\documentclass[11pt,a4paper]{article}
\usepackage{jcappub}

\usepackage[utf8]{inputenc}
\usepackage[T1]{fontenc}
\usepackage[normalem]{ulem}
\usepackage{amsmath,amssymb,amsfonts,mathtools,siunitx,xspace}
\usepackage{fontawesome}
\usepackage{graphicx}
\usepackage{tikz}
\usepackage{booktabs,array,tabularx,multirow,listings,makecell,nicematrix}
\usepackage[table]{xcolor}
\usepackage[inline]{enumitem}
\usepackage[skins,listings,breakable,raster]{tcolorbox}
\usepackage{hyperref}
\usepackage[nameinlink,noabbrev]{cleveref}

\makeatletter
\gdef\@fpheader{} 
\makeatother

\hypersetup{
  colorlinks=true,
  linkcolor=blue!55!black,
  citecolor=blue!55!black,
  urlcolor=blue!55!black
}

\graphicspath{{plots/}}

\DeclareSIUnit{\yr}{year}  
\numberwithin{equation}{section}

\AddToHook{cmd/appendix/before}{%
  \crefalias{section}{appendix}%
  \crefalias{subsection}{subappendix}%
  \crefalias{subsubsection}{subsubappendix}%
}
\crefname{lstlisting}{listing}{listings}
\Crefname{lstlisting}{Listing}{Listings}
\crefname{equation}{eq.}{eqs.}

\usetikzlibrary{tikzmark,calc,decorations.pathreplacing}
    
\newcommand{\alphath}[1][]{%
  \ensuremath{\alpha_{\bar\theta}%
    \if\relax\detokenize{#1}\relax
    \else^{#1}%
    \fi}\xspace%
}
\newcommand{\alphavac}[1][]{%
  \ensuremath{\alpha_{\rm vac}%
    \if\relax\detokenize{#1}\relax
    \else^{#1}%
    \fi}\xspace%
}
\newcommand{\alphashift}[1][]{%
	\ensuremath{\bar\alpha%
		\if\relax\detokenize{#1}\relax
		\else_{#1}%
		\fi}\xspace%
}
\newcommand*{\alphaP}{\ensuremath{\alpha_+}\xspace} 
\newcommand*{\alphaN}{\ensuremath{\alpha_N}\xspace} 
\newcommand*{\alphav}[2]{\ensuremath{\alphath[]\!\left(#1, #2\right)}\xspace} 
\newcommand*{\alpharw}[2]{\ensuremath{\alphaN\!\left(#1, #2\right)}\xspace}
\newcommand*{\rwdir}[1][]{%
  \ensuremath{r_w^{>}%
    \if\relax\detokenize{#1}\relax
    \else\!\left(#1\right)%
    \fi}\xspace}
\newcommand*{\rwinv}[1][]{%
  \ensuremath{r_w^{<}%
    \if\relax\detokenize{#1}\relax
    \else\!\left(#1\right)%
    \fi}\xspace}
\newcommand*{\xiw}{\ensuremath{\xi_w}\xspace}
\newcommand*{\vw}{\ensuremath{v_w}\xspace}
\newcommand*{\xish}{\ensuremath{\xi_{\text{sh}}}\xspace}
\newcommand*{\vwallframe}{v}
\newcommand*{\vp}{\ensuremath{\vwallframe_+}\xspace}
\newcommand*{\vm}{\ensuremath{\vwallframe_-}\xspace}
\newcommand*{\vpm}{\ensuremath{\vwallframe_\pm}\xspace}
\newcommand*{\vmp}{\ensuremath{\vwallframe_\mp}\xspace}
\newcommand*{\ccritDH}{\ensuremath{c_{-,\text{crit}}^{\text{DH}^+}}\xspace}
\newcommand*{\ccritID}{\ensuremath{c_{-,\text{crit}}^{\text{ID}^+}}\xspace}
\newcommand*{\ccritMH}{\ensuremath{c_{-,\text{crit}}^{\text{MH}^-}}\xspace}
\newcommand*{\cexID}{\ensuremath{c_{-,\text{ex}}^{\text{ID}^+}}\xspace}
\newcommand*{\cexMH}{\ensuremath{c_{-,\text{ex}}^{\text{MH}^-}}\xspace}

\newcommand*{\czeroID}{\ensuremath{c_{-,0}^{\text{ID}^+}}\xspace}
\newcommand*{\czeroIH}{\ensuremath{c_{-,0}^{\text{IH}}}\xspace}
\newcommand*{\alphaNfoldDH}{\ensuremath{\alpha_{N,\text{min}}^{\text{fold,DH}^+}}\xspace}
\newcommand*{\alphaNfoldID}{\ensuremath{\alpha_{N,\text{min}}^{\text{fold,ID}^+}}\xspace}
\newcommand*{\alphaNmaxID}{\ensuremath{\alpha_{N,\text{max}}^{\text{ID}^+}}\xspace}
\newcommand*{\alphaNfoldMH}{\ensuremath{\alpha_{N,\text{min}}^{\text{fold,MH}^-}}\xspace}
\newcommand*{\alphaNmaxMH}{\ensuremath{\alpha_{N,\text{max}}^{\text{MH}^-}}\xspace}

\newcommand*{\dd}{\mathrm{d}}
\newcommand*{\hydrodex}{\texttt{Hydr\'{o}Dex}\xspace}
\newcommand*{\cpp}{\texttt{C++}\xspace}
\newcommand*{\python}{\texttt{Python}\xspace}

\newcommand{\tikzunderbrace}[3]{\tikz[remember picture,overlay]{
  \draw[
    decorate,
    decoration={brace,mirror,amplitude=5pt},
    line width=1pt
  ]
  ([yshift=-4pt]#1.south west) -- ([yshift=-4pt]#2.south east)
  node[midway,below=4pt,font=\scriptsize,align=center] {#3};
}}
\newcommand{\tikzoverbrace}[3]{\tikz[remember picture,overlay]{
  \draw[
    decorate,
    decoration={brace,amplitude=5pt},
    line width=1pt
  ]
  ([yshift=4pt]#1.north west) -- ([yshift=4pt]#2.north east)
  node[midway,above=4pt,font=\scriptsize,align=center] {#3};
}}
\newcommand{\bracenode}[2]{\tikzmarknode{#1}{\vphantom{\xiw}#2}}
\newcommand{\fullstrut}{\rule[-32pt]{0pt}{60pt}}
\newcommand{\bottomstrut}{\rule[-32pt]{0pt}{0pt}}

\DeclareMathOperator{\Si}{si}
\DeclareMathOperator{\Ci}{ci}

\definecolor{codebg}{RGB}{247,247,247}
\definecolor{codecomment}{RGB}{0,120,0}
\definecolor{codekeyword}{RGB}{0,0,180}
\definecolor{codestring}{RGB}{160,30,30}
\definecolor{codenumber}{RGB}{100,100,100}
\definecolor{codelang}{RGB}{64,64,64}
\definecolor{latexcomment}{RGB}{128,128,128}

\lstdefinestyle{code}{
  basicstyle=\ttfamily\small,
  backgroundcolor=\color{codebg},
  frame=single,
  rulecolor=\color{black},
  framerule=0.4pt,
  framesep=4pt,
  numbers=left,
  numberstyle=\tiny\color{codenumber},
  numbersep=8pt,
  keywordstyle=\color{codekeyword},
  commentstyle=\color{codecomment},
  stringstyle=\color{codestring},
  showstringspaces=false,
  breaklines=true,
  breakatwhitespace=true,
  columns=fullflexible,
  keepspaces=true,
  tabsize=2,
  xleftmargin=1.5em,
  xrightmargin=0.5em,
  captionpos=b,
  escapeinside={(*@}{@*)},
}

\lstdefinestyle{python}{style=code, language=Python}
\lstdefinestyle{cpp}{style=code, language=C++}
\lstdefinestyle{shell}{style=code, language=bash}
\lstdefinestyle{inlinecpp}{style=cpp,basicstyle=\ttfamily}
\lstdefinestyle{inlinepy}{style=python,basicstyle=\ttfamily}

\newlength{\codeframesep}
\newlength{\codecaptionskip}
\newcommand{\codeleftskip}{\dimexpr 1.5em-\codeframesep-0.4pt\relax}
\newcommand*{\codecomment}[1]{\normalfont\color{latexcomment}#1}
\makeatletter
\let\code@PrintLabel\lst@SkipOrPrintLabel
\gdef\lst@SkipOrPrintLabel{%
  \hbox to\z@{\code@PrintLabel\hss}%
  \lst@numberfirstlinefalse%
}
\newcommand*{\lstnonumber}{%
  \leavevmode
  \setbox\z@\lastbox
  \ifdim\wd\z@=\z@ \ifdim\ht\z@=\z@ \setbox\z@\lastbox \fi\fi
  \ifvoid\z@
  \else\ifdim\wd\z@=\z@
    \global\advance\c@lstnumber-\lst@advancelstnum
  \else
    \box\z@
  \fi\fi
}
\makeatother

\tcbset{
  codebox/.style 2 args={
    enhanced, breakable,
    listing only,
    listing options={style=#1, frame=none, backgroundcolor={},
                     xleftmargin=0pt, xrightmargin=0pt,
                     aboveskip=0pt, belowskip=0pt},
    colback=codebg, colframe=black,
    boxrule=0.4pt, arc=0pt, outer arc=0pt,
    boxsep=0pt,
    left=\codeframesep, right=\codeframesep,
    top=\codeframesep, bottom=\codeframesep,
    left skip=\codeleftskip,
    right skip=\dimexpr 0.5em-\codeframesep-0.4pt\relax,
    codeskips,
    overlay unbroken and first={
      \node[anchor=north east, inner sep=3pt,
            font=\sffamily\scriptsize\bfseries,
            fill=codelang, text=white]
        at (frame.north east) {#2};
    },
  },
  codeskips/.style={before skip=\medskipamount, after skip=\medskipamount},
  codecaption/.style 2 args={after={\listingcaption{#1}{#2}}},
}

\makeatletter
\newcommand{\listingcaption}[2]{%
  \par\nopagebreak
  {\setlength{\abovecaptionskip}{\codecaptionskip}\setlength{\parindent}{0pt}%
   \refstepcounter{lstlisting}%
   \addcontentsline{lol}{lstlisting}{\protect\numberline{\thelstlisting}#1}%
   \@makecaption{\textbf{\lstlistingname~\thelstlisting}}{#1\label{#2}}}%
  \vskip\medskipamount}
\makeatother

\newtcblisting{pythoncode}[1]{codebox={python}{Python}, #1}
\newtcblisting{cppcode}[1]{codebox={cpp}{C++}, #1}
\newtcblisting{shellcode}[1]{codebox={shell}{Shell}, #1}

\usepackage{fontawesome}

\title{\hydrodex: gotta match 'em all!}

\author[a,b]{Giulio Barni}
\author[a,c,d]{, Eric Madge}

\affiliation[a]{Instituto de F\'isica Te\'orica IFT-UAM/CSIC, Cantoblanco, E-28049 Madrid, Spain}
\affiliation[b]{Deutsches Elektronen-Synchrotron DESY, Notkestr.~85, 22607 Hamburg, Germany}
\affiliation[c]{Departamento de F\'isica Te\'orica, Universidad Aut\'onoma de Madrid, Cantoblanco, E-28049 Madrid, Spain}
\affiliation[d]{Institute for Particle Physics Phenomenology, Durham University, South Road, Durham DH1\,3LE, United Kingdom}

\emailAdd{giulio.barni@desy.de}
\emailAdd{eric.madge-pimentel@durham.ac.uk}

\abstract{
    First-order phase transitions can source a stochastic gravitational-wave background within reach of LISA and other future detectors, with sound waves in the plasma typically providing the dominant contribution. Predicting the resulting gravitational-wave signal requires determining the stationary fluid profiles around expanding bubbles.
    We present a complete hydrodynamic classification of these profiles in the constant-sound-speed model, which generalizes the bag model by allowing the sound speeds in front of and behind the wall, \(c_+\) and \(c_-\), to differ. Solving the matching conditions at the wall and evolving the surrounding plasma, we recover the known direct and inverse detonations, deflagrations, and hybrids, and uncover mixed hybrids, a new class with rarefaction waves on both sides of the wall and oppositely directed fluid flows. For a sufficiently pronounced hierarchy \(c_+>c_-\), multiple stationary profiles can correspond to the same wall velocity \(\xiw\) and asymptotic transition strength \(\alphaN\). This non-uniqueness arises because this parametrization discards thermodynamic information and is resolved once the underlying microphysics is specified. We compute the kinetic-energy budget and acoustic gravitational-wave spectra within the sound-shell model, finding that mixed hybrids introduce no additional spectral feature and can be mimicked by ordinary direct or inverse hybrids with substantially different fluid profiles. Finally, we release \hydrodex, a public code implementing the complete classification, profile construction, energy-budget calculation, and gravitational-wave prediction.
 \par\vspace{1cm}
 \begin{center}
   \includegraphics[width=0.4\textwidth]{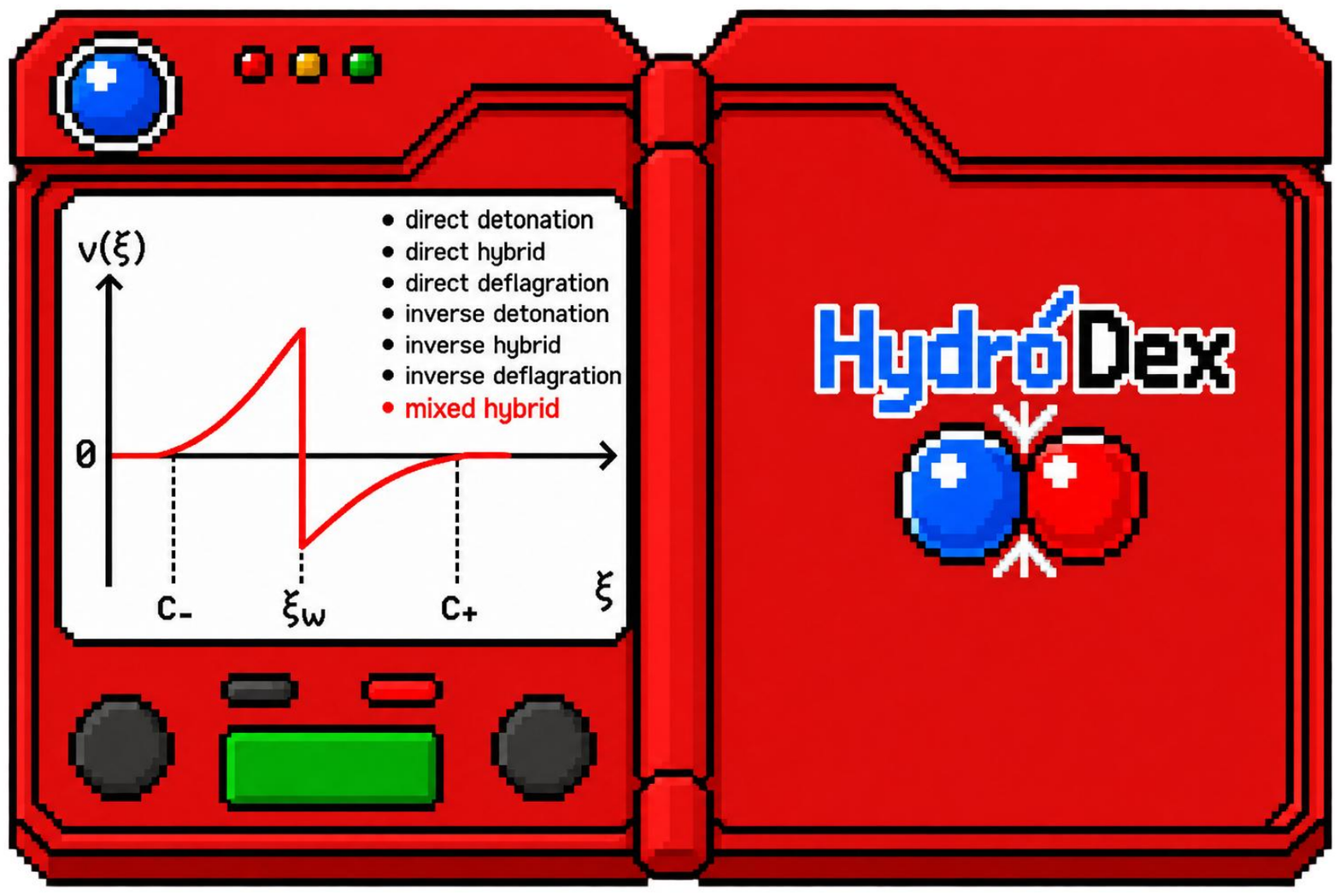}
 \end{center}
}

\begin{document}
\begin{flushright}
DESY-26-161\\
IFT-CSIC/UAM-26-165\\
IPPP/26/83
\end{flushright}
\vspace{0.75cm}
\maketitle

%
\section{Introduction}
\label{sec:introduction}
%

Cosmological phase transitions are generic in theories where the state of the early Universe changes as symmetries are broken or restored.
First-order phase transitions are especially important because they proceed through the nucleation and expansion of bubbles of the new phase~\cite{Coleman:1977py,Linde:1980tt,Linde:1981zj}.
They arise in many extensions of the Standard Model and can have several cosmological consequences.
They can provide the out-of-equilibrium environment required for baryogenesis~\cite{Kuzmin:1985mm,Dine:1992wr,Riotto:1998bt,Morrissey:2012db,Konstandin:2013caa}, affect dark-matter production~\cite{Baker:2019ndr,Hong:2020est,Azatov:2021ifm,Giudice:2024tcp}, generate primordial black holes~\cite{Baker:2021nyl,Jung:2021mku,Lewicki:2023ioy}, seed primordial magnetic fields~\cite{Baym:1995fk,Grasso:2000wj,Caprini:2009yp}, and source a stochastic gravitational-wave background~\cite{Witten:1984rs,Hogan:1986dsh,Kosowsky:1991ua,Kosowsky:1992rz,Kosowsky:1992vn,Kamionkowski:1993fg,Caprini:2015zlo,Caprini:2019egz,Schmitz:2020syl,Athron:2023xlk}.

A central ingredient in all these phenomena is the hydrodynamic response of the plasma to the expanding bubble wall.
Once a bubble has nucleated, energy--momentum conservation across the interface relates the fluid states immediately ahead of and behind the wall, while shocks, compression waves, and rarefactions connect these local states to the unperturbed plasma and to the bubble centre.
For a wall propagating at approximately constant velocity, the resulting flow becomes self-similar and determines how the energy released or absorbed by the transition is distributed between reheating and bulk motion~\cite{Steinhardt:1981ct,Ignatius:1993qn,Laine:1993ey,Kurki-Suonio:1995rrv,Espinosa:2010hh}.
A stationary regime requires microscopic friction on the wall to balance the driving force before bubble collisions~\cite{Moore:1995ua,Moore:1995si,Bodeker:2009qy,Bodeker:2017cim,Ai:2023see}.

In cosmological transitions, these single-bubble profiles provide the initial fluid configuration from which the gravitational-wave source develops.
Gravitational radiation can receive contributions from the scalar-field configuration, acoustic motion of the plasma, and, at later times, turbulent flows.
For non-runaway thermal transitions, numerical simulations have shown that the acoustic contribution typically dominates the spectrum~\cite{Hindmarsh:2013xza,Hindmarsh:2015qta,Hindmarsh:2017gnf,Caprini:2015zlo,Caprini:2019egz}.
Unlike the short-lived collision stage, the resulting sound waves can continue to source gravitational radiation after the transition has been completed, although their effective lifetime may be limited by non-linear evolution and shock formation~\cite{Ellis:2020awk,RoperPol:2023dzg}.
The sound-shell model~(SSM) turns this single-bubble description into a prediction for the velocity power spectrum, and hence for the acoustic gravitational-wave spectrum, using the self-similar fluid profile and the kinetic energy stored in the plasma~\cite{Hindmarsh:2016lnk,Hindmarsh:2019phv,Giombi:2024kju,Barni:2025gnm}.
Recent studies have refined this connection between single-bubble profiles and the velocity spectrum and have also shown that multi-bubble effects, including the slow-down of deflagrations and hybrids near percolation, can modify the acoustic source~\cite{Bhusal:2026iuw}. 

The standard analytic classification of bubble hydrodynamics is based on the bag equation of state, for which both phases are conformal and the speed of sound is fixed to \(c_s=1/\sqrt3\). 
In this limit, one obtains the familiar detonation, deflagration, and hybrid solutions~\cite{Steinhardt:1981ct,Gyulassy:1983rq,Kajantie:1986hq,Ignatius:1993qn,Laine:1993ey,Kurki-Suonio:1995rrv,Espinosa:2010hh}. 
This framework has been the basis for most computations of fluid profiles and efficiency factors.
However, it is a special limit.
In realistic thermal systems, the two phases need not have the same microscopic content, the same number of light degrees of freedom, or the same thermodynamic response.
As a result, the sound speed can differ across the wall and can deviate from the conformal value.
Such effects are relevant in hidden sectors, QCD-like sectors, strongly coupled transitions, and models where the masses or abundances of particles change substantially between the two phases
\cite{Giese:2020rtr,Giese:2020znk,Wang:2020nzm,Wang:2021dwl,Wang:2022lyd,Tenkanen:2022tly,Athron:2023xlk,Bea:2021zsu,Cline:2025bwe}.

The minimal analytic framework that keeps this thermodynamic information while preserving control over the hydrodynamic equations is the constant-speed-of-sound (CSS) model~\cite{Leitao:2014pda,Giese:2020rtr}.
Each phase is assigned a constant sound speed, but the two values are allowed to differ, \(c_+ \neq c_-\).
The bag equation of state is recovered as the special case \(c_+=c_-=1/\sqrt3\).
The CSS model therefore provides a simple setting in which to study how unequal sound speeds modify the wall matching conditions, the allowed self-similar profiles, and the energy budget of the transition.

The classification includes three generations of solutions.
The first generation (Gen.~I) consists of the ordinary direct solutions, in which fluid and bubble wall move in the same direction.
In the bag model these correspond to bubbles expanding towards a phase of lower vacuum energy.
The second generation (Gen.~II) consists of the inverse solutions, in which fluid and wall move in opposite directions.
In the bag model these expand, driven by thermal effects, against the vacuum-energy difference. 
Their hydrodynamics and energy budget were studied in Refs.~\cite{Barni:2024lkj,Bea:2024bxu}, and they have since been realized in a variety of microscopic scenarios~\cite{Barni:2025mud,Barni:2025ced,Azzola:2026mbq,Bigazzi:2026ppu,Cervantes:2026aji}.
We further point out the existence of a third generation (Gen.~III) of so-far unexplored mixed direct--inverse hybrids, in which the flow direction changes across the wall, and which appear only when the two sound speeds are unequal.

The reason unequal sound speeds induce additional hydrodynamic profiles is simple.
The sound speeds control whether the plasma supports rarefaction or compression waves and shock fronts on the respective side of the wall.
For \(c_+\neq c_-\), the wall velocity \xiw or the wall-frame fluid velocities \vpm ahead of or behind the phase front can lie between the two sound speeds --- i.e.\ subsonic with respect to one phase but supersonic with respect to the other phase --- allowing profiles which have no analogue in the equal-sound-speed limit.
In particular, in the interval \(c_- < \xiw < c_+\), rarefaction waves can form on both sides of the wall.
We refer to these solutions as mixed hybrids,
as they combine direct-like (fluid dragged by the wall) and inverse-like (fluid aspirated into the wall) hydrodynamic structures across the wall.

We provide a classification in terms of the wall velocity \xiw and the pseudo-trace phase-transition strength \alphaN evaluated in the unperturbed plasma at nucleation far ahead of the bubble.
This is a natural choice of physical input parameters as \alphaN can be computed directly from the finite-temperature effective potential once the nucleation temperature is determined.
For the purpose of constructing the profile solutions, though, the fluid velocities at each side of the phase front in the bubble wall's rest frame, \vp and \vm, provide a more convenient choice of variables as they enter the matching conditions across the wall and determine the subsequent fluid evolution.
Within the CSS model, they can be related to the pseudo-trace strength parameter \alphaP evaluated immediately in front of the wall.
However, \alphaP does not coincide with \alphaN unless the fluid ahead of the wall remains at rest --- the two strength parameters are related through the evolution of enthalpy (or temperature) between the phase front and the unperturbed plasma.
In general, the relation between \alphaP and \alphaN can only be established once the fluid profile solution is constructed.

The distinction between \alphaP and \alphaN becomes particularly important when the two phases have different sound speeds.
In the bag model, and more generally as long as the hierarchy between \(c_+\) and \(c_-\) is moderate, the map from the local wall state to the unperturbed plasma remains one-to-one.
Although different branches may overlap in the \((\xiw,\alphaP)\) plane, each point \((\xiw,\alphaN)\) corresponds to a unique completed fluid profile.
For sufficiently large \(c_+/c_-\), however, this changes qualitatively.
The heating or cooling of the fluid ahead of the wall can make \alphaN vary non-monotonically with \alphaP; the map becomes non-injective, developing a fold.
The same value of \alphaN can then be reached from different local wall states, so that several stationary profiles coexist at the same physical inputs \((\xiw,\alphaN)\).
In the same regime, the sign of \alphaN also ceases to provide a universal distinction between direct and inverse flows.
This happens for any \(c_+>c_-\) as mixed solutions arise.

This is a non-uniqueness of the reduced hydrodynamic parametrization in terms of the parameters \((\xiw,\alphaN, c_+, c_-)\) rather than necessarily of the underlying microscopic theory.
Additional thermodynamic information and the entropy production at the wall are needed to select the physical solution, while its dynamical realization also requires that the corresponding stationary state can actually be reached~\cite{Ai:2021kak,Ai:2023see,Ai:2024shx,Krajewski:2024gma,Krajewski:2024zxg,Krajewski:2026kcm,Laurent:2026jvx}.

We also revisit the energy budget and the kinetic-energy fraction.
In the bag model, the vacuum-energy parameter \(\epsilon\), or equivalently \alphaP, plays a double role.
Its sign determines whether vacuum energy acts as an energy source for the transition or as part of the final energetic cost, while at the same time it is tied to the hydrodynamic strength. 
In both the bag and CSS models, the pseudo-trace strength \(\alphaN\) controls the hydrodynamic matching, while the signed zero-temperature vacuum-energy difference determines whether vacuum energy is released during the transition or stored in the final phase.
For equal sound speeds, the signs of these two quantities are correlated, so that the hydrodynamic character of the solution tracks the energetic role of the vacuum contribution.
Once \(c_+\neq c_-\), this correspondence is lost, and the two notions become independent.
Thus, for unequal sound speeds, the hydrodynamic character of the solution need not reflect whether the vacuum contribution acts as an energy source or as a final-state energy cost \cite{Barni:2024lkj}.

Finally, we investigate the gravitational-wave signatures of the complete CSS hydrodynamic classification within the sound-shell model~\cite{Hindmarsh:2016lnk,Hindmarsh:2019phv}.
Building on previous extensions accounting for the cosmological expansion and
the finite lifetime of the acoustic source~\cite{Guo:2020grp,RoperPol:2023dzg}, and on studies of non-conformal sound speeds~\cite{Wang:2021dwl,Giombi:2024kju}, we keep separate the effective sound speed \(c_s\) governing the propagation of the acoustic perturbations from the equation-of-state parameter \(\omega\) controlling the homogeneous expansion.
This is relevant, for instance, when the transition occurs in a dark sector that constitutes only part of the total energy density.
The coupling of such a sector to the visible plasma during the sound-wave era can further modify the signal~\cite{Finetti:2026ool}.

Extending our previous analysis of the gravitational-wave signal of direct and inverse phase transitions~\cite{Barni:2025gnm}, we find that the dependence on the sound speed in the phase to which the Universe transitions is qualitatively common to both classes.
Increasing \(c_s=c_-\) generally shifts the gravitational-wave peak towards larger wavenumbers and enhances its amplitude, although peak frequency and amplitude can decrease again for \(c_-\to1\), depending on the branch.
We further present the first results for gravitational-wave signals from mixed-hybrid profiles.
The corresponding spectra are similar to those of direct and inverse solutions.
Despite the difference of the flow direction across the wall, their kinetic spectra exhibit a single acoustic peak.
Mixed hybrids can in fact be closely mimicked by ordinary hybrids with markedly different fluid profiles and even opposite signs of \alphaN, illustrating that the gravitational-wave spectrum alone does not resolve the hydrodynamic branch.

Together with the analytic classification, we release the public code \href{https://github.com/hydrodex-pt/hydrodex}{\faGithub\ \hydrodex} (formerly \href{https://github.com/eric-madge/inverse_pt}{\faGithub~\texttt{InversePTCalculator}}).
Given the thermodynamic parameters of a CSS transition and a wall velocity, \hydrodex determines the allowed hydrodynamic solution, constructs the corresponding fluid profile, evaluates how efficiently energy is transferred to bulk motion, and computes the resulting gravitational-wave spectrum within the sound-shell model.
In this way, the code provides a direct pipeline from the thermodynamic description of the transition to its hydrodynamic and gravitational-wave phenomenology.
The code is written in \cpp and interfaced to \python.
The repository includes the documentation and examples needed to reproduce the results presented in this work and to explore other regions of parameter space.

Several public tools address complementary parts of the same problem.
These include \texttt{PTtools}~\cite{pttools-git}, \texttt{CosmoGW}~\cite{cosmogw-git}, and \texttt{HydroGrav}~\cite{Linton:2026fpj,hydrograv-git}, which construct hydrodynamic profiles and/or the associated gravitational-wave signal.
More recently, \texttt{SNOBEX}~\cite{Mateos:2026moc} was developed to solve single-bubble hydrodynamics for general non-conformal equations of state, including state-dependent sound speeds, and to identify the corresponding wall and flow obstructions.  
To our knowledge, none of these public codes currently implements the complete hydrodynamic classification of all three generations of hydrodynamic solutions
(direct, inverse and mixed) presented here.
In particular, the mixed hybrids and the associated overlap regions discussed in this work have not appeared in any published hydrodynamic implementation.\footnote{%
    An independent unpublished implementation by B.~Laurent in \href{https://github.com/benoitlaurent96/HydrodynamicsFOPT}{\faGithub\ \texttt{HydrodynamicsFOPT}}~\cite{hydrodynamicsfopt-git} also contains direct and inverse hydrodynamic branches, including mixed hybrids.%
}
We have cross-checked \hydrodex against these public implementations, comparing the hydrodynamic solutions wherever their domains of applicability overlap, as well as their numerical accuracy and computational performance.
These comparisons provide an independent validation of our implementation.
Within the CSS framework considered here, \hydrodex offers the most complete coverage of the hydrodynamic solution space, including direct, inverse, and mixed branches.
It also provides the best performance in our benchmarks --- with the exception of the Numba JIT-compiled \texttt{PTtools} implementation, which is only moderately faster --- making it particularly well suited for dense parameter scans and for the subsequent computation of the energy budget and gravitational-wave signal.

The paper is organized as follows.
In \cref{sec:hydro} we review the relativistic hydrodynamic equations and the construction of self-similar bubble profiles.
In \cref{sec:CSS} we introduce the CSS equation of state, derive the wall and shock matching conditions, and relate the local strength \alphaP to the asymptotic strength \alphaN, including the thermodynamic and entropy constraints.
In \cref{sec:classification} we give the complete classification of direct, inverse, and mixed hydrodynamic solutions.
In \cref{sec:hydrodynamic-fold} we study when the map from local wall states to \((\xiw,\alphaN)\) becomes non-injective, leading to regions of the \((\xiw,\alphaN)\) plane in which multiple stationary profiles coexist, and discuss how the missing thermodynamic information lifts this degeneracy.
In \cref{sec:efficiency} we revisit the energy budget and kinetic-energy conversion in the CSS model, while in \cref{sec:ssm} we compute the corresponding acoustic gravitational-wave spectra within the sound-shell model.
\Cref{sec:code} provides an overview of the structure and implementation of the \hydrodex code, along with minimal examples demonstrating its usage workflow.
We summarize our results in \cref{sec:conclusions}.
The appendices discuss additional kinematic branches and their admissibility, provide analytic results for the properties of the hydrodynamic projection, and document the cross-validation and performance of \hydrodex.
A concise account of our main results and their physical implications is presented in the companion paper~\cite{Barni:2026ecg}.

%
\section{Hydrodynamics and self-similar profiles}
\label{sec:hydro}
%

The macroscopic evolution of a first-order phase transition can be described in terms of relativistic hydrodynamics once the microscopic wall thickness is much smaller than the length scales over which the plasma varies.
In this limit the wall is treated as a moving interface separating two locally equilibrated phases, while the plasma away from the wall obeys energy-momentum conservation.
This description is standard in studies of expanding bubbles and their associated fluid profiles~\cite{Steinhardt:1981ct,Kurki-Suonio:1995rrv,Kamionkowski:1993fg,Espinosa:2010hh,Megevand:2013yua,Megevand:2014yua,Hindmarsh:2017gnf}.

\subsection{Energy-momentum conservation}

The starting point is therefore the local conservation of the total energy and momentum carried by the scalar field and by the plasma.
On hydrodynamic scales the detailed microscopic structure of the wall enters only through matching conditions across the interface, while the bulk evolution is governed by the conservation of the total stress tensor.
We start from
\begin{equation}
  \nabla_\mu T^{\mu\nu}=0,
  \qquad
  T^{\mu\nu}=T^{\mu\nu}_\phi+T^{\mu\nu}_{\rm f}.
\end{equation}
The scalar order parameter contributes
\begin{equation}
  T^{\mu\nu}_\phi=
  (\partial^\mu\phi)(\partial^\nu\phi)
  -g^{\mu\nu}
  \left[
    \frac12(\partial\phi)^2+V(\phi)
  \right],
\end{equation}
whereas the thermal plasma is modelled, away from the wall, as a perfect fluid,
\begin{equation}
  T^{\mu\nu}_{\rm f}
  =
  w_{\rm f} u^\mu u^\nu+p_{\rm f} g^{\mu\nu},
  \qquad
  w_{\rm f}=e_{\rm f}+p_{\rm f},
  \qquad
  u^\mu=\gamma(1,\mathbf v),
  \qquad
  \gamma=(1-v^2)^{-1/2}.
\end{equation}
Here \(e_{\rm f}\), \(p_{\rm f}\), and \(w_{\rm f}\) are the local energy density, pressure, and enthalpy density of the plasma. 
The vacuum-energy contribution from the scalar potential is included in the phase-dependent equation of state, defining the effective pressure and energy density  \(p = p_{\rm f} - V_{\rm ph}\) and \(e = e_{\rm f} + V_{\rm ph}\), where \(V_{\rm ph}\) is vacuum-energy in each phase, so that, on scales larger than the wall thickness, the system is described by the effective perfect-fluid stress tensor
\begin{equation}
  T^{\mu\nu}
  =
  w u^\mu u^\nu+p g^{\mu\nu},
  \qquad
  w = e + p.
  \label{eq:perfect-fluid-tensor}
\end{equation}
The microscopic scalar dynamics then enters only through the matching conditions across the wall, while the regions on either side are governed by hydrodynamics.

Projecting \(\nabla_\mu T^{\mu\nu}=0\) parallel and orthogonal to the fluid four-velocity gives the relativistic energy and Euler equations,
\begin{subequations}
\begin{align}
  u_\nu \nabla_\mu T^{\mu\nu}=0
  &\quad\Longrightarrow\quad
  u^\mu\nabla_\mu e+w\nabla_\mu u^\mu=0,
  \label{eq:energy-projected}
  \\
  \Delta^\alpha{}_{\nu}\nabla_\mu T^{\mu\nu}=0
  &\quad\Longrightarrow\quad
  w u^\mu\nabla_\mu u^\alpha+\Delta^{\alpha\mu}\nabla_\mu p=0,
  \label{eq:momentum-projected}
\end{align}
\end{subequations}
where \(\Delta^{\alpha\nu}=g^{\alpha\nu}+u^\alpha u^\nu\) is the projector orthogonal to \(u^\mu\).
These equations provide the starting point for constructing the self-similar velocity, enthalpy, and temperature profiles around an expanding bubble~\cite{Landau:FluidMechanics,Steinhardt:1981ct,Ignatius:1993qn,Espinosa:2010hh,Hindmarsh:2017gnf}.

\subsection{Self-similar equations}

The hydrodynamic profiles around an expanding bubble simplify considerably when the wall has reached an approximately constant velocity.
In this regime there is no additional macroscopic length scale in the problem.
Distances from the bubble centre and times since nucleation enter only through their ratio.
The natural variable is therefore the self-similar coordinate \(\xi\equiv r/t\), and all fluid quantities are taken to depend only on \(\xi\).
This is the standard late-time description of bubble expansion in relativistic hydrodynamics~\cite{Ignatius:1993qn,Kurki-Suonio:1995rrv,Espinosa:2010hh,Hindmarsh:2017gnf}.

In \(D\) spatial dimensions the similarity form of \cref{eq:energy-projected,eq:momentum-projected} is
\begin{subequations}
\begin{align}
  (\xi-v)\frac{\dd e}{\dd\xi}
  &=
  w\left[
  \frac{D-1}{\xi}v
  +\gamma^2(1-\xi v)\frac{\dd v}{\dd\xi}
  \right],
  \label{eq:generalD-energy}
  \\
  (1-\xi v)\frac{\dd p}{\dd\xi}
  &=
  w\gamma^2(\xi-v)\frac{\dd v}{\dd\xi}.
  \label{eq:generalD-momentum}
\end{align}
\end{subequations}
These equations encode energy conservation and radial momentum conservation for a self-similar relativistic flow.
The term proportional to \((D-1)/\xi\) is purely geometrical and distinguishes planar, cylindrical, and spherical expansion.

Introducing the adiabatic sound speed \(c_{s}^2\equiv \frac{\dd p}{\dd e} \), and the velocity of the fluid measured in the frame moving with \(\xi\),
\begin{equation}
  \mu(\xi,v)\equiv \frac{\xi-v}{1-\xi v},
  \label{eq:mu-def}
\end{equation}
one obtains
\begin{align}
  \frac{D-1}{\xi}v
  &=
  {\gamma^2}(1-\xi v)
  \left[
  \frac{\mu(\xi,v)^2}{c_s^2}-1
  \right]
  \frac{\dd v}{\dd\xi},
  \label{eq:v-ode-general}\\
  \frac{\dd w}{\dd\xi}
  &=
  w\gamma^2\mu(\xi,v)
  \left(1+\frac{1}{c_s^2}\right)
  \frac{\dd v}{\dd\xi}.
  \label{eq:w-ode-general}
\end{align}
In a general equation of state, the sound speed is not necessarily constant.
It can depend on the local thermodynamic state, for instance on the temperature or on other control parameters.
The hydrodynamic equations must then be solved as a coupled system for the velocity and one thermodynamic variable, such as the enthalpy or the temperature.  

The constant-speed-of-sound (CSS) approximation used below provides a useful simplification: within each phase, \(c_s\) is fixed.
The velocity equation then becomes autonomous, while the temperature and enthalpy profiles can be reconstructed afterwards by quadrature.
For spherical bubbles\footnote{%
    The planar limit \(D=1\) is special.
    The geometrical source term in \cref{eq:v-ode-general} vanishes, so smooth regions are either constant-velocity plateaus, \(\dd v/\dd\xi=0\), or simple waves satisfying \(\mu^2=c_s^2\).
    In spherical symmetry the term \(2v/\xi\) forbids a smooth constant-velocity region with \(v\ne0\).
    Any nonzero radial flow must have a gradient.
    This is why a planar detonation may have a constant-velocity region behind the wall, whereas in a spherical profile the rarefaction wave starts directly at the wall, apart from the trivial region at rest near the bubble centre.%
}, \(D=3\), and \cref{eq:v-ode-general} becomes
\begin{equation}
  \frac{2v}{\xi}
  =
  \gamma^2(1-\xi v)
  \left[
  \frac{\mu(\xi,v)^2}{c_s^2}-1
  \right]
  \frac{\dd v}{\dd\xi}.
  \label{eq:v-ode-3d}
\end{equation}

Therefore, \cref{eq:v-ode-3d} is a closed equation for \(v(\xi)\).
Once the velocity profile has been determined, the temperature and enthalpy profiles follow from
\begin{equation}
  \frac{\dd\ln T}{\dd\xi}
  =
  \gamma^2\mu(\xi,v)\frac{\dd v}{\dd\xi},
  \qquad
  \frac{\dd\ln w}{\dd\xi}
  =
  \left(1+\frac{1}{c_s^2}\right)
  \frac{\dd\ln T}{\dd\xi}.
  \label{eq:T-w-quadrature}
\end{equation}
Equivalently,
\begin{equation}
  \ln\frac{w(\xi_2)}{w(\xi_1)}
  =
  \int\limits_{\xi_1}^{\xi_2}\dd\xi\,
  \left(1+\frac{1}{c_s^2}\right)
  \gamma^2(\xi)\mu(\xi,v(\xi))
  \frac{\dd v}{\dd\xi}
  =
  \int\limits_{v(\xi_1)}^{v(\xi_2)}\dd v\,
  \left(1+\frac{1}{c_s^2}\right)
  \gamma^2 \mu(\xi(v),v)
  .
  \label{eq:enthalpy-ratio}
\end{equation}
Thus, for the CSS equation of state, the construction of a smooth self-similar profile can be organized in two steps: first solve the velocity equation, and then reconstruct the thermodynamic variables.

\subsection{Hydrodynamic matching at the bubble wall and shock front}

The hydrodynamic fields are smooth away from the bubble wall, but they may be discontinuous across the wall itself.
In the thin-wall approximation, the microscopic wall thickness is much smaller than the hydrodynamic length scale, so the wall can be treated as an infinitesimally thin interface separating the `\(+\)' and `\(-\)' phases.
The matching conditions across this interface are the relativistic Rankine--Hugoniot conditions, obtained by integrating energy-momentum conservation across an infinitesimal volume enclosing the wall.
Equivalently, they express the continuity of the energy and momentum fluxes normal to the interface~\cite{Landau:FluidMechanics,Steinhardt:1981ct,Gyulassy:1983rq,Kamionkowski:1993fg,Ignatius:1993qn,Kurki-Suonio:1995rrv}.

Throughout the paper \(\vp\) and \(\vm\) denote the wall-frame fluid velocities immediately in front of and behind the wall.  They are taken to be positive quantities in the wall frame.
For a stationary planar wall, the relevant components of the perfect fluid stress tensor are
\begin{equation}
  T^{0z}=w\gamma^2 v,
  \qquad
  T^{zz}=w\gamma^2 v^2+p ,
\end{equation}
where \(z\) is the direction normal to the wall.
Continuity of the energy flux \(T^{0z}\) and of the momentum flux \(T^{zz}\) gives
\begin{subequations}
\label{eq:wall-energy-momentum-flux}
\begin{align}
  w_+\gamma_+^2 \vp
  &=
  w_-\gamma_-^2 \vm,
  \\
  w_+\gamma_+^2 \vp^2+p_+
  &=
  w_-\gamma_-^2 \vm^2+p_- .
\end{align}
\end{subequations}
These conditions express local energy and momentum conservation at the interface.
The wall velocity is not fixed by the stationary hydrodynamic classification alone.
It is determined dynamically by the coupled scalar-field and plasma equations, through the balance between the driving pressure and microscopic friction from particles crossing the wall~\cite{Moore:1995ua,Moore:1995si,Konstandin:2014zta,Dorsch:2018pat,Laurent:2022jrs,DeCurtis:2022hlx,DeCurtis:2023hil,DeCurtis:2024hvh,Branchina:2025adj,Ekstedt:2024fyq,Ai:2025bjw}.
Once the wall velocity is specified, however, the admissible hydrodynamic discontinuities are completely determined by \cref{eq:wall-energy-momentum-flux}~\cite{Espinosa:2010hh,Hindmarsh:2017gnf}.
The same conditions can be written in the symmetric form
\begin{equation}
  \vp \vm
  =
  \frac{p_+-p_-}{e_+-e_-},
  \qquad
  \frac{\vp}{\vm}
  =
  \frac{e_-+p_+}{e_++p_-}.
  \label{eq:symmetric-wall-matching}
\end{equation}
This form is often useful because it makes explicit that the jump in velocity is fixed by the jump in thermodynamic quantities across the wall.
Conversely, for a given equation of state, the matching conditions can be used to express the local strength of the transition directly in terms of the two wall-frame velocities.

In addition to discontinuities at the bubble wall, the matching conditions also allow for discontinuities within the same phase. 
If the fluid on both sides of the discontinuity satisfies the same equation of state, \(p = c_s^2 e + \text{const.}\), the fluid can encounter a shock front, at which the first condition in \cref{eq:symmetric-wall-matching} simplifies to 
\begin{equation}
    \vwallframe_{\uparrow}^{\rm sh} \vwallframe_{\downarrow}^{\rm sh}=c_s^2 ,
    \label{eq:shock-matching}
\end{equation}
where \(c_s=c_\pm\) depending on whether the shock occurs ahead of (in the `\(+\)' phase) or behind (in the `\(-\)' phase) the wall.
We here use `\(\uparrow\)' and `\(\downarrow\)' as indices for up- and downstream quantities instead of `+' and `-' to emphasize that the velocities are measured in the shock rest frame and to avoid confusion with the phase index.

\subsection{Fluid profile solutions}
\label{sec:solutions_general}

Before classifying the different branches, it is useful to state how a physical hydrodynamic profile is constructed.
The solution must satisfy the appropriate asymptotic conditions both at the centre of the bubble and far from the wall, while the fluid configuration between these regions can be assembled from the elementary building blocks introduced above: constant states, compression waves, rarefaction waves, and shocks.
A physical branch is then obtained by connecting these pieces consistently to the wall-matching solution and imposing the required regularity, causality, and entropy conditions.
In this subsection we apply this construction to determine the admissible global profiles associated with each branch.

\begin{figure}
    \centering
    \includegraphics{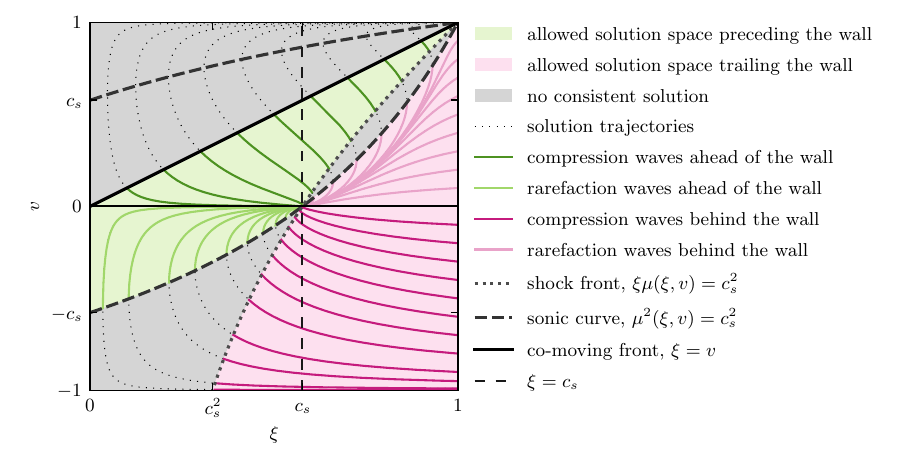}
    \caption{
        Different types of solutions for the fluid motion in front (green region) and behind (pink region) the bubble wall: compression waves (dark green or dark pink) end at a shock front (grey dotted line), whereas rarefaction waves go to \(v=0\) at \(\xi=c_s\).
        The grey region does not support consistent solutions.
        For simplicity, the figure assumes that the speed of sound \(c_s\) is constant.
    }
    \label{fig:solutions}
\end{figure}

Therefore, in combination with the boundary conditions --- in the plasma frame, the fluid velocity has to vanish in the bubble centre and tend to zero for \(\xi\to 1\) due to causality --- the fluid evolution \cref{eq:v-ode-3d} supports three different solutions on each side of the bubble wall, depicted in \cref{fig:solutions} (assuming a constant speed of sound \(c_s\)):%
\footnote{%
    Note that we here assume that the speed of sound is constant within each phase. We however expect that the conclusions in the general case with \(c_s = c_s(\xi)\) remain qualitatively similar, provided e.g.\ that \(\xi = c_s(\xi)\) does not admit multiple solutions for \(\xi \in (0,1)\).%
}
\begin{enumerate*}[label={\bfseries(\roman*)}]
    \item {rarefaction waves} and
    \item {compression waves}, as well as the trivial solution,
    \item {the fluid remains at rest}.
\end{enumerate*}

Solutions for the fluid in front of the wall (upstream) must lie below the \(\xi=v\) line (otherwise \(\vpm<0\), which also requires \(\xiw<0\) for consistent solutions) as well as above the shock front \(\xi \mu(\xi,v) = c_s^2\) and the sonic curve \(\mu(\xi,v) = c_s\) along which the equation of motion \cref{eq:v-ode-3d} possesses a critical point, whereas solutions for the plasma behind the wall (downstream) must lie below the shock front and the sonic curve.
The corresponding regions are shaded in \cref{fig:solutions} in light green and light pink, respectively.
The grey shaded regions do not admit consistent solutions.%
\footnote{%
    In principle, additional mathematical solutions exist between the sonic curve and the shock front, with the fluid accelerating from the bubble wall towards the discontinuity.
    They do not, however, correspond to physical hydrodynamic branches as the discontinuity is crossed as a rarefaction shock rather than a compressive shock, violating the entropy condition and hence the second law of thermodynamics.
    We discuss these unphysical solutions in \cref{app:exotic-branches}.%
    \label{fn:other-solutions}%
}

Compression waves are the ones evolving between the bubble wall and a shock front, where the velocity jumps to zero.
Ahead of the wall (dark green curves), they have positive fluid velocities, \(v(\xi)>0\) --- starting at a bubble wall with \(\max(0,\mu(\xiw,{c_s^2}/{\xiw})) < v_w < \xiw\).
Behind the wall (dark pink curves), they form with \(v(\xi)<0\) and require \(\xiw > c_s^2,\ v_w < \min(0,\mu(\xiw, {c_s^2}/{\xiw}))\).
Rarefaction waves, on the other hand, smoothly evolve between the wall and \(v=0\) at \(\xi=c_s\).
They can form for \(\xiw > c_s,\ 0<v_w\leq\mu(\xiw,c_s)\) with positive velocities, \(v(\xi)>0\), behind the wall (light pink curves), or for \(\xiw < c_s,\ \mu(\xiw,c_s)\leq v_w<0\) with negative velocities, \(v(\xi)<0\) ahead of the wall (light green curves).
Here, \(v_w = v(\xiw)\) denotes the fluid velocity either directly in front or directly behind the wall, depending on the context.

Combinations of these three solutions on each side of the wall give rise to eight different solution types for the full fluid profiles --- excluding the trivial solution with the fluid at rest on both sides of the wall.
The stability of the hydrodynamic solution however further requires the fluid velocities in the wall rest frame to be either both supersonic (\(\vpm\geq c_\pm\)) or both subsonic (\(\vpm\leq c_\pm\)) with respect to the respective speed of sound~\cite{Megevand:2013yua,Megevand:2014yua,Barni:2024lkj}.
This eliminates solutions with compression waves on both sides of the wall, reducing the number of possible types of solutions for the fluid profiles to seven.
Solutions with the fluid at rest on one side of the wall and a rarefaction or compression wave on the other side are called detonations and deflagrations, respectively, whereas solutions with the plasma in motion on both sides of the wall are called hybrids.\footnote{%
    We follow the convention of Ref.~\cite{Barni:2024lkj} in which ``detonation'' and ``deflagration'' denote the two branches of the hydrodynamic matching conditions, characterized by \(\vp>\vm\) and \(\vm>\vp\) in the wall frame, respectively.%
}
Direct and inverse solutions exhibit positive and negative fluid velocities in the plasma frame, respectively, while in mixed solutions, the fluid flow direction changes across the wall.
We hence obtain the following types of fluid profiles:
\begin{description}
    \item[direct detonations] Behind the wall, a rarefaction wave forms, while in front of the wall, the plasma is at rest.
    \item[inverse detonations] The fluid behind the wall is at rest, and develops a rarefaction wave in front of the wall.
    \item[direct deflagrations] The fluid remains unperturbed behind the wall and forms a compression wave ending at a shock front ahead of the wall.
    \item[inverse deflagrations] A compression wave propagates from the shock front behind the wall towards the wall, with the velocity jumping to zero in front of the wall.
    \item[direct hybrids] A rarefaction wave forms behind the wall with \(\vm=c_-\), whereas ahead of the wall, the plasma develops a compression wave ending at the shock front.
    \item[inverse hybrids] The profile evolves through a compression wave from a shock front towards the wall.
    Ahead of the wall, the velocity jumps to \(\vp=c_+\) (in the wall rest frame) and a rarefaction wave forms.
    \item[mixed direct--inverse hybrids] These solutions exhibit rarefaction waves on both sides of the wall, with either \(\vp=c_+\) or \(\vm=c_-\) in the wall rest frame.
    The fluid is moving with \(v(\xi)>0\) for \(\xi<\xiw\) and \(v(\xi)<0\) for \(\xi>\xiw\). 
    Mixed hybrids only exist for \(c_+ > c_-\).
\end{description}

While the first generation, direct solutions, has been discussed in the literature extensively~\cite{Gyulassy:1983rq,Espinosa:2010hh}, and the second generation, inverse solutions~\cite{Barni:2024lkj,Bea:2024bxu,Barni:2025mud}, is gaining traction, we here, for the first time, explicitly point out the existence of the third generation, mixed solutions.

%
\section{The Constant-Sound-Speed Model}
\label{sec:CSS}
%

To close the hydrodynamic equations, one must specify the equation of state of the two phases.
A widely used analytic approximation is the bag equation of state, in which the thermal component is radiation-like and the speed of sound is fixed to \(c_s^2=1/3\).
This provides an excellent description whenever both phases contain many relativistic degrees of freedom and their effective number changes only moderately across the transition, as is typically the case for electroweak-like cosmological phase transitions \cite{Steinhardt:1981ct,Kamionkowski:1993fg,Espinosa:2010hh,Hindmarsh:2017gnf}.

The bag equation of state becomes less appropriate when the microscopic properties of the two phases differ substantially, leading to sizeable changes in the number of active degrees of freedom or to a nontrivial dependence of the speed of sound on the thermodynamic state.
This occurs, for example, in QCD-like transitions and in matter at finite baryon chemical potential, as encountered in effective models of dense nuclear matter~\cite{Kajantie:1986hq,Giese:2020rtr,Giese:2020znk,Bea:2024bxu,Bea:2024bls,Cline:2025bwe}.

Hydrodynamic profiles beyond the bag equation of state can be constructed once a specific thermodynamic model is given, and several works have already studied non-conformal effects in the energy budget, in the sound-shell description, or in concrete models with nontrivial sound speed~\cite{Giese:2020rtr,Giese:2020znk,Wang:2020nzm,Wang:2021dwl,Wang:2022lyd,Tenkanen:2022tly,Wang:2023jto,Tian:2024ysd,Si:2025vdt,Mateos:2026moc}.
A fully general analytic classification for an arbitrary state-dependent equation of state, however, is much less tractable.
We therefore follow the constant-speed-of-sound framework used in Refs.~\cite{Giese:2020rtr,Giese:2020znk,Tenkanen:2022tly}, which is the minimal extension of the bag model that keeps the sound speeds as independent thermodynamic inputs while preserving analytic control over the hydrodynamic matching.
In this framework each phase is characterized by its own constant sound speed, \(c_+ \neq c_-\), together with a phase-dependent vacuum-energy contribution.
This provides the simplest setting in which the qualitative consequences of unequal sound speeds can be understood systematically.
Although the explicit solutions presented below are obtained within the CSS approximation, the resulting hydrodynamic classification relies only on local matching conditions and causality, and is therefore expected to remain qualitatively valid for more general equations of state with state-dependent sound speeds.

\subsection{CSS equation of state}

We now specify the thermodynamic input used to close the hydrodynamic equations.
In each phase we take
\begin{equation}
  e_\pm(T)
  =
  a_\pm T^{q_\pm}+\epsilon_\pm,
  \quad
  p_\pm(T)
  =
  c_\pm^2 a_\pm T^{q_\pm}-\epsilon_\pm,
  \quad
  w_\pm(T)
  =
  e_\pm(T)+p_\pm(T)
  =
  (1+c_\pm^2)a_\pm T^{q_\pm}.
  \label{eq:css-eos}
\end{equation}
The constants \(\epsilon_\pm\) are the vacuum-energy contributions in the two phases, while \(a_\pm\) parametrizes the size of the thermal contribution, or equivalently the effective number of degrees of freedom.
The exponent \(q_\pm\) is fixed by requiring the adiabatic sound speed to be constant
\begin{equation}
  c_\pm^2
  =
  \frac{\partial p_\pm}{\partial e_\pm}
  =
  \frac{
    \partial p_\pm/\partial T
  }{
    \partial e_\pm/\partial T
  }
  =
  \frac{1}{q_\pm-1},
  \qquad
  q_\pm>2 , \qquad \to \qquad q_\pm=1+\frac{1}{c_\pm^2}.
  \label{eq:css-sound-speed}
\end{equation}
The standard bag equation of state is recovered for \(q_\pm=4\), with \(c_\pm^2=\frac{1}{3}\). 
In that limit the thermal energy and pressure scale as \(T^4\), as appropriate for a relativistic radiation bath \cite{Steinhardt:1981ct,Espinosa:2010hh,Giese:2020rtr}.
The entropy density follows from the thermodynamic identity
\begin{equation}
  s_\pm(T)
  =
  \frac{\partial p_\pm(T)}{\partial T}
  =
  \frac{w_\pm(T)}{T}
  =   c_\pm^2 q_\pm a_\pm T^{q_\pm-1}.
  \label{eq:entropy-css}
\end{equation}

We further define \(T_+\) and \(T_-\) as the temperatures immediately in front (outside the bubble) and immediately behind (inside the bubble) the wall, respectively.
Whenever the temperature is not specified explicitly, all quantities in \cref{eq:css-eos,eq:entropy-css} are understood to be evaluated at the corresponding temperature adjacent to the wall, e.g.\ \(w_+ = w_+(T_+)\), \(p_- = p_-(T_-)\), etc.

It is useful to separate the vacuum-energy difference from the thermal part of the equation of state.
We define
\begin{equation}
  \alphavac[+]
  \equiv
  \alphavac(T_+)\,,
  \qquad
  \alphavac(T) \equiv
  \frac{\epsilon}{a_+ T^{q_+}},
  \qquad
  \epsilon
  \equiv
  \epsilon_+-\epsilon_-.
  \label{eq:alpha_vac}
\end{equation}
Here \alphavac[+] measures the zero-temperature vacuum-energy difference in units of the thermal energy scale immediately in front of the wall. 

For a conformal plasma, the trace of the stress-energy tensor, \(\theta\equiv T_\mu^\mu=e-3p\), isolates the departure from a purely radiation-like equation of state.
Accordingly, the strength of a first-order phase transition is often characterized in terms of the trace anomaly.
Away from the conformal limit, however, different definitions of the transition strength are no longer equivalent.
Their relation, and the corresponding generalization to arbitrary equations of state, has been discussed in Refs.~\cite{Giese:2020rtr,Giese:2020znk,Ai:2023see}.

Following these works, when the sound speed differs from \(1/\sqrt{3}\) the ordinary trace is no longer the most convenient thermodynamic quantity for the CSS equation of state.
We therefore introduce the pseudo-trace
\begin{equation}
  \bar\theta_\pm(T)
  =
  e_\pm(T)-\frac{p_\pm(T)}{c_-^2},
  \label{eq:pseudo-trace-def}
\end{equation}
where the reference sound speed is chosen to be that of the `\(-\)' phase.
In the limit \(c_-^2=1/3\), \(\bar\theta\) reduces to the ordinary trace \(\theta\).

The corresponding local strength parameter, evaluated just in front of the wall, is
defined as\footnote{%
    For notational convenience, we use \alphaP rather than \alphath[+] for the local pseudo-trace strength throughout this work --- not to be confused with the strength \alphavac[+] associated with the vacuum-energy difference.%
}
\begin{equation}
  \alphaP
  \equiv
  \alphath(T_+)
  \,,\qquad
  \alphath(T) 
  \equiv
  \frac{\Delta\bar\theta(T)}{3\,w_+(T)},
  \qquad
  \Delta\bar\theta(T)
  \equiv
  \bar\theta_+(T)-\bar\theta_-(T).
  \label{eq:alphath-def}
\end{equation}
Using \cref{eq:css-eos,eq:pseudo-trace-def}, this gives
\begin{equation}
  \alphaP
  =
  \alpha_0
  + \frac{1 + c_-^2}{3 c_-^2 (1+ c_+^2)}\,\alphavac[+],
  \qquad
  \alpha_0
  \equiv
  {\lim_{\epsilon\to 0} \alphaP =}
  \frac{c_-^2-c_+^2}{3c_-^2(1+c_+^2)}
  .
  \label{eq:alphath-local}
\end{equation}
This is the natural generalization of the usual bag-model strength parameter to the case \(c_+^2\neq c_-^2\).
Indeed, when \(c_+^2=c_-^2=\frac{1}{3}\), \cref{eq:alphath-local} reduces to \(\alphaP =\alphavac[+]\).
For different sound speeds, instead, the strength receives two contributions: one from the vacuum-energy difference, proportional to \(\alphavac[+]\), and one from the change in the
thermal equation of state across the interface.

\subsection{Hydrodynamic matching at the bubble wall}

\begin{figure}
    \centering
    \includegraphics[width=\textwidth]{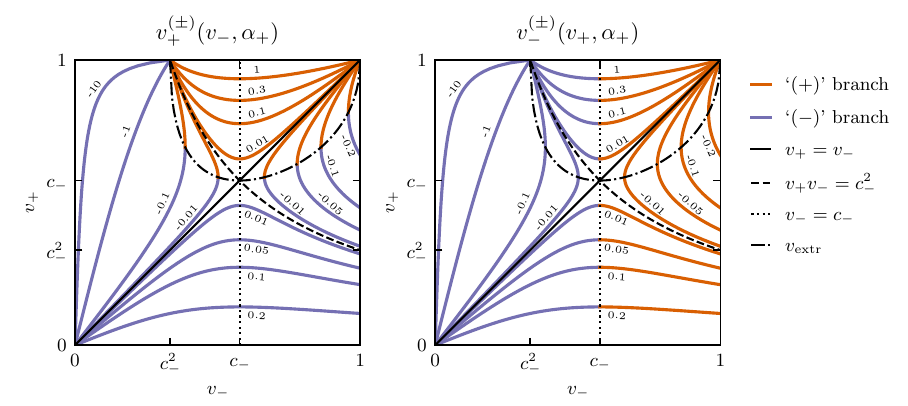}
    \caption{%
        Fluid velocities \(\vp\) and \(\vm\) ahead of and behind the wall in the wall rest frame for different values of \alphaP (see contour labels). 
        The `\((+)\)' and `\((-)\)' branches are shown in orange and purple, respectively, with the branch cuts occurring along the dash-dotted contour tracing the extremal values of \(\vp(\vm,\alphaP)\) and the dotted \(\vm=c_-\) line.
        The solid and dashed black lines indicate the \(\vp=\vm\) and \(\vp \vm = c_-^2\) contours across which the sign of \alphaP changes.
    }
    \label{fig:matching_contours}
\end{figure}

In the CSS model, using the equation of state in \cref{eq:css-eos} and the definition of the pseudo-trace strength in \cref{eq:alphath-def}, the matching conditions imply
\begin{subequations}
\label{eq:vplusminus_css}
\begin{align}
    \vp^{(\pm)}(\vm, \alphaP) &= \chi_w^{-1} \left[
	\frac{\vm}{2} + \frac{c_-^2}{2 \vm} \pm \sqrt{
		\left(\frac{\vm}{2} + \frac{c_-^2}{2 \vm}\right)^2 
		- c_-^2 \chi_w \chi_p
	}
	\right]\,,
    \label{eq:vplus_css}
    \\
    \vm^{(\pm)}(\vp, \alphaP) &=
    \frac{\chi_w \vp}{2}+ \frac{c_-^2\chi_p}{2 \vp}
    	\pm \sqrt{\left(\frac{\chi_w \vp}{2} + \frac{c_-^2 \chi_p}{2 \vp}\right)^2 - c_-^2}\,,
\end{align}
\end{subequations}
where \(\chi_w \equiv 1 + 3\, c_-^2 \alphaP\) and \(\chi_p \equiv 1 - 3\, \alphaP\).
The corresponding contours for different values of \alphaP are depicted in \cref{fig:matching_contours}.
For \(\alphaP>0\), the two branches of \(\vp^{(\pm)}(\vm,\alphaP)\) have a minimum or a maximum, respectively, at \(\vm=c_-\), with the `\((-)\)' branch disappearing for \(\alphaP>\frac{1}{3}\). 
The branches of \(\vm^{(\pm)}(\vp,\alphaP)\) exhibit extrema for negative \(0>\alphaP > - \frac{1-c_-^2}{6 c_-^2}\) while for smaller (larger negative) \alphaP, the `\((+)\)' branch disappears and the `\((-)\)' branch grows monotonically towards \(\vm = c_-^2\) at \(\vp=1\).
The position of the extrema can be parametrized as a function of \vp or \vm as
\begin{subequations}
\begin{align}
    \label{eq:vextr}
    \vwallframe_{+,\text{extr}}(\vm) &= 
    \frac{1}{c_-^2 + \vm^2} \left[(1+c_-^2)\, \vm - \sqrt{(1-\vm^2)(\vm^2-c_-^4)}\right]
    &\text{for}\ c_-^2 < \vm < 1\,,
    \\
    \vwallframe_{-,\text{extr}}^{(\pm)}(\vp) &= 
    \frac{1}{1+\vp^2} \left[(1+c_-^2)\,\vp \pm \sqrt{(\vp^2-c_-^2)(1-c_-^2\vp^2)}\right]
    &\text{for}\ c_- < \vp < 1\,,
\end{align}
\end{subequations}
and is depicted by the dash-dotted curve in \cref{fig:matching_contours}.

Similarly, for the front-side pseudo-trace strength at the wall, \alphaP, \cref{eq:css-eos,eq:alphath-def} imply
\begin{equation}
  \alphaP
  =
  \alphav{\vp}{\vm}
  =
  \frac{
    (\vp-\vm)(\vp \vm - c_-^2)
  }{
    3c_-^2\vm(1-\vp^2)
  } .
  \label{eq:alphath-vplus-vminus}
\end{equation}
Note that this relation divides the \((\vp,\vm)\) plane into four regions based on the sign of \alphaP (and the hierarchy of \vp and \vm), controlled by two simple curves (cf. dashed grey lines in \cref{fig:solutions_vp_vm}):
\(\vp=\vm\) --- separating solutions in which the fluid is accelerated across the wall from those in which it is decelerated --- and \(\vp\vm=c_-^2\) --- the sonic shock-condition associated with the `\(-\)' phase sound speed.

While the jump in the fluid velocity across the bubble wall is determined by the local pseudo-trace strength \alphaP evaluated immediately in front of the wall, the physical input parameter for the fluid-profile calculation is the pseudo-trace strength in the unperturbed plasma at nucleation, \(\alphaN\),\footnote{%
    As for the local strength, we use the short-hand notation \(\alphaN\) rather than \(\alphath[N]\) for the pseudo-trace strength at nucleation, which is different from the vacuum-energy strength \(\alphavac[N] = \alphavac(T_N)\) of \cref{eq:alpha_vac}.%
    \label{fn:alpha-vacuum-N}%
}
which is directly obtained from the effective potential evaluated at the nucleation temperature.
It is related to the local strength via the evolution of enthalpy in front of the wall.
From the CSS expression for the pseudo-trace strength, \cref{eq:alphath-local}, one finds
\begin{equation}
  \alpharw{\alphaP}{r_w}
  =
  \alpha_0 + r_w (\alphaP-\alpha_0),
  \qquad
  r_w
  \equiv
  \frac{w_+}{w_N}\, ,
  \label{eq:alphaN-rw-map}
\end{equation}
where \(r_w\) is the enthalpy ratio in front of the wall, and \(w_N \equiv w_+(T_N)\) is the enthalpy in the `\(+\)' phase at nucleation.

\subsection{Enthalpy evolution ahead of the wall}
\label{sec:enthalpy}

In order to obtain a classification of solutions in terms of transition strength \alphaN in the unperturbed plasma, we need to connect \alphaN to the corresponding strength parameter \alphaP at the phase front through \cref{eq:alphaN-rw-map}, i.e.\ via the evolution of the enthalpy density between the bubble wall and the unperturbed state far ahead of the wall.
If the fluid outside the bubble remains at rest, as for direct detonations and inverse deflagrations, then the ratio of the enthalpy densities directly in front and far ahead of the wall is \(r_w=w_+/w_N = 1\) and \(\alphaN=\alphaP\).
Otherwise, \(r_w\) has to be calculated from the self-similar profile in the \(+\) phase using \cref{eq:enthalpy-ratio}, once the velocity profile \(v(\xi)\) has been obtained from \cref{eq:v-ode-3d}.
In the following, we use superscripts `\(\gtrless\)' to denote enthalpy ratios of solutions with \(v\gtrless 0\) ahead of the wall, i.e.\ direct compression waves (`\(>\)') and inverse rarefaction waves (`\(<\)'), respectively.

If the profile features an inverse rarefaction wave ahead of the wall, as in inverse detonations as well as inverse and mixed hybrids, the profile starts with \(v_w = \mu(\xiw,\vp) < 0\) at \(\xi=\xiw<c_+\) and reaches the unperturbed state at \(\xi=c_+\) with \(v(c_+)=0\). 
The enthalpy ratio then becomes \(r_w = \rwinv[\vp]\) with\footnote{%
    We here suppress the \xiw dependence of \(r_w\) for notational brevity.%
}
\begin{equation}
  \rwinv[\vp]
  \equiv
  \exp\!\left[
    - q_+
    \int\limits_{v_w}^{0}
    \dd v\,
    \gamma^2(v)\,
    \mu(\xi(v;v_w),v)
  \right],
  \label{eq:rw-inverse}
\end{equation}
where \(\xi(v; v_w)\) is the solution of \cref{eq:v-ode-3d} with the initial condition \(\xi(v_w)=\xiw\).

If, on the other hand, the plasma develops a direct compression wave ending in a shock in front of the wall, like in the case of direct deflagrations and direct hybrids, the enthalpy density exhibits a jump at the shock front.
The enthalpy jump is fixed by energy-flux conservation, \(w\gamma^2 v=\text{const.}\), and the shock matching condition \cref{eq:shock-matching}, where, for a front-side shock, the unperturbed plasma ahead of the shock has \(\vwallframe_{\uparrow}^{\rm sh}=\xi_{\rm sh}\), so \(\vwallframe_{\downarrow}^{\rm sh}=c_+^2/\xi_{\rm sh}\).
The plasma-frame fluid velocity \(v_\text{sh}=\mu(\xi_\text{sh},\vwallframe_{\downarrow}^\text{sh})\) and the enthalpy ratio at the shock are thus
\begin{align}
  v_\text{sh}
  &= \frac{\xi_\text{sh}^2 - c_+^2}{\xi_\text{sh} (1-c_+^2)} \,,
  &
  \frac{w_{\downarrow}^\text{sh}}{w_{\uparrow}^\text{sh}}
  &=
  \frac{\xi_{\rm sh}^2-c_+^4}
       {c_+^2(1-\xi_{\rm sh}^2)} \,.
  \label{eq:front-shock-enthalpy-jump}
\end{align}
As \(w_{\uparrow}^\text{sh}\) can be identified with \(w_N\), one obtains
\begin{equation}
  \rwdir[\vp]
  \equiv \frac{w_{\downarrow}^\text{sh}}{w_N} \frac{w_+}{w_{\downarrow}^\text{sh}} 
  = 
  \frac{\xi_{\rm sh}^2-c_+^4}{c_+^2(1-\xi_{\rm sh}^2)} \,
  \exp\!\left[
    q_+
    \int\limits_{v_\text{sh}}^{v_w}
    \dd v\,
    \gamma^2(v)\,
    \mu(\xi(v;v_w),v)
  \right],
  \label{eq:rw-direct}
\end{equation}
where \(\xi_\text{sh}\) depends on the initial conditions, i.e.\ on \xiw and \(v_w\).

It can be shown (cf.\ \cref{app:projection-monotonicity}) that, at fixed \xiw, \(r_w\) decreases monotonically with \vp.
Since \alphaP is also a monotonic function of \vp\ within each solution branch, \(r_w\) is a monotonic function of \alphaP\ on each branch.
Nonetheless, since \alphaN\ depends on the product \(r_w(\alphaP-\alpha_0)\), the mapping \cref{eq:alphaN-rw-map} between \alphaP and \alphaN can become non-monotonic, so that some values of \((\xiw,\alphaN)\) can correspond to more than one solution.
In particular, this occurs for solutions on the `\((+)\)' branch of \cref{eq:vplus_css} once \(c_-\) falls below a critical value \(c_{-,\rm crit}(c_+)<c_+\).

\subsection{Thermodynamic completion and entropy constraints}
\label{sec:entropy-checks}

The construction described above determines which self-similar profiles can be assembled consistently from the wall matching conditions, continuous fluid flows, and shock fronts.
This is not yet sufficient, however, for a profile to be globally physical.
The final requirement is the second law of thermodynamics: entropy must not decrease across any discontinuity appearing in the solution.
This provides an additional selection criterion on top of energy--momentum conservation and can exclude otherwise perfectly admissible solutions of the hydrodynamic equations; see, e.g., Refs.~\cite{Steinhardt:1981ct,Barni:2024lkj,Eriksson:2025owh,Mateos:2026moc}.

There are two equivalent ways of implementing this last step.
If the complete microscopic equation of state is known, one can reconstruct the entropy density directly on the two sides of each discontinuity and check the corresponding entropy flux.
Alternatively, within the CSS description used here, the same information can be supplied through the relative thermodynamic normalization of the two phases
--- i.e.\ \(a_\pm\) in \cref{eq:css-eos} ---
which, in contrast to the velocity and enthalpy profiles, is not fixed by the reduced hydrodynamic problem.
In physical terms, this missing information encodes the change in the effective thermal degrees of freedom across the phase boundary.

A convenient parametrization of this thermodynamic information is the enthalpy ratio at the nucleation temperature~\cite{Ai:2023see,Sanchez-Garitaonandia:2023zqz},
\begin{equation}
  \Psi_N
  \equiv
  \frac{w_-(T_N)}{w_+(T_N)}
  =
  \frac{(1+c_-^2)a_-T_N^{q_-}}
       {(1+c_+^2)a_+T_N^{q_+}} .
  \label{eq:PsiN-def}
\end{equation}
For a microscopic model, \(\Psi_N\) is fixed by the thermodynamics of the two phases and therefore carries, in particular, information about the effective degrees of freedom participating in the transition.
Conversely, when working directly with the CSS template, it can be regarded as the additional input required to complete a hydrodynamic solution thermodynamically.
This separation between the reduced hydrodynamic variables and the enthalpy jump is also useful in model-independent descriptions of bubble dynamics~\cite{Ai:2023see,Sanchez-Garitaonandia:2023zqz}.

Indeed, once the velocity and enthalpy profiles have been constructed, the temperatures in the two phases follow from
\begin{equation}
  \frac{T_+}{T_N}
  =
  \left(\frac{w_+}{w_N}\right)^{1/q_+},
  \qquad
  \frac{T_-}{T_N}
  =
  \left[
    \frac{1}{\Psi_N}
    \frac{w_-}{w_N}
  \right]^{1/q_-},
  \label{eq:temperature-reconstruction-Psi}
\end{equation}
while the corresponding entropy densities are
\begin{equation}
  \frac{s_+}{s_N}
  =
  \left(\frac{w_+}{w_N}\right)^{1-1/q_+},
  \qquad
  \frac{s_-}{s_N}
  =
  \Psi_N^{1/q_-}
  \left(\frac{w_-}{w_N}\right)^{1-1/q_-},
  \label{eq:entropy-reconstruction-Psi}
\end{equation}
where \(s_N=w_N/T_N\).
Thus \(\Psi_N\) does not affect the velocity or enthalpy profile itself; it enters only when the reduced hydrodynamic solution is promoted to a complete thermodynamic one.

We can now impose the second law of thermodynamics at the phase boundary.
Across the bubble wall, the entropy-current flux must not decrease in the physical direction of the transition, {i.e.\ from the `\(+\)' to the `\(-\)' phase}.
With our conventions this requires
\begin{equation}
  s_-\gamma_-\vm
  \ge
  s_+\gamma_+\vp .
  \label{eq:wall-entropy-condition}
\end{equation}
Using \cref{eq:entropy-reconstruction-Psi}, this becomes a lower bound on the thermodynamic completion,
\begin{equation}
  \Psi_N
  \ge
  \Psi_{N,\min}
  \equiv
  \left[
  \frac{
    \left(w_+/w_N\right)^{1-1/q_+}\gamma_+\vp
  }{
    \left(w_-/w_N\right)^{1-1/q_-}\gamma_-\vm
  }
  \right]^{q_-}.
  \label{eq:PsiN-entropy-bound}
\end{equation}
Hence a kinematically admissible hydrodynamic profile corresponds to a physical phase-transition solution only if the underlying thermodynamics provides a value of \(\Psi_N\) satisfying \cref{eq:PsiN-entropy-bound}.
Equivalently, in a microscopic model this condition can be checked directly from the reconstructed entropy densities.
This illustrates explicitly how information about the actual thermodynamic degrees of freedom can further restrict the space of hydrodynamic solutions, as emphasized in Refs.~\cite{Ai:2023see,Sanchez-Garitaonandia:2023zqz,Mateos:2026moc}.

The same principle applies to shocks, although in this case both sides of the discontinuity belong to the same phase and no relative normalization such as \(\Psi_N\) is required.
The Rankine--Hugoniot conditions alone admit solutions crossing the shock front in either direction, with increasing or decreasing fluid speed, but the second law only allows for compressive shocks, for which the entropy flux increases from the upstream to the downstream state.
The opposite orientation is a rarefaction shock corresponding to a formal solution of the matching conditions but not a physical shock~\cite{Steinhardt:1981ct,Barni:2024lkj,Mateos:2026moc}.
This criterion is precisely what excludes the exotic kinematic candidates discussed in \cref{app:exotic-branches}.

The construction of a physical global profile can therefore be viewed as a sequence of increasingly restrictive steps. 
The local wall matching first selects the allowed states immediately adjacent to the phase boundary, the self-similar fluid equations and the available elementary waves then determine whether these states can be connected to the required asymptotic conditions, and finally, the second law selects among these globally constructed profiles those that admit a consistent thermodynamic completion.
In the CSS framework this last step is encoded by the entropy condition at the wall, \cref{eq:PsiN-entropy-bound}, which involves the enthalpy ratio \(\Psi_N\), and by the entropy condition on every shock.

\subsection{Chapman--Jouguet conditions}
\label{sec:ChapmanJouguet}

The stability of the hydrodynamic solutions requires the phase front to be evolutionary, meaning that small perturbations of the front do not introduce unstable modes~\cite{Landau:FluidMechanics,Megevand:2013yua,Megevand:2014yua,Barni:2024lkj}.
This excludes strong detonations and strong deflagrations for which the upstream and downstream flows lie on opposite sides of their respective sonic points, i.e.\ \(\vp >c_+\) and \(\vm < c_-\) or \(\vp<c_+\) and \(\vm>c_-\).
In order to allow for solutions with the plasma perturbed on both sides of the wall, hybrid solutions need to satisfy the Chapman--Jouguet (CJ) condition on one side of the wall,
\begin{equation}
    \vm = c_- \quad\text{(direct/mixed hybrids)},
    \qquad\text{or}\qquad
    \vp = c_+ \quad\text{(inverse/mixed hybrids)}.
    \label{eq:jouguet-condition}
\end{equation}
Note that the side of the wall satisfying the CJ condition has to be connected to a rarefaction wave, as compression waves encounter a shock before reaching the sonic front.

From the CJ conditions, we can define the corresponding Jouguet strengths and velocities for direct and inverse transitions as
\begin{equation}
\begin{aligned}
    \label{eq:jouguet-strenghts-and-velocities}
    \alpha_\text{CJ}^\text{dir}(\xiw) &\equiv \alphav{\xiw}{c_-} 
    \,, \qquad&
    \alpha_\text{CJ}^\text{inv}(\xiw) &\equiv \alphav{c_+}{\xiw}
    \,,
    \\
    \xi_\text{CJ}^\text{dir}(\alphaP) &\equiv \vp^{(+)}(c_-,\alphaP)\,,
    \qquad&
    \xi_\text{CJ}^\text{inv}(\alphaP) &\equiv \vm^{(-)}(c_+,\alphaP)\,.
\end{aligned}
\end{equation}
At the sonic front with fixed wall velocity \xiw, the local strengths \alphaP for detonation and hybrid solutions translate to the same value of the strength parameter \alphaN at nucleation, provided that the mapping between \alphaN and \alphaP remains monotonic (cf.\ discussion in \cref{sec:hydrodynamic-fold}).
Hence, in the standard bag model, these Jouguet boundaries also coincide with the usual limiting detonation or inverse-detonation velocities or strength in the \((\xiw,\alphaN)\) plane~\cite{Steinhardt:1981ct,Kamionkowski:1993fg,Espinosa:2010hh,Hindmarsh:2017gnf,Barni:2025gnm}.
In other words, \(\xi_\text{CJ}^\text{dir}\) and \(\alpha_\text{CJ}^\text{dir}\) correspond to the slowest/strongest direct detonation or fastest/weakest direct hybrid, whereas \(\xi_\text{CJ}^\text{inv}\) and \(\alpha_\text{CJ}^\text{inv}\) (once converted to \alphaN) are the fastest/strongest inverse detonation and slowest/weakest inverse hybrid solutions.

This interpretation does not necessarily carry over to the case of unequal sound speeds.
In the standard case, the CJ point identifies both the limiting detonation solution and, equivalently, the boundary beyond which the wall state can be completed by attaching a shock in front of the wall, thereby giving a hybrid profile.
These two notions coincide when \(c_+=c_-\), but they need not coincide in the general CSS model.

Indeed, while the usual interpretation still holds for \(c_->c_+\), for \(c_+>c_-\) direct and inverse detonations must additionally satisfy \(\xiw>c_+\) and \(\xiw<c_-\), respectively.
More importantly, as we will see in the next section, the usual identification between the CJ point and the limiting detonation/hybrid solution can fail on both the direct and inverse sides.
The local wall-matching conditions may admit states that would normally be completed into a hybrid profile by attaching a compressive shock --- in front of the wall for direct solutions and behind the wall for inverse ones --- while the corresponding shock can no longer be constructed because the wall velocity has crossed the relevant CJ kinematic bound.
These configurations give rise to the mixed-hybrid branches.

Therefore, in the general CSS model the CJ quantities are best viewed as the boundaries separating wall states that can or cannot be completed by a shock on the corresponding side of the wall.
Their familiar interpretation as the slowest detonation or the fastest hybrid, and analogous for inverse solutions, is recovered only in the standard case, where these notions happen to coincide.
For \(c_-<c_+\), this identification can break down completely, as discussed in \cref{sec:hydrodynamic-fold}.

%
\section{A complete index of hydrodynamic solutions}
\label{sec:classification}
%

We now classify the hydrodynamic solutions of the CSS model.
To make direct contact with the thermodynamics of the phase transition obtained from tunnelling calculations, we use as primary input parameters the transition strength, \alphaN, defined with respect to the unperturbed plasma far in front of the bubble, and the bubble-wall velocity, \xiw.
While \xiw can in principle be determined as part of the matching problem by imposing local thermal equilibrium or, more generally, a fixed entropy change across the wall~\cite{BarrosoMancha:2020fay,Ai:2021kak,Ai:2023see,Krajewski:2024zxg,Munzenberg:2025nwa,Krajewski:2026kcm, Ai:2024btx},
we treat it here as an external input parameter.

\subsection{Hydrodynamic classification}

\begin{figure}
  \centering
  \includegraphics{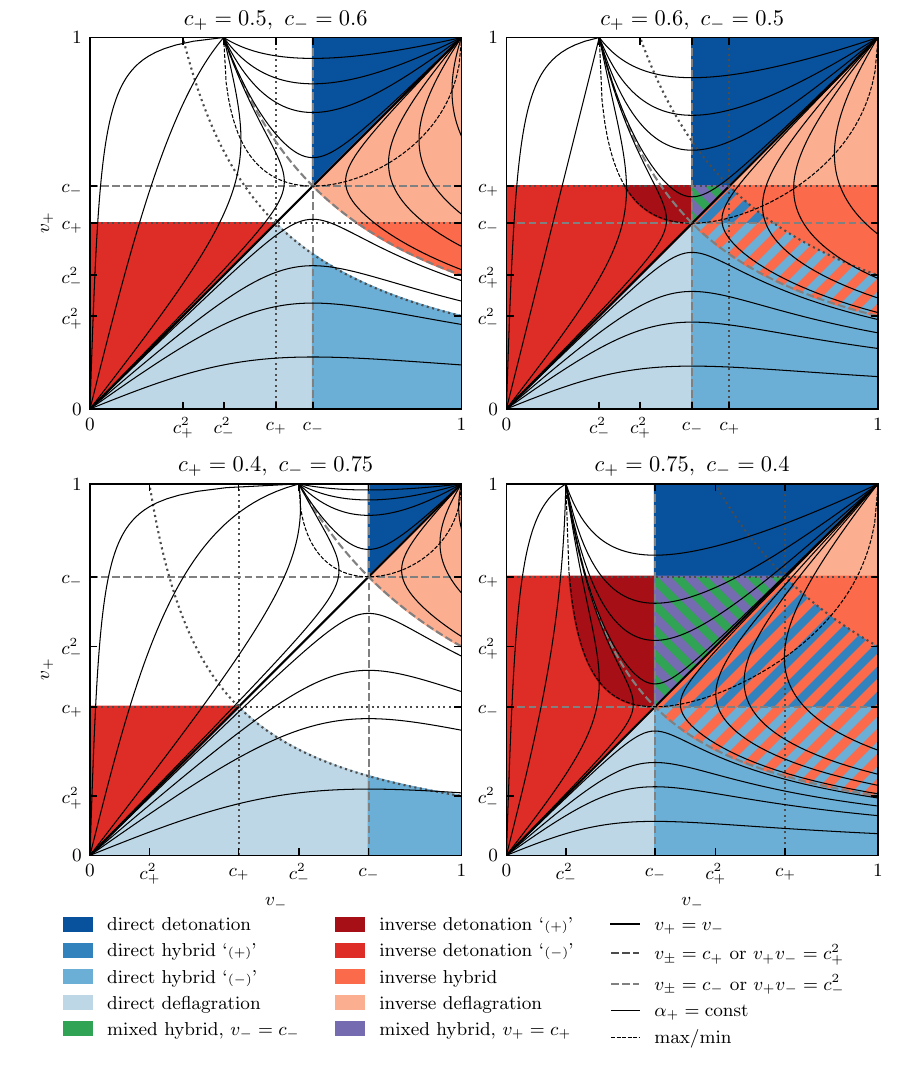}
  \caption{%
    Classification of solutions in terms of \(\vm\) and \(\vp\).
    Direct solutions are shown in shades of blue and inverse solutions in shades of red (from dark to light: detonation, hybrid, deflagration).
    The purple and green regions correspond to mixed direct--inverse hybrids with \(\vp=c_+\) or \(\vm=c_-\), respectively.
    For all hybrid solutions (direct, inverse, and mixed) either \(\vp=c_+\) or \(\vm=c_-\), and the corresponding axis is plotted in terms of the wall velocity \xiw rather than \(\vm\) or \(\vp\).
    The dashed light-grey lines correspond to \(\vpm = c_-\) and \(\vp \vm = c_-^2\), whereas the dark-grey ones indicate \(\vpm = c_+\) and \(\vp \vm = c_+^2\).}
  \label{fig:solutions_vp_vm}
\end{figure}

From the hydrodynamic point of view, the classification is most naturally done in terms of the fluid velocities at both sides of the wall, \(\vp\) and \(\vm\), as well as the wall velocity \xiw\ --- i.e.\ in terms of the local Rankine--Hugoniot conditions derived in \cref{eq:wall-energy-momentum-flux} --- which uniquely determine the corresponding velocity profiles. 
We hence start from the corresponding classification in the \((\vm,\vp)\) plane shown in \cref{fig:solutions_vp_vm}, where regions shaded in blue and red support direct and inverse solutions, respectively, while the purple and green hatched region corresponds to mixed hybrids with \(\vp=c_+\) or \(\vm=c_-\).

In general, we distinguish direct, inverse and mixed solutions, depending on whether the fluid velocity profile in the cosmic frame is positive (direct, \(v(\xi)\geq0\), i.e.\ \(\xiw>\vpm\)), negative (inverse, \(v(\xi)\leq 0\), i.e.\ \(\xiw<\vpm\)) or changes sign (mixed, \(\vpm>\xiw>\vmp\)).
These classes are further subdivided into detonations, deflagrations and hybrid solutions.
Detonations satisfy \(\vp>\vm\), whereas deflagrations have \(\vm>\vp\) in the wall rest frame.
Stable solutions are hydrodynamically weak, i.e.\ the flow is either supersonic (\(\vpm>c_\pm\)) or subsonic (\(\vpm<c_\pm\)) on both sides of the wall.
For direct solutions this corresponds to supersonic detonations and subsonic deflagrations, whereas for inverse solutions the roles are reversed.
The wall velocity \(\xiw\) can be identified with the larger or the smaller of the two wall-frame fluid velocities, depending on whether the solution is direct or inverse.
Finally, we denote detonations and deflagrations of Chapman--Jouguet type with either \(\vm=c_-\) or \(\vp=c_+\) as hybrid solutions.
As hybrid solutions have either \(\vp\) (inverse and mixed) or \(\vm\) (direct and mixed) fixed to the respective speed of sound, the corresponding axis in \cref{fig:solutions_vp_vm} then displays the value of the wall velocity \xiw. 
The hatched blue--orange and green--purple regions hence do not physically overlap, since they correspond to different values of \xiw. 
  
In addition to the solution types introduced in \cref{sec:solutions_general}, for direct hybrid and inverse detonations, we here differentiate between solutions on the different branches of the velocity \(\vp^{(\pm)}(\vm,\alphaP)\) in front of the wall, \cref{eq:vplus_css}.%
\footnote{%
    No analogous distinction is required for \(\vm^{(\pm)}(\vp,\alphaP)\), as the branches separate at \(\vm=c_-\), i.e.\ \(\vm^{(-)}<c_-\) and \(\vm^{(+)} > c_-\).%
}
This differentiation is, however, a ramification of our choice of parametrizing the matching in terms of the pseudo-trace strength transition parameter \alphath and, in contrast to the distinction between the types listed in \cref{sec:solutions_general}, not physical.

\begin{figure}
  \centering
  \includegraphics{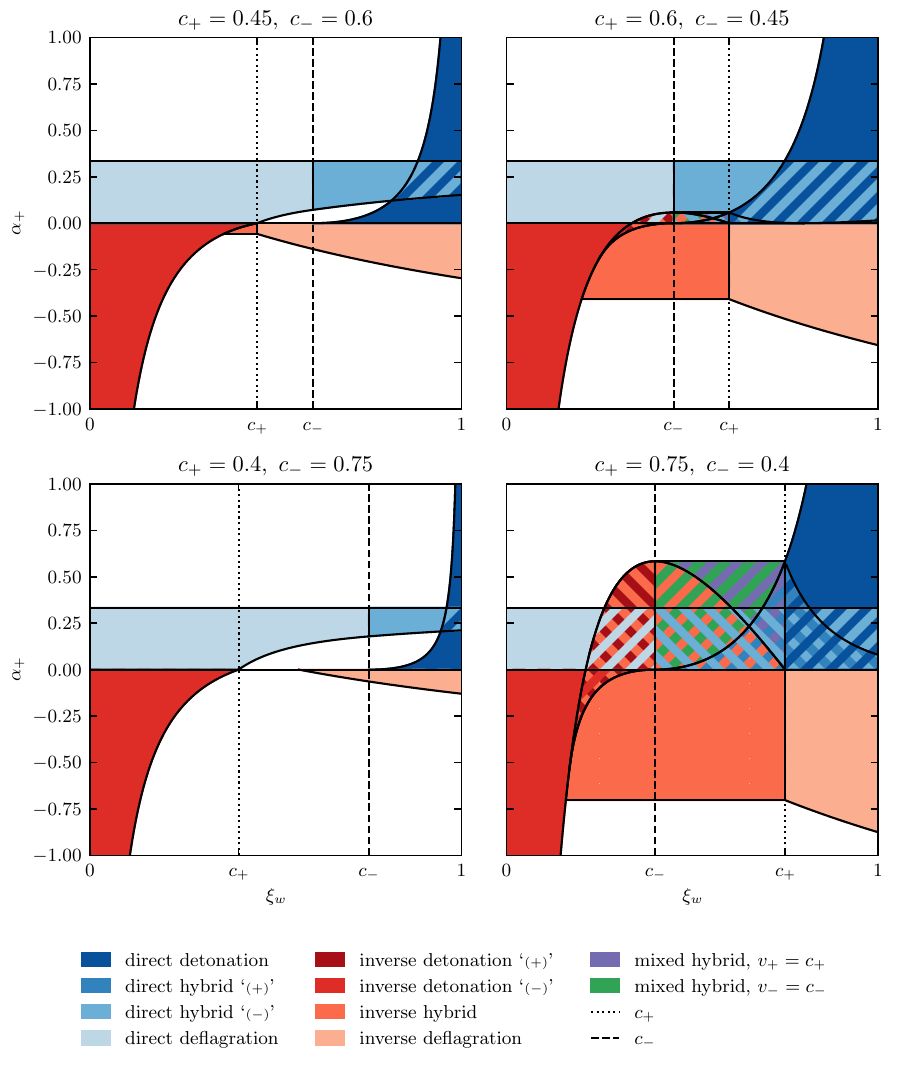}
  \caption{%
    Classification of possible solutions in terms of \(\xiw\) and \alphaP.
    Direct solutions are shown in shades of blue (from dark to light: detonation, hybrid with \vp on the `\((+)\)' branch, hybrid with \vp on the `\((-)\)' branch, deflagration), whereas inverse solutions are red (from dark to light: detonation with \(\vp\) on the `\((+)\)' branch, detonation with \(\vp\) on the `\((-)\)' branch, hybrid, deflagration).
    The purple and green regions correspond to mixed hybrids with \(\vp=c_+\) and \(\vm=c_-\), respectively.
    The dashed and dotted lines indicate \(\xiw=c_-\) and \(\xiw=c_+\).
    Hatched regions allow for more than one possible solution.
  }
  \label{fig:xi_alpha+}
\end{figure}

To translate the classification to the physical \((\xiw,\alphaN)\) parameter plane, we first calculate the local transition strength \alphaP in front of the bubble wall from the wall-frame fluid velocities \(\vp\) and \(\vm\).
The corresponding classification in the \((\xiw,\alphaP)\) plane is depicted in \cref{fig:xi_alpha+}.
As this reduces the number of input parameters, the hatched regions in the figure now correspond to overlaps between the different solution branches.
To obtain a unique solution for the fluid profile, we hence need to specify the corresponding solution branch, which determines which of the wall-frame velocities can be identified with the wall velocity or the respective speed of sound for hybrid solutions.

\begin{figure}
  \centering
  \includegraphics{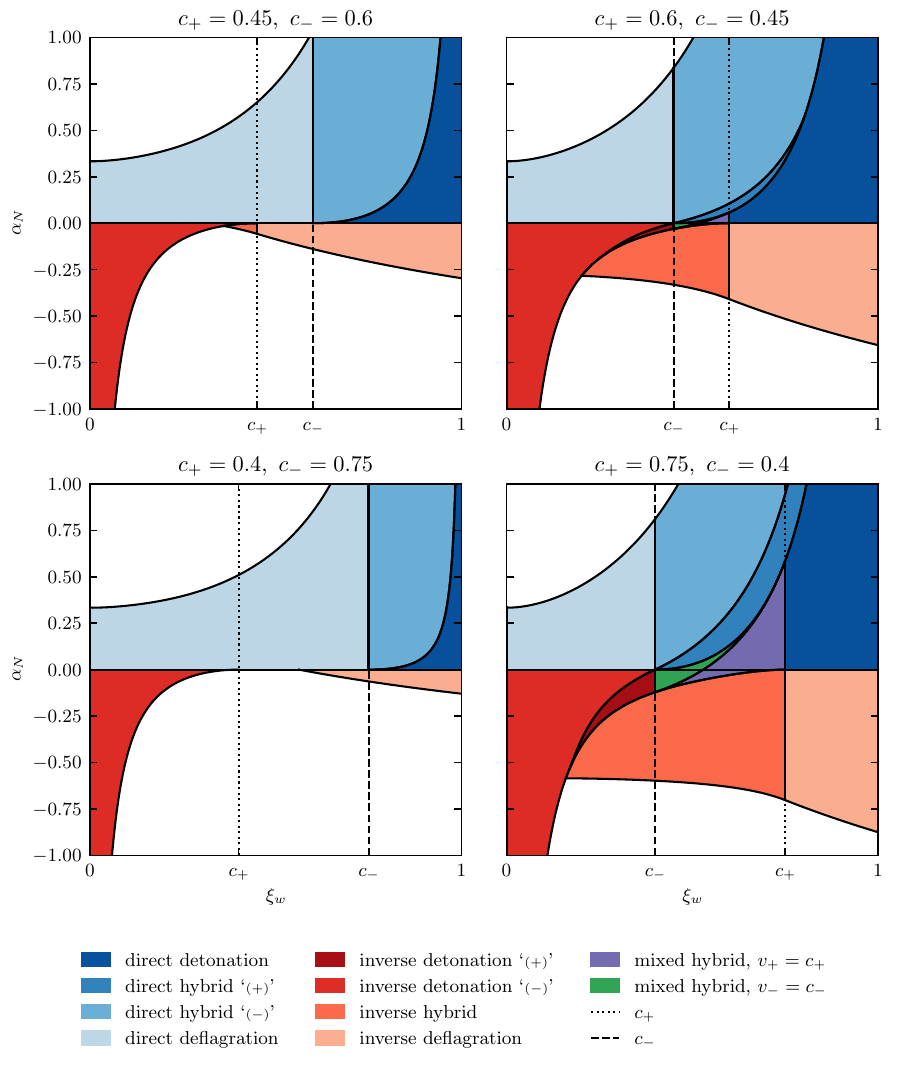}
  \caption{%
    Classification of possible solutions in terms of \xiw and \alphaN.
    Direct solutions are shown in shades of blue [from dark to light: detonation, hybrid with \vp on the `\((+)\)' branch, hybrid with \vp on the `\((-)\)' branch, deflagration], whereas inverse solutions are red [from dark to light: detonation with \(\vp\) on the `\((+)\)' branch, detonation with \(\vp\) on the `\((-)\)' branch, hybrid, deflagration].
    The purple and green regions correspond to mixed hybrids with \(\vp=c_+\) and \(\vm=c_-\), respectively.
    The dashed and dotted lines indicate \(\xiw=c_-\) and \(\xiw=c_+\).
  }
  \label{fig:xi_alphaN}
\end{figure}

Finally, the physical representation is obtained by replacing the local strength \alphaP with the nucleation strength \alphaN through \cref{eq:alphaN-rw-map}, with the mapping controlled by the enthalpy ratio \(r_w\) discussed in \cref{sec:enthalpy}.
The resulting classification in the \((\xiw,\alphaN)\) plane is shown in \cref{fig:xi_alphaN}.

This projection is generally more discriminating than the wall-local \((\xiw,\alphaP)\) representation, because different branches that overlap locally can evolve differently between the wall and the unperturbed plasma far in front of it, and therefore correspond to different values of \(r_w\) and hence of \alphaN.
In most of the parameter space this removes the degeneracies that are artefacts of using wall-local variables.
However, as discussed in \cref{sec:hydrodynamic-fold}, for sufficiently hierarchical sound speeds, in particular for \(c_+\gg c_-\), the map from \alphaP to \alphaN can itself become non-injective, so that distinct hydrodynamic solutions may again project onto the same point in the physical \((\xiw,\alphaN)\) plane.

In the next section we nevertheless introduce and discuss the different hydrodynamic branches one by one, without attempting to resolve possible overlaps at this stage.
As shown in \cref{sec:hydrodynamic-fold}, such non-injectivity occurs only under specific conditions and does not affect the general branch classification.

\subsection{Classification in the CSS model}

\begin{figure}
\centering
\includegraphics[width=1\linewidth]{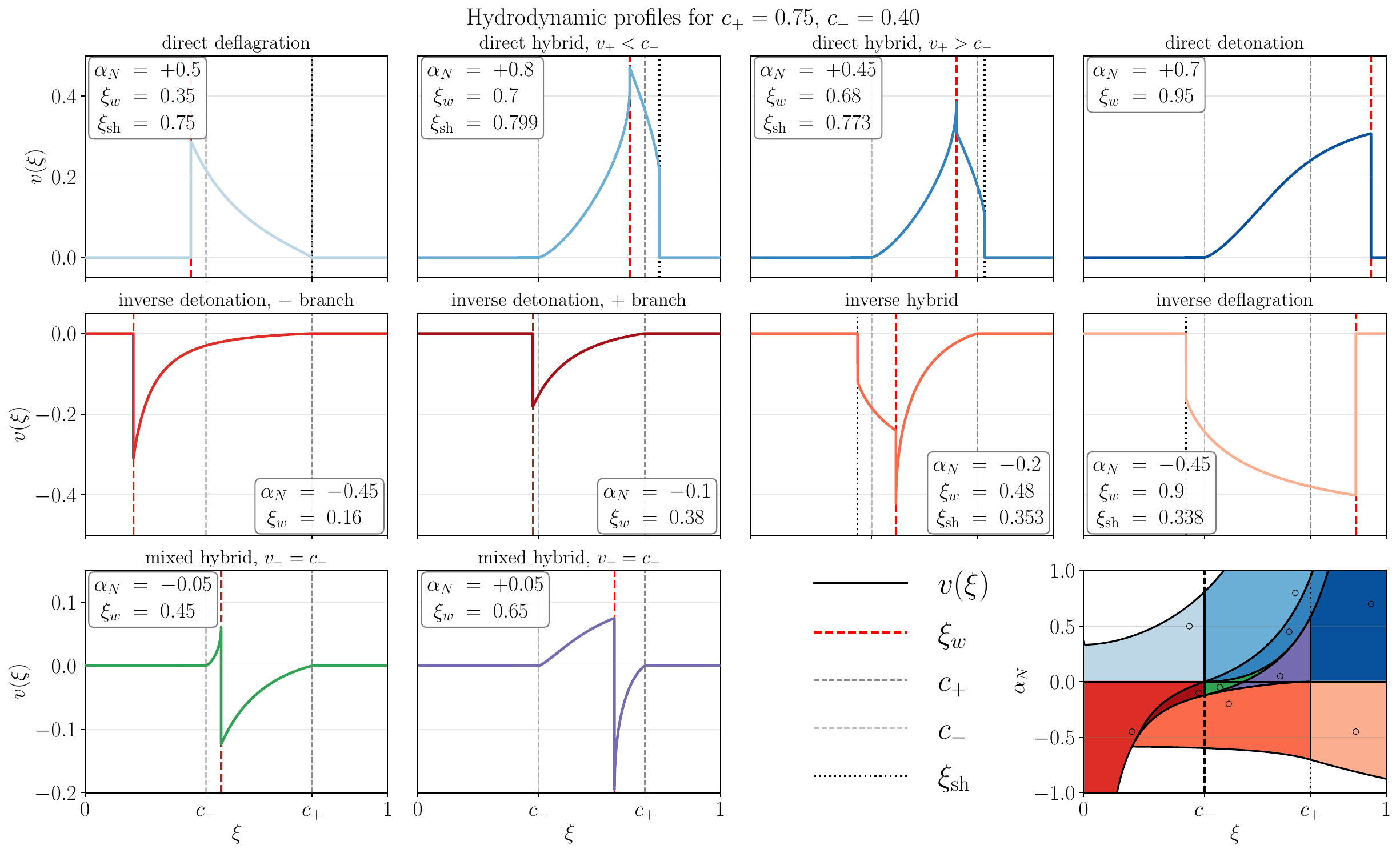}
\includegraphics[width=1\linewidth]{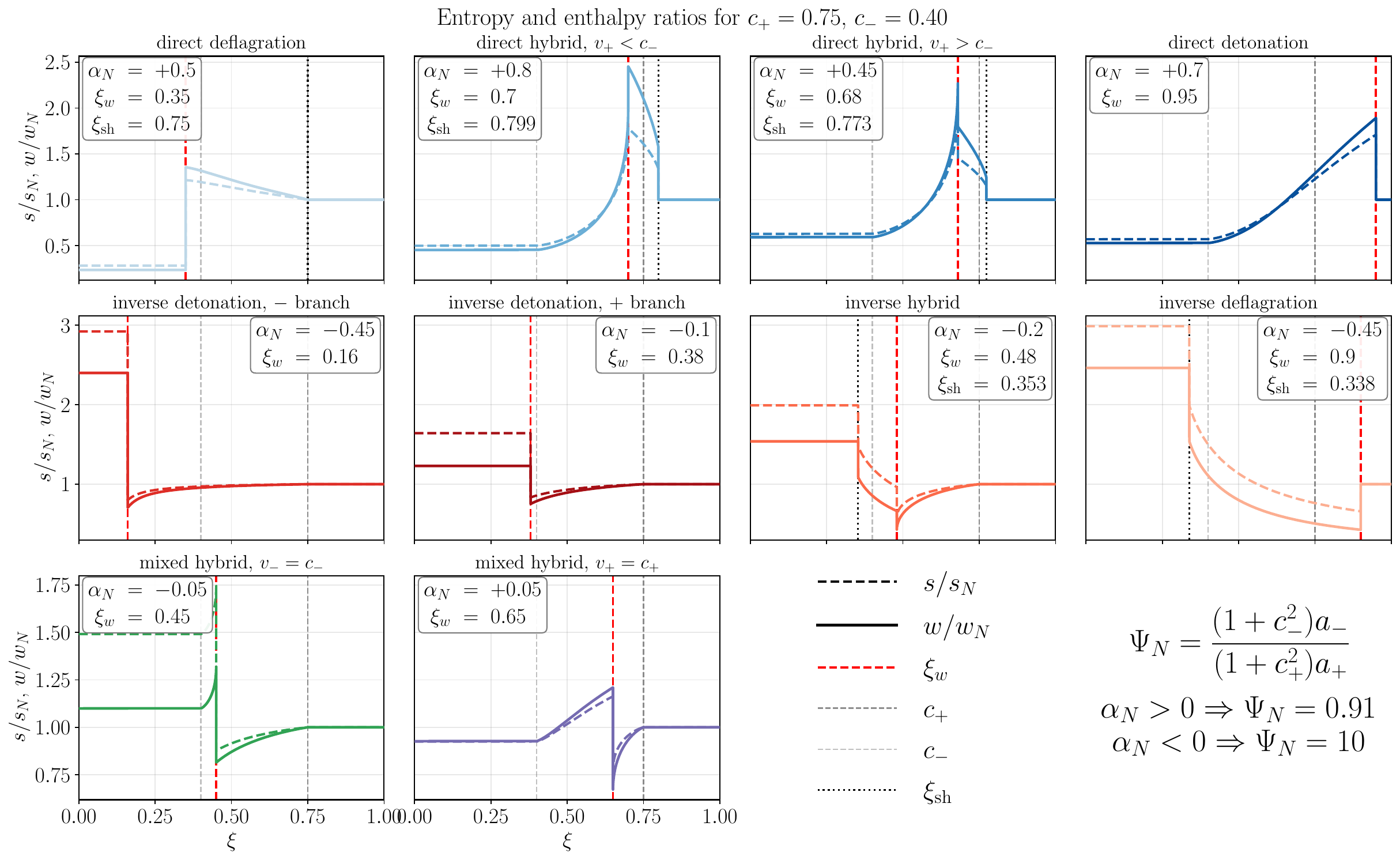}
\caption{%
    Representative hydrodynamic profiles for \(c_+=0.75\) and \(c_-=0.40\).
    The top table shows the velocity profiles \(v(\xi)\) for the different direct, inverse, and mixed branches.
    The bottom table shows the corresponding entropy and enthalpy ratios, \(s/s_N\) (dashed) and \(w/w_N\) (solid).
    Vertical lines mark \(\xiw\) (red dashed), \(c_+\) (dark-grey short-dashed), \(c_-\) (light-grey short-dashed), and, when present, \(\xish\) (black dotted).
    The value of \(\Psi_N\) has been chosen so that all displayed solutions are entropic, namely, such that the entropy-current flux increases across the wall.
}
\label{fig:all-branches}
\end{figure}

In the following, we present the classification of the different solution types in terms of the wall-frame fluid velocities \(\vp\) and \(\vm\), as well as the wall velocity \xiw and the local or nucleation strength parameters \alphaP or \alphaN.
The complete set of profiles is summarized in \cref{fig:all-branches}.
The different panels show the ordinary direct and inverse detonations, deflagrations and hybrids, together with the mixed hybrids that appear when the two sound speeds differ.  

\subsubsection{Direct detonations}

Direct detonations are the branches with positive fluid velocities in the cosmic frame for which the fluid in front of the wall is unperturbed --- i.e.\ the wall-frame velocity \(\vp\) ahead of the wall coincides with the wall velocity --- while behind the wall, a rarefaction wave forms.
In the wall frame they therefore satisfy
\begin{equation}
  \bracenode{detxiw}{\xiw}=\bracenode{detvp}{\vp}>\bracenode{detvm}{\vm}>\bracenode{detcm}{c_-},
  \qquad
  \bracenode{detvpcp}{\vp>c_+} .
  \fullstrut
  \tikzunderbrace{detxiw}{detvp}{direct\\(\(\xiw\!\geq\!\vpm\))}
  \tikzoverbrace{detvp}{detvm}{detonation}
  \tikzunderbrace{detvm}{detcm}{rarefaction\\behind wall}
  \tikzunderbrace{detvpcp}{detvpcp}{stability}
\end{equation}
Expressed in terms of the wall velocity and the pseudo-strength parameter, direct detonations form for
\begin{equation}
  \xiw>\max(c_+,c_-),
  \qquad
  0<\alphaP = \alphaN <\alphav{\xiw}{c_-},
  \label{eq:direct-det}
\end{equation}
where the upper and lower bound on \alphaP correspond to \(\vm\to\xiw\) and \(\vm\to c_-\), respectively, and the velocity behind the wall is given by \(\vm=\vm^{(+)}(\xiw,\alphaP)\).
Since the fluid ahead of the wall remains at rest, the strength parameter \alphaP at the bubble wall and \alphaN in the unperturbed plasma coincide. 
The upper bound on \alphaN is the direct Chapman--Jouguet strength in \cref{eq:jouguet-strenghts-and-velocities} and can be translated into a lower bound \(\xiw > \max\left[\xi_\text{CJ}^\text{dir}(\alphaN), c_+\right]\) on the wall velocity.

\subsubsection{Direct deflagrations}

For direct deflagrations, on the other hand, the plasma is unperturbed behind the wall and develops a compression wave ending at a shock front ahead of the wall. 
Hence,
\begin{equation}
  \bracenode{defvpcp}{\vp<c_+},
  \qquad
  \bracenode{defcm}{c_-} > \bracenode{defxiw}{\xiw} = \bracenode{defvm}{\vm} > \bracenode{defvp}{\vp},
  \qquad
  \bracenode{defshock}{\vp\vm<c_+^2} .
  \fullstrut
  \tikzoverbrace{defcm}{defxiw}{stability}
  \tikzunderbrace{defxiw}{defvm}{direct\\(\(\xiw\!\geq\!\vpm\))}
  \tikzoverbrace{defvm}{defvp}{deflagration}
  \tikzunderbrace{defvpcp}{defvpcp}{compression\\ahead of wall}
  \tikzunderbrace{defshock}{defshock}{shock front\\ahead of wall}
\end{equation}
For \(\xiw<c_+\), the shock condition is automatically satisfied, thus \(0<\vp<\xiw\), while for \(c_->c_+\), the wall velocity can further lie within the interval \(c_+<\xiw<c_-\), for which the additional condition \(\vp\vm<c_+^2\) instead imposes \(0 < \vp < c_+^2/\xiw\), giving
\begin{equation}
    \begin{rcases}
        \text{for}\ \xiw < c_+:
        \qquad&
        0
        \\
        \text{for}\ c_+ < \xiw < c_-:
        \qquad&
        \alphav{\frac{c_+^2}{\xiw}}{\xiw}
    \end{rcases}
    < \alphaP < 
    \frac{1}{3} \,.
\end{equation}
In either case, the upstream wall-frame velocity is given by \(\vp=\vp^{(-)}(\xiw,\alphaP)\).
The physical upper boundary is reached for \(\vp\to0\).  The lower boundary projects to \(\alphaN=0\), either when the shock disappears or when the wall coincides with the shock front.

To obtain the corresponding bounds on \alphaN using \cref{eq:alphaN-rw-map}, we need to take into account the enthalpy evolution from the wall to the shock as well as matching conditions at the shock front.
At the lower boundary, we either have \(\vp \to \xiw < c_-\) or \(v(\xiw) \to 0\) --- so that \(r_w \to 1\) --- or \(\vp\) sitting directly at the shock front --- so that only the shock-front matching contributes to \(r_w\). 
In the latter case, the two terms in \cref{eq:alphaN-rw-map} cancel against each other, so that, in either case, we obtain \(\alphaN > 0\). 
For the upper bound, we need to take into account the enthalpy evolution across the full trajectory from \(v(\xiw) \to\xiw\) to the shock front using \cref{eq:rw-direct}.
In \cref{app:projection-monotonicity}, we  show that the relation between \alphaP and \alphaN remains monotonic for \(0<\vp<\xiw<c_-\).
Hence, the corresponding range of \alphaN is 
\begin{equation}
  \xiw<c_-,
  \qquad
  0<\alphaN<
  \alpharw{\frac13}{\rwdir[0]} .
  \label{eq:direct-def-merged}
\end{equation}

\subsubsection{Direct hybrids}
\label{sec:direct-hybrids}

Direct hybrids are direct branches satisfying the Chapman--Jouguet condition \cref{eq:jouguet-condition} behind the wall,
\begin{equation}
  \bracenode{hybxiwv}{\xiw > \vpm},
  \qquad
  \bracenode{hybvmcm}{\vm=c_-},
  \qquad
  \bracenode{hybvpcp}{\vp<c_+},
  \qquad
  \bracenode{hybshock}{\xiw \vp<c_+^2} .
  \bottomstrut
  \tikzunderbrace{hybxiwv}{hybxiwv}{direct}
  \tikzunderbrace{hybvmcm}{hybvmcm}{CJ condition\\behind wall}
  \tikzunderbrace{hybvpcp}{hybvpcp}{compression\\ahead of wall}
  \tikzunderbrace{hybshock}{hybshock}{shock front\\ahead of wall}
  \label{eq:direct_hybrid_vp_vm}
\end{equation}
They contain a rarefaction wave in the `\(-\)' phase and a compression wave ending in a shock in front of the wall.  
As \cref{eq:direct_hybrid_vp_vm} does not impose any hierarchy between \(\vp\) and \(c_-\), when translating the classification into the \((\xiw, \alphaP)\) plane, \(\vp\) can lie on either branch of \(\vp^{(\pm)}(\vm,\alphaP)\), where the `\((+)\)' branch can only be populated for \(c_+ > c_-\).

\paragraph{\boldmath Direct hybrids on the `\((-)\)' branch:}
For \(\vp=\vp^{(-)}(c_-,\alphaP) < c_-\), the range of \alphaP becomes 
\begin{equation}
    \begin{rcases}
        \text{for}\ c_- < \xiw < \frac{c_+^2}{c_-}:
        \qquad&
        0
        \\
        \text{for}\ \xiw > \max\left(c_-, \frac{c_+^2}{c_-}\right):
        \qquad&
        \alphav{\frac{c_+^2}{\xiw}}{c_-}
    \end{rcases}
    < \alphaP < \frac{1}{3}\,,
\end{equation}
where the two cases arise depending on whether the upper bound on \(\vp\) is determined by \(\vp<c_-\) or \(\vp<c_+^2/\xiw\).
As before, for hybrids starting directly at the shock front (\(\vp=c_+^2/\xiw\)), the enthalpy evolution reduces to the matching factor at the shock front.
It turns out that \alphav{{c_+^2}/{\xiw}}{c_-} matches to \(\alphaN=\alphav{\xiw}{c_-}\), while the enthalpy ratio in the \(\vp\to 0\) and \(\vp\to c_-\) limits does not admit an analytic form.
Since \(\vp<c_-\) on this branch, as in the direct deflagration case, \alphaN is a monotonic function of \alphaP or \vp (cf. \cref{sec:hydrodynamic-fold}).
We therefore obtain
\begin{equation}
    \begin{rcases}
        \text{for}\ c_-<\xiw<\frac{c_+^2}{c_-}: \qquad&
        \alpharw{0}{\rwdir[c_-]}
        \\
        \text{for}\ \xiw>\max\!\left(c_-,\frac{c_+^2}{c_-}\right): \qquad&
        \alphav{\xiw}{c_-}
    \end{rcases}
    < \alphaN <
    \alpharw{\frac{1}{3}}{\rwdir[0]} .
\end{equation}

\paragraph{\boldmath Direct hybrids on the `\((+)\)' branch:}
For \(\vp=\vp^{(+)}(c_-,\alphaP)>c_-\) --- which, again, can occur only if \(c_+>c_-\) --- the bounds \(\vp < \xiw\) and \(\vp<c_+^2/\xiw\) translate to
\begin{equation}
    0 < \alphaP < \begin{cases}
        \alphav{\xiw}{c_-}
        \qquad&\text{for}\ 
        c_- < \xiw < c_+
        \,,\\
        \alphav{\frac{c_+^2}{\xiw}}{c_-}
        \qquad&\text{for}\ 
        c_+ < \xiw < \frac{c_+^2}{c_-} \,.
    \end{cases}
\end{equation}
At the upper end of the \alphaP range, the enthalpy ratio is either unity (for \(\vp=\xiw\)) or reduces to the shock matching factor (for \(\vp=c_+^2/\xiw\)), with \alphaP mapping to \(\alphaN=\alphav{\xiw}{c_-}\) in both cases.   
Provided that the mapping between \alphaP and \alphaN remains injective, this then translates into a lower bound on \alphaN.
However, as we will show in \cref{sec:hydrodynamic-fold}, for \(c_-\ll c_+\), \alphaN becomes non-injective as a function of \alphaP and develops a minimum
\begin{equation}
  \alphaNfoldDH(\xiw)
  \equiv
  \min_{\substack{c_-<\vp<\xiw\\\vp\;<\;{c_+^2}/{\xiw}}} \left\{
  \alpharw{\alphav{\vp}{c_-}}{\rwdir[\vp]}
  \right\} 
  \;\leq\; \alphav{\xiw}{c_-}
  \label{eq:alphaN-fold-direct-hybrid-plus}
\end{equation}
that does not correspond to the upper or lower bound of the respective \alphaP range.
The corresponding critical value \(\ccritDH(c_+)\) below which non-injectivity sets in is given in \cref{eq:ccrits} and discussed in \cref{sec:hydrodynamic-fold}.
The range of \alphaN on the `(\(+\))' branch of direct hybrids is thus
\begin{equation}
  c_-<\xiw<\frac{c_+^2}{c_-},
  \qquad
  \alphaNfoldDH(\xiw)
  <
  \alphaN
  <
  \alpharw{0}{\rwdir[c_-]} .
  \label{eq:alphaN-range-direct-hybrid-plus}
\end{equation}

\subsubsection{Inverse detonations}

Inverse detonations leave the plasma behind the wall unperturbed and develop a rarefaction wave with negative cosmic-frame velocities in front of the wall, reaching \(v(\xi)=0\) at \(\xi=c_+\), i.e.
\begin{equation}
  \bracenode{idetrf}{\vm<c_-},
  \qquad
  \bracenode{idetcp}{c_+} > \bracenode{idetvp}{\vp} > \bracenode{idetvm}{\vm} = \bracenode{idetxiw}{\xiw}
  .\fullstrut
  \tikzunderbrace{idetrf}{idetrf}{stability}
  \tikzunderbrace{idetcp}{idetvp}{rarefaction\\ahead wall}
  \tikzoverbrace{idetvp}{idetvm}{detonation}
  \tikzunderbrace{idetvm}{idetxiw}{inverse\\(\(\xiw \leq \vpm\))}
\end{equation}
As in the case for direct hybrids, when expressing the classification in terms of \alphaP, the front-side velocity can be either branch of \(\vp^{(\pm)}(\xiw,\alphaP)\).
For \(c_-\geq c_+\), only the `\((-)\)' branch corresponds to consistent solutions, with \(\xiw<\vp<c_+\).
If \(c_+>c_-\), however, both branches become accessible and split for \(\xiw>\xi_c\) with \(c_-^2 < \xi_c < c_-\) at the critical point \(\vp = v_c(\xiw)\) (along the dash-dotted curve in \cref{fig:matching_contours,fig:solutions_vp_vm}), where, using \cref{eq:vextr}
\begin{align}
    \xi_c &\equiv \vwallframe_{-,\text{extr}}^{(-)}(\max(c_+,c_-)) = \begin{cases}
        \vwallframe_{-,\text{extr}}^{(-)}(c_+) & \text{for}\ c_+ > c_-\,,\\
        c_- & \text{for}\ c_- \geq c_+\,,
    \end{cases}
    \qquad&
    v_c(\xi) &\equiv \vwallframe_{+,\text{extr}}(\xi) > c_- \,.
    \label{eq:xic-and_vc-inverse-detonation}
\end{align}

In either case, the enthalpy ratio between the state immediately ahead of the wall and the unperturbed fluid is given by \(r_w = \rwinv[\vp]\).

\paragraph{\boldmath Inverse detonations on the `\((-)\)' branch:}
As stated above, for \(c_-\geq c_+\), the wall-frame fluid velocity ahead of the wall satisfies \(\xiw <\vp<c_+\).
The same \vp range applies for \(c_+> c_-\) as long as \(\xiw < \xi_c\). 
For \(\xi_c < \xiw < c_-\), however, the `\((-)\)' branch ends before reaching \(c_+\), so that \(\xiw < \vp < v_c(\xiw)\) in this case. 
The corresponding range of the local strength \alphaP is hence
\begin{equation}
    \begin{rcases}
        \text{for}\ \xiw < \min(c_+,\xi_c): \qquad& \alphav{c_+}{\xiw}
        \\
        \text{for}\ \xi_c < \xiw < c_-: \qquad& \alphav{v_c(\xiw)}{\xiw} 
    \end{rcases}
    < \alphaP < 0 \,,
\end{equation}
where the second interval \(\xi_c<\xiw<c_-\) only exists for \(c_+>c_-\) (as otherwise \(\xi_c=c_-\)).

We will show in \cref{sec:hydrodynamic-fold} that the relation between \alphaN and \alphaP is always monotonic on this branch.
The corresponding physical range is therefore
\begin{equation}
    \begin{rcases}
        \text{for}\ \xiw < \min(c_+,\xi_c): \qquad& \alpharw{\alphav{c_+}{\xiw}}{\rwinv[c_+]}
        \\
        \text{for}\ \xi_c < \xiw < c_-: \qquad& \alpharw{\alphav{v_c(\xiw)}{\xiw}}{\rwinv[v_c(\xiw)]}
    \end{rcases}
    < \alphaN < 0 \,.
\end{equation}

\paragraph{\boldmath Inverse detonations on the `\((+)\)' branch:}
For \(c_+>c_-\) and \(\xiw > \xi_c\), inverse detonations can further populate the `\((+)\)' branch \(\vp^{(+)}(\xiw,\alphaP)\).
From \cref{fig:solutions_vp_vm}, one can see that the corresponding range of \vp is given by \(v_c(\xiw)<\vp<c_+\).
Thus,
\begin{align}
    \xi_c < \xiw < c_- 
    \,,\qquad& \alphav{v_c(\xiw)}{\xiw} < \alphaP < \alphav{c_+}{\xiw}\,.
\end{align}

For this sheet, however, the corresponding \alphaN range cannot in general be obtained by simply projecting the two endpoints of the local \(\alphaP\) interval.
As discussed in \cref{sec:hydrodynamic-fold}, the projection develops an internal minimum for \(c_-\) below the critical value \(c_- < \ccritID(c_+)\). 
Above \(c_->\ccritID(c_+)\), the upper and lower end of the \alphaN range are simply given by the projections of \(\alphav{v_c(\xiw)}{\xiw}\) and \(\alphav{c_+}{\xiw}\), respectively, while below \(c_-<\ccritID(c_+)\), the lower bound on \alphaN is given by the internal minimum.
At even lower values of \(c_-\), the ordering of the projections of the \alphaP limit changes, and the projection of \(\alphav{c_+}{\xiw}\) becomes the upper bound on \alphaN. 
Thus, defining
\begin{equation}\begin{aligned}
  \alphaNfoldID(\xiw)
  &\equiv
  \min_{v_c(\xiw)<\vp<c_+} \left\{
  \alpharw{\alphav{\vp}{\xiw}}{\rwinv[\vp]}
  \right\} ,\\
  \alphaNmaxID(\xiw) &\equiv
  \max\!\left(
    \alpharw{\alphav{c_+}{\xiw}}{\rwinv[c_+]},
    \alpharw{\alphav{v_c(\xiw)}{\xiw}}{\rwinv[v_c(\xiw)]}
  \right) ,
  \label{eq:alphaN-fold-min-IDplus}
\end{aligned}\end{equation}
inverse detonations on the `(\(+\))' branch lie within the \alphaN range
\begin{align}
  \xi_c < \xiw < c_- 
  \,,\qquad& 
  \alphaNfoldID(\xiw) < \alphaN < \alphaNmaxID(\xiw)\,.
  \label{eq:inv-det-plus-true-range}
\end{align}
See \cref{sec:hydrodynamic-fold} for further discussion. 

\subsubsection{Inverse deflagrations}

In inverse deflagrations, the fluid remains at rest ahead of the wall and undergoes a compression wave with negative fluid-frame velocities propagating from an inverse shock front to the wall,
\begin{equation}
  \bracenode{idefvmvm}{\vm>c_-},
  \qquad
  \bracenode{idefvm}{\vm} > \bracenode{idefvp}{\vp} = \bracenode{idefxiw}{\xiw} > \bracenode{idefcp}{c_+},
  \qquad
  \bracenode{idefshock}{\vp\vm>c_-^2} .
  \fullstrut
  \tikzunderbrace{idefvmvm}{idefvmvm}{compression\\behind wall}
  \tikzoverbrace{idefvm}{idefvp}{deflagration}
  \tikzunderbrace{idefvp}{idefxiw}{inverse\\(\(\xiw\leq \vpm\))}
  \tikzoverbrace{idefxiw}{idefcp}{stability}
  \tikzunderbrace{idefshock}{idefshock}{shock front\\behind wall}
\end{equation}
As the plasma in front of the wall is unperturbed, \alphaP and \alphaN coincide. Thus,
\begin{equation}
  \xiw>\max(c_+,c_-^2),
  \qquad
  \alphav{\xiw}{1}<\alphaP=\alphaN<0 \,.
  \label{eq:inv-def-merged}
\end{equation}
The corresponding downstream fluid velocity in the wall frame is given by \(\vm^{(+)}(\xiw,\alphaP)\).

\subsubsection{Inverse hybrids}

Inverse hybrids satisfy the Chapman--Jouguet condition in \cref{eq:jouguet-condition} in front of the wall,
\begin{equation}
  \bracenode{ihybxiwv}{\xiw < \vpm},
  \qquad
  \bracenode{ihybvpcp}{\vp=c_+},
  \qquad
  \bracenode{ihybvmcm}{\vm>c_-},
  \qquad
  \bracenode{ihybshock}{\xiw \vm>c_-^2} .
  \bottomstrut
  \tikzunderbrace{ihybxiwv}{ihybxiwv}{inverse}
  \tikzunderbrace{ihybvpcp}{ihybvpcp}{CJ condition\\ahead of wall}
  \tikzunderbrace{ihybvmcm}{ihybvmcm}{compression\\behind wall}
  \tikzunderbrace{ihybshock}{ihybshock}{shock front\\behind wall}
\end{equation}
They develop a compression wave starting at an inverse shock front behind the wall, and a rarefaction wave ahead of the wall, approaching \(v(\xi)=0\) at \(\xi=c_+\) from below.
Note that inverse hybrids exist only for \(c_+>c_-^2\), and disappear when \(c_+\leq c_-^2\).

The upper bound on \alphaP is set by either \(\vm>c_-^2/\xiw\) or \(\vm>\xiw\), depending on whether \(c_-^2<\xiw<\min(c_-,c_+)\) or \(c_-<\xiw<c_+\), but as \(\alphav{c_+}{c_-^2/\xiw}=\alphav{c_+}{\xiw}\), the parameter range for inverse hybrids in terms of \xiw and \alphaP becomes
\begin{equation}
  c_-^2<\xiw<c_+,
  \qquad
  \alphav{c_+}{1}<\alphaP<\alphav{c_+}{\xiw},
\end{equation}
where the wall-frame velocity behind the wall is \(\vm = \vm^{(+)}(c_+,\alphaP)\).
As \(\vp=c_+\) is fixed, the enthalpy evolution factor \(r_w\) for a fixed wall velocity \xiw is independent of \alphaP and given by \(\rwinv[c_+]\), so that the relation between \alphaP and \alphaN is linear.
In terms of \alphaN, we hence obtain
\begin{equation}
  c_-^2<\xiw<c_+,
  \qquad
  \alpharw{\alphav{c_+}{1}}{\rwinv[c_+]}
  <
  \alphaN
  <
  \alpharw{\alphav{c_+}{\xiw}}{\rwinv[c_+]}.
  \label{eq:inv-hyb-merged}
\end{equation}
Notice that, while the upper bound on \alphaN is negative for \(c_-\geq c_+\) and most values \(c_+\gtrsim c_-\), it can become positive for \(c_+ \gg c_-\), and hence invade regions one would naively attribute to direct transitions.

\subsubsection{Mixed direct--inverse hybrids}

Mixed direct--inverse hybrids are new types of solutions that arise only for
\(c_+>c_-\) and are hence absent in the equal-sound-speed limit \(c_+=c_-\), including the bag model.
They occur in the wall-velocity interval \(c_-<\xiw<c_+\), where the wall is supersonic with respect to the phase behind it, but subsonic with respect to the phase in front.  
The plasma then supports rarefaction waves on both sides of the phase-front, with the fluid moving towards the wall on both sides.
The mixed branches satisfy
\begin{equation}
  \bracenode{mhybcm}{c_-} \leq \bracenode{mhybvm}{\vm} < {\xiw} < \bracenode{mhybvp}{\vp} \leq \bracenode{mhybcp}{c_+} .
  \fullstrut
  \tikzunderbrace{mhybcm}{mhybvm}{rarefaction\\behind wall}
  \tikzoverbrace{mhybvm}{mhybvp}{mixed: \(\mu(\xiw,\vpm)\lessgtr 0\)}
  \tikzunderbrace{mhybvp}{mhybcp}{rarefaction\\ahead of wall}
\end{equation}
The local wall discontinuity is evolutionary only when one side is sonic at the wall, i.e.\ the stability of this hydrodynamic solution requires either \(\vm=c_-\) or \(\vp=c_+\), satisfying the Chapman--Jouguet condition \cref{eq:jouguet-condition} behind or in front of the wall, respectively.

\paragraph{\boldmath Mixed hybrids with \(\vp=c_+\):}

For the mixed branch with \(\vp=c_+\), the corresponding range of \alphaP is
\begin{equation}
  c_- < \xiw < c_+,
  \qquad
  \alphav{c_+}{\xiw}
  <\alphaP
  <\alphav{c_+}{c_-},
\end{equation}
and \(\vm = \vm^{(+)}(c_+,\alphaP)\).
As for inverse hybrids, the enthalpy ratio \(r_w = \rwinv[c_+]\) in this case only depends on the wall velocity \xiw and the sound speed \(c_+\) in the `\(+\)' phase but not on \(\alphaP=\alphav{c_+}{\vm}\).
Hence,
\begin{equation}
  c_- < \xiw < c_+,
  \qquad
  \alpharw{\alphav{c_+}{\xiw}}{\rwinv[c_+]}
  <
  \alphaN
  <
  \alpharw{\alphav{c_+}{c_-}}{\rwinv[c_+]}.
  \label{eq:mixed-plus-merged}
\end{equation}

\paragraph{\boldmath Mixed hybrids with \(\vm=c_-\):}

For the mixed branch with \(\vm=c_-\), \(\vp\in(\xiw,c_+)\), giving
\begin{equation}
  c_- < \xiw < c_+,
  \qquad
  \alphav{\xiw}{c_-}
  < \alphaP
  < \alphav{c_+}{c_-}
\end{equation}
with \(\vp=\vp^{(+)}(c_-,\alphaP)\).

The endpoint \(\vp=\xiw\) has unperturbed fluid in front of the wall and therefore projects trivially, while the endpoint \(\vp=c_+\) is connected to the unperturbed state by the inverse rarefaction wave and projects with \(\rwinv[c_+]\).
However, in contrast to the \(\vp=c_+\) case, \vp and hence also \rwinv[\vp] now depend on \alphaP.
As for the `(\(+\))' branches of direct hybrids and inverse detonations, the projected range is not in general obtained by simply projecting these two endpoints.
The relation between \alphaN and \alphaP becomes non-injective for \(c_- < \ccritMH(c_+)\), where the projection develops an internal minimum.
At \(c_->\ccritMH(c_+)\), the relation is monotonic and \(\alpharw{\alphav{c_+}{c_-}}{\rwinv[c_+]} < \alphaN < \alphav{\xiw}{c_-}\), while below \(c_-<\ccritMH(c_+)\), the minimal \alphaN is given by the internal minimum, and as \(c_-\) further decreases, the order of the projections of the \alphaP endpoints is interchanged.
We therefore define
\begin{equation}\begin{aligned}
  \alphaNfoldMH(\xiw)
  &\equiv
  \min_{\xiw<\vp<c_+} \left\{
  \alpharw{\alphav{\vp}{c_-}}{\rwinv[\vp]}
  \right\} ,\\
  \alphaNmaxMH(\xiw) &\equiv
  \max\!\left(
    \alphav{\xiw}{c_-},
    \alpharw{\alphav{c_+}{c_-}}{\rwinv[c_+]}
  \right) .
  \label{eq:alphaN-fold-min-mixed-minus}
\end{aligned}\end{equation}
The physical image of the mixed hybrid branch with \(\vm=c_-\) is then bounded by
\begin{align}
  c_- < \xiw < c_+
  \,,\qquad& 
  \alphaNfoldMH(\xiw) < \alphaN < \alphaNmaxMH(\xiw)
  \,.
  \label{eq:mixed-minus-true-range}
\end{align}

\begin{table}[th!]
\small
\centering
\begin{NiceTabular*}{\linewidth}{@{}>{\centering\arraybackslash}m{2em}@{\extracolsep{\fill}}ccc}[cell-space-limits=3pt]
    \toprule
    \rowcolor{gray!25}
    & {\bfseries\boldmath\xiw range}
    & {\bfseries\boldmath\alphaP range}
    & {\bfseries\boldmath\alphaN range}
    \\
    \midrule
    \Block[fill=gray!25]{11-1}{%
        \centering\rotatebox{90}{\bfseries Generation I}%
    }
    & \Block[fill=gray!10]{1-3}{%
        {\bfseries direct detonation:}
        \qquad\(\vp=\xiw,\qquad \vm=\vm^{(+)}(\xiw,\alphaP)\)
    } \\
    & \(\xiw > \max[c_\pm]\)
    & \(\left[0,\alphav{\xiw}{c_-}\right]\)
    & \(\left[0,\alphav{\xiw}{c_-}\right]\)
    \\

    & \Block[fill=gray!10]{1-3}{%
        {\bfseries\boldmath direct hybrid, \(\vp>c_-\):}
        \qquad\(\vp=\vp^{(+)}(c_-,\alphaP),\qquad \vm=c_-\)
    } \\
    & \(c_- < \xiw < c_+\)
    & \(\left[0,\alphav{\xiw}{c_-}\right]\)
    & \Block{2-1}{
        \(\left[
        \alphaNfoldDH\!\!(\xiw),
        \alpharw{0}{\rwdir[c_-]}
        \right]\)
    }
    \\
    & \(c_+ < \xiw < \frac{c_+^2}{c_-}\)
    & \(\left[
        0,\alphav{\frac{c_+^2}{\xiw}}{c_-}
        \right]
    \)
    &
    \\
    & \Block[fill=gray!10]{1-3}{%
        {\bfseries\boldmath direct hybrid, \(\vp<c_-\):}
        \qquad\(\vp=\vp^{(-)}(c_-,\alphaP),\qquad \vm=c_-\)
    } \\
    & \(c_- < \xiw < \frac{c_+^2}{c_-}\)
    & \(\left[0,\frac{1}{3}\right]\)
    & \(\left[
        \alpharw{0}{\rwdir[c_-]},
        \alpharw{\frac{1}{3}}{\rwdir[0]}
        \right]\)
    \\
    & \(\xiw > \max\left[c_-,\frac{c_+^2}{c_-}\right]\)
    & \(\left[
        \alphav{\frac{c_+^2}{\xiw}}{c_-},\frac{1}{3}
        \right]\)
    & \(\left[
        \alphav{\xiw}{c_-},
        \alpharw{\frac{1}{3}}{\rwdir[0]}
        \right]
    \)
    \\
    & \Block[fill=gray!10]{1-3}{%
        {\bfseries\boldmath direct deflagration:}
        \qquad\(\vp=\vp^{(-)}(\xiw,\alphaP),\qquad \vm=\xiw\)
    } \\
    & \(\xiw < \min[c_\pm]\)
    & \(\left[0,\frac{1}{3}\right]\)
    & \Block{2-1}{
        \(\left[
        0,\alpharw{\frac{1}{3}}{\rwdir[0]}
        \right]\)
    }
    \\
    & \(c_+ < \xiw < c_-\)
    & \(\left[
        \alphav{\frac{c_+^2}{\xiw}}{\xiw},\frac{1}{3}
        \right]
    \)
    &
    \\
    \midrule
    \Block[fill=gray!25]{9-1}{%
        \rotatebox{90}{\bfseries Generation II}%
    }
    & \Block[fill=gray!10]{1-3}{%
        {\bfseries\boldmath inverse detonation, \(\vp^{(-)}\):}
        \qquad\(\vp=\vp^{(-)}(\xiw,\alphaP),\qquad \vm=\xiw\)
    } \\
    & \(\xiw < \min\left[c_+,\xi_c\right]\)
    & \(\left[\alphav{c_+}{\xiw},0\right]\)
    & \(\left[
        \alpharw{\alphav{c_+}{\xiw}}{\rwinv[c_+]},0
        \right]\)
    \\
    & \(\xi_c < \xiw < c_-\)
    & \(\left[\alphav{v_c(\xiw)}{\xiw},0\right]\)
    & \(\left[
        \alpharw{\alphav{v_c(\xiw)}{\xiw}}
        {\rwinv[v_c(\xiw)]},0
        \right]\)
    \\
    & \Block[fill=gray!10]{1-3}{%
        {\bfseries\boldmath inverse detonation, \(\vp^{(+)}\):}
        \qquad\(\vp=\vp^{(+)}(\xiw,\alphaP),\qquad \vm=\xiw\)
    } \\
    & \(\xi_c < \xiw < c_-\)
    & \(\left[
        \alphav{v_c(\xiw)}{\xiw},\alphav{c_+}{\xiw}
        \right]\)
    & \(\left[
        \alphaNfoldID\!\!(\xiw),\alphaNmaxID(\xiw)
        \right]
    \)
    \\
    & \Block[fill=gray!10]{1-3}{%
        {\bfseries\boldmath inverse hybrid:}
        \qquad\(\vp=c_+,\qquad \vm=\vm^{(+)}(c_+,\alphaP)\)
    } \\
    & \(c_-^2 < \xiw < c_+\)
    & \(\left[
        \alphav{c_+}{1},\alphav{c_+}{\xiw}
        \right]\)
    & \Block{1-1}{%
        \([\alpharw{\alphav{c_+}{1}}{\rwinv[c_+]},
        \)\hspace*{2em}\\\hspace*{2em}\(
        \alpharw{\alphav{c_+}{\xiw}}{\rwinv[c_+]}]\)
    }
    \\
    & \Block[fill=gray!10]{1-3}{%
        {\bfseries\boldmath inverse deflagration:}
        \qquad\(\vp=\xiw,\qquad \vm=\vm^{(+)}(\xiw,\alphaP)\)
    } \\
    & \(\xiw > \max[c_+,c_-^2]\)
    & \(\left[\alphav{\xiw}{1},0\right]\)
    & \(\left[\alphav{\xiw}{1},0\right]\)
    \\
    \midrule
    \Block[fill=gray!25]{4-1}{%
        \rotatebox{90}{\bfseries Generation III}%
    }
    & \Block[fill=gray!10]{1-3}{%
        {\bfseries\boldmath mixed hybrid, \(\vp=c_+\):}
        \qquad\(\vp=c_+,\qquad \vm=\vm^{(+)}(c_+,\alphaP)\)
    } \\
    & \(c_- < \xiw < c_+\)
    & \(\left[
        \alphav{c_+}{\xiw},\alphav{c_+}{c_-}
        \right]\)
    & \Block{1-1}{%
        \([\alpharw{\alphav{c_+}{\xiw}}{\rwinv[c_+]},
        \)\hspace*{2em}\\\hspace*{2em}\(
        \alpharw{\alphav{c_+}{c_-}}{\rwinv[c_+]}]\)
    }
    \\
    & \Block[fill=gray!10]{1-3}{%
        {\bfseries\boldmath mixed hybrid, \(\vm=c_-\):}
        \qquad\(\vp=\vp^{(+)}(c_-,\alphaP),\qquad \vm=c_-\)
    } \\
    & \(c_- < \xiw < c_+\)
    & \(\left[
        \alphav{\xiw}{c_-},\alphav{c_+}{c_-}
        \right]\)
    & \(\left[
        \alphaNfoldMH\!(\xiw),\alphaNmaxMH(\xiw)
        \right]
    \)
    \\
    \bottomrule
\end{NiceTabular*}

\caption{%
    Summary of the hydrodynamic solutions with the respective \xiw ranges and the corresponding ranges of \alphaP and \alphaN.
    Cf.\ \cref{eq:vplusminus_css,eq:alphath-vplus-vminus,eq:alphaN-rw-map,eq:rw-direct,eq:rw-inverse,eq:xic-and_vc-inverse-detonation,eq:alphaN-fold-direct-hybrid-plus,eq:alphaN-fold-min-IDplus,eq:alphaN-fold-min-mixed-minus}.
}
\label{tab:classification}
\end{table}

\subsection{Brief summary}

\begin{figure}
  \includegraphics[width=\textwidth]{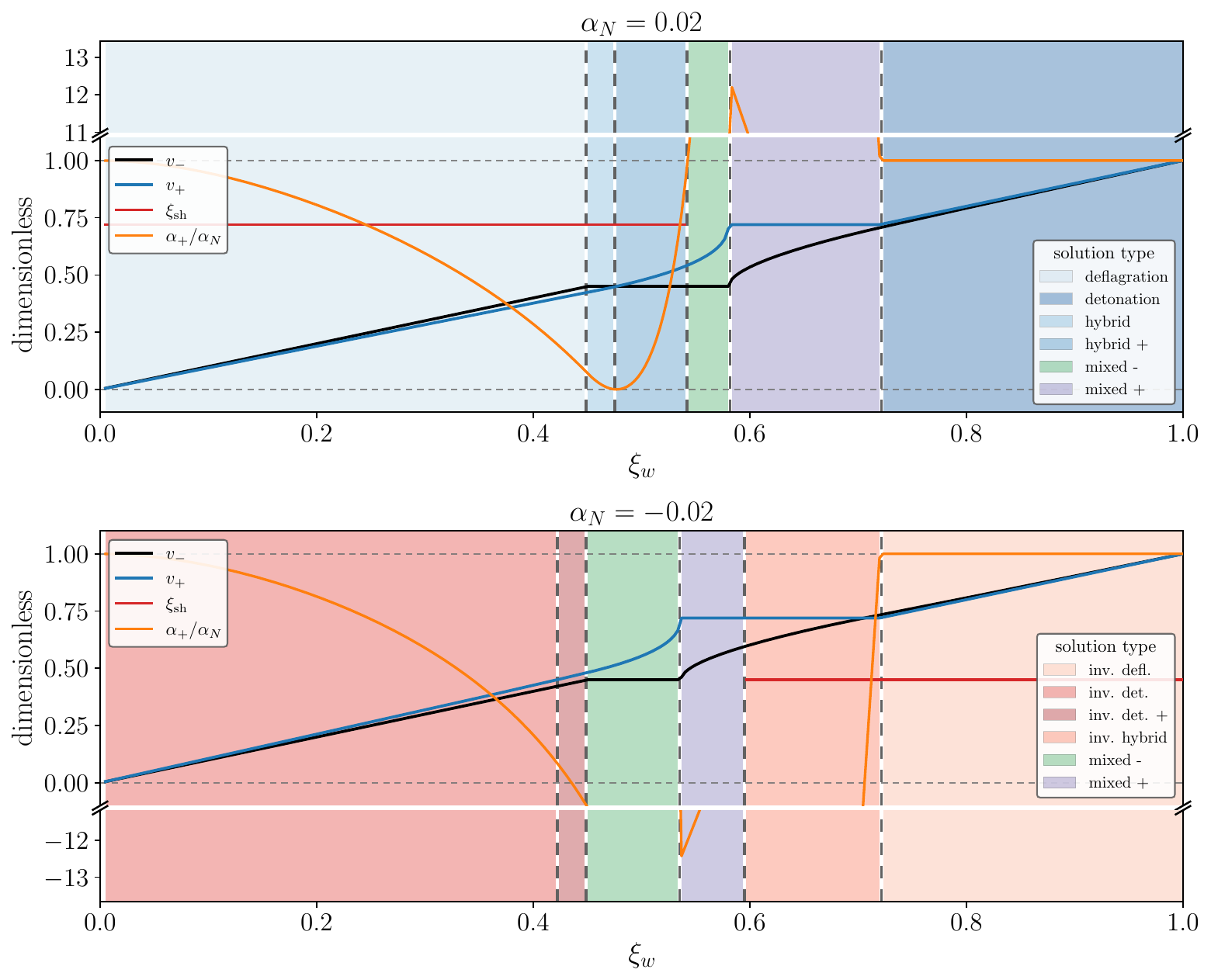}
  \caption{%
      Evolution of the wall-frame fluid velocities \vm\ and \vp\ (black and blue lines, respectively), the shock front position \xish (red), and the ratio \(\alphaP/\alphaN\) (orange) as a function of \xiw, for fixed \(\alphaN=0.02\) [top] and \(\alphaN=-0.02\) [bottom] and for \((c_+, c_-)=(0.72, 0.45)\).
      The colour shading of the background indicates the corresponding hydrodynamic branch.
      The plot illustrates that the local strength \alphaP can differ significantly from the physical strength \alphaN, and
      can even have the opposite sign.
  }
  \label{fig:solutions-evolution}
\end{figure}

A compact summary of the classification is given in \cref{tab:classification}, while \cref{fig:solutions-evolution} illustrates the relation between the wall-local strength \alphaP and the asymptotic strength \alphaN.
Representative solutions at fixed positive and negative \alphaN are followed as \xiw varies, together with the corresponding wall-frame velocities \vpm, shock position, and ratio \(\alphaP/\alphaN\).
In general, \alphaP and \alphaN need not have the same sign.
For instance, along inverse branches with \(\alphaN<0\), the local strength \alphaP can become positive.

For \(c_-\geq c_+\), the sign of \alphaN distinguishes direct (\(\alphaN>0\)) from inverse (\(\alphaN<0\)) solutions.
This identification fails for \(c_-<c_+\), where mixed solutions appear and, for sufficiently large sound-speed hierarchies, inverse detonations on the `(\(+\))' branch and inverse hybrids can extend into the \(\alphaN>0\) region.
The parametrization in terms of \((\xiw,\alphaN)\) nevertheless remains unique as long as \(c_->\ccritID(c_+)\).
Each point determines a single hydrodynamic solution and unique local quantities \alphaP and \vpm, as shown in \cref{fig:xi_alphaN}.
In this regime, the projection to \alphaN removes the apparent degeneracies visible in \cref{fig:xi_alpha+}, which arise from using the wall-local quantity \alphaP as an input.
For \(c_-<\ccritID(c_+)\), instead, the map between \alphaP and \alphaN becomes non-injective, leading to genuine overlaps in the \((\xiw,\alphaN)\) plane, as discussed in \cref{sec:hydrodynamic-fold}.

%
\section{Uniqueness of the hydrodynamic classification}
\label{sec:hydrodynamic-fold}
%

The previous section showed that \((\xiw,\alphaN)\) provides the natural physical parametrization of the hydrodynamic classification.
While the local strength \alphaP is evaluated in the fluid state immediately in front of the wall, which may have been heated, compressed, or rarefied by the hydrodynamic profile, \alphaN is defined in the unperturbed plasma far ahead of the bubble.

At first sight, one would expect the projection to \((\xiw,\alphaN)\) to make the hydrodynamic classification unique, and in most of parameter space this is indeed the case. 
This is also the assumption usually made when working with the physical strength \alphaN.  
However, as stated in \cref{sec:classification}, the projection is not always one-to-one.
For sufficiently different sound speeds, distinct wall states can be mapped to the same strength, so that a hydrodynamic branch folds in the \((\xiw,\alphaN)\) plane
and multiple stationary profiles coexist at fixed input values.

This phenomenon is restricted to \(c_+>c_-\) and to a small subset of the hydrodynamic branches.
Below we identify the branches on which folds can occur, determine the corresponding critical sound-speed hierarchies, and study when the resulting degeneracies become macroscopically relevant.
The formal monotonicity analysis and proofs are given in \cref{app:projection-monotonicity}.

\subsection{Degeneracies from non-injectivity}

At fixed wall velocity, the projection \(\alphaP \mapsto \alphaN\) is \cref{eq:alphaN-rw-map}, where the enthalpy ratio \(r_w=w_+/w_N\) is fixed by the fluid profile in front of the wall.
A solution branch becomes non-injective when \alphaN develops an internal turning point as a function of \alphaP at fixed \xiw, i.e.\ if
\begin{equation}
    \left.\frac{1}{r_w} \frac{\partial\alphaN}{\partial\alphaP}\right|_{\xiw} = 1 + (\alphaP-\alpha_0) \left.\frac{\partial \ln r_w}{\partial \vp}\right|_{\xiw} \left[\frac{\partial\alphaP}{\partial\vp}\right]_{\xiw}^{-1}
    = 0\,.
\end{equation}
\Cref{app:rw-monotonicity} shows that \(\partial_{\vp}\ln r_w|_{\xiw} < 0\).
Furthermore, within each solution branch, \alphaP is a monotonic function of \vp, where \(\partial_{\vp}\alphaP|_{\xiw} \gtrless 0\) for solutions using \(\vp=\vp^{(\pm)}(\vm,\alphaP)\), i.e.\ the map can become degenerate if \(\alphaP \gtrless \alpha_0\), respectively.
The fold appears when the logarithmic variation of \(r_w\) cancels the logarithmic variation of the shifted local strength \(\alphaP-\alpha_0\).

The complete branch-by-branch analysis of the projection \(\alphaP\mapsto\alphaN\) is presented in detail in \cref{app:projection-monotonicity}.  
For branches on which the fluid in front of the wall remains unperturbed, namely direct detonations and inverse deflagrations, the projection is trivial, \(\alphaN=\alphaP\).
These branches are therefore manifestly one-to-one.
The non-trivial projection arises when the state immediately in front of the wall must be connected to the asymptotic plasma by a compression wave or an inverse rarefaction.  
For inverse hybrids and mixed hybrids with \(\vp=c_+\), the fluid velocity ahead of the wall is fixed to the speed-of-sound in the wall frame, i.e.\ the enthalpy ratio is independent of \vp and injectivity follows from the monotonicity of \alphaP.
For direct deflagrations as well as the `(\(-\))' branches of direct hybrids and inverse detonations, monotonicity of the \(\alphaP\mapsto\alphaN\) map is shown analytically in \cref{app:projection-monotonicity}.

The situation changes on the remaining solution classes, which are the `(\(+\))' branches of direct hybrids (DH\textsuperscript{+}) and inverse detonations (ID\textsuperscript{+}) as well as mixed hybrids with \(\vm=c_-\) (MH\textsuperscript{\(-\)}).
These solutions only exist for \(c_-<c_+\) and satisfy \(\partial_{\vp}\alphaP|_{\xiw} > 0\) as well as \(\alphaP > \alpha_0\) throughout their respective parameter ranges.
Indeed, \(\partial_{\alphaP}\alphaN|_{\xiw}\) crosses zero in parts of the \xiw range once \(c_-<c_{-,\text{crit}}(c_+)<c_+\) falls below some critical value which depends on \(c_+\) as well as the respective solution type.
Above \(c_- > c_{-,\text{crit}}(c_+)\), the projection remains injective.
While the critical value for direct hybrids can be derived in a closed algebraic form, the remaining critical sound speeds can only be calculated numerically.
As shown in \cref{app:projection-monotonicity}, they satisfy
\begin{equation}
    \frac{3-c_+^2 - \sqrt{\left(1-c_+^2\right)\left(9-c_+^2\right)}}{2c_+} =
	\ccritDH(c_+) < \ccritMH(c_+) < \ccritID(c_+) < \frac{1-\sqrt{1-c_+^2}}{c_+} \,.
    \label{eq:ccrits}
\end{equation}

Below the corresponding critical values of \(c_-\), these non-monotonic branches generate degeneracies between multiple solutions, i.e.\ there are points in the \((\xiw,\alphaN)\) plane that can correspond to more than one solution type.
The parameter region of the inverse detonations' `(\(+\))' branch develops an overlap with the inverse hybrid range, the direct-hybrid `(\(+\))' branch overlaps with the mixed-hybrid regime, and mixed hybrids with \(\vm=c_-\) reach into the parameter space of inverse and mixed hybrids with \(\vp=c_+\).
However, this does not immediately imply a sizeable ambiguity in the \xiw vs.\ \alphaN plane.
When a branch becomes non-monotonic, the corresponding internal minimum might lie very close to a projected endpoint, so that the non-injective interval can be extremely small, and the physical classification diagram is still effectively single-valued.
Non-injectivity and its practical relevance are hence two distinct questions.

\begin{figure}
    \centering
    \includegraphics[width=\linewidth]{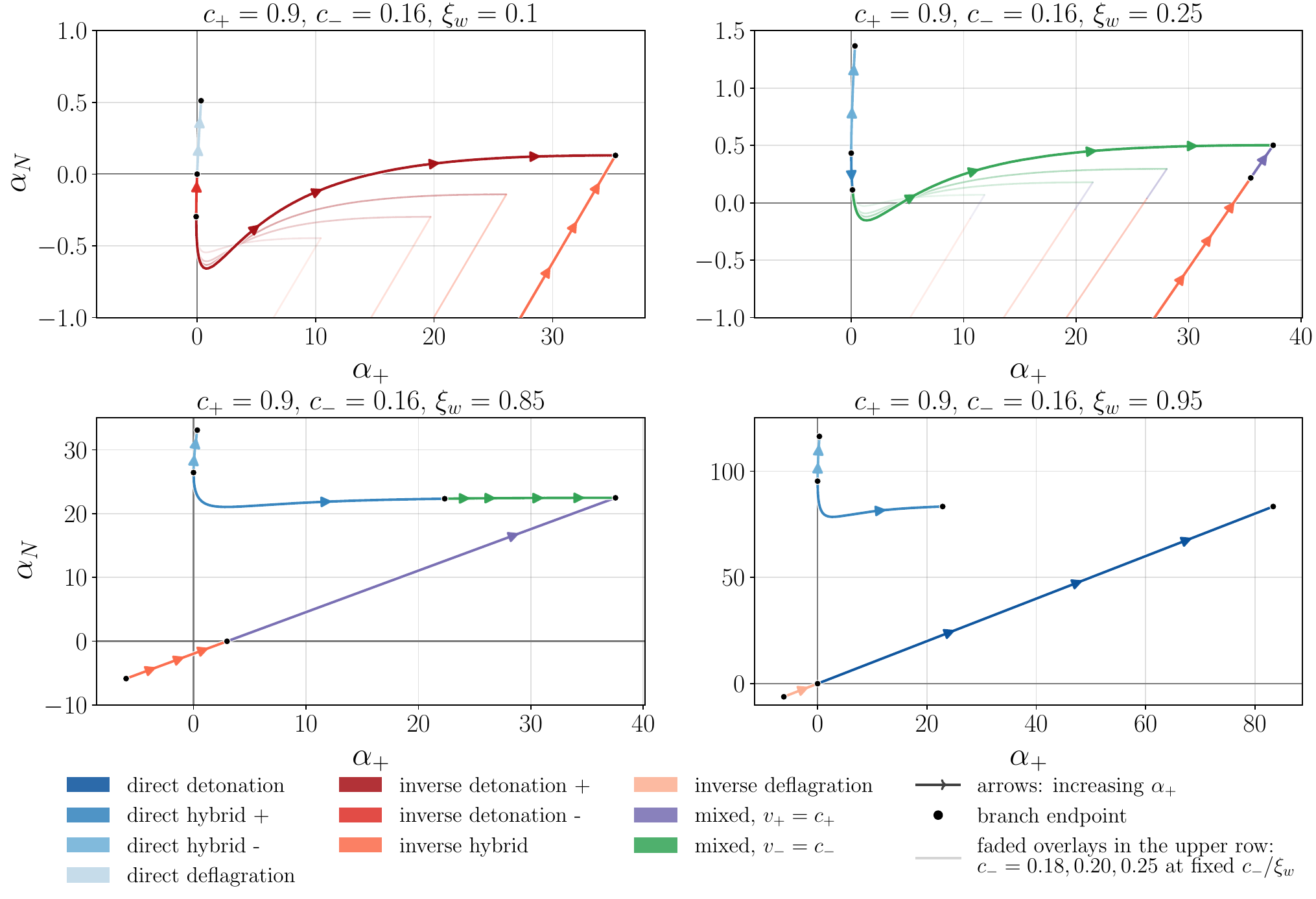}
    \caption{%
    Projected maps \(\alphaP\mapsto\alphaN\) for \(c_+=0.9\) and \(c_-=0.16\) with internal minima on the inverse detonation (upper left) and mixed hybrid with \(\vm=c_-\) branch (upper right), as well as direct hybrids (lower panels).
    Arrows indicate increasing \(\alphaP\), black dots mark branch endpoints, and the faded curves in the upper panels show larger values of \(c_-\) at fixed \(c_-/\xiw\).}
    \label{fig:alphaN-alpha-plus-folding}
\end{figure}

\Cref{fig:alphaN-alpha-plus-folding} depicts \alphaN as a function of \alphaP at different values of \xiw for \(c_+=0.9\) and \(c_-=0.16\), well within the region where all three of these branches develop a fold (\(\ccritDH(0.9)=0.52\)).
The upper-left panel displays the internal minimum of the inverse detonation projection, while the upper-right panel shows its continuation onto the mixed branch with \(\vm=c_-\).
The lower panels illustrate the fold of the direct hybrid sheet.
In each case, the physical lower edge of the projected branch is set by the internal minimum rather than by the image of either local endpoint.

\subsection{Additional degeneracies}

In addition, further degeneracies arise when the ordering of the \alphaN projections of the end points of the \alphaP range of a branch changes.
Once a branch folds, its lower boundary is set by the internal minimum rather than by either endpoint, while its upper boundary is given by the larger of the two endpoint images.
Close to the fold onset the ordering of these endpoints is typically the same as in the monotonic regime.
As the hierarchy \(c_+/c_-\) increases, however, the \alphaN values corresponding to the maximal and minimal value of \alphaP can exchange their ordering. 

The order of end points of the \(\vp^{(+)}\) direct-hybrid branch could change if \(\rwdir[c_-] < 1 + \alphav{\xiw}{c_-}/|\alpha_0|\).
We however find numerically (cf.\ \Cref{app:projection-monotonicity}) that \(\rwdir[c_-]\) always remains above this value, so that the ordering is preserved.
For the remaining non-injective branches, we can define the values of \(c_-\) below which the order of the end-points is reversed by
\begin{equation}\begin{aligned}
    1 &= \min_{\xi_c<\xiw<c_-} \left\{
        \frac{\rwinv[v_c(\xiw)]}{\rwinv[c_+]}\,
        \frac{\alphav{v_c(\xiw)}{\xiw}-\alpha_0}{\alphav{c_+}{\xiw}-\alpha_0}
    \right\}_{c_-=\cexID(c_+)} ,\\
    1 &= \min_{c_-<\xiw<c_+} \left\{
        \frac{1}{\rwinv[c_+]}\,
        \frac{\alphav{\xiw}{c_-}-\alpha_0}{\alphav{c_+}{c_-}-\alpha_0}
    \right\}_{c_-=\cexMH(c_+)} .
    \label{eq:cexs}
\end{aligned}\end{equation}
In other words, if \(c_-<\cexID\), for at least one \(\xiw\) in the inverse-detonation domain with \(\vp=\vp^{(+)}(\vm,\alphaP)\), the \alphaN boundary shared with the \(\vp=\vp^{(-)}(\vm,\alphaP)\) sub-branch at the extremum of \alphaP and the sonic point \(\vp=c_+\) bordering the inverse hybrids intersect.
Similarly, the \alphaN boundaries of mixed hybrids with \(\vm=c_-\) to the \(\vp=c_+\) mixed hybrids and direct hybrids intersect for \(c_-<\cexMH\).
As this crossing requires non-monotonicity of \alphaN in terms of \alphaP, we necessarily have \(\cexID\leq\ccritID\) and \(\cexMH\leq\ccritMH\).

\begin{figure}
    \centering
    \includegraphics[width=\linewidth]{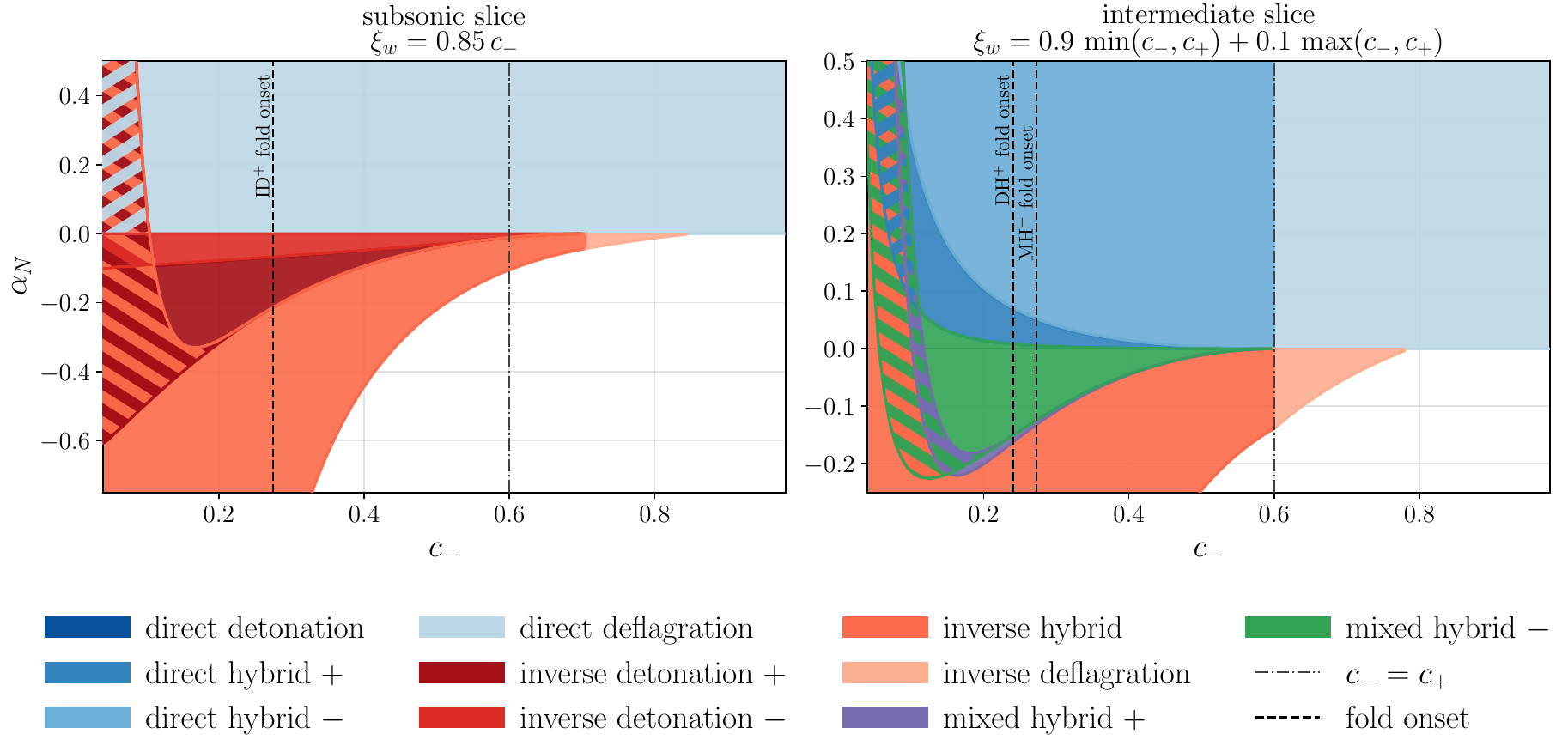}
    \caption{%
    Projected branch intervals in the \((c_-,\alphaN)\) plane at fixed \(c_+=0.6\), for two slices at fixed wall velocity.
    The vertical dot-dashed line marks \(c_-=c_+\).
    The black dashed lines indicate the fold onset for the inverse detonations on the `(\(+\))' branch (left) and mixed hybrids with \(\vm=c_-\) and direct hybrids on the `(\(+\))' branch (right).
    Moving to the left of these lines corresponds to increasing the hierarchy \(c_+/c_-\), where the non-injectivities worsen.  \textbf{Left:}
    Subsonic slice \(\xiw=0.85\,c_-\). 
    The \(\alphaP-\alphaN\) relation for inverse detonations on the `(\(+\))' branch develops an internal minimum and its projected endpoint shared with the inverse-hybrid side moves upward as \(c_-\) decreases.
    For a sufficiently strong hierarchy, different solution types overlap (hatched regions) and the inverse-detonation parameter space can extend to positive \(\alphaN\).
    \textbf{Right:}
    Intermediate slice \(\xiw=0.9 \min(c_-,c_+) + 0.1 \max(c_-,c_+)\).
    For \(c_+>c_-\), this slice corresponds to supersonic wall velocities with respect to \(c_-\).
    The mixed branch with \(\vm=c_-\) develops a fold, leading to visible overlaps between mixed, inverse, and direct branches.
    }
    \label{fig:phase-plane-all-branches-slices}
\end{figure}

This effect leads to a three-way overlap between different solution types, as the crossing boundaries also correspond to the lower or upper bounds of the neighbouring branches.
For \(c_-<\cexID\), we hence obtain a region in the \(\xiw,\alphaN\) plane that can correspond to inverse detonations on either branch of \(\vp^{(\pm)}(\vm,\alphaP)\) as well as inverse hybrids, while \(c_-<\cexMH\) leads to degeneracies between mixed hybrids with either \(\vpm=c_\pm\) and the `(\(+\))' sub-branch of direct hybrids.
This is illustrated in \cref{fig:phase-plane-all-branches-slices}, where the \alphaN range of the different solution types and their overlap are shown as a function of \(c_-\) at fixed \(c_+\) for two slices in the wall velocity \xiw.

\Cref{fig:phase-plane-all-branches-slices} further reveals an additional feature that leads to degeneracies.
At still larger hierarchies \(c_-\ll c_+\), the \alphaN range of inverse hybrids and the `\((+)\)' branch of inverse detonations can extend into the \(\alphaN>0\) space usually only occupied by direct or hybrid solutions.
In contrast to the sources of overlaps discussed above, this effect is independent of the monotonicity of \alphaN on the branch.

Like the other sources of degeneracies, these zero crossings arise only for \(c_- \ll c_+\).
As \(\alpha_0 < 0\) in this regime, and \(\rwinv < 1\), \cref{eq:alphaN-rw-map} implies that \(\alphaN>0\) also requires \(\alphaP>0\).
As further discussed in \cref{app:zero-crossing}, at fixed \(c_+\), the crossing first occurs for inverse hybrids at the critical value \(c_-=\czeroIH\) implicitly defined by the condition \(\alphaN=0\), or, equivalently,
\begin{equation}
    \max\limits_{c_-<\xiw<c_+}
    \left\{\rwinv[c_+]\left(1-\frac{\alphav{c_+}{\xiw}}{\alpha_0}\right)\right\}_{c_-=\czeroIH} = 1\,,
    \label{eq:czero-IH}
\end{equation}
with the zero attained at a wall velocity in the range \(c_-<\xiw<c_+\).
Subsequently, the upper \alphaN bound of the `\((+)\)' branch of inverse detonations crosses zero directly at the upper end of its \xiw range, \(\xiw=c_-\), at a critical value \(c_- = \czeroID < \czeroIH\).
The equivalent of \cref{eq:czero-IH} for inverse detonations can then be written as
\begin{equation}
    \rwinv[c_+]\Big|_{\xiw=c_-=\czeroID}
    = \frac{(1-c_+^2)(c_++c_-)}{2c_+(1-c_+c_-)}\bigg|_{c_-=\czeroID}\,.
    \label{eq:czero-ID}
\end{equation}

The fold onset, endpoint exchange, and crossing through \(\alphaN=0\) are separate thresholds, probing progressively larger departures from the injective regime.
The fold appears first, the endpoint ordering changes only later, and the crossing through \(\alphaN=0\), first for inverse hybrids and then for inverse detonations, requires an even more pronounced hierarchy.
Once sufficiently pronounced, they generate broad overlaps between mixed, inverse, and ordinary direct stationary branches, and the sign of \alphaN ceases to provide a reliable distinction between direct and inverse hydrodynamic solutions.
This progression is illustrated in \cref{fig:phase-plane-all-branches-slices}, where \(c_+\) is kept fixed while \(c_-\) is decreased.
The left panel shows a subsonic slice through the inverse-detonation region.
At mild hierarchy the projection is monotonic, while deeper in the \(c_+>c_-\) region the inverse-detonation `(\(+\))' sheet develops a visible internal minimum. 
The minimum then moves below the \alphaN projections of the \alphaP endpoints, and the endpoint shared with the inverse-hybrid branch moves upward until the ordering is reversed.
For still stronger hierarchies, part of the inverse-detonation region eventually extends to positive \alphaN.
This does not change the hydrodynamic character of the solution, but shows that the sign of \alphaN alone is no longer a reliable direct--inverse classifier in this regime.
The right panel probes the interval between the two sound speeds, with \(\xiw = 0.9\min(c_+,c_-) + 0.1\max(c_+,c_-)\).
For \(c_+>c_-\), this choice guarantees \(c_-<\xiw<c_+\), i.e.\ the slice can support mixed-hybrid solutions.
The same progressive deformation is visible as the hierarchy increases, with increasingly broad overlaps with neighbouring direct and inverse branches appearing.

\subsection{Physically motivated sound speeds}
\label{subsec:overlap-sound-speeds}

\begin{figure}
  \centering
  \includegraphics[width=\linewidth]{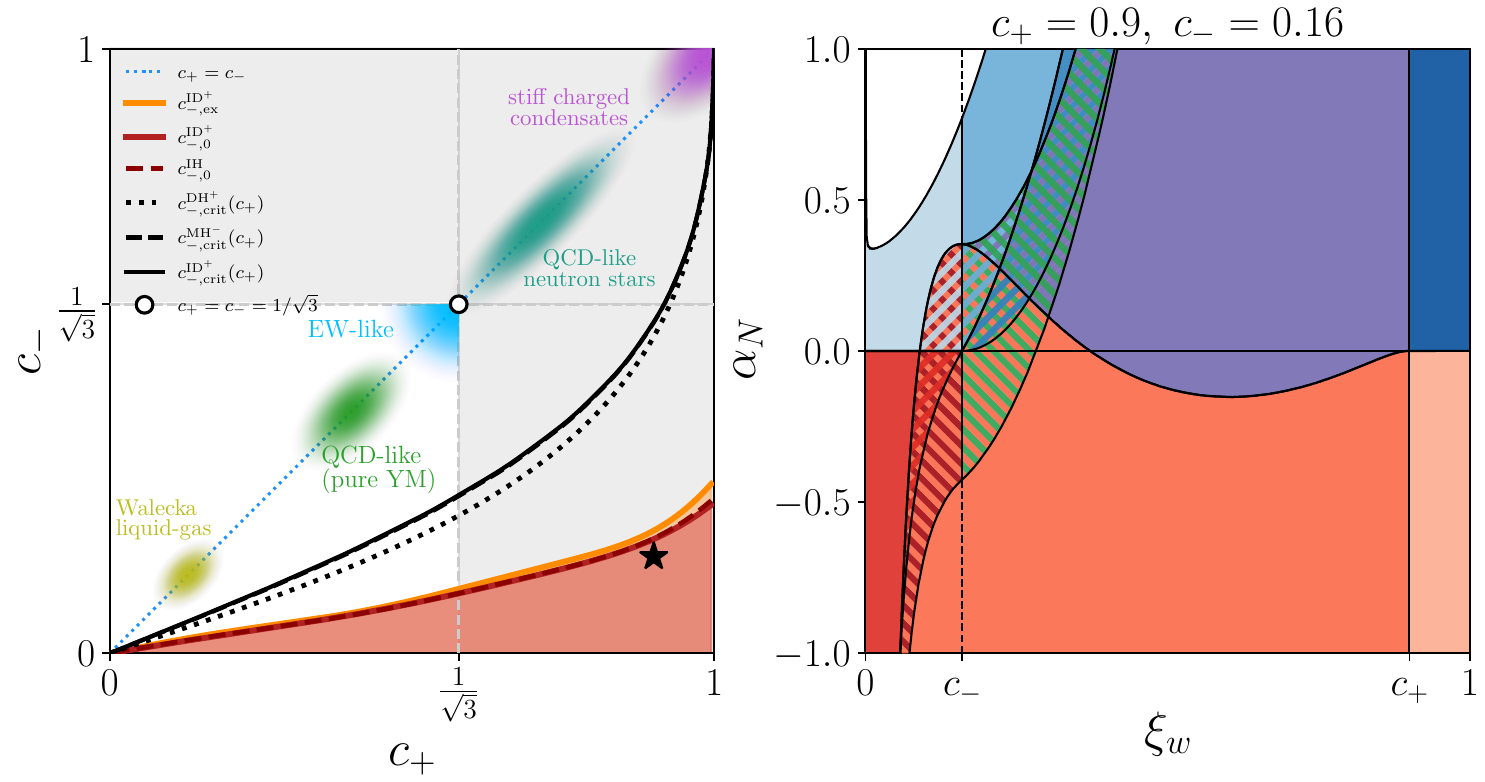}
  \caption{%
    \textbf{Left}:
    sound-speed plane showing physically motivated parameter regions and the thresholds controlling the loss of injectivity.
    The curves \(\ccritDH(c_+)\), \(\ccritMH(c_+)\), and \(\ccritID(c_+)\) mark the onset of folds in the respective branches, cf.\ \cref{eq:ccrits}.
    The threshold \(\cexID(c_+)\) marks the subsequent exchange in the ordering of the projected branch endpoints, cf.\ \cref{eq:cexs}, while \(\czeroIH(c_+)\) marks the  crossing of the upper inverse-hybrid boundary through \(\alphaN=0\), cf.\ \cref{eq:czero-IH}.
    The corresponding threshold \(\czeroID(c_+)\) for inverse detonations, \cref{eq:czero-ID}, lies slightly below.
    The dotted blue line denotes \(c_+=c_-\), and the dashed grey lines \(c_\pm=1/\sqrt{3}\); the grey background indicates the region where at least one sound speed exceeds the conformal value.
    Coloured blobs schematically indicate physically motivated equations of state, while the black star marks the hierarchical benchmark used in the right panel.
    \textbf{Right}: 
    hydrodynamic classification in the \((\xiw,\alphaN)\) plane for \(c_+=0.9\) and \(c_-=0.16\), with the same colour coding as \cref{fig:solutions_vp_vm,fig:xi_alpha+,fig:xi_alphaN}.
    Hatched regions indicate overlaps in which distinct stationary profiles correspond to the same \((\xiw,\alphaN)\).
}
  \label{fig:hydro-fold-overlap}
\end{figure}

It is useful to compare the regions identified above with the sound speeds expected in physically motivated equations of state.
The left panel of \cref{fig:hydro-fold-overlap} illustrates the approximate ranges of sound-speeds in these scenarios (shaded blobs) together with the corresponding sound-speed thresholds of \cref{eq:ccrits,eq:cexs,eq:czero,eq:czero-ID} in the \(c_+\) vs.\ \(c_-\) plane.
The coloured regions associated with microscopic models should be understood only schematically as theoretically motivated parameter space.

For an approximately conformal relativistic plasma, \(c_s^2\simeq1/3\), and electroweak-like transitions (cyan) with many relativistic degrees of freedom typically remain close to this value~\cite{Steinhardt:1981ct,Kamionkowski:1993fg,Espinosa:2010hh,Hindmarsh:2017gnf}.
They therefore populate a region close to the diagonal \(c_+\simeq c_-\simeq 1/\sqrt{3}\), far from the pronounced hierarchy \(c_+\gg c_-\) for which the folds identified above become macroscopically important.

Strongly coupled thermal plasmas can instead be substantially softer, with direct consequences for bubble dynamics~\cite{Bea:2021zsu,Li:2023xto}.
A useful example is pure Yang--Mills theory (green), whose confinement--deconfinement transition is first order for \(SU(3)\).
Recent lattice calculations based on density-of-states methods have resolved the thermodynamics in the critical region, including the metastable and unstable branches associated with the transition~\cite{Lucini:2023irm,Mateos:2026moc}.
Around the critical temperature, the sound speed is substantially suppressed relative to the conformal value, while the deconfined phase approaches \(c_s^2=1/3\) only gradually at higher temperature~\cite{Khaidukov:2018lor,Giusti:2025fxu,Mateos:2026moc}.
Pure Yang--Mills theory therefore provides a microscopic realization in which both phases close to the transition can be considerably softer than an electroweak-like relativistic plasma.
The green region in \cref{fig:hydro-fold-overlap} is intended only as a schematic representation of this QCD-like behaviour, rather than as a constant-\(c_s\) approximation to the full Yang--Mills equation of state.
Similarly, nuclear matter close to the liquid--gas transition (olive) can be considerably softer.
Mean-field Walecka-type equations of state display a strong variation of the sound speed with temperature and density in the vicinity of the coexistence and spinodal regions~\cite{Walecka:1974qa,Serot:1984ey,Kapusta:2006pm,He:2022yrk}.
These examples demonstrate that sound speeds well below the conformal value arise naturally, although they do not by themselves generate a large hierarchy between the two phases.

At finite baryon density, the equation of state can instead become considerably stiffer than the conformal expectation.
In particular, sound speeds above \(1/\sqrt{3}\) are compatible with neutron-star phenomenology and can occur over a sizeable range of supranuclear densities~\cite{Bedaque:2014sqa,Altiparmak:2022bke,Ecker:2022xxj}, see the teal region in \cref{fig:hydro-fold-overlap}.
First-order transitions proceeding through bubble nucleation in such dense environments have recently been studied in holographic models~\cite{Bea:2024bxu,Bea:2024bls} and in (proto-)neutron stars~\cite{Bleau:2026ala,Ecker:2026ixk}.
This statement can also be checked directly in microscopic and phenomenological neutron-star equations of state tabulated in the CompOSE database~\cite{CompOSE:web,CompOSECoreTeam:2022ddl}.
For example, CompOSE contains widely used relativistic mean-field equations of state such as HS(DD2)~\cite{Typel:2009sy} and SFHo~\cite{Steiner:2012rk}, which provide representative realizations of the range of stiffnesses encountered in dense hadronic matter.
In particular, the cold beta-equilibrated HS(DD2) equation of state develops a sound speed above the conformal value over part of the density range relevant for neutron-star interiors~\cite{Ecker:2022xxj}.
A systematic study of these equations of state and of their implications for relativistic hydrodynamics at finite density will be presented in Ref.~\cite{Barni:2026FiniteDensity}.

The neutron-star example should therefore be understood as evidence that \(c_s>1/\sqrt{3}\) is physically realizable in dense matter, rather than as suggesting that neutron-star equations of state naturally generate the hierarchy \(c_+\gg c_-\).  
Indeed, representative dense matter equations of state typically vary continuously through the stiff part of the \((c_+,c_-)\) plane and remain much closer to the diagonal than the strongly hierarchical benchmark considered below.

A more extreme realization is provided by stiff charged condensates (purple). In scalar theories with a conserved charge and a chemical potential, the charge-carrying condensate can drive the equation of state towards the stiff regime \(p\simeq e\), while the adiabatic sound speed \(c_s^2=dp/de\) can correspondingly approach unity.
This behaviour appears in the \(O(N)\) scalar model studied in Ref.~\cite{Moore:2024gmt} and is closely related to the finite-isospin QCD example of Son and Stephanov~\cite{Son:2000xc,Son:2000by}.
Such systems can naturally populate the large-\(c_s\) part of the plane, but their hydrodynamics may contain additional slow Goldstone modes and transport effects that are not captured by the single-fluid CSS parametrization~\cite{Moore:2024gmt}.

The black curves in \cref{fig:hydro-fold-overlap} mark the onset of non-injectivity for the `(\(+\))' branches of inverse detonations and direct hybrids, as well as for mixed hybrids with \(\vm=c_-\), indicated by solid, dotted, and dashed lines, respectively.
The orange region identifies the stronger endpoint-exchange condition for inverse detonations derived in \cref{eq:cexs}, while the red region corresponds to the still more hierarchical regime in which inverse hybrids, and for slightly stronger hierarchies also inverse detonations, extend to \(\alphaN>0\), cf.\ \cref{eq:czero,eq:czero-ID}.
The dotted blue line denotes \(c_+=c_-\), and the dashed grey lines indicate the conformal value \(c_s=1/\sqrt{3}\).
The grey background highlights the part of the plane in which at least one of the two phases has \(c_s>1/\sqrt{3}\).%
\footnote{%
    The conformal value \(c_s^2=1/3\) is not a fundamental upper bound on the speed of sound, but rather the value approached by an approximately conformal relativistic plasma.
    Interacting and finite-density equations of state can consistently satisfy \(c_s^2>1/3\).%
}

Hence, both rather soft and superconformal equations of state can arise in well-motivated microscopic systems, but the hydrodynamic degeneracies become sizeable only when the two phases have very different stiffnesses, corresponding to a pronounced hierarchy \(c_+\gg c_-\).
Realizing such a contrast across the same phase transition is considerably more restrictive than obtaining either a small or a large sound speed separately.
Accordingly, for conventional cosmological phase transitions the map to \((\xiw,\alphaN)\) is expected to remain approximately injective, whereas extended overlap regions arise only sufficiently deep in the hierarchical regime.

The black star in the left panel of \cref{fig:hydro-fold-overlap} marks the deliberately hierarchical benchmark \(c_+=0.9\), \(c_-=0.16\).
The corresponding physical \((\xiw,\alphaN)\) phase diagram is shown in the right panel.
At this point in the sound-speed plane, the folds are sufficiently developed that extended overlaps occur between distinct stationary hydrodynamic branches, providing a concrete example of the regime in which the loss of injectivity can no longer be neglected.
Moreover, sufficiently deep in the hierarchical regime, inverse branches can project across \(\alphaN=0\), so that the sign of \(\alphaN\) alone cannot be used to distinguish direct and inverse solutions, even in the parameter regions where no mixed solutions arise.
These degeneracies should be interpreted as a non-uniqueness of stationary hydrodynamic solutions, rather than as an inconsistency of the construction.
The matching conditions and self-similar fluid equations classify the admissible profiles, but do not by themselves select which one is dynamically realized.
In the overlapping region, the fluid profiles cannot be characterized by \((\xiw,\alphaN,c_+,c_-)\) alone without an additional selection principle.
Such a principle may depend on microscopic friction, stability, initial conditions, or the full time-dependent evolution of the wall~\cite{Ai:2024shx,Krajewski:2024gma,Laurent:2026jvx}.

\subsection{Thermodynamic lifting of the hydrodynamic degeneracy}
\label{sec:fold-thermodynamic-lifting}

The non-injectivity discussed above is a property of the reduced hydrodynamic problem.
At fixed \((\xiw,\alphaN)\), several distinct local wall states can project onto the same asymptotic parameters. 
A complete thermodynamic model, however, carries additional information through the enthalpy ratio \(\Psi_N\), and the entropy condition associates with each hydrodynamic solution a minimum value \(\Psi_{N,\min}\), given in \cref{eq:PsiN-entropy-bound}.

This provides a simple way of testing whether the hydrodynamic degeneracy survives once the thermodynamic completion is specified.
In the regions where the projection is non-injective, we reconstruct all physically distinct hydrodynamic solutions of the same \((\xiw,\alphaN)\).
At most three such solutions can coexist at any point.
Each of them defines a different entropy threshold, \(\Psi_{N,\min}^{(i)}\), \(i\in\{1,2,3\}\), which can be ordered locally as \(\Psi_{N,\min}^{(1)} < \Psi_{N,\min}^{(2)} < \Psi_{N,\min}^{(3)}\).
To quantify how well separated the competing thermodynamic completions are, we define
\begin{equation}
  \mathcal D_{\rm LTE}(\xiw,\alphaN)
  \equiv
  \min_{i<j}
  \left|
    \ln
    \frac{\Psi_{N,\min}^{(i)}}
         {\Psi_{N,\min}^{(j)}}
  \right| .
  \label{eq:D-LTE}
\end{equation}
For two solutions this simply measures their relative separation, while for three solutions it selects the closest pair in the hierarchy.
A vanishing value, \(\mathcal D_{\rm LTE}=0\), would therefore imply that two genuinely distinct hydrodynamic states require the same thermodynamic completion.

\begin{figure}
  \centering
  \includegraphics[width=0.62\textwidth]{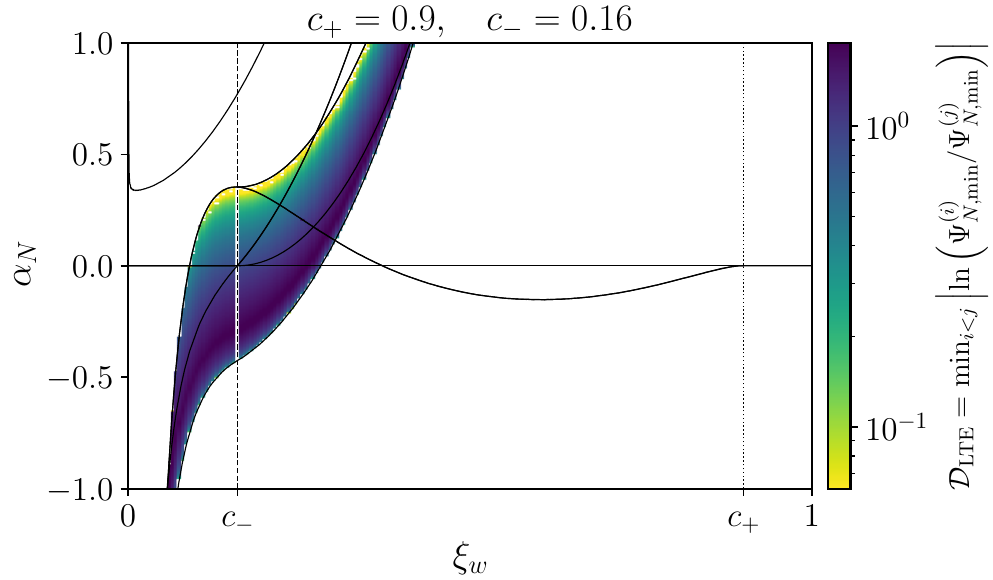}
  \caption{%
      Thermodynamic splitting of the hydrodynamically degenerate solutions for \(c_+=0.9\) and \(c_-=0.16\).
      The colour scale shows \(\mathcal D_{\rm LTE}\), defined in \cref{eq:D-LTE}, only in the regions where at least two physically distinct hydrodynamic solutions coexist at the same \((\xiw,\alphaN)\).
      A vanishing value would signal two distinct solutions requiring the same \(\Psi_N\) in LTE.
      No such degeneracy is found in the interior of the overlap regions within the numerical resolution of the scan, showing that the thermodynamic completion lifts the hydrodynamic non-injectivity.
  }
  \label{fig:lte-splitting}
\end{figure}

This is particularly relevant in local thermal equilibrium (LTE).
LTE corresponds to saturation of the entropy bound, \(\Psi_N=\Psi_{N,\min}\), so that two degenerate hydrodynamic solutions could both satisfy LTE for the same microscopic model only if their values of \(\Psi_{N,\min}\) coincided.
As shown in \cref{fig:lte-splitting}, this does not occur in the interior of the non-injective regions for the benchmark \(c_+=0.9\), \(c_-=0.16\).
The different hydrodynamic solutions exhibit a definite hierarchy in \(\Psi_{N,\min}\), and \(\mathcal D_{\rm LTE}\) remains non-vanishing within our numerical resolution.
Hence, at fixed \((\xiw,\alphaN)\), a given thermodynamic model can satisfy the LTE condition for at most one of the competing hydrodynamic solutions.

As emphasized in \cref{sec:entropy-checks}, \(\Psi_N\) is not an extra parameter that can be freely adjusted once the microscopic theory has been specified.
It fixes the relative normalization of the two equations of state and therefore encodes, in particular, the change in the effective thermal degrees of freedom between the two phases.
Choosing different values of \(\Psi_N\) generally corresponds to considering different thermodynamic models.
The fact that two hydrodynamically degenerate solutions require different values of \(\Psi_{N,\min}\) therefore means that they cannot represent LTE solutions of the same underlying theory.

More generally, the same reasoning applies to any closure that fixes the entropy production across the wall.
Using \cref{eq:entropy-reconstruction-Psi,eq:PsiN-entropy-bound}, the entropy-flux ratio can be written as
\begin{equation}
  \frac{s_-\gamma_-\vm}{s_+\gamma_+\vp}
  =
  \left(
    \frac{\Psi_N}{\Psi_{N,\min}}
  \right)^{1/q_-}.
  \label{eq:entropy-flux-ratio-Psi}
\end{equation}
Thus fixing the entropy production, or equivalently fixing this ratio, again maps the different values of \(\Psi_{N,\min}\) into distinct thermodynamic completions. 
By contrast, the weaker requirement of non-negative entropy production alone, \(\Psi_N\geq\Psi_{N,\min}\), does not by itself imply uniqueness.

%
\section{Efficiency and energy budget in the CSS model}
\label{sec:efficiency}
%

The classification of the stationary fluid profiles determines whether a transition is hydrodynamically direct, inverse or mixed. 
The amount of energy converted into gravitational waves, however, depends on the amount of energy transferred to bulk fluid motion.
In this context, direct and inverse transitions are more naturally defined in terms of the zero-temperature vacuum energy difference \(\epsilon \equiv \epsilon_+-\epsilon_-\), cf.\ \cref{eq:alpha_vac}, or the corresponding strength parameter \alphavac[N] defined in \cref{fn:alpha-vacuum-N}, i.e.\ \(\alphavac[N]\gtrless 0\) for energetically direct and energetically inverse transitions, respectively.
The sign of \(\epsilon\) determines the role played by the vacuum contribution in the energy budget.
For \(\epsilon>0\), vacuum energy is released during the transition, while for \(\epsilon<0\), vacuum energy must instead be produced at the expense of the thermal bath. 
In the CSS model the sign of \(\epsilon\), and therefore of \(\alphavac[N]\), need not coincide
with the sign of the pseudo-trace strength \(\alphaN\) at nucleation or the sign of the fluid velocity in the cosmic frame.
Only in the conformal bag limit, \(c_+^2=c_-^2=1/3\), they are guaranteed to coincide as \(\alphaN=\alphavac[N]\).
In this section, we drop the qualifier `energetically' and simply refer to these two cases as `direct' and `inverse', respectively.

For the CSS equation of state, the vacuum-energy strength parameter \alphavac[N] can be written in terms of the thermal contribution \(e_{\rm th}\) to the energy density as
\begin{equation}
    e_{{\rm th},\pm}(T) = a_\pm T^{q_\pm} = \frac{w_\pm(T)}{1+c_\pm^2}
    \qquad\Longrightarrow\qquad
    \alphavac[N] = \frac{\epsilon}{e_{{\rm th}, N}} = (1+c_+^2) \frac{\epsilon}{w_N},
    \quad
    e_{{\rm th},N} = e_{{\rm th},+}(T_N)
    \,.
    \label{eq:eth-css-efficiency}
\end{equation}
In the conformal bag equation of state, \(c_\pm^2=1/3\), so that \(e_{\rm th}=3w/4\) and \(\alphavac[N]=4\epsilon/(3w_N)\), reproducing the usual bag-model normalization.

The energy budget of a phase transition is usually described in terms of an efficiency factor or, equivalently once the normalization is specified, a kinetic-energy fraction \cite{Kamionkowski:1993fg,Espinosa:2010hh,Hindmarsh:2017gnf}.
This quantity also provides the normalization of the acoustic source in the sound-shell description \cite{Hindmarsh:2017gnf,Barni:2025gnm}.
While the kinetic energy itself is unambiguous once the hydrodynamic profile is known,
the energy reservoir with respect to which it is normalized requires some care.
In a direct transition, the vacuum energy released by the phase conversion is an additional source of energy.
In an inverse transition the situation is reversed, and part of the initial thermal energy must instead be used to increase the vacuum energy of the converted phase \cite{Barni:2024lkj}.

The kinetic part of the fluid energy follows directly from \(T^{00}= w\gamma^2-p = e+w\gamma^2v^2\), so that the energy stored in bulk motion is weighted by \(w\gamma^2v^2\).
The energy balance is then given by
\begin{subequations}
\label{eq:energy-budget}
\begin{align}
  \textbf{direct}:& \quad
  \underbrace{\vphantom{\int_0^{\xi_f}}\frac{\xiw^3}{3}\,\epsilon}_{\substack{E_{\rm vac}/(4\pi t^3)\\
  \text{vacuum energy}\vphantom{l}}}
  +
  \underbrace{\int_0^{\xi_f}
  \frac{w_N}{1+c_+^2}\,\xi^2\,\dd\xi}_{\substack{E_{\rm ini}/(4\pi t^3)\\
  \text{initial thermal energy}}}
  =
  \underbrace{\int_0^{\xi_f}
  \gamma^2 v^2 w\,\xi^2\,\dd\xi}_{\substack{E_K/(4\pi t^3)\\
  \text{fluid motion}}}
  +
  \underbrace{\int_0^{\xi_f}
  \frac{w(\xi)}{1+c_s^2(\xi)}\,\xi^2\,\dd\xi}_{\substack{E_{\rm fin}/(4\pi t^3)\\
  \text{final thermal energy}}},
  \label{eq:energy-budget-standard}
  \\
  \textbf{inverse}:& \quad
  \underbrace{\int_0^{\xi_f}
  \frac{w_N}{1+c_+^2}\,\xi^2\,\dd\xi}_{\substack{E_{\rm ini}/(4\pi t^3)\\
  \text{initial thermal energy}}}
  =
  \underbrace{\vphantom{\int_0^{\xi_f}}\frac{\xiw^3}{3}\,|\epsilon|}_{\substack{E_{\rm vac}/(4\pi t^3)\\
  \text{vacuum-energy cost}\vphantom{l}}}
  +
  \underbrace{\int_0^{\xi_f}
  \gamma^2 v^2 w\,\xi^2\,\dd\xi}_{\substack{E_K/(4\pi t^3)\\
  \text{fluid motion}}}
  +
  \underbrace{\int_0^{\xi_f}
  \frac{w(\xi)}{1+c_s^2(\xi)}\,\xi^2\,\dd\xi}_{\substack{E_{\rm fin}/(4\pi t^3)\\
  \text{final thermal energy}}},
  \label{eq:energy-budget-inverse}
\end{align}
\end{subequations}
where \(c_s(\xi)=c_-\) inside the bubble and \(c_s(\xi)=c_+\) outside, and
\(\xi_f\equiv\max\{\xiw,c_+,\xi_{\rm sh}\}\)
is the outer edge of the perturbed region, i.e.\ \(\xi_f=\xiw\) when the plasma ahead of the wall is unperturbed, \(\xi_f=\xi_{\rm sh}\) for a compression profile in front of the wall ending in a shock, and \(\xi_f=c_+\) when the perturbation ahead of the wall terminates at the sound horizon.\footnote{%
    Note that \(w(\xi)=w_N\) is constant for \(\xi\geq \xi_f\), so that the corresponding contributions to \(E_{\text{ini}}\) and \(E_{\text{fin}}\) cancel against each other.%
}
In the direct case, both the initial thermal energy and the released vacuum energy are available and are redistributed between bulk motion and the final thermal state.
In the inverse case, by contrast, the only initial reservoir is the thermal bath.
Part of it is converted into bulk motion and part is necessarily spent increasing the vacuum energy.

From \cref{eq:energy-budget}, we can define the kinetic-energy fractions \(K\equiv \rho_\text{kin}/\rho_{\text{tot}}^N\) normalized to the total energy initially available, and the corresponding efficiency factors \(\kappa\).
For an energetically direct transition, this is the thermal reservoir plus the released vacuum energy, while for an energetically inverse transition, the available initial reservoir is instead given by the thermal energy alone.
Thus,
\begin{subequations}
\begin{align}
  K_{\rm dir} 
  &\equiv \frac{E_{K}}{E_{\rm ini} + E_{\rm vac}}
  \qquad\text{and}\qquad
  \kappa_\text{direct} = \frac{E_K}{E_{\rm vac}}
  \qquad
  &\text{for}\
  \alphavac[N]>0\,,
  \label{eq:K-direct}
  \\
  K_{\rm inv} &\equiv \kappa_\text{inverse}
  \equiv \frac{E_{K}}{E_{\rm ini}}
  \qquad
  &\text{for}\
  \alphavac[N]<0 \,,
\end{align}
\end{subequations}
where we follow Ref.~\cite{Barni:2024lkj} regarding the convention of the efficiency factor for inverse transitions.
The numerator is always the kinetic energy contained in the full disturbed region, while the denominator is the energy reservoir chosen according to the budget in \cref{eq:energy-budget}. 
In particular, inverse transitions are not normalized to the magnitude of a vacuum-energy difference which acts as an energy cost rather than as an available source.

The initial thermal energy can be written as \(E_\text{ini} = \tfrac{4\pi}{3} \xi_f^3 t^3 e_{{\rm th},N}\).
We hence define the average kinetic and final thermal energy densities as \(\bar{e}_i \equiv 3 E_i / (4 \pi \xi_f^3 t^3)\), \(i=K,\text{fin}\).
This convention averages the energy over the region actually affected by the hydrodynamic disturbance, rather than over the smaller bubble volume or over the entire causal region.\footnote{%
    Other normalization conventions are possible and change only the definition of
    the efficiency, not the physical kinetic energy \(E_K\).%
}
Further defining the ratio \(r_V \equiv \xiw^3/\xi_f^3\) of the bubble volume to the perturbed fluid volume, the kinetic energy fractions can be written in terms of the energy densities as
\begin{align}
  \label{eq:K-perturbed-volume}
  K_{\rm dir} 
  = \frac{\rho_{\rm kin}}{e_{{\rm th},N} + r_V \epsilon}
  =\frac{1-\bar e_{\rm fin}/e_{{\rm th},N}+r_V\alphavac^N}{1+r_V\alphavac[N]},
  \qquad
  K_{\rm inv}
  = \frac{\rho_{\rm kin}}{e_{{\rm th},N}}
  = 1 - \frac{\bar e_{\rm fin}}{e_{{\rm th},N}} - r_V\left|\alphavac[N]\right|.
\end{align}

The energy fractions for direct and inverse transitions exhibit different upper limits.  Neglecting the final thermal remnant,
\begin{equation}
  K_{\rm dir}\longrightarrow 1\,,
  \qquad
  K_{\rm inv}\longrightarrow
  1-r_V|\alphavac[N]|<1 \,,
  \label{eq:K-budget-limits}
\end{equation}
i.e.\ direct transitions can reach kinetic fractions up to unity, whereas the maximal inverse kinetic fraction lies below unity (note that \(\bar e_{\rm fin}\) cannot be neglected in the \(\alphavac[N]\to 0\) limit),
as for inverse transitions, the vacuum-energy increase is counted as one of the products of the transition, on the same footing as the kinetic and final thermal energies.
Even in the idealized limit of vanishing final thermal energy, part of the initial thermal reservoir must be converted into vacuum energy.\footnote{%
    One could instead account for the vacuum-energy cost in the initial energy and normalize the inverse kinetic energy to the net energy remaining after creating the new vacuum,
    \begin{equation*}
      \widetilde K_{\rm inv}
      \equiv
      \frac{E_K}{E_{\rm ini}-E_{\rm vac}}
      =
      1-
      \frac{E_{\rm fin}}{E_{\rm ini}-E_{\rm vac}} .
    \end{equation*}
    With this convention \(\widetilde K_{\rm inv}\to1\) whenever the final thermal remnant tends to zero.
    The difference between \(K_{\rm inv}\) and \(\widetilde K_{\rm inv}\) is therefore purely one of interpretation and normalization.
    The physical kinetic energy \(E_K\) is of course the same.
    Our choice follows Ref.~\cite{Barni:2024lkj} and makes explicit what fraction of the \emph{original thermal reservoir} is ultimately converted into bulk motion.%
}

Alternatively, the literature often uses the initial thermal energy stored within the bubble volume, \(E_\text{ini}^* = \tfrac{4\pi}{3} \xiw^3 t^3 e_{{\rm th},N}\), rather than the thermal energy \(E_\text{ini}\) within the perturbed fluid volume.
In the bag limit, the kinetic energy fraction in \cref{eq:K-direct} can then be written as \(K_\text{dir} = \kappa_\text{direct}\alphaN/(1+\alphaN)\) as in Ref.~\cite{Espinosa:2010hh}.
For very strong direct transitions, \(\alphaN\gg1\), the vacuum contribution dominates the available energy and \(K_\text{dir}\) can approach unity, while for inverse transitions \(K_\text{inv}\to 1-|\alphaN|<1\).

The maximal kinetic fraction for inverse transitions is determined by the limiting hydrodynamic profile, given by an inverse deflagration with \(\xiw \to 1\),
\begin{equation}
  \alphaN^{\rm min}
  = \lim_{\xiw\to 1}
  \alphaP(\xiw,1)
  =
  -\,\frac{1-c_-^2}{6c_-^2},
  \qquad
  \alphavac[N,\rm min]
  =
  -\,\frac{1-c_+^2}{2}.
  \label{eq:inverse-endpoint-strengths-css}
\end{equation}
At this endpoint the perturbed region coincides with the bubble volume, \(r_V=1\).
Defining \(c_{\rm min}=\min(c_+,c_-)\), the limiting profile retains the thermal and kinetic fractions
\begin{equation}
  \left.\frac{\bar e_{\rm fin}}{e_{{\rm th},N}}\right|_{\rm max}
  =
  \frac{c_+(c_+-c_{\min})}{2} 
  \qquad\text{and}\qquad
  K_{\rm inv}^{\rm max}
  =
  \frac{1+c_+c_{\rm min}}{2}\,.
  \label{eq:K-inverse-max-css}
\end{equation}
Thus the upper bound depends on both sound speeds through the limiting hydrodynamic profile.
In the conformal bag model, \(c_+=c_-=1/\sqrt{3}\), this reduces to \(K_{\rm inv}^{\rm max}=2/3\), in agreement with Ref.~\cite{Barni:2024lkj}. 
This should not be interpreted as a fundamental upper bound on the fraction of the \emph{net} energy that can go into motion.
It is the maximal fraction of the original thermal reservoir converted into kinetic energy under the normalization adopted here. 
The remaining \(1/3\) is precisely the energy required to create the higher-vacuum-energy phase.

\begin{figure}
  \centering
  \includegraphics[width=\textwidth]{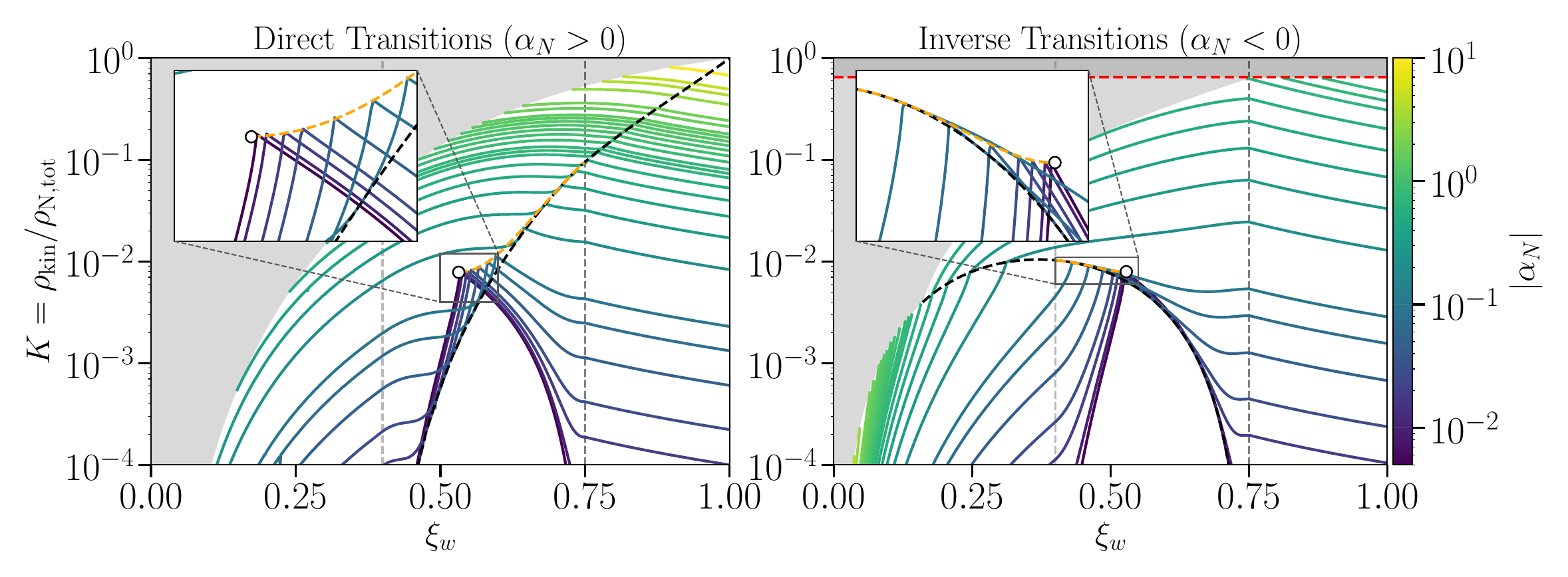}
    \caption{%
        Kinetic-energy fraction \(K=\rho_{\rm kin}/\rho_{\rm tot}\) for \((c_+, c_-)=(0.75, 0.4)\), marked by grey vertical dashed lines.
        The left and right panels show \(K\) for direct and inverse solutions, with colour indicating \(|\alphaN|\).
        The black dashed curves show \(K\) along the ordinary CJ boundaries \(\alpha_{\rm CJ}^{\rm dir}(\xiw)\) and \(\alpha_{\rm CJ}^{\rm inv}(\xiw)\) defined in \cref{eq:jouguet-strenghts-and-velocities}.
        The orange dashed curves extend the same sonic conditions, \(\vm=c_-\) and \(\vp=c_+\), onto the corresponding mixed-hybrid branches, as discussed in \cref{sec:ChapmanJouguet}; open circles mark the endpoints of the displayed mixed-hybrid segments.
        In the inverse panel, the horizontal red dashed line marks \(K_{\rm inv}^{\rm max}\) of \cref{eq:K-inverse-max-css}, equal to \((1+c_+c_-)/2\) for this benchmark.
        Grey-shaded regions are kinematically excluded.%
    }
    \label{fig:efficiency-css}
\end{figure}

The kinetic-energy fractions \(K\) are shown in \cref{fig:efficiency-css} as a function of the wall velocity \xiw for different values of \alphaN, using the perturbed-volume normalization introduced above.
The left panel corresponds to energetically direct transitions, whereas the right panel depicts energetically inverse transitions.
For large values of \(|\alphaN|\) and \(\xiw \to 1\), direct transitions reach \(K=1\), whereas inverse transitions satisfy \(K<K_{\text{inv}}^{\text{max}} = 0.65\) for \(c_+=0.75\) and \(c_-=0.4\). 

It should be noted that the quantity \(K\) defined here is derived from the profile of a single bubble.
Its role in the gravitational-wave calculation is to characterize the amount of energy initially stored in fluid motion.
Ultimately, however, the single-bubble efficiency discussed above should be regarded as a bookkeeping prescription rather than as a fundamental observable.
The physically relevant kinetic-energy fraction is the one generated by the full many-bubble dynamics, including the conversion of energy during bubble expansion, collisions, and the subsequent formation of the acoustic fluid.
This distinction is particularly relevant for the sound-shell model used in the next section.
There the quantity entering the gravitational-wave signal is the kinetic-energy fraction of the acoustic fluid after the ensemble of bubbles has nucleated and collided, obtained from the resulting velocity power spectrum rather than from a choice of single-bubble normalization.
The isolated-bubble hydrodynamic profile remains the microscopic input from which this spectrum is constructed, but the normalization relevant for gravitational-wave production is ultimately a property of the resulting many-bubble fluid configuration~\cite{Hindmarsh:2017gnf,Barni:2025gnm}.

%
\section{Gravitational wave production}
\label{sec:ssm}
%

We now turn to the gravitational-wave signal generated by the single-bubble hydrodynamic solutions of the CSS model, focusing on the sound-wave contribution from weak to moderately strong phase transitions.
As confirmed by numerical simulations, in this regime, the bulk fluid motion generated by the expanding bubbles remains acoustic for a time much longer than the vacuum bubble collision time and thus typically provides the dominant contribution to the gravitational-wave signal~\cite{Hindmarsh:2013xza,Hindmarsh:2015qta,Hindmarsh:2017gnf,Caprini:2024gyk}.
The sound-shell model (SSM)~\cite{Hindmarsh:2016lnk,Hindmarsh:2019phv} provides a semi-analytic description of the corresponding spectra.
The self-similar velocity and energy-density profiles around individual bubbles determine the initial conditions for sound waves after bubble overlap, while the stochastic superposition of freely-propagating shells\footnote{%
    Cf.\ Ref.~\cite{Cai:2023guc} for the forced propagating contribution from the initial vacuum bubbles.%
} nucleated at different positions and times fixes the unequal-time correlator of the fluid stress~\cite{Hindmarsh:2016lnk,Hindmarsh:2019phv}.
The framework has subsequently been extended to include the cosmological expansion and a finite acoustic lifetime~\cite{Guo:2020grp,RoperPol:2023dzg}, non-conformal sound speeds and equations of state~\cite{Wang:2021dwl,Giombi:2024kju,Maki:2025ykv}, next-to-leading order corrections in the bubble size over the Hubble radius~\cite{Giombi:2025tkv}, as well as signals from inverse phase transitions within the bag model~\cite{Barni:2025gnm}, and to sound waves leaking from a dark sector into a weakly coupled visible plasma~\cite{Finetti:2026ool}.
Recent work has further refined the mapping from the expanding-bubble profiles to the initial velocity spectrum~\cite{Caprini:2026nnk}, constructed accurate fitting formulae and emulators for SSM spectra~\cite{Guo:2024gmu,Hajkarim:2025efr,Tian:2025zlo}, and implemented different parts of this calculation in public tools such as \texttt{PTtools}~\cite{pttools-git}, \texttt{CosmoGW}~\cite{cosmogw-git}, and \texttt{HydroGrav}~\cite{Linton:2026fpj,hydrograv-git}.
The SSM has also been used directly in forecasts and reconstruction studies for space-based interferometers~\cite{Gowling:2021gcy,Boileau:2022ter,Gowling:2022pzb}.

We use the CSS hydrodynamic profiles constructed in the previous sections as the single-bubble input of the SSM.
This extends our previous implementation for direct and inverse transitions~\cite{Barni:2025gnm} to the complete CSS classification, including the two mixed-hybrid branches, and is implemented in the public code \hydrodex~\cite{github-page}.
In our review of the SSM in \cref{sec:ssm-review}, we further keep separate the equation-of-state parameter \(\omega\) controlling the expansion of the total cosmological background and the effective sound speed \(c_s\) controlling the propagation of the acoustic perturbations.
This distinction makes the formalism applicable when the sound-carrying plasma is only a sub-component of the total energy density.
If the transition instead occurs in a dark sector weakly coupled to the visible plasma, energy and momentum exchange during the sound-wave era can also induce acoustic waves in the visible sector, generally suppressing the resulting gravitational-wave signal~\cite{Finetti:2026ool}, however, we do not consider this effect here.
For the numerical analysis in \cref{sec:ssm-results}, we follow the single-fluid setup of Ref.~\cite{Giombi:2024kju} and set \(\omega=c_s^2\).

\subsection{The sound-shell model at constant speed of sound}
\label{sec:ssm-review}

In this section we review the SSM and generalize its cosmological evolution to allow the background equation-of-state parameter \(\omega\) to differ from that of the fluid undergoing the phase transition.
This can occur, for instance, if an additional component contributes significantly to the total
stress-energy tensor and controls the expansion rate without participating in the transition.
The acoustic perturbations generated by the transition, however, propagate in the post-transition `\(-\)' phase and are therefore characterized by its sound speed, \(c_s=c_-\).

We use the metric convention \(d s^2  = a^2(\tau) \left[ - d\tau^2 + \left(\delta_{ij} + h_{ij}(\tau,\vec{x})\right) dx_i dx_j\right]\), where \(\tau\) denotes conformal time, \(a\) is the scale factor and \(h_{ij}(\tau,\vec{x})\) are the gravitational-wave metric perturbations in the transverse-traceless gauge.
Since, in the context of inverse phase transitions, the energy content after the transition has completed will in general also contain a vacuum-energy component, we model the total homogeneous background and the separately conserved acoustic fluid by independent equations of state,
\begin{align}
    p_{\rm tot} &= \omega\rho_{\rm tot},
    &
    \bar p &= c_s^2\bar\rho,
    &
    \bar w &\equiv \bar\rho+\bar p = (1+c_s^2)\bar\rho,
    \label{eq:ssm-two-eos}
\end{align}
where the average energy and pressure densities \(\bar\rho\) and \(\bar{p}\) are potentially only a subcomponent of the total densities \(\rho_\text{tot}\) and \(p_\text{tot}\) of the Universe.
For simplicity, both \(\omega\) and \(c_s\) are taken to be constant during the acoustic stage, which is a good approximation as long as the acoustic source is short-lived compared to the Hubble time.
The corresponding background scalings are
\begin{align}
    \rho_{\rm tot} &\propto a^{-3(1+\omega)},
    &
    \bar\rho &\propto a^{-3(1+c_s^2)},
    &
    a &\propto \tau^{\frac{2}{1+3\omega}},
    &
    \mathcal H &= \frac{2}{1+3\omega}\frac{1}{\tau},
    \label{eq:ssm-background-scaling}
\end{align}
where \(\mathcal{H} \equiv \frac{a'}{a} = \sqrt{\frac{8 \pi G}{3}\,a^2 \rho_\text{tot}}\) is the comoving Hubble rate with primes denoting derivatives with respect to conformal time.
The usual radiation-dominated case is recovered for \(c_s^2=\omega=1/3\).
Keeping these two parameters independent is useful, for example, in the presence of spectator species or a homogeneous vacuum component that affects the expansion rate without participating in the acoustic perturbations.
For later convenience, let us further define
\begin{align}
    \sigma \equiv \frac{1-3\omega}{1+3\omega},
    \quad
    \bar{\sigma} \equiv \frac{1-3 c_s^2}{1+3\omega}
    \qquad\Longrightarrow\qquad
    \rho_\text{tot} \propto \tau^{-4-2 \sigma},
    \quad
    \bar\rho \propto \tau^{-4 (1+\sigma) + 2 \bar{\sigma}},
    \quad
    a \propto \tau^{1+\sigma}.
    \label{eq:sigma-def}
\end{align}

The tensor perturbations satisfy the linearized Einstein equation~\cite{RoperPol:2023dzg,Giombi:2024kju}
\begin{align}
    \left(\partial_\tau^2 + k^2 - \frac{a''(\tau)}{a(\tau)}\right) a(\tau) \,h_{ij}(\tau,\vec{k}) = 16 \pi G\,a^3(\tau)\,{\bar w}(\tau)\, \Pi_{ij}(\tau,\vec{k}) \,,
    \label{eq:ssm-einstein}
\end{align}
where \(\Pi_{ij}(\tau,\vec{k})= \Lambda_{ij,mn}(\hat{k})\, T_{mn}(\tau,\vec{k}) / \bar{w}(\tau)\) is the dimensionless anisotropic stress tensor normalized to the mean enthalpy-density \(\bar w\) of the source, with the transverse-traceless projector \(\Lambda_{ij,mn}(\vec{k})\).
The corresponding Green-function solution \(G_k(\tau,\eta)\) is
\begin{align}
    h_{ij}(\tau,\vec{k}) = \frac{16 \pi G}{a(\tau)} \,\int\limits_{\tau_*}^{\tau} \dd\eta \,
    a^3(\eta)\,\bar{w}(\eta) \,\Pi_{ij}(\eta,\vec{k})\,G_k(\tau,\eta)\,.
    \label{eq:ssm-green-solution}
\end{align}
The gravitational-wave spectrum is defined using the Isaacson formula for the high-frequency effective stress tensor~\cite{Isaacson:1968hbi,Isaacson:1968zza},
\begin{align}
    \Omega_{\rm GW}(\tau,k) \equiv
    \frac{1}{\rho_{\rm tot}(\tau)} \frac{d\rho_{\rm GW}}{d\log k},
    \qquad
    \rho_{\rm GW}(\tau) =
    \frac{\langle h'_{ij}(\tau,\vec{x}) h'_{ij}(\tau,\vec{x})\rangle}{32 \pi G a^2(\tau)}.
    \label{eq:ssm-omega-gw-def}
\end{align}

In the non-relativistic and irrotational limit relevant to the SSM, the dimensionless anisotropic stress of the source can be written as
\begin{align}
    \Pi_{ij}(\tau,\vec{k}) = \int\!\frac{\dd^3 p}{(2\pi)^3}\,\Lambda_{ij,mn}(\hat{k}) \,v_m(\tau,\vec{p}) \,v_n(\tau, \vec{k}-\vec{p}) \,,
    \label{eq:ssm-pi-velocity}
\end{align}
where \(\vec{v}(\tau,\vec{k})\) is the fluid velocity field.
We define the anisotropic stress and kinetic unequal-time correlators~(UETCs) by
\begin{align}
    \label{eq:PiUETC}
    \langle\Pi_{ij}(\tau_1,\vec{k}) \,\Pi_{ij}^*(\tau_2,\vec{k}')\rangle &= (2 \pi)^3\, \delta^{(3)}(\vec{k}-\vec{k}')\,\frac{2\pi^2}{k^2}\,E_\Pi(\tau_1,\tau_2,k) \,,
    \\
    \langle v_i(\tau_1,\vec{k})\,v_j^*(\tau_2,\vec{k}')\rangle 
    &= (2\pi)^3\,\delta^{(3)}(\vec{k}-\vec{k}') \, \hat{k}_i \hat{k}_j \,\frac{4 \pi^2}{k^2}\, E_\text{kin}(\tau_1,\tau_2,k)\,.
    \label{eq:ssm-kinetic-uetc-def}
\end{align}
Assuming Gaussian statistics, Wick's theorem reduces the velocity four-point correlator in \cref{eq:PiUETC} to a convolution of two kinetic two-point correlators.
With \(\widetilde p \equiv \sqrt{k^2+p^2-2kpz},\) one obtains
\begin{align}
    E_\Pi(\tau_1,\tau_2,k) = 2 k^2 \int\limits_0^\infty\!\dd p\, \int\limits_{-1}^1\!\dd z\, \frac{p^2\,(1-z^2)^2}{\tilde{p}^4}\,E_\text{kin}(\tau_1,\tau_2,p)\,E_\text{kin}(\tau_1,\tau_2,\tilde{p}) \,.
    \label{eq:ssm-pi-uetc}
\end{align}
The square on \((1-z^2)\) is the angular factor generated by the transverse-traceless projection.

The time dependence of the kinetic correlator follows from the linearized conservation equations of the acoustic fluid.
Splitting energy density and pressure into a background component and fluctuations, \(\rho=\bar\rho+\delta \rho\) and \(p=\bar{p}+\delta p\), and defining the density contrast \(\lambda(\tau,\vec{k}) \equiv \delta\rho(\tau,\vec{k})/\bar{w}(\tau)\), the background conservation law and the spatial component of energy-momentum conservation can be written as
\begin{align}
    \frac{\bar w'}{\bar w} = -3(1+c_s^2)\mathcal{H}\,,
    \qquad
    (\bar w v_i)' + 4\mathcal H\bar w v_i + \partial_i\delta p = 0
    \,.
\end{align}
Using \(\delta p = c_s^2 \delta\rho\) and inserting \cref{eq:ssm-background-scaling}, the equations of motion for the Fourier modes of the velocity field and the density contrast become, using \(\bar{\sigma}\) defined in \cref{eq:sigma-def},
\begin{align}
    {\lambda}'(\tau,{\vec{k}}) - i\,\vec{k}\cdot\vec{v}(\tau,{\vec{k}}) = 0 \,, \qquad
    \vec{v}'(\tau,{\vec{k}})  - i\, c_s^2 \vec{k}\,{\lambda}(\tau,{\vec{k}}) + \frac{2 \bar{\sigma}}{\tau}\,\vec{v}(\tau,{\vec{k}})= 0 \,.
    \label{eq:ssm-continuity-euler}
\end{align}
The numerator of the dissipation term is fixed by the momentum dilution of the acoustic fluid, whereas the denominator is fixed by Hubble expansion.
For sub-horizon modes, \(c_s k\tau\gg1\), the leading solution is a damped acoustic wave,
\begin{align}
    \vec v(\tau,\vec k)
    \simeq
    \left(\frac{\tau}{\tau_*}\right)^{-\bar{\sigma}}
    \left[
        \vec v_{\vec k}\,e^{-ic_s k(\tau-\tau_*)}
        +
        \vec v_{-\vec k}^{\,*}\,e^{ic_s k(\tau-\tau_*)}
    \right].
    \label{eq:ssm-velocity-wave}
\end{align}
The resulting plane wave solutions are damped by a factor \(\propto \tau^{-\bar{\sigma}}\) compared to the conformal \(c_s^2=\frac{1}{3}\) case.

Neglecting the rapidly oscillating cross-correlators between the two frequency branches, the kinetic UETC becomes
\begin{align}
    E_\text{kin}(\tau_1, \tau_2, k) \simeq E_\text{kin}(k)\,\left(\frac{\tau_1 \tau_2}{\tau_*^2}\right)^{\!-\bar{\sigma}} \!\cos\left[c_s k(\tau_1-\tau_2)\right]\, .
    \label{eq:ssm-kinetic-uetc}
\end{align}
For \(c_s^2=\omega\), this reduces to the result of Ref.~\cite{Giombi:2024kju} with \(\bar{\sigma}=\sigma=(1-3\omega)/(1+3\omega)\).

The time-independent spectrum \(E_\text{kin}(k)\) is fixed by the single-bubble velocity and energy-density profiles and by the bubble nucleation history~\cite{Hindmarsh:2016lnk,Hindmarsh:2019phv,RoperPol:2023dzg}.
Schematically, it is given in terms of the two acoustic mode amplitudes by
\begin{align}
    E_\text{kin}(k) = \frac{k^2}{2\pi^2 \beta^6 R_*^3} \int\limits_0^\infty\!\dd\tilde{T}\,\tilde{T}^6 \nu(\tilde{T}) |\mathcal{A}_+(\tilde{T} k/\beta)|^2\,,
    \qquad
    \mathcal{A}_\pm(\chi)
    =
    -\frac{i}{2}
    \left[
        f'(\chi)
        \pm i c_s\ell(\chi)
    \right],
    \label{eq:ssm-single-bubble-amplitudes}
\end{align}
where \(R_*\) is the mean bubble separation, \(\beta = (8\pi)^{1/3}\xiw/R_*\) the inverse duration of the transition, \(\tilde{T}=T \beta\), and \(f'(\chi)\equiv \dd f(\chi)/\dd\chi\).
The bubble lifetime distribution function \(\nu(\tilde{T})\) is given by \(\nu(\tilde{T}) = \exp(-\tilde{T})\) for exponential nucleation and \(\nu(\tilde{T}) = \tilde{T}^2/2 \exp(-\tilde{T}^3/6)\) for simultaneous nucleation, respectively.
The functions \(f\) and \(\ell\) are Fourier kernels constructed from the self-similar velocity and normalized energy-density profiles,
\begin{align}
    f(\chi)
    &=
    \frac{4\pi}{\chi}
    \int_0^\infty \dd\xi\,
    v_{\rm ip}(\xi)\,
    \sin(\chi\xi),
    \qquad&
    \ell(\chi)
    &=
    \frac{4\pi}{\chi}
    \int_0^\infty \dd\xi\,
    \xi\,\lambda_{\rm ip}(\xi)\,
    \sin(\chi\xi).
    \label{eq:ssm-f-and-l}
\end{align}
Here \(v_{\rm ip}(\xi)\) and \(\lambda_{\rm ip}(\xi)\) are, respectively, the single-bubble self-similar velocity and normalized energy-density perturbation profiles at the onset of the acoustic stage.
The subsequent average over bubble centres and lifetimes determines \(E_{\rm kin}(k)\).
For later convenience we define the kinetic spectral shape function \(\zeta_\text{kin}(k R_*)\), the mean-squared fluid velocity \(\bar{U}_f^2\), and the kinetic energy fraction \(\Omega_K = \rho_\text{kin}/\rho_\text{tot}\) as 
\begin{align}
   E_\text{kin}(k) &= \bar{U}_f^2\,R_*\,\zeta_\text{kin}(k\,R_*) 
   \,,&
   \bar{U}_f^2 &= 2\,\int\limits_0^\infty\!\dd k\, E_\text{kin}(k)
   \,,&
   \Omega_K &= \bar\Omega_*\,\bar\Gamma\, \bar{U}_f^2\,,
    \label{eq:ssm-kinetic-normalization}
\end{align}
where \(\bar\Omega_*\equiv\bar\rho_*/\rho_{{\rm tot},*}\) and \(\bar\Gamma\equiv\bar w_*/\bar\rho_*=1+c_s^2\) are the fractional energy density and the mean adiabatic index of the constant-\(c_s\) fluid, respectively, at the time \(\tau_*\) of gravitational-wave production.
Thus the hydrodynamic profile fixes both the kinetic normalization \(\Omega_K\) and the shape \(\zeta_{\rm kin}\).

Combining \cref{eq:ssm-green-solution,eq:ssm-pi-uetc,eq:ssm-kinetic-uetc}, the spectrum can be written in the form
\begin{align}
    \Omega_\text{GW}(\tau, k) &= 
    3 \, \bar\Gamma^2 \,\bar\Omega_*^2\,k^3 
    \int\limits_0^\infty\!\dd p\, \int\limits_{-1}^1\!\dd z\, \frac{p^2\,(1-z^2)^2}{\widetilde{p}^4}\,E_\text{kin}(p)\,E_\text{kin}(\widetilde{p}) \,
    \Delta(\tau,\tau_*,k,p,\widetilde{p}) \,.
    \label{eq:ssm-final-spectrum-dimensional}
\end{align}
The time integration kernel \(\Delta(\tau,\tau_*,k,p,q)\) is
\begin{align}\begin{aligned}
    \Delta(\tau, \tau_*, k, p, q) 
    &= 2\, (\mathcal{H}_*\tau_*)^2\,\left(\frac{\tau}{\tau_*}\right)^{\!\!-2\, \sigma}
    \int\limits_{\tau_*}^\tau\! \frac{\dd\tau_1}{\tau_*}\, \int\limits_{\tau_*}^\tau \!\frac{\dd\tau_2}{\tau_*}\,
    \left(\frac{\tau_*^2}{\tau_1\tau_2}\right)^{\!1+\sigma}
    \times\\&\qquad\times
    G_k'(\tau,\tau_1)\,
    G_k'^*(\tau,\tau_2)
    \cos\left[c_s p(\tau_1-\tau_2)\right]
    \cos\left[c_s q(\tau_1-\tau_2)\right],
    \label{eq:ssm-time-kernel}
\end{aligned}\end{align}
where \(G_k'(\tau,\eta)\equiv\partial_\tau G_k(\tau,\eta)\) and \(\sigma\) is defined in \cref{eq:sigma-def}.
Note that the final power in \cref{eq:ssm-time-kernel} is independent of \(c_s\).
Each source insertion contributes \((a_i^3\bar w_i)/(a_*^3\bar w_*) = (\tau_i/\tau_*)^{-(1+\sigma)+2\bar{\sigma}}\), while the two kinetic correlators contained in the stress UETC contribute \((\tau_1\tau_2/\tau_*^2)^{-2\bar{\sigma}}\), with the exponents combining to \(1+\sigma=2/(1+3\omega)\).
The sound speed therefore cancels from the power-law time weight, but remains in the acoustic phases, in the resonance condition, and in the initial spectrum \(E_{\rm kin}(k)\).

At leading order in the short-wavelength expansion \(\mathcal{H}_* R_* \ll 1\), the Green's function is given by~\cite{Giombi:2024kju}
\begin{align}
    G_k(\tau,\eta) = 
    \frac{\sin\left[k(\tau-\eta)\right]}{k}\,\Theta(\tau-\eta)
    + \mathcal{O}(\mathcal{H}_* R_*)
    \,.
    \label{eq:ssm-leading-green}
\end{align}
Corrections beyond this approximation have been derived in Ref.~\cite{Giombi:2025tkv} and are not included here.
After decomposing the products of cosines and averaging over the rapidly oscillating observation-time dependence, the kernel can be expressed as
\begin{align}\begin{aligned}
    \Delta(\tau, \tau_*, {k}, p, q) &= 
     (1+\sigma)^2\,\left(\frac{\tau}{\tau_*}\right)^{-2 \sigma}
    \sum\limits_{m,n=\pm} \Delta_{mn}(\tau,\tau_*, |k + c_s (n p + m q)|)\,,
    \label{eq:ssm-kernel-sum}
    \\
    \Delta_{mn}(\tau,\tau_*, \widehat{p}_{mn}) &= \frac{|\widehat{p}_{mn} \tau_*|^{2\,\sigma}}{4} \,\left[ 
        \Ci^2(\widehat{p}_{mn}\tau, \widehat{p}_{mn}\tau_*; -\sigma)
        + \Si^2(\widehat{p}_{mn}\tau, \widehat{p}_{mn}\tau_*; -\sigma)
    \right] ,
\end{aligned}\end{align}
where we used \(\mathcal{H}\tau = 1+\sigma\).
Our convention for the generalized trigonometric integrals is
\begin{align}
    \Si(x,x_0;s) &\equiv \int\limits_{x_0}^x\!\dd t\,\frac{\sin(t)}{t^{1-s}}\,, &
    \Ci(x,x_0;s) &\equiv \int\limits_{x_0}^x\!\dd t\,\frac{\cos(t)}{t^{1-s}}\,.
    \label{eq:ssm-generalized-cisi}
\end{align}

In the low-frequency limit, \(k\to 0\), the momenta of the kinetic spectra coincide, \(q=\tilde{p}\to p\), and the kernel simplifies to
\begin{align}\begin{aligned}
    \Delta_0(\tau,\tau_*,p) 
    \equiv& \lim\limits_{k\to 0} \Delta(\tau,\tau_*, k, p ,\tilde{p})
    \\
    =& (1+\sigma)^2\,\left(\frac{\tau}{\tau_*}\right)^{-2 \sigma} \left\{
        \frac{1}{2 \sigma^2} \left[1-\left(\frac{\tau}{\tau_*}\right)^{\!-\sigma}\right]^2
        + 2\, \Delta_{++}(\tau,\tau_*, 2 c_s p)
    \right\}\,,
    \label{eq:ssm-kernel-low-f}
\end{aligned}\end{align}
where the first term in the curly bracket becomes \(\ln^2(\tau/\tau_*)/2\) in radiation domination, i.e.\ for \(\sigma\to 0\).
In the limit of infinite source duration, on the other hand, \(\tau \gg \tau_*\) and \(k\gg (\tau-\tau_*)^{-1}\), the kernel simplifies to a \(\delta\) distribution~\cite{Giombi:2024kju}
\begin{align}
    \Delta_\text{high}(\tau,\tau_*,k,p,q) = 
    \frac{1+\sigma}{1+2\sigma} \left(\frac{\tau}{\tau_*}\right)^{\!-2 \sigma}
    \left[
        1 - \left(\frac{\tau_*}{\tau}\right)^{1+2\sigma}
    \right]
    \mathcal{H}_* \frac{\pi}{2}
    \delta\big(k-c_s(p+q)\big)\,,
    \label{eq:ssm-kernel-infinite-duration}
\end{align}
making the \(z\) integration  in \cref{eq:ssm-final-spectrum-dimensional} trivial.

Finally, introducing \(K=k R_*\), \(P=p R_*\), and \(\widetilde P=\sqrt{K^2+P^2-2KPz}\), and using the normalization of \cref{eq:ssm-kinetic-normalization}, the equal-time correlator \(E_\Pi(k) \equiv E_\Pi(\tau,\tau,k)\) of the anisotropic stress can be written as \(k E_\Pi(k) = 2 (k R_*)^3 \bar{U}_f^4 \,\mathcal{C}\, \zeta_\Pi(k R_*)\), where the normalization \(\mathcal{C}\) and the shape function \(\zeta_\Pi(K)\) are given by
\begin{align}
   \mathcal{C} = \frac{16}{15}\, \int\limits_0^\infty\! d K\, \frac{\zeta_\text{kin}^2(K)}{K^2}\,,
   \qquad
   \mathcal{C}\,\zeta_\Pi(K) = \int\limits_0^\infty\!d P\,P^2 \zeta_\text{kin}(P) \int\limits_{-1}^1\!d z \,(1-z^2)^2 \,\frac{\zeta_\text{kin}(\tilde{P})}{\tilde{P}^4}
   \,,
\end{align}
and the corresponding gravitational-wave spectrum in \cref{eq:ssm-final-spectrum-dimensional} can be written in terms of the dimensionless variables and spectral shape functions as
\begin{align}\begin{gathered}
    \Omega_\mathrm{GW} = 3\,\Omega_K^2\,K^3\,\int\limits_0^\infty\!\dd P \int\limits_{-1}^1\!\dd z\,
    \frac{P^2 (1-z^2)^2}{\tilde{P}^4}\, \zeta_\text{kin}(P) \zeta_\text{kin}(\tilde{P})\,\bar\Delta(\mathcal{H_*} \delta\tau_\text{fin},\mathcal{H}_* R_*, K, P, \tilde{P}) \,,
    \label{eq:ssm-final-spectrum}
    \\
    \bar\Delta(\mathcal{H_*} \delta\tau_\text{fin},\mathcal{H}_* R_*, K, P, \tilde{P}) = \Delta\left(\mathcal{H}_*\tau_* + \mathcal{H}_*\delta\tau_\text{fin}, \mathcal{H}_*\tau_*, \frac{K}{\mathcal{H}_*R_*}, \frac{P}{\mathcal{H}_*R_*}, \frac{\tilde{P}}{\mathcal{H}_*R_*}\right) ,
\end{gathered}\end{align}
where \(\mathcal{H}_*\) factors cancel in the ratios and products in \cref{eq:ssm-kernel-sum}.
The hydrodynamic profile determines the energy budget \(\Omega_K\) and the kinetic spectral shape \(\zeta_{\rm kin}\), whereas the background and the acoustic propagation after the bubbles have collided enter through the time evolution kernel \(\overline\Delta\).

\subsection{Results for direct, inverse, and mixed-hybrid spectra}
\label{sec:ssm-results}

We now apply the formalism of \cref{sec:ssm-review} to the complete CSS hydrodynamic classification.
We first study how the sound speed in the primordial plasma at the end of the transition affects the peak of the gravitational-wave spectrum, and then investigate whether the new mixed-hybrid profiles generate distinctive spectral features.
Unless stated otherwise, we use exponential nucleation, fix \(\mathcal H_*R_*\), and set the acoustic lifetime by the onset of non-linearities, \(\mathcal H_*\Delta\tau_{\rm nl} \simeq \mathcal H_* R_*/\sqrt{\Omega_K/\bar\Omega_*}\).
This estimate follows the prescription adopted in our previous analysis of inverse transitions~\cite{Barni:2025gnm} and in the characterization of the finite-duration sound-shell signal of Ref.~\cite{RoperPol:2023dzg}.

\paragraph{Sound-speed dependence of direct and inverse profiles.}

Ref.~\cite{Giombi:2024kju} showed that, for a fixed initial kinetic spectrum approximated as a broken power-law and a single fluid satisfying \(\omega=c_s^2\), decreasing the sound speed produces an approximately scale-independent suppression of the gravitational-wave signal and shifts its peak towards smaller wavenumbers, without changing the asymptotic power-law indices.
The displacement follows from the acoustic resonance, \(k_{\rm peak} \sim 2 c_s k_{\rm kin}\), where \(k_{\rm kin}\) is the characteristic scale of the kinetic spectrum.
The amplitude is simultaneously affected by the time evolution and coherence of the acoustic source.

Here we test whether the same qualitative behaviour persists for the complete set of direct and inverse CSS solutions.
Instead of assuming a broken power-law kinetic spectrum as in Ref.~\cite{Giombi:2024kju}, we here vary \(c_-\) self-consistently in both the hydrodynamic profile and the sound-wave propagation after collisions, identifying \(\omega = c_s^2 = c_-^2\).
Consequently, at fixed \((\xiw, \alphaN)\), changing \(c_-\) modifies not only the acoustic kernel, but also the initial self-similar profiles, the kinetic spectrum, and the kinetic-energy fraction.

\begin{figure}
    \centering
    \includegraphics[width=\textwidth]{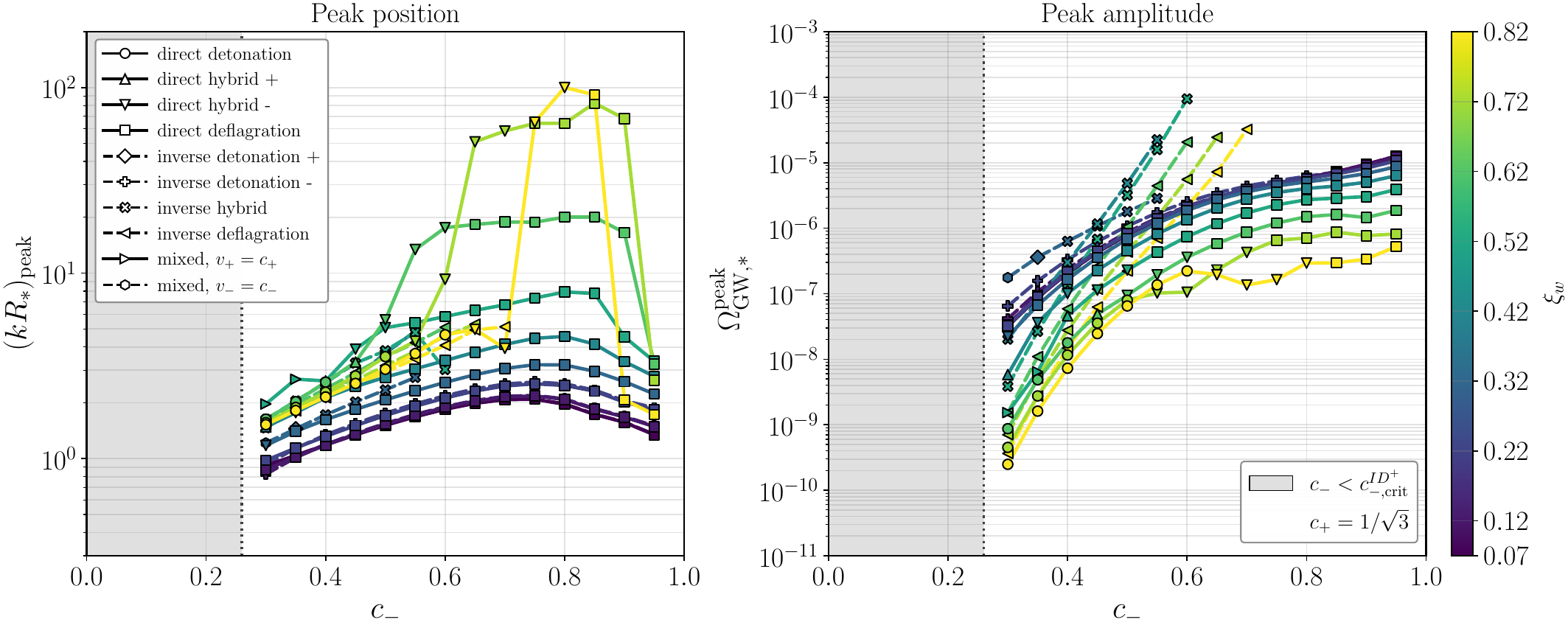}
    \caption{%
        Peak position (left) and peak amplitude (right) as functions of the sound speed \(c_s=c_-\), for fixed \(c_+=1/\sqrt{3}\) and \(\mathcal{H}_* R_*=0.1\).
        Solid lines denote direct solutions with \(\alphaN=+0.1\), while dashed lines denote inverse solutions with \(\alphaN=-0.1\).
        The colour indicates the wall velocity and the markers identify the hydrodynamic branch.
        The grey regions mark \(c_-<\ccritID\) where different solution branches can overlap in the \((\xiw,\alphaN)\) plane and they are not displayed.
    }
    \label{fig:ssm-cs-direct-inverse}
\end{figure}

The resulting peak position and amplitude are shown in \cref{fig:ssm-cs-direct-inverse}.
Solid lines correspond to direct solutions, while dashed lines denote inverse ones, with \(\alphaN=+0.1\) and \(\alphaN=-0.1\), respectively.
The overall dependence on the sound-speed \(c_s=c_-\) is consistent with the behaviour found in previous studies, in particular Ref.~\cite{Giombi:2024kju}.
For moderate values of \(c_-\), increasing \(c_-\) shifts the peak towards larger \(kR_*\), although this trend weakens and can reverse as \(c_-\to1\).
The peak amplitude shows a similar overall enhancement with increasing \(c_-\).

The same qualitative trend is found for both direct and inverse hydrodynamic solutions, showing that the dependence on \(c_-\) is not peculiar to direct transitions.
Along individual branches, however, the behaviour need not be monotonic.
In particular, the \(\vp^{(-)}\) sub-branches of direct hybrids and inverse detonations exhibit pronounced dips in the peak amplitude over part of the scan.
Such branch-dependent structure is expected because varying \(c_-\) changes not only the propagation speed of the acoustic modes, but also the self-consistent hydrodynamic profile and therefore the shape of the kinetic spectrum.
Moreover, the quantities shown in \cref{fig:ssm-cs-direct-inverse} are defined by the global maximum of the gravitational-wave spectrum.
If two local spectral peaks exchange hierarchy as \(c_-\) is varied, the identified peak switches from one to the other, leading to the sharp variations seen in the figure.
Despite these local features, the overall trend at low and moderate \(c_-\) remains that increasing the sound speed in the `\(-\)' phase shifts the signal towards larger wavenumbers and enhances its peak amplitude.

\paragraph{Mixed hybrids.}

\begin{figure}
    \centering
    \includegraphics[width=\textwidth]{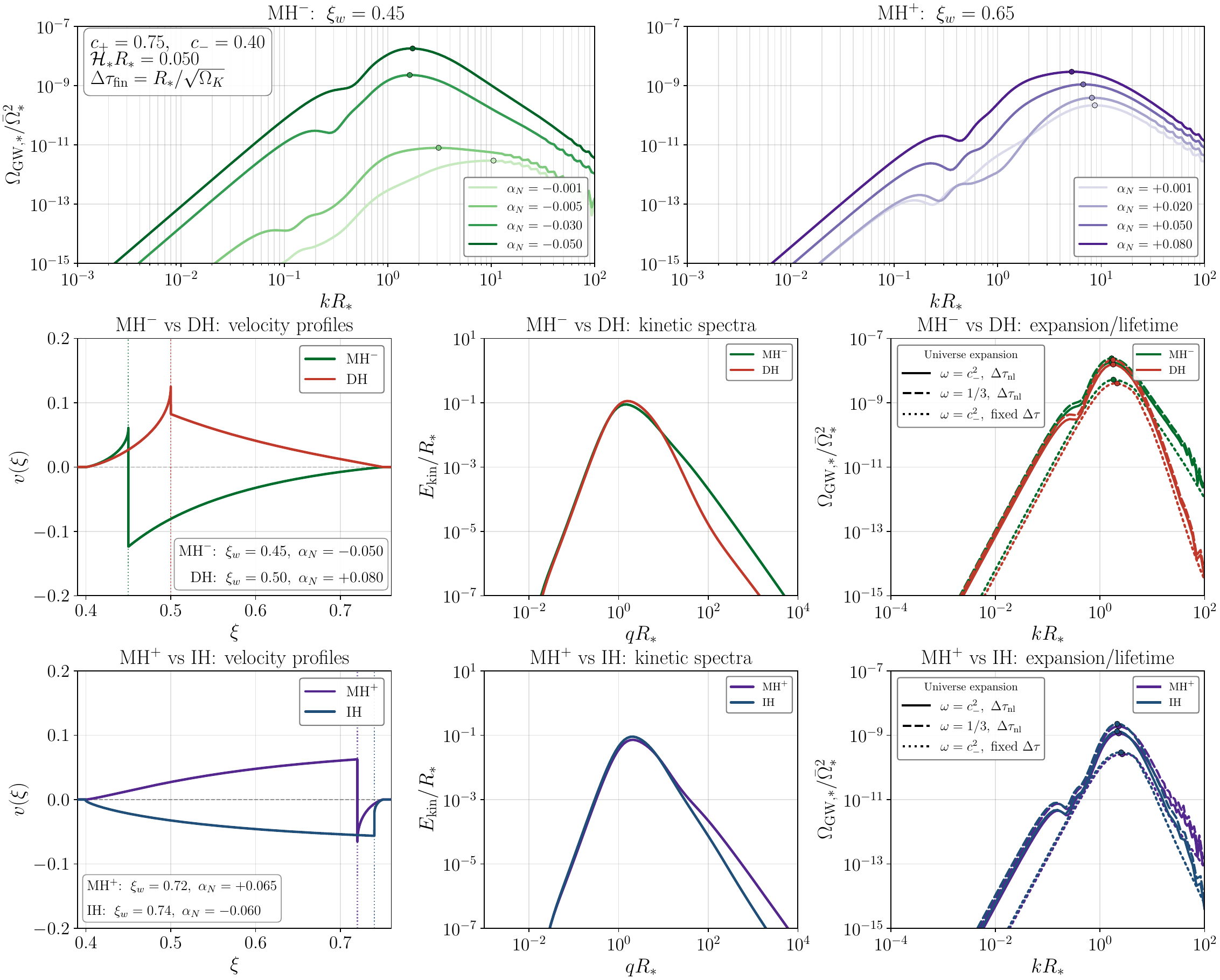}
    \caption{%
        Mixed-hybrid spectra and comparison with neighbouring ordinary branches for \(c_+=0.75\), \(c_-=0.40\), and \(\mathcal{H}_* R_*=0.05\).
        The upper panels show scans along the two mixed-hybrid branches at fixed wall velocity.
        The lower rows compare the branches to direct and inverse hybrids. From left to right, the velocity profiles, normalized kinetic spectra, and gravitational-wave spectra are shown.
        Despite the oppositely directed fluid motion on the two sides of the wall, the mixed-hybrid kinetic and gravitational-wave spectra closely resemble ordinary hybrids with different velocity profiles.
        Changing the background from \(\omega=c_-^2\) to \(\omega=1/3\) at fixed \(c_s\) and source duration has a comparatively modest effect, whereas changing the acoustic lifetime significantly modifies the amplitude and removes the low-\(k\) shoulder present for \(\Delta\tau_\text{nl}\).
    }
    \label{fig:ssm-mixed-hybrids}
\end{figure}

We next investigate whether the two mixed-hybrid branches lead to a qualitatively distinctive gravitational-wave signal.
The upper panels of \cref{fig:ssm-mixed-hybrids} show narrow scans along the two mixed hybrid branches for the benchmark choice \(c_+=0.75\) and \(c_-=0.40\), at fixed wall velocity.
Along both branches, the spectra evolve continuously as \alphaN is varied.
Their characteristic frequencies, overall amplitudes, and spectral widths remain of the same order as those of direct and inverse hybrids.
The middle and lower panels compare mixed hybrids with \(\vm=c_-\) to a direct hybrid, and mixed hybrids with \(\vp=c_+\) to an inverse hybrid, respectively.
Since mixed and ordinary hybrids do not coexist at the same \((\xiw,\alphaN)\), for this choice of sound speeds, we select for each mixed benchmark an ordinary hybrid whose gravitational-wave spectrum is as close as possible.
Although the corresponding real-space velocity profiles are markedly different, with oppositely directed flows and opposite signs of \alphaN in the lower panel, the resulting gravitational-wave spectra almost overlap.
At the level of the final spectrum, mixed hybrids are thus effectively indistinguishable from the direct- or inverse-hybrid branches to which they are hydrodynamically connected.

Moreover, the two lower right-hand panels separate the effects of the background expansion from those of the finite acoustic lifetime.
Solid lines show our baseline single-fluid scenario, in which the same CSS fluid both dominates the homogeneous expansion and carries the acoustic perturbations, so that \(\omega=c_-^2\), with the source assumed to remain active until the onset of non-linearities, \(\Delta\tau=\Delta\tau_{\rm nl}\).
Dashed lines instead correspond to a radiation-dominated Universe, \(\omega=1/3\), where the expansion is controlled by an additional conformal relativistic component that does not participate in the phase transition, while the sound waves still propagate with \(c_s=c_-\) and source gravitational waves up to \(\Delta\tau_{\rm nl}\).
To isolate the effect of the background expansion, both lower panels (right column) show the gravitational-wave spectrum normalized by \(\bar{\Omega}_*^2\). 
For the representative benchmarks considered here, changing the expansion history in this way produces only a modest modification of the spectrum, indicating that the position and shape of the peak are primarily controlled by the acoustic fluid rather than by the homogeneous background.

Dotted lines isolate the effect of the source lifetime.  They return to the single-fluid choice \(\omega=c_-^2\), but replace the nonlinear lifetime by the fixed duration
\(\mathcal H_*\Delta\tau=0.1\).
This produces a more visible change in both the overall normalization and the shape of the spectrum around the peak.
For \(\Delta\tau=\Delta\tau_{\rm nl}\), the spectra of both mixed and ordinary hybrids develop a shoulder at \(kR_*\sim \numrange{0.1}{0.3}\), below the main peak, which is absent for the shorter fixed duration. 
For mixed hybrids, as for the other solutions, the detailed gravitational-wave morphology is therefore controlled not only by the initial hydrodynamic profile, but also by the duration over which the resulting acoustic configuration remains an efficient source.

For the benchmarks considered here, the spectral shape is more sensitive to the acoustic lifetime than to the equation of state governing the homogeneous expansion.
The background parameter \(\omega\) mainly modifies the time-dependent kernel, whereas the shape of the spectrum is controlled primarily by the kinetic spectrum and by the duration of the acoustic source.
Overall, our analysis extends previous sound-shell studies by confirming the sound-speed dependence for inverse solutions, allowing the acoustic fluid and the cosmological background to obey different equations of state, and showing that mixed-hybrid configurations, despite their distinct hydrodynamics, yield gravitational-wave spectra that are degenerate with those of ordinary hybrids.

%
\section{The \texorpdfstring{\hydrodex}{HydróDex} code}
\label{sec:code}
%

This paper marks the public release of version~1.0 of the \hydrodex code --- a \cpp code for computing hydrodynamic fluid profiles and the corresponding gravitational-wave spectra from cosmological phase transitions.
\hydrodex provides a complete index of fluid profiles within the CSS model as well as routines for calculating the gravitational waves within the sound-shell model assuming different equations of state for the Universe and the fluid in which the sound-wave source propagates.
The code further provides a \python interface via \texttt{pybind11}~\cite{pybind11}.

The \hydrodex code implements the classes \texttt{FluidProfile} and \texttt{SoundShellSpectrum}, which provide an object-oriented interface to calculate the hydrodynamic profiles of the CSS model and the corresponding gravitational-wave spectra within the sound-shell model. 
The underlying functions used internally are implemented in the modules \texttt{profile\_calculator} and \texttt{sound\_shell\_model}, respectively.

\subsection{Installation and prerequisites}

The \hydrodex source code is available at \href{https://github.com/hydrodex-pt/hydrodex}{\faGithub\ \hydrodex}.
The following dependencies are required to build \hydrodex:
\begin{itemize}
    \item the \texttt{CMake} build system~\cite{CMake}, version~3.24 or higher;
    \item a \cpp17 compatible compiler;
    \item the Boost C++ libraries~\cite{boost}, in particular \texttt{Boost.Log};
    \item the GNU Scientific Library~\cite{Galassi:2019czg};
    \item the \texttt{OpenMP} API~\cite{Dagum:1998rbq};
    \item a \python interpreter, version 3.8 or higher (optional, required to build the \python interface);
    \item the \texttt{pybind11} library~\cite{pybind11}, version 3.0.0 or higher (optional, required to build the \python interface);
    \item the \texttt{NumPy} package~\cite{Harris:2020xlr} (optional, required for using the \python interface);
    and
    \item the \texttt{doxygen} documentation generator~\cite{doxygen}, version 1.9.7 or higher (optional, required to build the documentation).
    Building the \python documentation further requires \texttt{Sphinx}~\cite{Sphinx} and \texttt{Furo}~\cite{Furo}.
\end{itemize}
The requirements can be installed using a package manager, e.g.\ \texttt{apt} on Debian and \texttt{MacPorts}~\cite{MacPorts} or \texttt{Homebrew}~\cite{homebrew} on MacOS.
\texttt{CMake} will automatically find these dependencies if they are installed into a standard location.
Further instructions in case that \texttt{CMake} fails to detect them can be found in the \texttt{README.md} file of the \hydrodex source directory.

Once the prerequisites for the desired features are installed, \hydrodex can be built and installed using \texttt{CMake} by running
\begin{shellcode}{codecaption={Installation instructions}{lst:installation}}
cmake -S<root_dir> -B<build_folder> -DCMAKE_INSTALL_PREFIX=<install_path>
cmake --build <build_folder> -j <N>
cmake --install <build_folder>
\end{shellcode}
where \texttt{<root\_dir>} is the root of the source directory, \texttt{<build\_folder>} is the directory in which the code is built, \texttt{<install\_path>} is the path into which the code is installed, and \texttt{<N>} is the number of parallel jobs in the build process.
Additional options are documented in the \texttt{README.md} file.

After installation, the \texttt{<install\_path>} will contain the following files:
\begin{description}[font=\ttfamily,style=nextline]
    \item[lib/libhydrodex.so \text{or} lib/libhydrodex.dylib] 
        The shared \hydrodex library to which user code should be linked.
    \item[include/hydrodex/*.hpp]
        The header files for the \cpp library that need to be included to use the library.
    \item[bin/hydrodex/hydrodex] 
        A command-line-interface executable that calculates the fluid profiles and gravitational-wave spectrum for given input values (if \texttt{BUILD\_TEST=ON}).
    \item[lib/python<version>/site-packages/hydrodex.cpython-<version>-<platform>.so]
        The \python interface to the library (if \texttt{BUILD\_PYTHONLIB=ON}).
    \item[share/hydrodex/docs/] 
        The \texttt{HTML} and \LaTeX documentation generated with \texttt{Doxygen} (if \texttt{BUILD\_DOCS=ON}).
    \item[share/hydrodex/examples/]
        Example Jupyter notebooks with a tutorial and the code required to recreate the figures in this paper (if \texttt{BUILD\_PYTHONLIB=ON}).
\end{description}
The compilation of the optional components (test script, \python interface, and documentation) can be controlled by passing the corresponding argument \texttt{-D<OPTION>=<ON/OFF>} to the initial \texttt{cmake} call (first line of \cref{lst:installation}). 
Alternatively, the \python interface can be installed with \texttt{pip} using the \texttt{Scikit-build-core} backend~\cite{schreiner2024scikitbuildcore} via
\lstinline[style=shell, basicstyle=\ttfamily]{pip install <root_dir>}.

\subsection{Code structure and usage}

To use the \cpp library, the corresponding headers need to be included. 
For instance, the class \lstinline[style=inlinecpp]{FluidProfile} requires \lstinline[style=inlinecpp]{#include "hydrodex/fluid_profile.hpp"}. 
If the installation directory is not within the system's standard search paths, the user must either set the corresponding environment variables or pass the include and library directories to the compiler and linker when using the library. 

All \hydrodex classes and functions are defined in the \lstinline[style=inlinecpp]{hydrodex} namespace or one of its nested namespaces. 
The headers \lstinline[style=inlinecpp]{fluid_profile.hpp} and \lstinline[style=inlinecpp]{sound_shell_spectrum.hpp} declare the classes \lstinline[style=inlinecpp]{hydrodex::FluidProfile} and \lstinline[style=inlinecpp]{hydrodex::SoundShellSpectrum}, respectively, both directly in the \lstinline[style=inlinecpp]{hydrodex} namespace.
The namespaces \lstinline[style=inlinecpp]{hydrodex::profile_calculator} and \lstinline[style=inlinecpp]{hydrodex::sound_shell_model} containing the corresponding functions used internally to calculate the fluid profiles and gravitational-wave spectra are defined in the headers \lstinline[style=inlinecpp]{profile_calculator.hpp} and \lstinline[style=inlinecpp]{sound_shell_model.hpp}.
Additional helper functions and constants within the namespaces \lstinline[style=inlinecpp]{hydrodex::utils} and \lstinline[style=inlinecpp]{hydrodex::constants} are declared in the respective headers \lstinline[style=inlinecpp]{utils.hpp} and \lstinline[style=inlinecpp]{constants.hpp}.
The header \lstinline[style=inlinecpp]{settings.hpp} contains global default settings of the code (which can be overwritten in the user code) within the namespace \lstinline[style=inlinecpp]{hydrodex::settings}.

The \python interface can be used by importing the \lstinline[style=inlinepy]{hydrodex} module via \lstinline[style=inlinepy]{import hydrodex}.
As in \cpp, the classes \lstinline[style=inlinepy]{hydrodex.FluidProfile} and \lstinline[style=inlinepy]{hydrodex.SoundShellSpectrum} are available directly from the \lstinline[style=inlinepy]{hydrodex} module.
The same applies for the \lstinline[style=inlinepy]{hydrodex.UniformLinearInterpolator} class, which is part of the \lstinline[style=inlinecpp]{hydrodex::utils} namespace in \cpp.
The internally used functions and variables are defined within the respective submodules \lstinline[style=inlinepy]{hydrodex.profile_calculator} and \lstinline[style=inlinepy]{hydrodex.sound_shell_model}.
Global settings of \hydrodex can be accessed and changed via the corresponding getter and setter functions \lstinline[style=inlinepy]{hydrodex.get_<setting>} and \lstinline[style=inlinepy]{hydrodex.set_<setting>}.

\subsection{Fluid profile calculation}

The primary user interface for calculating fluid profiles from the wall velocity \xiw and transition strength parameter \alphaN is the \texttt{FluidProfile} class.
The class is declared in the \cpp header \texttt{hydrodex/fluid\_profile.hpp}.
The primary constructor to create a fluid profile from the wall velocity \xiw and the strength parameter \alphaN is given by
\begin{cppcode}{codecaption={Main constructor for the \lstinline[style=inlinecpp]{FluidProfile} \cpp class.}{lst:FluidProfile-cpp}}
hydrodex::FluidProfile(
    double xi_w,       (*@\codecomment{wall velocity \xiw}@*)
    double alpha_N,    (*@\codecomment{pseudotrace strength parameter \alphaN (far in front of the wall)}@*)
    double cp = hydrodex::constants::cs_bag,    (*@\codecomment{sound speed \(c_+\) ahead of the wall}@*)
    double cm = hydrodex::constants::cs_bag,    (*@\codecomment{sound speed \(c_-\) behind the wall}@*)
(*@\lstnonumber@*)    (*@\codecomment{step size for the fluid velocity (or coordinate) when solving \cref{eq:v-ode-3d} numerically}@*)
    double step_size = hydrodex::profile_calculator::profile_step_size,
(*@\lstnonumber@*)    (*@\codecomment{\texttt{ProfileSolveMode} switch determining which differential equation for the velocity profile is solved}@*)
    hydrodex::profile_calculator::ProfileSolveMode solve_mode 
        = hydrodex::profile_calculator::profile_solve_mode
);
\end{cppcode}
The constructor first classifies the corresponding solution type in terms of \xiw (\lstinline[style=inlinecpp]{xi_w}) and \alphaN (\lstinline[style=inlinecpp]{alpha_N}) using the \alphaN ranges in \cref{tab:classification}.
The values of the sound speeds \(c_\pm\) in the two phases of the CSS can be passed as optional arguments \lstinline[style=inlinecpp]{cp} and \lstinline[style=inlinecpp]{cm}, respectively, and default to the bag value \(c_\pm=1/\sqrt{3}\) if not provided.
The condition \(c_- < \ccritID(c_+)\), below which different solution types may overlap in the \((\xiw,\alphaN)\) plane, is tested based on a polynomial fit to the \(\ccritID(c_+)\) line, which is the largest of the critical sound speeds in \cref{eq:ccrits}.
\hydrodex then throws an exception, displays a warning or ignores the overlap depending on whether the global \lstinline[style=inlinecpp]{enum class hydrodex::settings::CheckMode} switch \lstinline[style=inlinecpp]{hydrodex::settings::noninjectivity_check} is set to \lstinline[style=inlinecpp]{CheckMode::ERROR}, \lstinline[style=inlinecpp]{CheckMode::WARNING} (default) or \lstinline[style=inlinecpp]{CheckMode::SILENT}, respectively.
The code currently only returns one solution type and only tests \alphaN values lying within the projections of the respective \alphaP endpoints.

Subsequently, the corresponding value of \alphaP is reconstructed using the TOMS 748 root-finding algorithm~\cite{TOMS748} implemented in \texttt{Boost}~\cite{boost} within the \alphaP range of the respective branch, and the wall-frame fluid-velocities \vpm on either side of the wall are determined using \cref{eq:vplusminus_css}.
Next, the velocity profile \(v(\xi)\) is calculated using a fourth-order Runge-Kutta method~\cite{Runge:1895hdo} with a step size controlled by the optional \lstinline[style=inlinecpp]{step_size} argument, where the default value controlled by \lstinline[style=inlinecpp]{hydrodex::profile_calculator::profile_step_size} is set to \num{e-4}.
The code integrates \cref{eq:v-ode-3d} either as \(\dd v/\dd\xi\) or \(\dd\xi/\dd v\) depending on the \lstinline[style=inlinecpp]{solve_mode} parameter.
Possible options for this parameter are
\begin{cppcode}{codecaption={\lstinline[style=inlinecpp]{enum class ProfileSolveMode} selector for solving \cref{eq:v-ode-3d}.}{lst:ProfileSolveMode}}
enum class hydrodex::profile_calculator::ProfileSolveMode {
    D_XI_D_V,            (*@\codecomment{integrate \(\dd\xi/\dd v\) as a function of \(v\)}@*)
    D_V_D_XI,            (*@\codecomment{integrate \(\dd v/\dd\xi\) as a function of \(\xi\)}@*)
    D_V_D_XI_FOR_SHOCKS  (*@\codecomment{integrate \(\dd v/\dd\xi\) for compression waves and \(\dd\xi/\dd v\) for rarefaction waves}@*)
};
\end{cppcode}
\noindent
where the default \lstinline[style=inlinecpp]{hydrodex::profile_calculator::profile_solve_mode} is set to always solve \(\dd\xi/\dd v\) (\lstinline[style=inlinecpp]{hydrodex::profile_calculator::ProfileSolveMode::D_XI_D_V}).
The enthalpy profile \(w(\xi)\) is computed from the velocity profile integrating \cref{eq:enthalpy-ratio} analytically, interpolating linearly on each segment in the velocity profile vector, assuming \(w_N=1\).

The fluid profiles with the corresponding self-similar coordinate \(\xi\) as well as the respective solution type can be obtained as \lstinline[style=inlinecpp]{std::vector<double>} from a \lstinline[style=inlinecpp]{hydrodex::FluidProfile} instance using the following member functions:
\begin{cppcode}{codecaption={Member functions of the \lstinline[style=inlinecpp]{hydrodex::FluidProfile} class for obtaining the profiles.}{lst:FluidProfile-functions}}
(*@\lstnonumber\codecomment{get the hydrodynamic solution branch:}@*)
hydrodex::profile_calculator::HydroSolutionType hydrodex::FluidProfile::get_solution_type() const;
(*@\lstnonumber\codecomment{get the self-similar coordinate vector \(\xi\):}@*)
const std::vector<double>& hydrodex::FluidProfile::get_xi_coordinate() const;
(*@\lstnonumber\codecomment{get the velocity profile \(v(\xi)\):}@*)
const std::vector<double>& hydrodex::FluidProfile::get_fluid_velocity() const;
(*@\lstnonumber\codecomment{get the enthalpy density profile \(w(\xi)\):}@*)
const std::vector<double>& hydrodex::FluidProfile::get_enthalpy_density() const;
(*@\lstnonumber\codecomment{get the energy density profile \(e(\xi)\):}@*)
const std::vector<double>& hydrodex::FluidProfile::get_energy_density() const;
(*@\lstnonumber\codecomment{get the pressure profile \(p(\xi)\):}@*)
const std::vector<double>& hydrodex::FluidProfile::get_pressure() const;
(*@\lstnonumber\codecomment{get the entropy density profile \(s(\xi)\):}@*)
std::vector<double> hydrodex::FluidProfile::get_entropy_density(
    double Psi_N,       (*@\codecomment{enthalpy ratio \(\Psi_N\) across the wall at nucleation}@*)
    double s_N = 1.0    (*@\codecomment{entropy density \(s_N\) at nucleation}@*)
) const;
(*@\lstnonumber\codecomment{get the temperature profile \(T(\xi)\):}@*)
std::vector<double> hydrodex::FluidProfile::get_temperature(
    double Psi_N,       (*@\codecomment{enthalpy ratio \(\Psi_N\) across the wall at nucleation}@*)
    double T_N = 1.0    (*@\codecomment{nucleation temperature \(T_N\)}@*)
) const;    
(*@\lstnonumber\codecomment{get the efficiency factor \(\kappa\) or kinetic energy fraction \(K\):}@*)
double hydrodex::FluidProfile::get_efficiency(
    hydrodex::profile_calculator::EfficiencyNormalization normalization
        = hydrodex::profile_calculator::EfficiencyNormalization::TotalEnergyDensity
) const;
\end{cppcode}
\noindent%
The functions \lstinline[style=inlinecpp]{get_entropy_density} and \lstinline[style=inlinecpp]{get_temperature} require the enthalpy ratio \(\Psi_N\) across the wall evaluated at the nucleation temperature, cf.\ \cref{eq:PsiN-def}. 
The values of the entropy density \(s_N\) and temperature \(T_N\) at nucleation (i.e.\ at \(\xi=1\)) can be provided as optional arguments \lstinline[style=inlinecpp]{s_N} and \lstinline[style=inlinecpp]{T_N}, respectively, and are set to \num{1} if omitted.
The function \lstinline[style=inlinecpp]{get_efficiency} returns the efficiency factor \(\kappa\) or kinetic energy fraction \(K\) discussed in \cref{sec:efficiency}.
By default, it normalizes the efficiency to the total energy density, although alternative normalizations can be obtained by passing the corresponding  \lstinline[style=inlinecpp]{hydrodex::profile_calculator::EfficiencyNormalization} enumerator option.\footnote{%
    The default normalization \lstinline[style=inlinecpp]{EfficiencyNormalization::TotalEnergyDensity} gives the kinetic-energy fraction \(K=\rho_{\rm kin}/\rho_{\rm tot}\) used in this work, with the full perturbed region as reference volume, while the choices \lstinline[style=inlinecpp]{EfficiencyNormalization::VacuumEnergy} and \lstinline[style=inlinecpp]{EfficiencyNormalization::AlphaPseudoTrace} use the vacuum-energy difference and \(\alpha_{\bar\theta}^N\), respectively.
    For inverse transitions, the latter two prescriptions automatically normalize to the initial thermal-energy reservoir.
    Alternatively, the option \lstinline[style=inlinecpp]{EfficiencyNormalization::BubbleVolume} normalizes to the bubble volume.%
}
The function \lstinline[style=inlinecpp]{get_solution_type} returns a  \lstinline[style=inlinecpp]{hydrodex::profile_calculator::HydroSolutionType} object which indicates the respective solution branch as follows:
\begin{cppcode}{codecaption={\lstinline[style=inlinecpp]{enum class HydroSolutionType} corresponding to the hydrodynamic classification in \cref{tab:classification}.}{lst:HydroSolutionType}}
enum class hydrodex::profile_calculator::HydroSolutionType { 
    Deflagration,             (*@\codecomment{direct deflagrations\(\vphantom{\vp^{(+)}}\)}@*)
    Detonation,               (*@\codecomment{direct detonations\(\vphantom{\vp^{(+)}}\)}@*)
    Hybrid,                   (*@\codecomment{direct hybrids with \(\vp=\vp^{(-)}(c_-,\alphaP)\) on the `(\(-)\)' branch\(\vphantom{\vp^{(+)}}\)}@*)
    HybridPlus,               (*@\codecomment{direct hybrids with \(\vp=\vp^{(+)}(c_-,\alphaP)\) on the `(\(+)\)' branch\(\vphantom{\vp^{(+)}}\)}@*)
    InverseDeflagration,      (*@\codecomment{inverse deflagrations\(\vphantom{\vp^{(+)}}\)}@*)
    InverseDetonation,        (*@\codecomment{inverse detonations with \(\vp=\vp^{(-)}(\xiw,\alphaP)\) on the `(\(-)\)' branch\(\vphantom{\vp^{(+)}}\)}@*)
    InverseDetonationPlus,    (*@\codecomment{inverse detonations with \(\vp=\vp^{(+)}(\xiw,\alphaP)\) on the `(\(+)\)' branch\(\vphantom{\vp^{(+)}}\)}@*)
    InverseHybrid,            (*@\codecomment{inverse hybrids\(\vphantom{\vp^{(+)}}\)}@*)
    MixedHybridMinus,         (*@\codecomment{mixed hybrids with \(\vm=c_-\)\(\vphantom{\vp^{(+)}}\)}@*)
    MixedHybridPlus           (*@\codecomment{mixed hybrids with \(\vp=c_+\)\(\vphantom{\vp^{(+)}}\)}@*)
};
\end{cppcode}

In the \python interface, the corresponding class is accessible as \lstinline[style=inlinepy]{hydrodex.FluidProfile} once the \lstinline[style=inlinepy]{hydrodex} module is imported.
The constructor in terms of the wall velocity \xiw and transition strength \alphaN is equivalent to
\begin{pythoncode}{codecaption={Main constructor for the \lstinline[style=inlinepy]{FluidProfile} \python class.}{lst:FluidProfile-py}}
class FluidProfile:
    def __init__(
        self,
        xi_w: float,       (*@\codecomment{wall velocity \xiw}@*)
        alpha_N: float,    (*@\codecomment{pseudotrace strength parameter \alphaN (far in front of the wall)}@*)
        cp: float = <default value>,    (*@\codecomment{sound speed \(c_+\) ahead of the wall}@*)
        cm: float = <default value>,    (*@\codecomment{sound speed \(c_-\) behind the wall}@*)
(*@\lstnonumber@*)        (*@\codecomment{step size for the fluid velocity (or coordinate) when solving \cref{eq:v-ode-3d} numerically}@*)
        step_size: float = <default value>,
(*@\lstnonumber@*)        (*@\codecomment{switch determining which differential equation for the velocity profile is solved}@*)
        solve_mode: hydrodex.profile_calculator.ProfileSolveMode = <default value>
    ):
        ...
\end{pythoncode}
\noindent%
where the default values are taken from the \cpp library.
They can be changed for the \lstinline[style=inlinepy]{step_size} and \lstinline[style=inlinepy]{solve_mode} parameters using the \lstinline[style=inlinepy]{hydrodex.profile_calculator.set_profile_step_size} and \lstinline[style=inlinepy]{hydrodex.profile_calculator.set_profile_solve_mode} functions, respectively. 

The \python \lstinline[style=inlinepy]{FluidProfile} class has the same member functions as the corresponding \cpp class listed in \cref{lst:FluidProfile-functions}.
Instead of a \lstinline[style=inlinecpp]{std::vector<double>}, the corresponding \python functions return a \lstinline[style=inlinepy]{numpy.array}~\cite{Harris:2020xlr}.
The function \lstinline[style=inlinepy]{hydrodex.FluidProfile.get_solution_type} returns the solution branch as a \lstinline[style=inlinepy]{hydrodex.profile_calculator.HydroSolutionType} \lstinline[style=inlinepy]{Enum} object as in \cref{lst:HydroSolutionType}.

\subsection{Gravitational wave spectra}

From the fluid profile, the corresponding gravitational-wave signal can be calculated within the SSM using the \lstinline[style=inlinecpp]{SoundShellSpectrum} class declared in the \lstinline[style=inlinecpp]{hydrodex/sound_shell_spectrum.hpp} header.
The class can be constructed either from a pre-computed \lstinline[style=inlinecpp]{hydrodex::FluidProfile} instance or directly from the values of the wall velocity \xiw and transition strength \alphaN.
\begin{cppcode}{codecaption={Constructors of the \lstinline[style=inlinecpp]{SoundShellSpectrum} \cpp class.}{lst:SoundShellSpectrum-cpp}}
hydrodex::SoundShellSpectrum(
    const hydrodex::FluidProfile& profile,    (*@\codecomment{fluid profile}@*)
    double R_H,    (*@\codecomment{mean bubble separation in units of the comoving Hubble radius, \(\mathcal{H}_* R_*\)}@*)
    hydrodex::sound_shell_model::NucleationHistory nhist    (*@\codecomment{nucleation history}@*)
        = hydrodex::sound_shell_model::NucleationHistory::EXPONENTIAL,
    hydrodex::sound_shell_model::ExpansionHistory exhist    (*@\codecomment{expansion history}@*)
        = hydrodex::sound_shell_model::ExpansionHistory::RADIATION_DOMINATION,
    double cs = hydrodex::constants::cs_bag,    (*@\codecomment{sound speed \(c_s\) in the fluid}@*)
    double omega = 1.0 / 3.0,    (*@\codecomment{constant background equation-of-state parameter \(\omega\)}@*)
    double Omega_ini = 1.0       (*@\codecomment{source-fluid energy-density fraction \(\bar\Omega_*\)}@*)
);

hydrodex::SoundShellSpectrum(
    double xi_w,       (*@\codecomment{wall velocity \xiw}@*)
    double alpha_N,    (*@\codecomment{pseudotrace strength parameter \alphaN}@*)
    double R_H,        (*@\codecomment{mean bubble separation in units of the comoving Hubble radius, \(\mathcal{H}_* R_*\)}@*)
    ...
    double Omega_ini = 1.0,    (*@\codecomment{source-fluid energy-density fraction \(\bar\Omega_*\)}@*)
    double cp = hydrodex::constants::cs_bag,    (*@\codecomment{sound speed \(c_+\) ahead of the wall}@*)
    double cm = hydrodex::constants::cs_bag,    (*@\codecomment{sound speed \(c_-\) behind the wall}@*)
(*@\lstnonumber@*)    (*@\codecomment{step size for the fluid velocity (or coordinate) when solving \cref{eq:v-ode-3d} numerically}@*)
    double step_size= hydrodex::profile_calculator::profile_step_size,
(*@\lstnonumber@*)    (*@\codecomment{\texttt{ProfileSolveMode} switch determining which differential equation for the velocity profile is solved}@*)
    hydrodex::profile_calculator::ProfileSolveMode solve_mode
        = hydrodex::profile_calculator::profile_solve_mode
);
\end{cppcode}
In addition to the \lstinline[style=inlinecpp]{hydrodex::FluidProfile} instance (or the wall velocity \xiw and strength \alphaN), the constructor requires the mean bubble separation \(R_* \mathcal{H}_*\) normalized to the comoving Hubble radius.
The nucleation history entering  \cref{eq:ssm-single-bubble-amplitudes} can be set with the optional \lstinline[style=inlinecpp]{nhist} parameter by passing the corresponding \lstinline[style=inlinecpp]{hydrodex::sound_shell_model::NucleationHistory} enumerator.
Possible options are \lstinline[style=inlinecpp]{NucleationHistory::EXPONENTIAL} for exponential nucleation (default) and \lstinline[style=inlinecpp]{NucleationHistory::SIMULTANEOUS} for simultaneous nucleation. 
Similarly, the expansion history of the Universe is controlled by passing a \lstinline[style=inlinecpp]{hydrodex::sound_shell_model::ExpansionHistory} enumerator as the optional \lstinline[style=inlinecpp]{exhist} parameter.
\hydrodex currently provides the following expansion history options:
\begin{cppcode}{codecaption={\lstinline[style=inlinecpp]{enum class ExpansionHistory} selector}{lst:ExpansionHistory}}
enum class hydrodex::sound_shell_model::ExpansionHistory{
    FLAT_UNIVERSE,           (*@\codecomment{for a flat universe without expansion}@*)
    RADIATION_DOMINATION,    (*@\codecomment{for a radiation dominated universe (\(\omega=1/3)\))}@*)
    CONSTANT_EOS,            (*@\codecomment{for an expanding universe with equation-of-state parameter \(\omega\)}@*)
    NONE                     (*@\codecomment{no time integration, for calculating the autocorrelation \(\zeta_\Pi\)}@*)
};
\end{cppcode}
\noindent
The arguments \lstinline[style=inlinecpp]{cs} and \lstinline[style=inlinecpp]{omega} control the speed-of-sound \(c_s\) in the sound-wave plasma and the equation-of-state parameter \(\omega\) in \cref{eq:ssm-two-eos}, respectively.
They default to \(c_s=1/\sqrt{3}\) and \(\omega=1/3\), respectively, where the value of \lstinline[style=inlinecpp]{omega} is ignored unless the expansion history is set to \lstinline[style=inlinecpp]{ExpansionHistory::CONSTANT_EOS}.
If the \lstinline[style=inlinecpp]{SoundShellSpectrum} is constructed from a precomputed \lstinline[style=inlinecpp]{FluidProfile}, \lstinline[style=inlinecpp]{cs} can be set to a negative value to automatically set \(c_s=c_-\) using the value of \(c_-\) stored in the fluid profile.
Finally, the argument \lstinline[style=inlinecpp]{Omega_ini} controls the value of the fractional energy-density \(\bar\Omega_*\) of the fluid in \cref{eq:ssm-kinetic-normalization}.
If the \lstinline[style=inlinecpp]{SoundShellSpectrum} is constructed passing \xiw and \alphaN directly, the user can further control the sound-speeds \(c_\pm\) on either side of the wall, the integration step size and the integration strategy as in the \lstinline[style=inlinecpp]{FluidProfile} constructor in \cref{lst:FluidProfile-cpp}.
Further getter and setter functions to access and modify these settings can be found in the documentation.

The SSM calculation then proceeds by subsequently computing the Fourier-transformed single-bubble profiles and shape function \(|\mathcal{A}_+|^2(\chi)\), the dimensionless kinetic spectrum \(E_\text{kin}(q R_*)/R_*\) and the gravitational-wave spectrum \(\Omega_\text{GW}(k R_*)\) using \cref{eq:ssm-single-bubble-amplitudes,eq:ssm-final-spectrum}.

For the shape function \(|\mathcal{A}_+|^2(\chi)\), the user first has to set the corresponding values of the \(\chi\) array with the \lstinline[style=inlinecpp]{set_chi} member function of the \lstinline[style=inlinecpp]{SoundShellSpectrum} class.
The function either takes a \lstinline[style=inlinecpp]{std::vector<double>} vector of \(\chi\) values or a minimal and maximal value and the number of points to be logarithmically spaced between these points (including the start and end value).
The shape function is then calculated calling \lstinline[style=inlinecpp]{calculate_A_squared} member function.
It can either be called without arguments, in which it uses the previously set \(\chi\) array, or with the same arguments that \lstinline[style=inlinecpp]{set_chi} accepts, in which case the \(\chi\) vector is set accordingly before calculation.
The function returns an \lstinline[style=inlinecpp]{std::pair} of \lstinline[style=inlinecpp]{std::vector<double>} vectors with the corresponding values of \(f'(\chi)\) and \(\ell(\chi)\) as defined in \cref{eq:ssm-f-and-l} and stores the values of \(|\mathcal{A}_+|^2(\chi)\) internally.
The Fourier integrals are evaluated analytically assuming that the fluid profile is piecewise linear between the respective \(\xi\) values.
The function \lstinline[style=inlinecpp]{get_shape_function} returns the vector of \(|\mathcal{A}_+|^2(\chi)\) values.
If \lstinline[style=inlinecpp]{calculate_A_squared} has not been called, it calls the function internally.

\begin{cppcode}{codecaption={Routines for calculating the shape function \(|\mathcal{A}_+|^2\)}{lst:A2-functions}}
(*@\lstnonumber\codecomment{set the values of \(\chi\) at which the shape function \(|\mathcal{A}_+|^2\) is evaluated (cf. \cref{eq:ssm-single-bubble-amplitudes}):}@*)
void hydrodex::SoundShellSpectrum::set_chi(double chi_min, double chi_max, size_t N);
void hydrodex::SoundShellSpectrum::set_chi(const std::vector<double>& chi_vec);
(*@\lstnonumber\codecomment{calculate \(f'(\chi)\), \(\ell(\chi)\) and \(|\mathcal{A}_+(\chi)|^2\) (cf.\ \cref{eq:ssm-single-bubble-amplitudes,eq:ssm-f-and-l}):}@*)
std::pair<std::vector<double>,std::vector<double>>
    hydrodex::SoundShellSpectrum::calculate_A_squared(...);
    // call without arguments or with same argument types as set_chi
(*@\codecomment{get the shape function \(|\mathcal{A}_+(\chi)|^2\):}@*)
const std::vector<double>& hydrodex::SoundShellSpectrum::get_shape_function();
\end{cppcode}

Similarly, to calculate the dimensionless kinetic spectrum \(E_\text{kin}(qR_*)/R_*\), the \(q R_*\) vector must be set using the \lstinline[style=inlinecpp]{set_qR} function or passed to the \lstinline[style=inlinecpp]{calculate_kinetic_spectrum} call, which accept the same types of arguments as \lstinline[style=inlinecpp]{set_chi} and \lstinline[style=inlinecpp]{calculate_A_squared}, respectively.
The kinetic spectrum is calculated from \cref{eq:ssm-single-bubble-amplitudes} evaluating the integral using the trapezoidal rule on the precalculated \((\chi,|\mathcal{A}_+|^2)\) values, and stored in the class instance.
It can be retrieved as a \lstinline[style=inlinecpp]{std::vector<double>} by calling the \lstinline[style=inlinecpp]{get_kinetic_spectrum} function of the \lstinline[style=inlinecpp]{SoundShellSpectrum} instance.
If not already calculated, \lstinline[style=inlinecpp]{calculate_A_squared} and \lstinline[style=inlinecpp]{calculate_kinetic_spectrum} are called internally.
The mean-squared fluid velocity \(\bar{U}_f^2\) and the kinetic energy density parameter \(\Omega_K\) in \cref{eq:ssm-kinetic-normalization} can be obtained integrating the kinetic spectrum using the trapezoidal rule by calling \lstinline[style=inlinecpp]{get_mean_squared_fluid_velocity} and \lstinline[style=inlinecpp]{get_kinetic_energy_density_parameter}, respectively.

\begin{cppcode}{codecaption={Routines for calculating the dimensionless kinetic spectrum \(E_\mathrm{kin}(q R_*)/R_*\)}{lst:Ekin-functions}}
(*@\lstnonumber\codecomment{set the values of \(q R_*\) at which the dimensionless kinetic spectrum \(E_\text{kin}(q R_*)/R_*\) is evaluated:}@*)
void hydrodex::SoundShellSpectrum::set_qR(...); // same argument types as set_chi
(*@\lstnonumber\codecomment{compute \(E_\mathrm{kin}(q R_*)/R_*\) (cf. \cref{eq:ssm-single-bubble-amplitudes}):}@*)
void hydrodex::SoundShellSpectrum::calculate_kinetic_spectrum(...);
    // same argument types as calculate_A_squared
(*@\lstnonumber\codecomment{get the kinetic spectrum \(E_\mathrm{kin}(q R_*)/R_*\):}@*)
const std::vector<double>& hydrodex::SoundShellSpectrum::get_kinetic_spectrum();
(*@\lstnonumber\codecomment{compute the mean-squared fluid velocity \(\bar{U}_f^2\):}@*)
double hydrodex::SoundShellSpectrum::get_mean_squared_fluid_velocity();
(*@\lstnonumber\codecomment{compute the kinetic energy density parameter \(\Omega_K\):}@*)
double hydrodex::SoundShellSpectrum::get_kinetic_energy_density_parameter();
\end{cppcode}

The momenta \(k R_*\) on which the gravitational-wave spectrum is evaluated can be set in units of the mean bubble distance using the \lstinline[style=inlinecpp]{set_kR} function. 
The function accepts the same types of arguments as \lstinline[style=inlinecpp]{set_chi} or \lstinline[style=inlinecpp]{set_qR} with an additional parameter \lstinline[style=inlinecpp]{N_qR} as the last argument.
If provided, the \(q R_*\) array on which the kinetic spectrum is calculated is set automatically from \(k R_*\) as logarithmically spaced with  \lstinline[style=inlinecpp]{N_qR} steps\footnote{%
	The range of \(Q=q R_*\) is chosen to extend by one logarithmic step and an additional factor \((1\mp c_s)/(2 c_s)\) below the minimum and above the maximum value of the \(K=k R_*\) range, i.e.\ \(N_Q\) logarithmic steps between \(Q_0 = K_0 \times (K_0/K_1) \times (1-c_s)/(2 c_s)\) and \(Q_{N_Q-1} = K_{N_K-1} \times (K_{N_K-1}/K_{N_K-2}) \times (1+c_s)/(2 c_s)\), where the \(c_s\)-dependent factors come from the integration limits of \(q\) in the inifinite source-duration limit \cref{eq:ssm-kernel-infinite-duration}.%
}.
If \lstinline[style=inlinecpp]{N_qR} is omitted or set to  \lstinline[style=inlinecpp]{-1} (or  \lstinline[style=inlinepy]{None} in the \python interface), \(q R_*\) needs to be set using  \lstinline[style=inlinecpp]{set_qR}, or after setting \(k R_*\) calling  \lstinline[style=inlinecpp]{set_qR_from_kR}.  

Furthermore, to evaluate the time integration in \cref{eq:ssm-final-spectrum} the source duration \(\mathcal{H}_* \delta \tau_\text{fin}\) in units of the Hubble time is required. 
It can be set using the  \lstinline[style=inlinecpp]{set_source_duration} function or estimated from the shock-formation time, \(\mathcal{H}_*\delta\tau = \mathcal{H}_* R_* /\sqrt{\Omega_K/\bar\Omega_*} = \mathcal{H}_* R_*/(\sqrt{\bar\Gamma}\,\bar{U}_f)\), cf.\ \cref{sec:ssm-results}, calling   \lstinline[style=inlinecpp]{calculate_source_duration}.
The latter is the default if \lstinline[style=inlinecpp]{set_source_duration} is not called explicitly.

Finally, the gravitational-wave spectrum at production in \cref{eq:ssm-final-spectrum} is calculated calling the routine \lstinline[style=inlinecpp]{calculate_gw_spectrum} (which accepts the same type of parameters as \lstinline[style=inlinecpp]{set_kR} or can be called without arguments) and retrieved as a \lstinline[style=inlinecpp]{std::vector<double>} using the function \lstinline[style=inlinecpp]{get_gravitational_wave_spectrum}.
The momentum integration is performed using the trapezoidal rule, with the kinetic spectrum interpolated linearly in logarithmic space on the precalculated values. 
The number of points in the \(P\) and \(z\) integration of \cref{eq:ssm-final-spectrum} can be set with \lstinline[style=inlinecpp]{set_correlator_integration_settings}.
The respective default values are controlled by \lstinline[style=inlinecpp]{sound_shell_model::N_pR_default} (5000) and \lstinline[style=inlinecpp]{sound_shell_model::N_z_default} (100).
The function further sets which form of the time-integration kernel in \cref{eq:ssm-kernel-sum} is used, passing the corresponding \lstinline[style=inlinecpp]{sound_shell_model::MomentumIntegration}  enumerator.
To evaluate the full form of the kernel (default), pass \lstinline[style=inlinecpp]{MomentumIntegration::FULL} as the third argument of \lstinline[style=inlinecpp]{set_correlator_integration_settings}, while passing \lstinline[style=inlinecpp]{MomentumIntegration::LOW_FREQUENCY} or \lstinline[style=inlinecpp]{MomentumIntegration::INFINITE_DURATION} corresponds to the low-frequency approximation and infinite-duration limit of the integral, \cref{eq:ssm-kernel-low-f,eq:ssm-kernel-infinite-duration},  respectively.%
\footnote{%
    The numerical evaluation differs slightly from the implementation discussed in Ref.~\cite{Giombi:2024kju}.
    There, the generalized sine and cosine integrals are expressed in terms of incomplete Gamma and confluent hypergeometric functions, with special care required near vanishing argument, where the individual generalized functions can be singular although the combination entering the kernel remains finite.
    In \hydrodex we instead evaluate directly the finite differences appearing in \cref{eq:ssm-kernel-sum}, i.e.\ the oscillatory integrals between the two physical time arguments, using the \texttt{qawo} routine implemented in the GNU Scientific Library (GSL) \cite{Galassi:2019czg}.  
    This avoids constructing the generalized trigonometric functions separately and therefore also avoids the corresponding cancellation between their endpoint values.
    The radiation case \(\sigma=0\) is treated separately using the ordinary GSL sine and cosine integrals.%
}

\begin{cppcode}{codecaption={Routines for calculating the gravitational-wave spectrum at production, \(\Omega_\text{GW}(k R_*)\)}{lst:OmegaGW-functions}}
(*@\lstnonumber\codecomment{set the values of \(k R_*\) at which the  gravitational-wave spectrum \(\Omega_\text{GW}(k R_*)\) is evaluated:}@*)
void hydrodex::SoundShellSpectrum::set_kR(..., size_t N_qR=static_cast<size_t>(-1));
    // first arguments are the same as set_chi
(*@\lstnonumber\codecomment{set the values of \(q R_*\) at which the kinetic spectrum is calculated from the \(k R_*\) array:}@*)
void hydrodex::SoundShellSpectrum::set_qR_from_kR(size_t N);
(*@\lstnonumber\codecomment{set the source duration \(\mathcal{H}_*\delta\tau_\text{fin}\):}@*)
void hydrodex::SoundShellSpectrum::set_source_duration(double duration);
(*@\lstnonumber\codecomment{calculate the source duration from the shock-formation time:}@*)
void hydrodex::SoundShellSpectrum::calculate_source_duration();
(*@\lstnonumber\codecomment{specify the momentum integration settings:}@*)
void hydrodex::SoundShellSpectrum::set_correlator_integration_settings(
    size_t n_p, 
    size_t n_z, 
    MomentumIntegration integ = MomentumIntegration::FULL
);
(*@\lstnonumber\codecomment{compute the gravitational-wave spectrum  \(\Omega_\text{GW}(k R_*)\) at production:}@*)
void hydrodex::SoundShellSpectrum::calculate_gw_spectrum(...); 
    // call without arguments or with same argument types as set_kR
(*@\lstnonumber\codecomment{get the gravitational-wave spectrum  \(\Omega_\text{GW}(k R_*)\) at production:}@*)
const std::vector<double>& hydrodex::SoundShellSpectrum::get_gravitational_wave_spectrum()
(*@\lstnonumber\codecomment{get the values of \(kR_*\) at which the spectrum is evaluated:}@*)
const std::vector<double>& hydrodex::SoundShellSpectrum::get_kR() const;
\end{cppcode}

The \lstinline[style=inlinepy]{hydrodex.SoundShellSpectrum} class of the \python interface can be used in the same way as the corresponding \cpp class. 
As its \cpp version, it can be constructed from either a  \lstinline[style=inlinepy]{FluidProfile} instance or from the values of \xiw and \alphaN. 
For the former case, the constructor is equivalent to
\begin{pythoncode}{codecaption={Constructors of the \lstinline[style=inlinecpp]{SoundShellSpectrum} \python class.}{lst:SoundShellSpectrum-python}}
class SoundShellSpectrum:
  def __init__(
    self,
    profile: hydrodex.FluidProfile,    (*@\codecomment{fluid profile}@*)
    H_R: float,    (*@\codecomment{mean bubble separation in units of the comoving Hubble radius, \(\mathcal{H}_* R_*\)}@*)
    nhist: hydrodex.sound_shell_model.NucleationHistory = <default_value>, (*@\codecomment{nucl.\ hist.}@*)
    exhist: hydrodex.sound_shell_model.ExpansionHistory = <default_value>, (*@\codecomment{exp.\ hist.}@*)
    cs: float = <default_value>,    (*@\codecomment{sound speed \(c_s\) in the fluid}@*)
    omega: float = 1.0/3.0,         (*@\codecomment{constant background equation-of-state parameter \(\omega\)}@*)
    Omega_star: float = 1.0          (*@\codecomment{source-fluid energy-density fraction \(\bar\Omega_*\)}@*)
  ):
    ...
\end{pythoncode}
\noindent
The default values for the number of points in the momentum integration can be controlled with the corresponding setter functions \lstinline[style=inlinepy]{hydrodex.sound_shell_model.set_N_pR_default} and \lstinline[style=inlinepy]{hydrodex.sound_shell_model.set_N_z_default}, respectively.

\subsection{Minimal workflow for the gravitational-wave signal}

A minimal working example for calculating the stationary fluid profile and the corresponding gravitational-wave spectrum within the SSM starting from the macroscopic transition parameters \((\xiw,\alphaN,c_+,c_-)\) is given below.
The workflow in \cpp is:
\begin{cppcode}{codecaption={Minimal workflow for calculating the gravitational-wave spectrum in \cpp.}{lst:example-cpp}}
#include <vector>
#include "hydrodex/fluid_profile.hpp"
#include "hydrodex/sound_shell_model.hpp"
#include "hydrodex/sound_shell_spectrum.hpp"

using namespace hydrodex;
namespace ssm = hydrodex::sound_shell_model;
using vecd = std::vector<double>;

int main() {
// sound speeds (*@\codecomment{(\(c_+=0.75,\ c_-=0.4\))}@*)
double c_plus = 0.75, c_minus = 0.40;

//  calculate the fluid profile (*@\codecomment{for \(\xiw=0.6\) and \(\alphaN = 0.1\)}@*)
FluidProfile profile(0.6, 0.1, c_plus, c_minus);

// initialize the sound shell calculator
SoundShellSpectrum spec(
	profile,
	0.01,  // mean bubble distance, (*@\codecomment{\(\mathcal{H}_* R_*=0.01\)}@*)
	ssm::NucleationHistory::EXPONENTIAL,    // exponential nucleation
	ssm::ExpansionHistory::CONSTANT_EOS,    // expanding Universe
	c_minus,         // speed of sound in the fluid, (*@\codecomment{\(c_s = c_-\)}@*)
	c_minus*c_minus, // equation-of-state parameter of the Universe, (*@\codecomment{\(\omega = c_-^2\)}@*)
	1.0              // fractional energy density of the fluid, (*@\codecomment{\(\Omega_*=1\)}@*)
);

// set the source duration
spec.set_source_duration(0.1);    // source lifetime, (*@\codecomment{\(\mathcal{H}_* \delta\tau_\mathrm{fin}=0.1\)}@*)
// use the full time integration kernel with 100/48 steps in the P/z integration
spec.set_correlator_integration_settings(100, 48, ssm::MomentumIntegration::FULL);

// calculate A^2 for 10k log-spaced chi-values between 0.01 and 1e4
spec.calculate_A_squared(0.01, 1e4, 10000); 
// use 100 log-spaced kR-values between 0.01 and 5000, and 400 values for qR
spec.calculate_gw_spectrum(0.01, 5000., 100, 400);
vecd kR = spec.get_kR();    // get the values of k*R
vecd Omega_GW = spec.get_gravitational_wave_spectrum();    // GW spectrum

return 0;
}
\end{cppcode}
\noindent
If \hydrodex is installed in a standard location, the code can be compiled simply by running
\begin{shellcode}{}
$CXX -std=c++17 example.cpp -lhydrodex -o example
\end{shellcode}
\noindent
where \texttt{\$CXX} must be a \cpp17 compatible compiler.
Otherwise, if \hydrodex is installed in a custom location, the compilation command must be adapted to specify the corresponding include and library paths. 
\begin{shellcode}{}
$CXX -std=c++17 -I<install_path>/include -L<install_path>/lib -Wl,-rpath,<install_path>/lib example.cpp -lhydrodex -o example
\end{shellcode}
\noindent
Likewise, if \texttt{OpenMP} is not available in the default include search paths, the corresponding include flag for the \texttt{omp.h} header must be specified.

In \python, the same spectrum can be obtained as follows:
\begin{pythoncode}{codecaption={Minimal \python workflow.}{lst:minimal-model-interface-python}}
import numpy as np
from hydrodex import FluidProfile, SoundShellSpectrum
from hydrodex import sound_shell_model as ssm

# sound speeds (*@\codecomment{(\(c_+=0.75,\ c_-=0.4\))}@*)
c_plus, c_minus  = 0.75, 0.40

# calculate the fluid profile (*@\codecomment{for \(\xiw=0.6\) and \(\alphaN = 0.1\)}@*)
profile = FluidProfile(xi_w=0.6, alpha_N=0.1, cp=c_plus, cm=c_minus)

# initialize the sound shell calculator (*@\codecomment{for \(\mathcal{H}_* R_*=0.01\), \(c_s=c_-\), \(\omega=c_-^2\) and \(\Omega_* = 1.0\)}@*)
spec = SoundShellSpectrum(
	profile = profile,
	H_R = 0.01,
	nhist = ssm.NucleationHistory.EXPONENTIAL,
	exhist = ssm.ExpansionHistory.CONSTANT_EOS,
	cs = c_minus,
	omega = c_minus**2,
	Omega_star = 1.0,
)

# set the source duration (*@\codecomment{(\(\mathcal{H}_* \delta\tau_\mathrm{fin}=0.1\))}@*) and integration settings
spec.set_source_duration(0.1)
spec.set_correlator_integration_settings(100, 48, ssm.MomentumIntegration.FULL)

# calculate the spectra
kR  = np.logspace(np.log10(1.e-2), np.log10(5.e3), 100)
spec.calculate_A_squared(0.01, 1e4, 10000)
spec.calculate_gw_spectrum(kR, 400) # use 400 values for qR
Omega_GW = spec.get_gravitational_wave_spectrum()
\end{pythoncode}
\noindent
The resulting arrays contain the dimensionless production spectrum \(\Omega_{\rm GW,*}(kR_*)\), where \(R_*\) is the characteristic bubble separation.
The numerical grids entering the Fourier transforms and momentum convolutions should be increased until the spectrum is stable over the range of interest.

%
\section{Conclusions}
\label{sec:conclusions}
%

We have presented a complete classification of stationary self-similar hydrodynamic profiles for first-order phase transitions in the constant-speed-of-sound equation of state, allowing the two phases to have independent sound speeds \(c_+\) and \(c_-\).
Going beyond the equal-sound-speed limit does more than quantitatively deform the familiar detonation, deflagration, and hybrid solutions of  generations~I and~II, i.e.\ direct and inverse profiles.
It enlarges the space of possible fluid configurations by generation~III of mixed hybrids, in which the flow reverses direction across the wall.
These solutions arise naturally when the wall velocity lies between the two sound speeds, \(c_-<\xiw<c_+\), and disappear in the equal-sound-speed limit.

A central result of this work is that unequal sound speeds can also invalidate the usual two-parameter description of stationary bubble hydrodynamics.
The local matching problem is naturally formulated in terms of the strength \alphaP immediately ahead of the wall, whereas the physical input \alphaN is defined in the unperturbed plasma.
Heating, cooling, compression, or rarefaction ahead of the wall makes the relation between these two quantities depend on the entire fluid profile.
For a sufficiently large hierarchy \(c_+/c_-\), the map \(\alphaP\mapsto\alphaN\) becomes non-monotonic and ceases to be one-to-one.
Distinct wall states can then project onto the same \((\xiw,\alphaN)\), and we find regions in which up to three stationary profiles coexist for the same values of these macroscopic parameters.
Moreover, the identification of positive and negative \(\alphaN\) with direct and inverse hydrodynamics, respectively, which holds for the standard branches in the bag model and more generally when \(c_+\leq c_-\), no longer applies once \(c_+ \gg c_-\).
Mixed hybrids span both signs of \(\alphaN\) and are not purely direct or inverse. 
For sufficiently small \(c_-\), below \(c_{-,0}^{\mathrm{ID}^+}\) and \(c_{-,0}^{\mathrm{IH}}\), ordinary inverse solutions also extend to positive \(\alphaN\).
 
This non-uniqueness should not, however, be interpreted as implying that a fixed microscopic theory necessarily admits multiple physical stationary solutions.
The macroscopic parameter set \((\xiw,\alphaN,c_+,c_-)\) does not retain the relative thermodynamic normalization of the two phases or the entropy produced at the wall.
We have shown explicitly that, in local thermal equilibrium, the competing profiles in the overlap regions require different values of the thermodynamic normalization \(\Psi_N\).
Fixing the microscopic equation of state therefore selects at most one of them in the benchmark studied here.
More generally, specifying the entropy production provides the additional information needed to complete the stationary problem, whereas imposing only non-negative entropy production is not sufficient.  
The folds found here thus reflect the non-injectivity of the map from local wall states to the macroscopic variables \((\xiw,\alphaN)\); whether any physical degeneracy survives must instead be established within a complete microscopic model.

The sound-speed hierarchy required to reach the non-injective regime is, however, difficult to realize in realistic microscopic models.
Radiation-dominated and electroweak-like thermal transitions typically remain close to the conformal regime and are therefore expected to lie well outside the regions where the folds become macroscopically relevant.
Softer QCD-like or nuclear equations of state can explore a wider range of sound speeds, while finite-density systems and stiff charged condensates provide more natural routes towards the strongly hierarchical regime.
These examples should presently be regarded as thermodynamic targets rather than explicit realizations of the overlapping bubble solutions.
Nevertheless, they show that the relevant region of the sound-speed plane is not intrinsically unphysical.

We have also revisited the energy budget beyond the bag model.
In a general CSS equation of state, the hydrodynamic character of a solution and the energetic role of the vacuum contribution need no longer coincide.
The pseudo-trace strength \(\alphaN\) controls the hydrodynamic matching, whereas the signed vacuum-energy difference determines whether vacuum energy is released during the transition or must instead be supplied by the initial thermal reservoir. 
In the bag model the two coincide, \(\alphaN=\alphavac[N]\), and for equal sound speeds their signs remain correlated.
Once \(c_+\neq c_-\), the additional thermal contribution to the pseudo-trace can drive \(\alphaN\) through zero independently of the vacuum-energy difference.
The hydrodynamic classification and the energetic interpretation can therefore become genuinely decoupled, which is particularly relevant for inverse transitions.

The same fluid profiles provide the initial conditions for the acoustic gravitational-wave source. 
Extending the sound-shell calculation to the full CSS classification, we find that the dependence on the acoustic sound speed is qualitatively common to direct and inverse solutions. 
Increasing \(c_s\) moves the characteristic gravitational-wave scale to larger wavenumbers and enhances the signal.
Mixed hybrids exhibit spectral shapes similar to those of ordinary hybrids.
Despite their oppositely directed fluid flows, they introduce no additional features, and their gravitational-wave spectra can be reproduced by ordinary direct or inverse hybrids with different profiles and even opposite signs of \(\alphaN\).
The gravitational-wave signal alone therefore does not discriminate between these hydrodynamic branches.
For the benchmarks considered here, the spectral shape is more sensitive to the acoustic lifetime than to the equation of state controlling the homogeneous expansion. 

Finally, we release \hydrodex, which implements the complete CSS classification, constructs the corresponding self-similar profiles, computes their energy budget, and propagates them to the sound-shell gravitational-wave signal.
We have cross-validated the code against existing public implementations wherever their physical domains overlap.
The independent calculations agree once conventions and implemented approximations are matched, while \hydrodex provides the broadest hydrodynamic coverage and the overall best performance in the like-for-like benchmarks considered here.
It therefore provides a practical framework for systematic studies of phase-transition hydrodynamics beyond the conformal bag model.

\paragraph{Outlook.}

The most immediate extension is to move beyond the CSS approximation and ask whether the qualitative structure identified here survives for general equations of state with state-dependent sound speeds.
The non-injectivity of the map \((\alphaP,\xiw)\to(\alphaN,\xiw)\) originates from the competition between the local matching at the wall and the hydrodynamic evolution connecting the wall to the asymptotic plasma, and is therefore not obviously tied to constant sound speeds.

A natural generalization is to phase transitions at finite density, where the equation of state depends on both temperature and chemical potential and the matching conditions involve an additional conserved current.
This extra constraint modifies the set of allowed stationary solutions and may remove or reshape the overlap regions found here.
At the same time, dense and strongly interacting matter can realize much larger sound-speed hierarchies than radiation-dominated plasmas, which makes it the most promising setting for a sizeable effect.
The CSS hydrodynamics at finite density will be presented in Ref.~\cite{Barni:2026cssfiniteDensity}, while its application to realistic systems, from degenerate Fermi gases to neutron-star matter, will be discussed in Ref.~\cite{Barni:2026FiniteDensity}.

A second open question concerns the selection among stationary solutions.
Our LTE analysis shows that, in the examples studied here, the degeneracy is lifted once the thermodynamic normalization is fixed.
This does not exhaust the possible microscopic closures.
Acomplete treatment should include entropy production at the wall, the microscopic pressure and friction, and the stability of the corresponding steady states.
At finite density the additional conserved charge enters the LTE conditions as well, so it remains to be seen whether the degeneracy is still lifted in that case; this question will be addressed in Ref.~\cite{Barni:2026cssfiniteDensity}.

Ultimately, these questions should be addressed directly in microscopic models, both for conventional thermal transitions, where the effect is expected to be small, and for strongly coupled or finite-density theories with large sound-speed hierarchies~\cite{Barni:2026ecg}.
Such studies will determine to what extent the loss of universality found here has observable consequences for reheating, kinetic-energy production, and the resulting acoustic gravitational-wave signal.

\paragraph{Acknowledgements}
It is a pleasure to thank Simone Blasi, Carlo Branchina, Marco Finetti, Helena Kole\v{s}ov\'a, David Mateos, Jos\'e Miguel No, and Mikel Sanchez-Garitaonandia for helpful discussions and feedback on the manuscript.
Research at the IFT is supported by the Spanish Research Agency (Agencia Estatal de Investigaci\'on, AEI; MCIN/AEI/10.13039/501100011033) through the IFT Severo Ochoa centre of Excellence grant CEX2020-001007-S\@. 
E.M. acknowledges support from STFC under IPPP Consolidated and National Centre grants UKRI4227
and UKRI4534, as well as from grants CNS2023-14453 and PID2022-137127NB-I00, funded by AEI, the European Union NextGenerationEU/PRTR, and ESF+\@.
G.B. is supported by the Deutsche Forschungsgemeinschaft under Germany’s Excellence Strategy - EXC 2121 Quantum universe - 390833306, as well as from grant CNS2023-145069, funded by AEI and by the European Union NextGenerationEU/PRTR\@.

\paragraph{Data availability}
The custom code used for the hydrodynamic analysis and to generate the results presented in this work is publicly available at \href{https://github.com/hydrodex-pt/hydrodex}{\faGithub\ \hydrodex}.

\paragraph{AI disclosure.}
Standard AI-assisted programming tools were used during the development of the code underlying the results presented in this work, primarily to improve efficiency in tasks such as debugging and modifying existing code.
Large language models were also used to assist with editing and polishing the manuscript.
They did not independently determine the scientific results or conclusions, which were developed, checked, and verified by the authors.

\newpage
\appendix

%
\section{Exotic branches}
\label{app:exotic-branches}
%

\begin{figure}
    \centering
    \includegraphics[width=\linewidth]{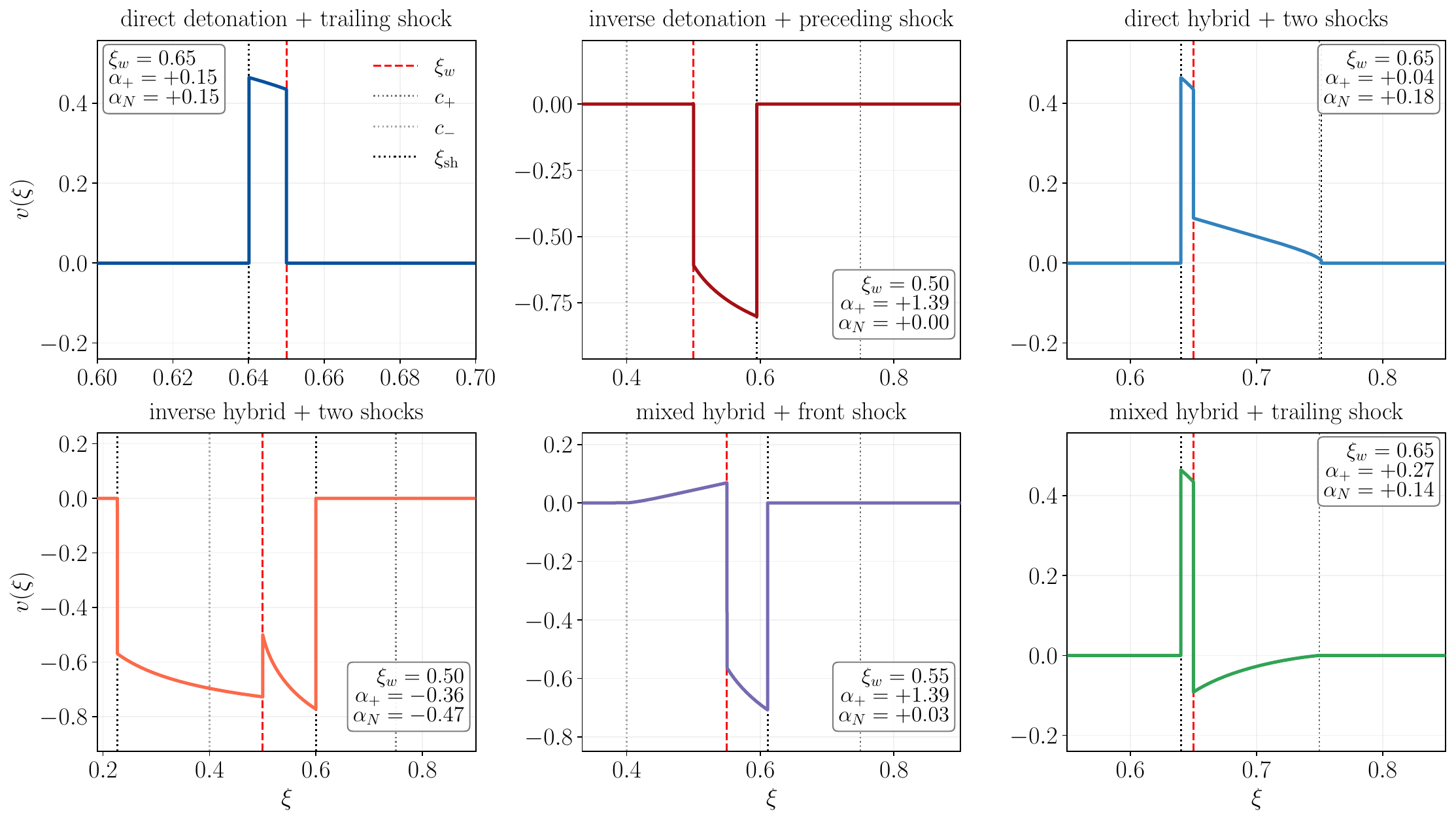}
    \caption{%
    Representative kinematic exotic CSS profiles for \(c_+=0.75\) and \(c_-=0.40\).
    The red dashed line marks the wall position \xiw, the grey dotted lines mark \(c_+\) and \(c_-\), and the black dotted lines mark the shock front positions.
    The boxes report the corresponding values of \xiw, \alphaP, and \alphaN.
    All configurations satisfy the local wall matching, but they do not define physical hydrodynamic branches.
    In each case at least one shock front is traversed as a rarefaction shock and therefore fails both the evolutionarity and entropy conditions.
    }
    \label{fig:exotic-profiles}
\end{figure}

In the main text we classified the self-similar hydrodynamic profiles under the (reasonable) assumption that the absolute value of the fluid velocity reaches its maximum at the bubble wall and decreases when moving away from the wall.
As noted in \cref{fn:other-solutions}, relaxing this assumption enlarges the set of kinematically possible profiles.
One can formally construct additional configurations in which the fluid is connected to the asymptotic state through one or more extra shocks.
Representative examples are shown in \cref{fig:exotic-profiles}.

These configurations solve the local matching conditions at the bubble wall and the Rankine--Hugoniot conditions at the additional discontinuities.
This is not, however, sufficient for them to define physical hydrodynamic branches. 
The bubble wall and every shock are independent discontinuities, and each of them must separately satisfy the appropriate evolutionarity and entropy conditions.

As we show below, the global completion of every candidate considered here necessarily contains at least one same-phase shock crossed in the rarefactive orientation, whereas all deflagration and hybrid solutions in \cref{sec:classification} cross the shock front with a compressive orientation.
By a compressive shock we mean a discontinuity across which the fluid is compressed as it passes from the upstream to the downstream state, with increasing temperature, pressure, and energy density.
A rarefactive shock, {on the other hand}, corresponds to the opposite orientation, in which these quantities decrease across the discontinuity.
For a CSS equation of state, the Rankine--Hugoniot conditions allow both orientations algebraically, but only compressive shocks satisfy the entropy condition.
The rarefactive orientation instead decreases the entropy flux and is therefore forbidden by the second law of thermodynamics \cite{Spillane:2015daa}.

This exclusion is specific to the CSS equation of state.
Rarefaction shocks are not forbidden in general; for sufficiently non-convex equations of state they can satisfy both the entropy and stability conditions \cite{Zamfirescu:2008rsw}.

\subsection{Kinematic candidates}
\label{app:exotic-kinematics}

The possible kinematic candidates are summarized in \cref{tab:exotic-branches}.
At the level of the bubble wall, these configurations are perfectly consistent.
The wall discontinuity obeys the same matching conditions as the ordinary branches and lies within the corresponding kinematically allowed domains.

The obstruction arises only when completing the full self-similar profile.
Whenever an additional shock is introduced, the two states across the discontinuity must satisfy the Rankine--Hugoniot conditions.
These conditions alone, however, do not determine which orientation of the jump is physically admissible.
The same pair of states can formally be connected in either direction, while the entropy condition selects the physical one.
As we show in \cref{app:shock-direction}, every candidate below necessarily contains at least one shock traversed in the entropy-decreasing orientation.

The candidates are:
\begin{enumerate}
  \item a direct detonation with a rarefaction wave ending at a trailing shock behind the wall instead of terminating smoothly at the sonic point \(\xi=c_-\);
  \item an inverse detonation with a rarefaction wave ending at a preceding shock in front of the wall instead of terminating smoothly at the sonic point \(\xi=c_+\);
  \item a direct hybrid completed by shocks on both sides of the wall;
  \item an inverse hybrid completed by shocks on both sides of the wall;
  \item a mixed hybrid, for \(c_+>c_-\), with a shock instead of smooth termination ahead of the wall;
  \item a mixed hybrid, for \(c_+>c_-\), with a shock instead of smooth
  termination behind the wall.
\end{enumerate}
The corresponding wall-frame fluid velocities listed in \cref{tab:exotic-branches} follow from the required ordering of \vp, \vm, \xiw, and the two sound speeds, together with the algebraic conditions required for the additional shocks to exist.

\begin{table}[t]
\centering
\footnotesize
\begin{NiceTabular}[cell-space-limits=3pt]{c@{\hspace{4em}}c@{\hspace{4em}}c}
    \toprule\rowcolor{gray!25}
    \textbf{Solution}
    & \textbf{Wall-frame domain}
    & \textbf{Excluding shock}
    \\\midrule
    \Block[fill=gray!10]{1-3}{\textbf{Direct-type exotic completions}}
    \\\hline
    \Block{1-1}{direct detonation\\with trailing shock}
    & \Block{1-1}{
        \(c_+>\xiw=\vp>c_->\vm\)\\
        \({c_-^2}/{\xiw}<\vm<c_-\)
    }
    & \Block{1-1}{
        rear (`\(-\)' phase)\\
        \(\vwallframe_{\uparrow}^{\rm sh} < c_- < \vwallframe_{\downarrow}^{\rm sh}\)
    }
    \\\hline
    \Block{1-1}{direct hybrid\\with two shocks}
    &
    \Block{1-1}{
        \(0<\vp<\min(c_+,\xiw,{c_+^2}/{\xiw})\)\\
        \({c_-^2}/{\xiw}<\vm<c_-\)
    }
    &
    \Block{1-1}{
        rear (`\(-\)' phase)\\
        \(\vwallframe_{\uparrow}^{\rm sh}<c_-<\vwallframe_{\downarrow}^{\rm sh}\)
    }
    \\\midrule
    \Block[fill=gray!10]{1-3}{\textbf{Inverse-type exotic completions}}
    \\\hline
    \Block{1-1}{inverse detonation\\with preceding shock}
    &
    \Block{1-1}{
        \(\vp>c_+>\vm=\xiw>c_-\)\\
        \(c_+<\vp<{c_+^2}/{\xiw}\)
    }
    &
    \Block{1-1}{
        front (`\(+\)' phase)\\[0.15em]
        \(\vwallframe_{\uparrow}^{\rm sh}<c_+<\vwallframe_{\downarrow}^{\rm sh}\)
    }
    \\\hline
    \Block{1-1}{inverse hybrid\\with two shocks}
    &
    \Block{1-1}{
        \(c_+<\vp<{c_+^2}/{\xiw}\)\\
        \(\max(c_-,\xiw,{c_-^2}/{\xiw})<\vm<1\)
    }
    &
    \Block{1-1}{
        front (`\(+\)' phase)\\[0.15em]
        \(\vwallframe_{\uparrow}^{\rm sh}<c_+<\vwallframe_{\downarrow}^{\rm sh}\)
    }
    \\\midrule
    \Block[fill=gray!10]{1-3}{\textbf{Mixed exotic completions (\(c_+>c_-\))}}
    \\\hline
    \Block{1-1}{mixed hybrid\\with front shock}
    &
    \Block{1-1}{
        \(c_-<\xiw<c_+\)\\
        \(c_+<\vp<{c_+^2}/{\xiw},\quad c_-<\vm<\xiw\)
    }
    &
    \Block{1-1}{
        front (`\(+\)' phase)\\
        \(\vwallframe_{\uparrow}^{\rm sh}<c_+<\vwallframe_{\downarrow}^{\rm sh}\)
    }
    \\\hline
    \Block{1-1}{mixed hybrid\\with trailing shock}
    &
    \Block{1-1}{
        \(c_-<\xiw<c_+\)\\
        \(\xiw<\vp<c_+,\quad {c_-^2}/{\xiw}<\vm<c_-\)
    }
    &
    \Block{1-1}{
        rear (`\(-\)' phase)\\[0.15em]
        \(\vwallframe_{\uparrow}^{\rm sh}<c_-<\vwallframe_{\downarrow}^{\rm sh}\)
    }
    \\\bottomrule
\end{NiceTabular}
\caption{%
    Additional kinematic CSS solutions obtained by relaxing the assumption that the fluid-velocity magnitude decreases away from the wall.
    The last column identifies, for each candidate, an additional shock that is necessarily traversed in the rarefactive orientation, \(\vwallframe_{\uparrow}^{\rm sh}<c_i< \vwallframe_{\downarrow}^{\rm sh}\), and therefore fails the stability and entropy conditions discussed in \cref{app:shock-direction}.
}
\label{tab:exotic-branches}
\end{table}

\subsection{Shock admissibility in the CSS model}
\label{app:shock-direction}

We now determine which orientation of crossing a shock front is physical.
As shocks occur within a single phase, the fluid obeys the same equation of state on the two sides of the discontinuity.
We hence omit the `\(\pm\)' phase index.
Following the notation introduced in \cref{eq:shock-matching}, we denote by \(\vwallframe_{\uparrow}^{\rm sh}\) and \(\vwallframe_{\downarrow}^{\rm sh}\) the fluid velocities in the shock rest frame, respectively upstream and downstream along the physical direction in which the fluid crosses the shock.
The arrows refer to up- and downstream states and should not be confused with the `\(\pm\)' phase labels.

For the CSS equation of state, within a fixed phase we may write
\begin{equation}
  e_i= a T_i^q + \epsilon,
  \qquad
  p_i = c_s^2 a T_i^q - \epsilon,
  \qquad
  w_i=(1+c_s^2) a T_i^q,
  \qquad
  q=1+\frac{1}{c_s^2},
  \qquad
  i=\uparrow,\downarrow \,,
  \label{eq:exotic-shock-css-X}
\end{equation}
where since the speed of sound \(c_s\), the vacuum energy \(\epsilon\) and the degree-of-freedom parameter \(a\) are the same on both sides of the shock, the matching condition reduces to \cref{eq:shock-matching},
\(\vwallframe_{\uparrow}^{\rm sh} \vwallframe_{\downarrow}^{\rm sh} = c_s^2\).
Energy-flux conservation across the shock gives 
\(
    T_{\downarrow}^q {\gamma_{\downarrow}^{\rm sh}}^2 \vwallframe_{\downarrow}^{\rm sh}
    = T_{\uparrow}^q {\gamma_{\uparrow}^{\rm sh}}^2 \vwallframe_{\uparrow}^{\rm sh}
\), so that
\begin{equation}
  T_{\downarrow}>T_{\uparrow}
  \quad\Longleftrightarrow\quad
  \vwallframe_{\downarrow}^{\rm sh}
  <
  c_s
  <
  \vwallframe_{\uparrow}^{\rm sh}
  \qquad\quad\text{or}\quad\qquad
  T_{\downarrow}<T_{\uparrow}
  \quad\Longleftrightarrow\quad
  \vwallframe_{\downarrow}^{\rm sh}
  >
  c_s
  >
  \vwallframe_{\uparrow}^{\rm sh},
  \label{eq:exotic-compressive-condition}
\end{equation}
i.e.\ the fluid enters a compressive shock (\(\vwallframe_\uparrow^\text{sh}>\vwallframe_\downarrow^\text{sh}\)) supersonically and leaves it subsonically with increasing temperature, whereas traversing the same algebraic discontinuity in the opposite direction, i.e.\ a rarefaction or expansion shock (\(\vwallframe_\uparrow^\text{sh}<\vwallframe_\downarrow^\text{sh}\)), results in a decrease of temperature.

Both orientations solve the Rankine--Hugoniot equations, but they are not equally admissible.
First, the entropy current must increase across the shock,
\begin{equation}
  s_{\downarrow}\gamma_{\downarrow}^{\rm sh}
  \vwallframe_{\downarrow}^{\rm sh}
  \geq
  s_{\uparrow}\gamma_{\uparrow}^{\rm sh}
  \vwallframe_{\uparrow}^{\rm sh}, \qquad\text{with}\qquad
  s_i \propto T_i^{q-1} \,.
  \label{eq:exotic-entropy-current}
\end{equation}
Energy-flux conservation and the entropy-flux condition together select precisely the compressive orientation in \cref{eq:exotic-compressive-condition}.
The reversed jump decreases the entropy flux and is therefore forbidden by the second law of thermodynamics.

The same distinction follows independently from the stability analysis of the discontinuity.
Applying the criterion discussed in Ref.~\cite{Barni:2024lkj} to the shock shows that the compressive orientation corresponds to the stable, evolutionary branch, whereas the reversed rarefactive orientation does not.
For a non-trivial CSS shock, the Rankine--Hugoniot relation requires the fluid velocity to lie above the sound speed on one side of the discontinuity and below it on the other, so the two orientations are obtained by interchanging which state is upstream and which is downstream.
Only the compressive ordering satisfies the stability requirement.
Thus, for a CSS shock, both the stability analysis and the entropy condition select the compressive orientation and rule out the rarefactive one.

For a direct compression shock in front of the wall, the upstream state is at rest, so that \(\vwallframe_{\uparrow}^{\rm sh}=\xish>c_s>\vwallframe_{\downarrow}^{\rm sh}\).
For an inverse compression shock behind the wall, the downstream state is at rest, so that
\(\vwallframe_{\uparrow}^{\rm sh}>c_s>\vwallframe_{\downarrow}^{\rm sh}=\xish\).
Thus, in both cases the physical shock is traversed from a supersonic upstream state to a subsonic downstream state, with the thermodynamic variables increasing across the discontinuity.
Placing the corresponding shock on the opposite side of the wall reverses this ordering and produces the rarefactive, entropy-decreasing orientation.

We can now apply this result to the candidates in \cref{tab:exotic-branches}.
In every case, completing the self-similar profile requires at least one additional shock on the side opposite to the corresponding physical compressive shock, and therefore forces that discontinuity to be traversed in the rarefactive, entropy-decreasing orientation.
Hence none of the configurations in \cref{fig:exotic-profiles,tab:exotic-branches} defines a physical CSS branch.
This does not contradict the ``shocked detonations'' of Ref.~\cite{Mateos:2026moc}, where the equation of state is genuinely non-conformal and the sound speed varies with the thermodynamic state.
In that case the same-phase shock matching condition involves
\(
    \vwallframe_\downarrow^\text{sh}\vwallframe_\uparrow^\text{sh}
    =(p_\downarrow-p_\uparrow)/(e_\downarrow-e_\uparrow)
\),
rather than a single constant \(c_s^2\), so the two states are not constrained to straddle the same fixed sonic point and an additional entropy-admissible shock can provide a physical continuation of the flow.
The CSS exclusion derived above therefore does not apply to those solutions.

%
\section{Proof of the projection properties}
\label{app:projection-monotonicity}
%

In this appendix, we provide the details of the results presented in \cref{sec:hydrodynamic-fold}.
We determine when the projection from the local strength \alphaP in front of the wall to the far-front  strength \alphaN at nucleation is one-to-one and when it develops an extremum between the limiting values of \alphaP leading to multiple hydrodynamic solutions at fixed \alphaN.
Since \alphaN depends not only on the local wall matching but also on the hydrodynamic profile connecting the wall to the asymptotic fluid, monotonicity must be established separately for each branch.

In the following, we work at fixed wall velocity \xiw. 
As \(\partial_{\vp} \alphaP\) has a definite sign on each branch (\(\partial_{\vp} \alphaP\gtrless 0\) if the solution branch uses \(\vp^{(\pm)}\)) and \(\partial_{\vp} \vw = \partial_{\vp} \mu(\xiw,\vp) < 0\), it is sufficient to show that any derivative of \alphaN at fixed \xiw with respect to \alphaP, \(\vw\) or \vp does not cross zero,
\begin{equation}
	\left(\frac{\partial \alphaN}{\partial \alphaP}\right)_{\!\!\xiw} 
	= \left(\frac{\partial\alphaP}{\partial \vp}\right)^{\!\!-1}_{\!\!\xiw}  \left(\frac{\partial \alphaN}{\partial \vp}\right)_{\!\!\xiw} 
	= \left(\frac{\partial\alphaP}{\partial \vp}\right)_{\!\!\xiw}^{\!\!-1} \left(\frac{\partial \vw}{\partial \vp}\right)_{\!\!\xiw} \left(\frac{\partial \alphaN}{\partial \vw}\right)_{\!\!\xiw} 
	\overset{!}{\neq} 0\,.
\end{equation}
In the remainder of this section, we will drop the \xiw subscript.

To simplify notation, we further introduce the shifted variable \(\alphashift[i] = \alpha_i - \alpha_0\), \(i\in\{+,N\}\), where \(\alphashift[i] \sim \alphavac[i]\) is the vacuum-energy strength up to a constant multiplicative factor, cf.\ \cref{eq:alpha_vac}.
\Cref{eq:alphath-local,eq:alphaN-rw-map} then yield
\begin{align}
	\alphashift[N] = r_w \alphashift[+]
	\qquad\Longrightarrow\qquad
	\frac{\partial\alphashift[N]}{\partial \vp} 
	= r_w \left[ \frac{\partial \alphashift[+]}{\partial \vp} + \alphashift[+] \frac{\partial \ln r_w}{\partial \vp}\right] 
	= \frac{\partial\alphaN}{\partial \vp} .
	\label{eq:alpha-shift}
\end{align}

We first establish a general monotonicity property of the enthalpy evolution between the wall and unperturbed fluid, which will be used repeatedly in the remainder of the section, and then examine each hydrodynamic branch separately.
As direct detonations and inverse deflagrations leave the plasma ahead of the wall unperturbed, \(\alphaN=\alphaP\) on these branches and the relation between the strengths is trivially one-to-one. 
Similarly, inverse hybrids and mixed hybrids with \(\vp=c_+\) fix the upstream wall-frame velocity \vp to the sound speed, so that the enthalpy ratio \(r_w\) only depends on \xiw in this case and \(\partial_{\alphaP} \alphaN = r_w > 0\).
These cases will hence not be discussed here.	
We will first provide the proof that for direct deflagrations as well as direct hybrids on the `\((-)\)' branch \(\vp^{(-)}(c_-,\alphaP)\), \(\alphaN(\alphaP)\) remains injective throughout the entire parameter space, while in the case of hybrids on the `\((+)\)' branch \(\vp^{(+)}(c_-,\alphaP)\), it develops a minimum within the \alphaP range for \(c_- < \ccritDH\).
Subsequently, we turn to the inverse case, where inverse detonations are injective on the `\((-)\)' branch \(\vp^{(-)}(\xiw,\alphaP)\) but non-injective on the `\((+)\)' branch \(\vp^{(+)}(\xiw,\alphaP)\) for \(c_-<\ccritID\).
Finally, for mixed hybrids with \(\vm=c_-\), \alphaN develops a minimum as a function of \alphaP if \(c_- < \ccritMH(c_+)\).

\subsection{Properties of the enthalpy ratio}
\label{app:rw-monotonicity}

\subsubsection{Monotonicity}

Consider the solution curves \(\xi(v;\vw)\) of \cref{eq:v-ode-3d} in the region ahead of the wall, with fixed wall velocity \(\xiw\), 
\begin{align}
	\frac{\dd}{\dd v}\xi(v;\vw) &= f(\xi, v) \equiv \frac{\xi \gamma^2 (1-\xi v)}{2 v} \left[\frac{\mu^2(\xi, v)}{c_+^2} - 1\right] ,
	&
	\xi(\vw; \vw) &= \xiw\,,
	\label{eq:xi-ode}
\end{align}
where \(f(\xi,v) \lessgtr 0\) along (direct) compression and (inverse) rarefaction waves, respectively.
Using \(\frac{\dd}{\dd \vw}\xi(\vw;\vw) = 0\) at the initial point, we obtain the derivative of the profile \(\xi(v;\vw)\) at fixed \(v\) with respect to the initial condition
\begin{equation}
	\frac{\dd}{\dd v} \frac{\partial \xi}{\partial \vw} = \frac{\partial f}{\partial \xi} \frac{\partial\xi}{\partial \vw}
	\qquad\Longrightarrow\qquad
	\frac{\partial \xi}{\partial \vw} = - f(\xiw, \vw) \exp\!\left[\,\int\limits_{\vw}^v \!\dd u\,\frac{\partial f}{\partial \xi}(\xi(u;\vw),u)\right] ,
\end{equation}
i.e.\ \(\partial_{\vw}\xi(v;\vw)\) is positive for compression solutions and negative for rarefactions.

Differentiating \cref{eq:rw-inverse}, we hence directly obtain
\begin{equation}
  \frac{\partial \ln \rwinv}{\partial \vw}
  =
  q_+
  \left[
    \frac{\vp}{1-\vw^2}
    -\int\limits_{\vw}^{0}
    \dd v\,
    \frac{\frac{\partial}{\partial \vw} \xi(v;\vw)}{\big(1-\xi(v;\vw)\,v\big)^2}
  \right]
   > 0,
  \label{eq:dlogrw-inverse-dvw}
\end{equation}
where the first term comes from the integration boundary and the second one from the deformation of the trajectory.
As both terms are positive, the enthalpy ratio \rwinv for inverse rarefaction waves is a monotonically increasing function of \(\vw\) at fixed wall velocity \xiw .

For the direct case, differentiating \cref{eq:rw-direct} in addition produces terms from the enthalpy jump factor at the shock front as well as \(v_\text{sh}\) in the integration limit.
Using \cref{eq:front-shock-enthalpy-jump}, these two terms combine into a single term proportional to the shift in the shock position, \(\partial_{\vw} \xi_\text{sh} > 0\). Hence,
\begin{equation}
  \frac{\partial \ln \rwdir}{\partial \vw}
  =
    q_+
    \left[
      \frac{\vp}{1-\vw^2}
      +\int\limits_{v_\text{sh}}^{\vw}
      \!\dd v\,
      \frac{\frac{\partial}{\partial \vw}\xi(v;\vw)}{\big(1-\xi(v;\vw)\,v\big)^2}
    \right]
    +\frac{(1-c_+^4)\,(\xi_\text{sh}^2-c_+^2)}
          {\xi_\text{sh}\,(\xi_\text{sh}^2-c_+^4)\,(1-\xi_\text{sh}^2)}\,
     \frac{\partial\xi_\text{sh}}{\partial \vw}
  > 0,
  \label{eq:dlogrw-direct-dvw}
\end{equation}
where, again, all terms are positive and, at fixed \xiw, \rwdir is monotonically increasing with \(\vw\).
Correspondingly, both \rwinv and \rwdir are monotonically decreasing functions of \vp.
In particular, dropping the integral and the shock contribution, we obtain for both cases
\begin{equation}
	\frac{\partial \ln r_w}{\partial \vp} < \frac{q_+ \vp}{1-\vw^2} \frac{\partial \vw}{\partial \vp} = - \frac{\vp (1+c_+^2)}{c_+^2 (1-\vp^2)} < 0 \,.
	\label{eq:dlogrw-dvp}
\end{equation}

\subsubsection{Limiting cases}

The enthalpy ratio can further be evaluated analytically in the thin profile limit, i.e.\ at \(\vw \to 0\) with \(\xiw<c_+\), as well as \(\vw\to\mu(\xiw,c_+)\) with \(\xiw<c_+\) for inverse rarefactions and \(\vw \to v_\text{sh} = \mu(\xiw, c_+^2/\xiw)\) with \(\xiw=\xi_\text{sh}>c_+\) for compression waves.
For \(\vw\to0\), \cref{eq:xi-ode} can be solved analytically,
\begin{equation}
	\frac{\dd\xi}{\dd v} = \frac{\xi (\xi^2-c_+^2)}{2 c_+^2 v}
	\qquad\Longrightarrow\qquad
	\xi(v;\vw) = c_+ \left(1 + \frac{v}{\vw} \frac{c_+^2-\xiw^2}{\xiw^2}\right)^{\!\!-\frac{1}{2}} \hspace{-1em},
\end{equation}
which, inserted into \cref{eq:dlogrw-inverse-dvw,eq:dlogrw-direct-dvw}, yields
\begin{equation}
	\lim_{\vw \to 0} \frac{\partial \ln r_w}{\partial \vw} = q_+ \left[\xiw - \frac{\xiw (\xiw-c_+)}{\xiw + c_+} \right]
	= \frac{2 \xiw (1+c_+^2)}{c_+ (\xiw+c_+)} \,.
	\label{eq:dlogrw-zero}
\end{equation}
Note that \(\xi_\text{sh}\to c_+\) in this limit, so that the part proportional to \(\partial_{\vw}\xi_\text{sh}\) in \cref{eq:dlogrw-direct-dvw} vanishes.

At the sonic front, \(\vp \to c_+\) or \(\vw \to \mu(\xiw,c_+)\), the derivative of the trajectory with respect to the initial condition, \cref{eq:xi-ode}, vanishes as \(f(\xiw,\vw)\to 0\) in this limit.
\Cref{eq:dlogrw-inverse-dvw} then becomes
\begin{equation}
	\lim_{\vp\to c_+} \frac{\partial \ln \rwinv}{\partial \vw} = \frac{c_+ q_+}{1-\mu^2(\xiw,c_+)} = \frac{(1+ c_+^2) (1 - c_+ \xiw)^2}{c_+ (1-c_+^2) (1-\xiw^2)} \,.
	\label{eq:dlogrw-cp}
\end{equation}

For compression waves, on the other hand, the contribution from the integral in \cref{eq:dlogrw-direct-dvw} vanishes at the shock front.
The derivative of the shock position, \(\partial_{\vw}\xi_\text{sh}\), can be evaluated from the condition \(\xi_\text{sh} \mu(\xi_\text{sh}, v_\text{sh}) = c_+^2\), where the fluid velocity for walls close to the shock is given by \(v_\text{sh} = \vw + {(\xi_\text{sh}-\xiw)}/{f_w}\) with \(f_w = f(\xiw,\vw)\).
Taking the derivative of the shock condition then yields
\begin{equation}
	\frac{\partial}{\partial \vw}\! \left[\xi_\text{sh} \mu(\xi_\text{sh}, v_\text{sh})\right] = 0 
	\qquad\Longrightarrow\qquad
	\lim_{\vw \to v_\text{sh}} \frac{\partial \xi_\text{sh}}{\partial \vw} = \frac{\xiw^2 (1-c_+^2)^2}{3 (\xiw^2-c_+^4)  + c_+^2 (1-\xiw^2)} \,.
\end{equation}
Hence, inserting this into \cref{eq:dlogrw-direct-dvw}, 
\begin{equation}
	\lim_{\vw \to v_\text{sh}} \frac{\partial\ln\rwdir}{\partial \vw}
	=
	\frac{2 \xiw (1-c_+^2) (1-c_+^4) [\xiw^2 (1 - c_+^2) + (\xiw^2 - c_+^4)]}{(1-\xiw^2)(\xiw^2-c_+^4)[3 (\xiw^2-c_+^4)  + c_+^2 (1-\xiw^2)]}
	\,.
	\label{eq:dlogrw-sh}
\end{equation}

\subsubsection{Upper bound for inverse rarefactions}
\label{app:rwinv-upper-bound}

Finally, for inverse rarefactions, \cref{eq:dlogrw-dvp} can be complemented by an upper bound,
\begin{equation}
	\frac{\partial \ln \rwinv}{\partial \vw} \leq \frac{2 \gamma_w^2 c_+ q_+ \vp }{c_+ + \vp}
	=
	\frac{2}{1-\vw^2}\,\frac{\vp}{c_+}\,\frac{1 + c_+^2}{c_++\vp} 
	\,,
	\label{eq:dlogrw-dvp-upper}
\end{equation}
which reproduces \cref{eq:dlogrw-zero,eq:dlogrw-cp} in the \(\vw\to 0\) (\(\vp\to\xiw\)) and \(\vp\to c_+\) limit, respectively.
To prove \cref{eq:dlogrw-dvp-upper}, we transport the wall along a solution trajectory of \cref{eq:v-ode-3d}.
Along such a trajectory, the total derivative with respect to the wall velocity and the commutator of the total and the partial derivative are given by
\begin{equation}
	\frac{\dd}{\dd \vw} = \left.\frac{\partial}{\partial \vw} \right|_{\xiw}+ f_w \left.\frac{\partial}{\partial \xiw}\right|_{\vw}
	\qquad\Longrightarrow\qquad
	\left[\frac{\dd}{\dd \vw}, \frac{\partial}{\partial \vw}\right] = - \frac{\partial f_w}{\partial \vw} \frac{\partial}{\partial \xiw} \,.
	\label{eq:transport-derivatives}
\end{equation}
Using the commutator as well as that, along the trajectory, \(\frac{\dd \ln r_w}{\dd \vw} = q_+ \vp \gamma_w^2\) from \cref{eq:w-ode-general}, all derivatives of \(\ln r_w\) can be expressed in terms of \(\partial_{\vw}\ln r_w\).
We then obtain the linear equation
\begin{equation}
	\left[\frac{\dd}{\dd \vw} - \frac{1}{f_w}\frac{\partial f_w}{\partial \vw}\right]\frac{\partial \ln r_w}{\partial \vw}
	=
	\left[\frac{\partial}{\partial \vw} - \frac{1}{f_w}\frac{\partial f_w}{\partial \vw}\right]\frac{q_+\vp}{1-\vw^2}
	\,.
	\label{eq:dlogrw-transport}
\end{equation}
Denoting the difference between the left-hand side and right-hand side of \cref{eq:dlogrw-dvp-upper} as
\(\Delta \equiv \partial_{\vw}\ln\rwinv - 2 \gamma_w^2 c_+ q_+ \vp/(\vp+c_+)\),
\cref{eq:dlogrw-transport} gives
\begin{equation}
	\left[\frac{\dd}{\dd \vw} - \frac{1}{f_w}\frac{\partial f_w}{\partial \vw}\right] \Delta
	=
	\frac{2 c_+ q_+ (1-\xiw^2)(c_+-\vp)}
	{(1-\vw^2)(1-\vw\xiw)^2 (\vp+c_+)^2}
	\geq 0 \,.
	\label{eq:dlogrw-transport-difference}
\end{equation}
Furthermore, for inverse rarefactions, we have \(f_w>0\) and, as \(\vp < c_+\) and \(\vw<0\),
\begin{equation}
	\frac{1}{f_w}\,\frac{\partial f_w}{\partial \vw}
	= \frac{2 \vp (1-c_+^2)}{(1-\vw^2)(c_+^2-\vp^2)} + \frac{2 \xiw}{1-\xiw \vw} - \frac{1}{\vw (1-\xiw \vw)}
	> 0 \,.
	\label{eq:dlogf-dv}
\end{equation}
Now, as \(\Delta\) vanishes at \(\vw\to0\) and \(\vp\to c_+\), its derivative \(\dd\Delta/\dd \vw\) along a solution trajectory has to change sign in between.
Combining \cref{eq:dlogrw-transport-difference,eq:dlogf-dv}, this is only possible if \(\Delta < 0\) (\(\dd\Delta/\dd \vw \geq 0\) if \(\Delta \geq 0\)), proving \cref{eq:dlogrw-dvp-upper}.

\subsection{Direct deflagrations}
\label{app:monotonicity-deflagrations}

Let us decompose the transition strength into energy and pressure contributions,%
\footnote{%
	We can define a corresponding decomposition of \(\alpha_i\) with \(\varepsilon_i=(e_+-e_-)/(3 w_i)\) and \(\pi_i=(p_+-p_-)/(3 w_i)\).
    The \(\lim_{\epsilon\to0}\) part comes from the \(\alpha_0\) shift.%
}
\begin{align}
	\alphashift[i] &= \bar\varepsilon_i - \frac{\bar\pi_i}{c_-^2}
	\,,&
	\bar\varepsilon_i &\equiv \left[1 - \lim_{\epsilon\to0}\right]\frac{e_+-e_-}{3 w_i}
	,&
	\bar\pi_i &\equiv \left[1 - \lim_{\epsilon\to0}\right]\frac{p_+-p_-}{3 w_i}
	\,,&
	i\in\{+,N\},
\end{align}
where \(\bar\varepsilon_i\) and \(\bar\pi_i\) are independent of \(c_-\) and satisfy relations equivalent to \cref{eq:alpha-shift}, i.e.\ \(\bar{X}_N = r_w \bar{X}_+\) for \(X\in\{\varepsilon,\pi\}\). 
From \cref{eq:alphath-local,eq:alphath-vplus-vminus}, we then obtain
\begin{align}
	\bar\varepsilon_+ 
	&= \frac{(1-\xiw \vp) (c_+^2 \vw - \vp )}{3 \xiw(1-\vp^2) (1+c_+^2)}
	\,,
	&
	\bar\pi_+ 
	&= \frac{(1-\xiw \vp) (\vp \vw - c_+^2)}{3 (1-\vp^2)(1+c_+^2)}  
	\,,
\end{align}
where we used that \(\vm=\xiw\) for direct deflagrations.

The pressure contribution \(\bar\pi_+ < 0\) is always negative as \(\vw < \xiw\) and \(\xiw \vp < c_+^2\) in the direct deflagration case.
Using \cref{eq:dlogrw-dvp} we directly obtain (recalling \(\vp<c_+\) and \(\xiw > \vp\))
\begin{equation}
	\frac{1}{r_w} \frac{\partial \bar\pi_N}{\partial \vp} 
	= \frac{\partial\bar\pi_+}{\partial \vp} + \bar\pi_+  \frac{\partial \ln r_w}{\partial \vp}
	>  \frac{(c_+^2-\vp^2)(\xiw-\vp)}{3 c_+^2 (1-\vp^2)^2} 
	> 0 \,.
	\label{eq:pi-monotonicity-deflagration}
\end{equation}
For the energy contribution, on the other hand, this simple argument does not work since \(\bar\varepsilon_+\) changes sign at \(\vp = c_+^2 \vw\). 
The derivative \(\partial_{\vp}\bar\varepsilon_+ <0\) is always negative, so that, using \cref{eq:dlogrw-dvp}, \(\partial_{\vp} \bar\varepsilon_N < 0\) for \(\bar\varepsilon_+ \geq 0\), i.e.\ for \(c_+^2 \vw \geq \vp\).
For \(\bar\varepsilon_+ < 0\) this estimate is, however, inconclusive.

To show that \(\partial_{\vw}\bar\varepsilon_N > 0\) (i.e.\ \(\partial_{\vp}\bar\varepsilon_N < 0\)) also for \(c_+^2 \vw < \vp\), we start by considering the derivative directly at the shock front, \(\xiw \vp = c_+^2\). 
From \cref{eq:dlogrw-sh}, we obtain%
\footnote{We are here using \(\partial_{\vw}\) rather than \(\partial_{\vp}\) as \(\partial_{\xiw}\) will be evaluated at fixed \vw in the following.}
\begin{equation}
	\lim_{\vw\to v_\text{sh}} \frac{1}{r_w} \frac{\partial \bar\varepsilon_N}{\partial \vw} = \frac{(1-c_+^2)^3 \xiw (\xiw^2-c_+^2)}{(\xiw^2-c_+^4)(1-\xiw^2)[3(\xiw^2-c_+^4) + c_+^2 (1-\xiw^2)]} > 0\,.
	\label{eq:epsilon-derivative-shock-front}
\end{equation}
To show that \(\partial_{\vw}\bar\varepsilon_N\) remains positive away from the shock front, as in \cref{app:rwinv-upper-bound}, we transport the wall along a solution trajectory of \cref{eq:v-ode-3d} and show that the derivative cannot change sign in the \((\xiw,\vw)\) region of interest.%
\footnote{%
	The trajectories are traversed across the entire compression wave region, potentially including \(\xiw > c_-\), where \(\bar\varepsilon_\pm\) and \(r_w\) remain well defined although the configuration strictly speaking does not correspond to a physical direct deflagration.%
}
Using \cref{eq:transport-derivatives,eq:w-ode-general}, we then obtain
\begin{align}
	\frac{\dd}{\dd \vw} \frac{1}{r_w} \frac{\partial \bar\varepsilon_N}{\partial \vw} = 
	\frac{\dd}{\dd \vw} \frac{\partial \bar\varepsilon_+}{\partial \vw}
	+ q_+ \bar\varepsilon_+ \left[ \frac{\partial}{\partial \vw} - \frac{1}{f_w}\frac{\partial f_w}{\partial \vw}\right] \gamma_w^2 \vp
	+ \left[\frac{\dd \bar\varepsilon_+}{\dd \vw} + \frac{\bar\varepsilon_+}{f_w} \frac{\partial f_w}{\partial \vw} \right] \frac{\partial \ln r_w}{\partial \vw}
	.
\end{align}

Now, suppose that \(\partial_{\vw} \bar\varepsilon_N\) crosses zero when shifting along the trajectory. 
At the zero crossing, the equivalent of \cref{eq:pi-monotonicity-deflagration} for \(\bar\varepsilon\) in terms of \(\vw\) gives \(\partial_{\vw}\ln r_w = - (\partial_{\vw}\bar\varepsilon_+)/\bar\varepsilon_+\), and hence
\begin{equation}
	\left.\frac{\dd}{\dd \vw} \frac{1}{r_w} \frac{\partial \bar\varepsilon_N}{\partial \vw}\right|_{\partial_{\vw}\bar\varepsilon_N=0} = 
	\frac{(1-c_+^2 \vw^2) (c_+^2-\vp^2)}{2 c_+^2 \vw (1-\vw^2)^2 ( \vp-c_+^2 \vw)}
	> 0 
	\qquad\text{for}\ \vp > c_+^2 \vw\,.
	\label{eq:transport}
\end{equation}
Therefore, when moving towards smaller \(\vw\) --- i.e.\ pushing the wall towards the shock front, which is the smallest \(\vw\) attained along the trajectory --- the derivative can cross zero only from \(\partial_{\vw} \bar\varepsilon_N>0\) to \(\partial_{\vw} \bar\varepsilon_N<0\) (unless \(\vp \leq c_+^2 \vw\), for which we already know that \(\partial_{\vw} \bar\varepsilon_N>0\)).
As \(\partial_{\vw}\bar\varepsilon_N>0\) at the shock front, cf.\ \cref{eq:epsilon-derivative-shock-front}, no such crossing can occur before  traversing the shock front, and the derivative remains positive along the entire trajectory.
Thus, for direct deflagrations we always have \(\partial_{\vp} \bar\varepsilon_N < 0\), and, with \cref{eq:pi-monotonicity-deflagration}, \(\partial_{\alphaP} \alphaN > 0\).

\subsection{Direct hybrids}

For direct hybrids, the fluid in the wall rest frame behind the wall moves at the speed of sound, \(\vm = c_-\), i.e.
\begin{align}
	\alphashift[+] 
	&= \frac{(\vp-c_-)^2}{3 c_-^2 (1-\vp^2)} + \frac{c_+^2-c_-^2}{3 c_-^2 (1+c_+^2)} 
	\,,&
	\frac{\partial\alphashift[+]}{\partial\vp} 
	&= \frac{2 (\vp-c_-)(1-c_-\vp)}{3 c_-^2 (1-\vp^2)^2}
	\,.
	\label{eq:alphashiftplus-hybrid-c-}
\end{align}
As discussed in \cref{sec:direct-hybrids}, hybrid solutions split into two sub-branches, corresponding to the `\((\pm)\)' branches  of \(\vp = \vp^{(\pm)}(c_-,\alphaP) \gtrless c_-\), where the `\((+)\)' branch is only accessible for \(c_+  > c_-\).

\subsubsection[\texorpdfstring{`\((-)\)'}{'(-)'} branch]{\boldmath `\((-)\)' branch}

As  \(\partial_{\vp}\alphashift[+] < 0\) is negative on the `\((-)\)' branch, we trivially obtain \(\partial_{\vp} \alphaN < 0\) from \cref{eq:dlogrw-dvp} as long as \(\alphashift[+]>0\) is positive, i.e.\ for
\begin{equation}
	c_- q_- = c_-+\frac{1}{c_-} \geq \frac{2 \vp (1+c_+^2)}{c_+^2+\vp^2} \equiv q_c \,.
\end{equation}
In particular, this condition is automatically satisfied for  \(c_+ \geq c_-\).
For the \(c_- q_- < q_c\) case, on the other hand, we can use the transport argument from \cref{app:monotonicity-deflagrations} to prove monotonicity, replacing \(\bar\varepsilon\) in \cref{eq:transport} by \alphashift.%
\footnote{%
	Note that \(q_c\) depends on \vp and hence varies along the trajectory, so that a given trajectory generally passes through both the \(c_- q_- < q_c\) and the \(c_- q_- > q_c\) region.
	As zero crossings are excluded in the latter, the argument proceeds as for direct deflagrations.%
}
With \cref{eq:dlogrw-sh}, the derivative \(\partial_{\vw}\alphashift[N]\) at the shock front evaluates to
\begin{equation}
	\lim_{\vw\to v_\text{sh}} \frac{1}{r_w} \frac{\partial \alphashift[N]}{\partial \vw}
	= \left.\frac{
		2 (1-c_+^2)^2 \xiw (\xiw-\vp)  [
		c_- (2-c_+^2-c_-\vp) + c_- - \vp
		]
	}{ 
		3 c_-^2 (1 - \vp^2) (1-\xiw^2) [3\xiw^2(1-\vp^2)+c_+^2(1-\xiw^2)]
	}\right|_{\vp=\frac{c_+^2}{\xiw}} 
	,
	\label{eq:alpha-derivative-direct-hybrid-shock}
\end{equation}
which is positive since \(\xiw > \vp\) and \(\vp<c_+<c_-\) (as the case \(c_+\geq c_-\) is proven trivially) for direct hybrids.
The transport derivative is given by
\begin{equation}\begin{gathered}
	\label{eq:transport-derivative-hybrid}
	\left.\frac{\dd}{\dd \vw} \frac{1}{r_w} \frac{\partial \alphashift[N]}{\partial \vw}\right|_{\partial_{\vw}\alphashift[N]=0}
	= 
	\frac{%
		2 (c_+^2-\vp^2) (\vp+\vw)(3-\vp^2)
		[c_-^2 q_-^2 \mathcal{A}_2 + c_- q_- \mathcal{A}_1 - 1 ]
	}{%
		9 c_+^2 c_-^2 
		\vw (1-\vw^2)^2 (1-\vp^2)^2
		\alphashift[+]
	} ,
	\\
	\text{with}\quad
	\mathcal{A}_1 = \frac{c_+^2 (1+\vp^3 \vw) + 3 \vp(\vp+\vw)}{(1+c_+^2)(\vp+\vw)(3-\vp^2)},
	\quad
	\mathcal{A}_2 = \frac{c_+^2 \vw (1-\vp^2) - \vp^2 (\vp+\vw)}{(1+c_+^2)(\vp+\vw)(3-\vp^2)}.
\end{gathered}\end{equation}
The numerator is negative at \(c_-q_-=q_c\).
Furthermore, its derivative with respect to \(c_- q_-\), \(2 c_- q_- \mathcal{A}_2 + \mathcal{A}_1\), is positive for \(c_- q_- \leq q_c\) (for \(\mathcal{A}_2\geq 0\) it is positive for all \(c_- q_-\), and for \(\mathcal{A}_2<0\) it is positive at \(q_c\) and crosses zero only at \(c_- q_- > q_c\)).%
\footnote{%
	\label{fn:direct-hybrid-transport-numerator}
	Writing \(\mathcal{N}(x)= \mathcal{A}_0 [x^2 \mathcal{A}_2 + x \mathcal{A}_1 - 1]\), where \(\mathcal{A}_0>0\) is the denominator of \(\mathcal{A}_{1,2}\), we have
	\begin{align*}\textstyle
		\mathcal{N}(q_c) &= - \frac{(1+c_+^2) (1-\vp^2) (c_+^2-\vp^2)}{(c_+^2+\vp^2)^2} [(c_+^2-\vp^2)(\vp+\vw) + 2 c_+^2 \vw (1-\vp^2)] < 0 \,,\\
		\textstyle
		\mathcal{N}'(q_c) &= \frac{
			(c_+^4-\vp^4) + 4 c_+^2 \vp^2 (1-\vp^2) 
			+ \vw \vp [ (1+c_+^2)(c_+^2-\vp^2) + c_+^2 (1-\vp^2)  (6 + 3 c_+^2 - \vp^2)]
		}{c_+^2+\vp^2}  > 0\,.
	\end{align*}%
}
Hence, as \(\alphashift[+]<0\) in this region, the transport derivative \cref{eq:transport-derivative-hybrid} is positive for  \(c_-q_-\leq q_c\), and, using the same argumentation as in the deflagration case, \(\partial_{\vw}\alphashift[N]>0\) remains positive.

Therefore, \(\partial_{\alphaP}\alphaN > 0\) is positive everywhere on the `\((-)\)' branch of direct hybrid solutions, i.e.\ \alphaN is monotonic.

\subsubsection[\texorpdfstring{`\((+)\)'}{'(+)'} branch]{\boldmath `\((+)\)' branch}

In contrast to the `\((-)\)' branch, the relation between \alphaN and \alphaP on the direct hybrid `\((+)\)' branch can indeed become non-monotonic.
This can be seen directly considering the boundaries of the solution region, recalling that the `\((+)\)' branch is only accessible for \(c_+>c_-\).

At the endpoint shared with the `\((-)\)'  branch, \(\vp=c_-\), both \alphaP and \(\partial_{\vp}\alphaP\) vanish.
Thus, using \(\alpha_0<0\) and \(\partial_{\vw} \ln \rwdir > 0\),
\begin{equation}
	\left.
	\frac{1}{r_w} \frac{\partial\alphashift[N]}{\partial \vw}
	\right|_{\vp=c_-}
	=
	- \alpha_0
	\left.
	\frac{\partial\ln \rwdir}{\partial \vw}
	\right|_{\vp=c_-}
	> 0.
	\label{eq:alpha-derivative-direct-hybrid-cminus}
\end{equation}

For \(c_-<\xiw<c_+\), the upper bound on \(\vp<\xiw\) is given by the \(\vw\to 0\) limit, so that, using \cref{eq:dlogrw-zero},
\begin{equation}
	\lim\limits_{\vw \to 0} \frac{1}{r_w} \frac{\partial \alphashift[N]}{\partial \vw} 
	= \frac{2 (c_+-\xiw) [
		c_+c_-
		+  2c_-\xiw
		- \xiw^2(1+c_-^2+c_+c_-)
		]}{
		3 c_-^2 c_+ (c_++\xiw) (1-\xiw^2)
	} .
	\label{eq:alpha-derivative-direct-hybrid-zero}
\end{equation}
In the \(c_+<\xiw<c_+^2/c_-\) region, on the other hand, the upper bound is given by the shock front \(\vp < c_+^2/\xiw\).
We hence again obtain \cref{eq:alpha-derivative-direct-hybrid-shock}.
The sign of \cref{eq:alpha-derivative-direct-hybrid-shock,eq:alpha-derivative-direct-hybrid-zero} is controlled by the respective square bracket in the numerator, which is monotonically increasing with \xiw in the shock case and decreasing in the \(\vw\to 0\) case. 
The \alphaP--\alphaN relation on the direct hybrid `\((+)\)' branch hence can only be monotonic if \cref{eq:alpha-derivative-direct-hybrid-shock,eq:alpha-derivative-direct-hybrid-zero} are positive for the entire respective \xiw range, i.e.\ in particular at \(\xiw\to c_+\).
Both equations then yield,
\begin{equation}
	\ccritDH(c_+)
	\equiv
	\frac{
		3-c_+^2
		-
		\sqrt{\left(1-c_+^2\right)\left(9-c_+^2\right)}
	}{
		2c_+
	},
	\label{eq:DHplus-critical-cminus}
\end{equation}
where \(c_- \geq \ccritDH(c_+)\) is a necessary condition for monotonicity. 
For \(c_- < \ccritDH(c_+)\), the interval develops an extremum between the limiting values of \alphaP.

Note that, in the remaining interval \(\ccritDH(c_+) < c_- < \vp\), the square bracket in the numerator of \cref{eq:transport-derivative-hybrid} is positive on both endpoints,%
\footnote{%
	Using the same notation as in \cref{fn:direct-hybrid-transport-numerator}, we obtain for \(c_-=\ccritDH\) and \(c_-=\vp\), respectively,
	\begin{align*}\textstyle
		\mathcal{N}\left(\frac{3 - c_+^2}{c_+}\right)
		&=
		\frac{(c_+-\vp)(1-\vw)}{c_+^2} [(c_+^2 + 4 c_+ \vp + 7 \vp^2)(1-c_+^2) + 2 (c_+-\vp)^2] 
		+ \frac{(1-c_+)(1+\vp) \vw}{c_+^2} 
		\times\\ &\phantom{=\ }
		\big[
			(c_+^2-\vp^2)(6+9 c_+-c_+^3) 
			+ 4 (1-c_+^2) c_+ \vp 
			+ 4 \vp (c_+ -\vp) 
			+ (1-c_+^2)^2 \vp (c_++\vp)
		\big] > 0,
		\\ \textstyle
		\mathcal{N}\left(\frac{1+\vp^2}{\vp}\right) &= \frac{(c_+^2-\vp^2)(1-\vp^2)^2 (\vp+\vw)}{\vp^2} > 0 \,.
	\end{align*}%
}
and hence on the entire interval.\footnote{%
	For \(\mathcal{A}_2<0\) this follows by concavity, and for \(\mathcal{A}_2>0\) because \(\mathcal{N}\) is then monotonically increasing (\(\mathcal{A}_1>0\)).%
}
With \(\alphashift[+]>0\) on the `\((+)\)' branch of direct hybrids, the transport derivative at any zero crossing of \(\partial_{\vw}\alphashift[N]\) is therefore positive, and  \(\partial_{\vw}\alphashift[N] =0\) can only be crossed in one direction.
Furthermore, along a trajectory \(\dd\vp/\dd\vw = \partial_{\vw}\mu + f_w \partial_{\xiw}\mu< 0\), so that every configuration on the `\((+)\)' branch is connected at decreasing \vw to the shock front, where \(\partial_{\vw}\alphashift[N] >0\) by \cref{eq:alpha-derivative-direct-hybrid-shock}.
Hence, using \(\partial_{\vp}\alphaP >0\) on the `\((+)\)' branch,  we obtain \(\partial_{\alphaP}\alphaN < 0\), i.e.\ the relation between \alphaN and \alphaP is monotonic for \(c_-> \ccritDH(c_+)\). 

\subsection{\boldmath Inverse detonations}

For inverse detonations, \(\vp>\vm=\xiw\), i.e.\ \(\vw < 0\).
Again, the solutions split into the `\((\pm)\)' branches of \(\vp = \vp^{(\pm)}(\xiw,\alphaP)\), where solutions on the `\((+)\)' branch only exist for \(c_+>c_-\).

\subsubsection[\texorpdfstring{`\((-)\)'}{'(-)'} branch]{\boldmath `\((-)\)' branch}

On this branch, the derivative \(\partial_{\vp}\alphashift[+]<0\) is negative.
Hence, with \cref{eq:dlogrw-dvp} we trivially obtain \(\partial_{\vp}\alphashift[N] < 0\) for \(\alphashift[+]\geq0\) from \cref{eq:alpha-shift}.
For \(\alphashift[+]<0\), on the other hand, which holds in most of the inverse-detonation parameter space, we can use the upper bound of \cref{eq:dlogrw-dvp-upper} to obtain
\begin{equation}
	\frac{1}{r_w}\frac{\partial \alphashift[N]}{\partial \vw}
	\geq
	\frac{\left(c_+-\vp\right)\left[c_+(1-\vw^2)(c_-^2+\xiw^2) + 2(\xiw-\vw)(c_-^2+\vw\xiw)\right]}
	{3c_-^2 c_+ \xiw (1-\vw^2)^2 (\vp+c_+)}
	> 0 \,,
	\label{eq:alpha-derivative-inverse-detonation}
\end{equation}
where \(c_-^2+\vw \xiw > c_-^2 - \vp\xiw > 0\) on this solution branch.
Thus, \(\partial_{\vw}\alphashift[N]>0\) on the entire branch, i.e.\ \(\partial_{\alphaP}\alphaN>0\) for inverse detonations on the `\((-)\)' branch.

\subsubsection[\texorpdfstring{`\((+)\)'}{'(+)'} branch]{\boldmath `\((+)\)' branch}

Similar to the `\((+)\)' branch of direct hybrids, the relation between \alphaN and \alphaP on the `\((+)\)' branch of inverse detonations also becomes non-injective.
Note that this branch only exists for \(c_+>c_-\), and that \(\partial_{\vp}\alphaP > 0\) as well as \(\alphaP > \alphav{v_c(\xiw)}{\xiw} > \alphav{c_+}{\xi_c} = \alpha_0\), i.e.\ \(\alphashift[+]>0\) on this branch, with \(c_-<v_c(\xiw)<\vp<c_+\).
Inserting \cref{eq:dlogrw-dvp} into \cref{eq:alpha-shift} then yields\footnote{%
    \label{fn:alpha-bar-plus}%
	In terms of \(v_*\), \alphashift[+] can be written as
	\(\displaystyle
	\alphashift[+](\vp,\vm) = \frac{c_+^2(1+c_-^2)}{3 c_-^2(1+c_+^2)} + \frac{(1+c_-^2) \vp (\vp-v_*(\vm))}{3 c_-^2 (1-\vp^2)} 
	\).%
}
\begin{equation}
	\frac{1}{r_w}\frac{\partial\alphashift[N]}{\partial\vp}
	\leq
	\frac{(c_+^2-\vp^2)(1+c_-^2)\,\big[\vp - v_*(\xiw)\big]}{3c_+^2 c_-^2(1-\vp^2)^2}\,,
	\qquad
	v_*(\xiw) \equiv \frac{c_-^2+\xiw^2}{\xiw\,(1+c_-^2)}\,.
	\label{eq:ID-plus-lower-bound}
\end{equation}
Hence \(\partial_{\vp}\alphashift[N]<0\) whenever \(\vp\leq v_*(\xiw)\), and any zero must lie strictly above \(v_*\).
On the branch, \(v_*\) decreases monotonically from \(v_*(\xi_c)=2c_+/(1+c_+^2)>c_+\) to \(v_*(c_-)=2c_-/(1+c_-^2)\).
In particular, if \(c_+\leq 2c_-/(1+c_-^2)\), then \(v_*>c_+\) for all \xiw and the right-hand side of \cref{eq:ID-plus-lower-bound} is negative on the entire branch, i.e.\ the map is monotonic in this case.

Along a solution trajectory, using \cref{eq:transport-derivatives,eq:dlogrw-transport}, the transport derivative at a zero crossing of \(\partial_{\vw}\alphashift[N]\) reads 
\begin{equation}\begin{gathered}
    \label{eq:transport-derivative-ID}
    \left.\frac{\dd}{\dd \vw} \frac{1}{r_w} \frac{\partial \alphashift[N]}{\partial \vw}\right|_{\partial_{\vw}\alphashift[N]=0}
    =
    \frac{(1-\xiw\vw)^3 (c_+^2-\vp^2) 
        \big[\mathcal{B}_2\,\vp^2 + \mathcal{B}_1\,\vp + \mathcal{B}_0\big]}
    {18\,c_+^2 (1+c_+^2)\,\xiw^2 (v_*-\xiw)^2\;\vw\,(1-\vw^2)^3\;\alphashift[+]}\,,
    \\
    \mathcal{B}_0 = -3(1+c_+^2)\,\xiw v_*^2 + c_+^2(3+\xiw^2)\,v_* + 2c_+^2\xiw\,,
    \\
    \mathcal{B}_1 = -2\big[2c_+^2 - (3-c_+^2)\,\xiw v_*\big]\,,
    \qquad
    \mathcal{B}_2 = (1+c_+^2)\,\xiw v_*^2 - (1+\xiw^2-2c_+^2)\,v_* - 2\xiw\,.
\end{gathered}\end{equation}
Since \(\vw<0\) and \(\alphashift[+]>0\), the transport derivative is positive if the polynomial in the square bracket is negative, which holds on the whole region \(v_*\leq\vp<c_+\) where zero crossings of \(\partial_{\vw}\alphashift[N]\) are possible by \cref{eq:ID-plus-lower-bound}.%
\footnote{%
	Only \xiw with \(v_*<\vp<c_+\) admit zero crossings of \(\partial_{\vw}\alphashift[N]\), which implies \(c_- < c_+/(1+ \sqrt{1-c_+^2}) < 2 c_+/(3 - c _+^2)\).	
	With \(\xiw<c_-\) we hence obtain \(\mathcal{B}_1<0\).
	Denoting \(\mathcal{P}(\vp) = \mathcal{B}_0 + \mathcal{B}_1 \vp + \mathcal{B}_2 \vp^2\), the endpoint values are \(\mathcal{P}(c_+) = - \tfrac{1}{2} \mathcal{B}_1 \big[(1+c_+^2)v_* - 2c_+\big] < 0\) and \(\mathcal{P}(v_*) = -(1-c_+^2)\,v_*^2 (1-\xiw v_*)(v_*-\xiw) - c_+^2 (1-v_*^2)\big[(1-\xiw^2)v_* - 2\xiw(1-v_*^2)\big] < 0\) as \(\xiw < c_-\).
	Finally \(\mathcal{P}'=2\mathcal{B}_2\vp+\mathcal{B}_1<0\) for \(\mathcal{B}_2\leq0\), so that \(\mathcal{P}\leq\mathcal{P}(v_*)<0\), while for \(\mathcal{B}_2>0\) the parabola is convex and bounded by \(\max[\mathcal{P}(v_*),\mathcal{P}(c_+)]<0\).%
}
Thus, moving along a trajectory towards larger \vw and, as \(f_w>0\), larger \xiw, i.e.\ towards the exit of the branch at \(\xiw=c_-\), \(\partial_{\vw}\alphashift[N]\) can cross zero only from negative to positive values.
Trajectories enter the branch either through \(\vp=v_c\) or through \(\vp=c_+\).
At \(\vp=v_c\), \(\partial_{\vp}\alphaP=0\) by definition, so that, as \(\partial_{\vp} \ln r_w < 0\), we directly obtain \(\partial_{\vp} \alphashift[N]|_{\vp=v_c} < 0\).
At \(\vp=c_+\), on the other hand, the derivative vanishes, \(\partial_{\vp} \alphashift[N]|_{\vp=c_+} = 0\), as the change in the enthalpy ratio, \cref{eq:dlogrw-cp}, compensates the change in \alphaP at the sonic front.
The sign of \(\partial_{\alphaP}\alphaN\) close to the sonic point is therefore controlled by the second derivative, and monotonicity on the branch requires \(\partial_{\vp}\alphashift[N]<0\) on the whole interval \(v_c<\vp<c_+\), i.e.\  \(\alphashift[N]\) must approach its endpoint value at \(\vp=c_+\) from above.
Hence, \(\partial_{\vp}^2\alphashift[N]|_{\vp=c_+} < 0\) --- or, equivalently,  \(\partial_{\vw}^2\alphashift[N]|_{\vp=c_+} < 0\) --- implies non-monotonicity.

\begin{figure}
    \centering
    \includegraphics[width=0.62\textwidth]{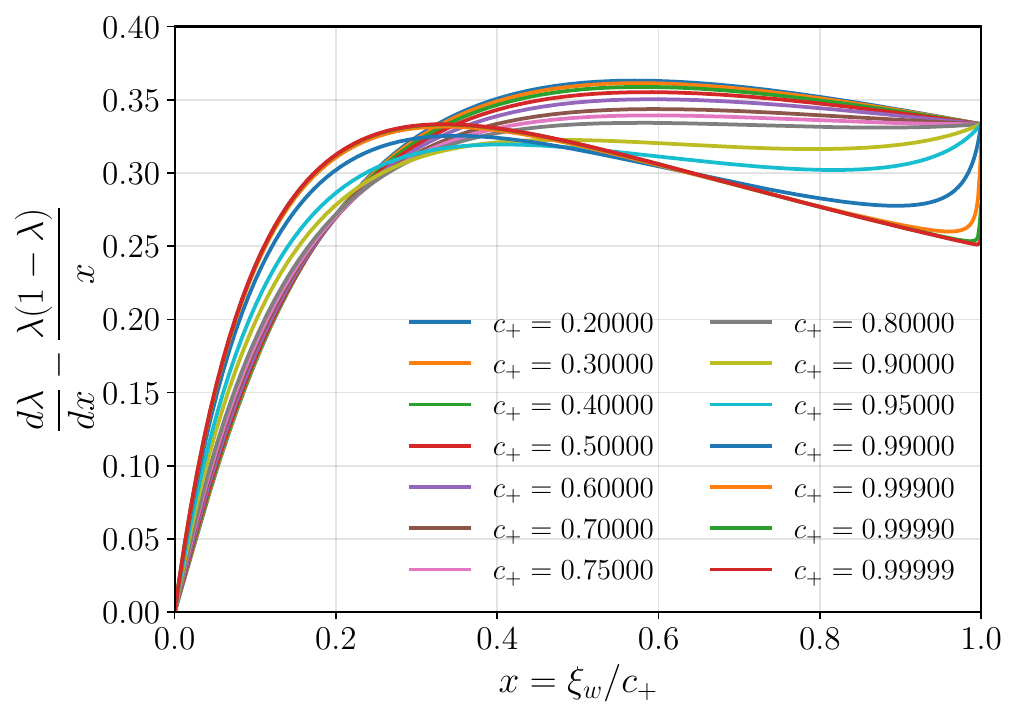}
    \caption{
    Numerical test of the monotonicity of \(\lambda(c_+,\xiw)\).
    We plot \(d\lambda/dx-\lambda(1-\lambda)/x\), with \(x=\xiw/c_+\), for representative values of \(c_+\).
    The quantity remains positive throughout \(0<x<1\), implying \(\partial_{\xiw}\lambda>0\).
    }
    \label{fig:lambda-monotonicity}
\end{figure}

Thus, consider the second derivative 
\begin{equation}
	\frac{1}{r_w} \frac{\partial^2 \alphashift[N]}{\partial \vw^2} = \frac{1}{r_w} \frac{\partial \ln r_w}{\partial \vw}  \frac{\partial \alphashift[N]}{\partial \vw} + \frac{\partial^2 \alphashift[+]}{\partial \vw^2} +  \frac{\partial \ln r_w}{\partial \vw} \frac{\partial \alphashift[+]}{\partial \vw} + \alphashift[+] \frac{\partial^2 \ln r_w}{\partial \vw^2} \,,
\end{equation}
where the first term vanishes at the sonic front.
From \cref{eq:xi-ode,eq:w-ode-general} and using  \(f_w\to 0\) for \(\vp\to c_+\), we obtain
\(\partial_{\vw}^2 \ln \rwinv|_{\vp=c_+} = [\partial_{\vw}(q_+\gamma_w^2 \vp) - (\partial_{\vw}f_w) \partial_{\xiw} \ln\rwinv]_{\vp=c_+}\).
By comparison to \cref{eq:dlogrw-inverse-dvw}, and noting that \(\frac{\dd}{\dd \vw} \xi(\vw) = 0\) along a solution trajectory, i.e.\ \(\partial_{\vw}\xi = -f_w \partial_{\xiw} \xi\), one further finds
\begin{equation}
	\frac{\partial \ln \rwinv}{\partial \xiw} 
	= - q_+ \int\limits_{\vw}^{0} \dd v\, \frac{\frac{\partial}{\partial\xiw} \xi(v;\vw)}{\big(1-\xi(v;\vw)\,v\big)^2}
	= - q_+ \int\limits_{\vw}^{0} \dd v\, \frac{e^{\,\int_{\vw}^v \!\!\dd u\,\frac{\partial f}{\partial \xi}(\xi(u;\vw),u)}}{\big(1-\xi(v;\vw)\,v\big)^2} < 0 \,.
	\label{eq:dlogrw-dxiw}
\end{equation}
The integral in \cref{eq:dlogrw-dxiw} can, in general, not be computed analytically, and no closed-form expression can be found in the \(\vp\to c_+\) limit.
Normalizing \(\partial_{\xiw} \ln\rwinv\) to its lower bound derived from \cref{eq:dlogrw-dvp-upper}, 
\begin{equation}
	\label{eq:lambda-def}
	\lambda(\vp, \xiw) \equiv 
	- \frac{(c_++\vp)^2 \xiw (1-\xiw^2)}{2 (1+c_+^2) \vp (\vp-\xiw)}
	\frac{\partial \ln\rwinv}{\partial \xiw},
	\qquad 
	0 < \lambda(\vp,\xiw) < 1
	\,,
\end{equation}
the second derivative of \alphaN with respect to \vp at the sonic front can be written as
\begin{equation}
	\lim_{\vp\to c_+} \frac{1}{r_w} \frac{\partial^2 \alphaN}{\partial \vp^2} = 
	\frac{(1+c_-^2)[4 - (1+c_+^2) \lambda(c_+,\xiw)]}{6 c_-^2 c_+ (1-c_+^2)^2} \left[
		v_*(\xiw) - \frac{ c_+ (4 - 2 \lambda(c_+,\xiw))}{4 - (1+c_+^2) \lambda(c_+,\xiw)}
	\right]\!.
	\label{eq:d2logrw-d2v-cp}
\end{equation}
Numerically, we furthermore find that \(\lambda(c_+,\xiw)\) is strictly increasing throughout \(0<\xiw<c_+\).
In fact, as shown in \cref{fig:lambda-monotonicity}, the stronger inequality
\begin{equation}
    \frac{\partial\lambda(c_+,\xiw)}{\partial\xiw}
    >
    \frac{\lambda(c_+,\xiw)\,[1-\lambda(c_+,\xiw)]}{\xiw}
    >0 \ ,
\end{equation}
is satisfied for all values of \(c_+\) that we have tested.
Hence \(\partial_{\alphaP}\alphaN<0\) close to the \(\vp=c_+\) line --- and using \cref {eq:transport-derivative-ID} in the entire \vp range --- whenever \(v_*(\xiw)>c_+\), while \(\alphaN\) develops a minimum inside the branch whenever \(v_*(\xiw)<2c_+/(3-c_+^2)\). Only in the window \(2c_+/(3-c_+^2)<v_*<c_+\) does the sign depend on the actual slope of the inverse-hybrid enthalpy ratio.

Note that, at \(\xiw=c_-\), the derivative of \cref{eq:d2logrw-d2v-cp} with respect to \xiw at a zero crossing of \cref{eq:d2logrw-d2v-cp} has the same sign as \(\partial_{\xiw}\lambda(c_+,\xiw)>0\) (since \(\partial_{\xiw} v_*(c_-)=0\)), i.e.\ the onset of non-injectivity does not occur at the endpoint \(\xiw=c_-\), but must be determined by the numerical minimization over the full interval \(\xi_c<\xiw<c_-\).
We therefore conclude that the `\((+)\)' branch of inverse detonations is injective for every \xiw if and only if
\begin{equation}
	\min\limits_{\xi_c<\xiw<c_-} \left[\frac{1}{r_w} \frac{\partial^2 \alphaN}{\partial \vp^2}\right]_{\vp=c_+} \geq\ 0 
	\qquad\Longrightarrow\qquad
	\min\limits_{\xi_c<\xiw<c_-} \left[\frac{1}{r_w} \frac{\partial^2 \alphaN}{\partial \vp^2}\right]_{\substack{
			\vp=c_+\hphantom{\ccritID()}\\
			c_-=\ccritID(c_+)
	}} = 0 \,,
	\label{eq:ID-critical-cminus}
\end{equation}
which defines the critical sound speed \(\ccritID(c_+)\).
While the critical value needs to be calculated numerically, it is bounded by
\begin{equation}
	\ccritDH(c_+) = \frac{3-c_+^2-\sqrt{(1-c_+^2)(9-c_+^2)}}{2c_+}
	\;<\;\ccritID(c_+)\;<\;
	\frac{1-\sqrt{1-c_+^2}}{c_+}\,,
	\label{eq:ID-critical-bracket}
\end{equation}
since \(0<\lambda<1\) and as the lowest value of \(v_*(\xiw)\) is obtained at \(\xiw=c_-\).

\subsection[Mixed hybrids with \texorpdfstring{\(\vm=c_-\)}{v-=c-}]{\boldmath Mixed hybrids with \(\vm=c_-\)}

Mixed hybrid solutions with \(\vm=c_-\) only exist for \(c_+>c_-\).
They satisfy \(c_-<\xiw<\vp <c_+\) with \(\alphashift[+]>0\) given by \cref{eq:alphashiftplus-hybrid-c-} and have the enthalpy ratio \rwinv{} of inverse detonations.
\Cref{eq:dlogrw-dvp,eq:dlogrw-dvp-upper} then imply
\begin{equation}
	\frac{2 \mathcal{C} \, (c_+-\vp) (\vp-v_0^+)(\vp-v_0^-)}{3 c_+ c_-^2 (c_++\vp) (1-\vp^2)^2}
	\leq
	\frac{1}{r_w} \frac{\partial \alphashift[N]}{\partial \vp} 
	\leq
	\frac{(c_+^2-\vp^2) (1+c_-^2) \left[  \vp - v_*(c_-)\right]}{3 c_+^2 c_-^2 (1-\vp^2)^2} \,,
	\label{eq:MH-dalpha-bounds}
\end{equation}
where \(v_*(c_-)= 2 c_-/(1+c_-^2)\) as defined in \cref{eq:ID-plus-lower-bound} and 
\begin{equation}
	v_0^\pm = \frac{c_- \pm \sqrt{c_-^2 + \mathcal{C}\,c_+ c_- }}{\mathcal{C}}
	\qquad\text{with}\qquad
	\mathcal{C} = {1+c_-c_++c_-^2} > 1\,,
\end{equation}
with \(v_0^- < 0\) and \(v_0^+ > c_-\).
Hence,  the map is monotonic whenever \(c_+\leq 2c_-/(1+c_-^2)\), and zeros of \(\partial_{\vp}\alphashift[N]\) require \(v_* < \vp < v_0^+\).
Here and in the following, we use the short-hand notation \(v_* = v_*(c_-)\).

Using \cref{eq:transport-derivatives} together with \(\dd\ln r_w/\dd\vw=q_+\gamma_w^2\vp\) along a trajectory, the definition of \(\lambda(\vp,\xiw)\) in \cref{eq:lambda-def} is equivalent to
\begin{equation}
	\frac{\partial\ln\rwinv}{\partial\vp}
	= - \frac{q_+\vp}{1-\vp^2}\left[1+\lambda(\vp,\xiw)\,\frac{c_+-\vp}{c_++\vp}\right],
	\qquad 0<\lambda<1\,,
	\label{eq:dlogrw-lambda}
\end{equation}
which interpolates between \cref{eq:dlogrw-dvp} at \(\lambda=0\) and \cref{eq:dlogrw-dvp-upper} at \(\lambda=1\).
Inserting \cref{eq:dlogrw-lambda} into \cref{eq:alpha-shift} gives
\begin{equation}
	\frac{1}{r_w} \frac{\partial\alphashift[N]}{\partial \vp}
	=
	(1-\lambda) \frac{(c_+^2-\vp^2) (1+c_-^2) (\vp - v_*)}{3 c_+^2 c_-^2 (1-\vp^2)^2} 
	+ \lambda\, \frac{2 \mathcal{C} \, (c_+-\vp) (\vp-v_0^+)(\vp-v_0^-)}{3 c_+ c_-^2 (c_++\vp) (1-\vp^2)^2}
	\,,
\end{equation}
where all \xiw dependence enters through \(\lambda(\vp,\xiw)\).
The derivative vanishes for
\begin{equation}
	\lambda(\vp,\xiw) = \lambda_c(\vp)
	\equiv
	\frac{(c_++\vp)^2\,(\vp-v_*)}{\vp\,\big[\vp^2-(1+c_+^2)v_*\vp+c_+^2\big]}\,.
	\label{eq:MH-crossing}
\end{equation}
We can thus define the critical value \(\ccritMH(c_+)\) as the largest value of \(c_-\) for which the difference \(\lambda(\vp,\xiw)-\lambda_c(\vp)\) vanishes somewhere in the domain \(c_-<\xiw<c_+\) and \(\max(v_*,\xiw)<\vp<\min(v_0^+,c_+)\) in which \cref{eq:MH-dalpha-bounds} allows for non-monotonicity.

Moreover, it can be proven analytically that, throughout the region in which a zero of \cref{eq:MH-crossing} is possible,
\begin{equation}
    \frac{\partial}{\partial{\vp}}
    \big[\lambda(\vp,\xiw)-\lambda_c(\vp)\big] < - \frac{1}{c_+} <0 .
    \label{eq:MH-crossing-monotonicity}
\end{equation}
We here outline the proof, since it implies that \cref{eq:MH-crossing} can have at most one solution at fixed \xiw.

Using \cref{eq:transport-derivatives} and \(\dd\ln r_w/\dd\vw=q_+\gamma_w^2\vp\), the definition \cref{eq:lambda-def} gives the linear equation
\begin{equation}
	\frac{\dd\lambda}{\dd\xiw} = 
    -\frac{(c_++\vp)(1-\vp^2)}{\vp(c_+-\vp)(1-\xiw^2)} 
    + \frac{(1-\vp^2)\left[4c_+^3(\vp-\xiw)+\xiw(c_++\vp)^3\right]}{\vp\xiw(c_+-\vp)(c_++\vp)^2(1-\xiw^2)} \lambda\,,
	\label{eq:lambda-transport}
\end{equation}
where \(\dd/\dd\xiw = f_w^{-1}\dd/\dd\vw\) denotes the derivative along a trajectory, and the linear coefficient is positive.
All trajectories end at \((\xiw,\vp)\to(c_+,c_+)\), where expanding \cref{eq:xi-ode,eq:dlogrw-dxiw} gives \(\lambda\to1\) and \(\partial_{\vp}\lambda\to-1/(3c_+)\).

First, \cref{eq:lambda-transport} implies the stronger lower bound \(\lambda>\xiw/\vp\).
Indeed, \(\lambda-\xiw/\vp\) obeys the same equation with the same linear coefficient and a negative inhomogeneous term,
\begin{equation}\begin{multlined}
	\frac{\dd}{\dd\xiw}\left(\lambda-\frac{\xiw}{\vp}\right)
	= \frac{(1-\vp^2)\left[4c_+^3(\vp-\xiw)+\xiw(c_++\vp)^3\right]}{\vp\xiw(c_+-\vp)(c_++\vp)^2(1-\xiw^2)} \left(\lambda-\frac{\xiw}{\vp}\right) \\
	- \frac{(\vp-\xiw)\left\{2c_+(1-c_+^2)(\vp+3c_+) + (c_+^2-\vp^2)\left[\vp^2+3c_+\vp+6c_+^2+\xiw(c_++\vp)\right]\right\}}{\vp(c_+-\vp)(c_++\vp)^2(1-\xiw^2)} \,,
\end{multlined}\end{equation}
so that it can only cross zero from positive to negative values along the trajectory.
As \(\lambda-\xiw/\vp\to0\) at the endpoint, it is positive throughout.

Second, the partial derivative of \(\lambda\) with respect to \vp is bounded by \(\partial_{\vp}\lambda(\vp,\xiw) < 1/c_+\).
Differentiating \cref{eq:lambda-transport} with respect to \vp, \(\partial_{\vp}\lambda-1/c_+\) obeys a linear equation whose inhomogeneous term is affine in \(\lambda\) and positive at both \(\lambda=\xiw/\vp\) and \(\lambda=1\), hence for all \(\xiw/\vp<\lambda<1\).
Hence \(\partial_{\vp}\lambda-1/c_+\) can only cross zero from negative to positive values along the trajectory, which is excluded as it approaches \(-4/(3c_+)\) at the endpoint.
Thus, \(\partial_{\vp}\lambda<1/c_+\) throughout.

In the window \(\max(v_*,\xiw)<\vp<\min(v_0^+,c_+)\), where \(0<\lambda_c<1\), differentiating \cref{eq:MH-crossing} then gives
\begin{equation}\begin{aligned}
	\lambda_c' 
	&= (1-\lambda_c)\frac{(c_++\vp)^2}{c_+^2\vp(1-\vp^2)} 
	+ \lambda_c\frac{4}{(c_++\vp)(1-\vp^2)} 
	+ \lambda_c(1-\lambda_c)\frac{\vp(c_+-\vp)(1+c_+^2)}{c_+^2(c_++\vp)(1-\vp^2)}\\
	&> \frac{4}{(c_++\vp)(1-\vp^2)} > \frac{2}{c_+} \,,
\end{aligned}\end{equation}
where we used \((c_++\vp)^3-4c_+^2\vp = c_+^2(c_+-\vp)+3c_+\vp^2+\vp^3>0\).
Together with \(\partial_{\vp}\lambda<1/c_+\), this yields \cref{eq:MH-crossing-monotonicity}.

With \cref{eq:MH-crossing-monotonicity}, and using the monotonicity \(\partial_{\xiw}\lambda(c_+,\xiw)>0\) shown in \cref{fig:lambda-monotonicity}, together with \(\lambda>\lambda_c\) at the lower end of the \vp window,
\footnote{%
	For \(\xiw\geq v_*\), \(\lambda(\xiw,\xiw)=1\) while \(\lambda_c<1\) for \(\vp<v_0^+\). 
	For \(\xiw<v_*\), \(\lambda_c(v_*)=0\) while \(\lambda>0\).%
}
\cref{eq:MH-crossing} is most restrictive at \(\vp=c_+\) and \(\xiw=c_-\), so that \ccritMH can be defined by
\begin{equation}
	\frac{2c_-}{1+c_-^2}\,\Big[4-(1+c_+^2)\,\lambda(c_+,c_-)\Big] = c_+\Big[4-2\lambda(c_+,c_-)\Big]
	\qquad\text{for}\quad c_-=\ccritMH(c_+)\,.
	\label{eq:MH-critical-cminus}
\end{equation}

Finally, \cref{eq:MH-critical-cminus} coincides with the definition of \(\ccritID(c_+)\) from \cref{eq:ID-critical-cminus} evaluated at \(\xiw=c_-\), i.e.\ at the shared junction between the two solution types. 
We therefore obtain the hierarchy
\begin{equation}
    \frac{
		3-c_+^2
		-
		\sqrt{\left(1-c_+^2\right)\left(9-c_+^2\right)}
	}{
		2c_+
	} =
	\ccritDH(c_+) < \ccritMH(c_+) < \ccritID(c_+) < \frac{1-\sqrt{1-c_+^2}}{c_+} \,,
\end{equation}
where the `(\(+\))' branch of inverse detonations is the first solution type to develop non-monotonicity as \(c_-\) decreases with respect to \(c_+\).

\subsection[Crossing of inverse detonations and inverse hybrids to positive \texorpdfstring{\(\alphaN>0\)}{alphaN>0}]{\boldmath Crossing of inverse detonations and inverse hybrids to \(\alphaN>0\)}
\label{app:zero-crossing}

Independently of becoming non-monotonic, for sufficiently large hierarchies \(c_- \ll c_+\), inverse solution branches can also cross into the \(\alphaN > 0\) regions ordinarily associated with direct (or mixed) solutions.
These zero crossings can only occur for the `(\(+\))' branch of inverse detonations or for inverse hybrids, and only for \(c_-<c_+\) (for \(c_->c_+\), the upper border of inverse hybrids is shared with the `(\(-\))' branch of inverse detonations, which is monotonic and has \(\alphaN<0\)).
Since \(\alpha_0<0\) and \(\rwinv<1\), \cref{eq:alphaN-rw-map} implies that \(\alphaN>0\) also requires \(\alphaP>0\).
On both branches, the upper end of the \alphaP range is \(\alphav{c_+}{\xiw}\), which is positive only for \(c_-^2/c_+<\xiw<c_+\), while the lower \alphaP bound is always negative.\footnote{%
    For the `(\(+\))' branch of inverse detonations, the lower bound \(\alphav{v_c(\xiw)}{\xiw}\) approaches zero from below for \(\xiw\to c_-\), while for inverse hybrids it is given by \(\alphav{c_+}{1}<0\), independently of \xiw.%
}
As the \(\vp=v_c\) end of the `(\(+\))' branch hence projects onto \(\alphaN<0\), positive \alphaN on this branch requires the \(\vp=c_+\) end to lie above it, so that the zero crossing of inverse detonations can only occur after their endpoints have exchanged, i.e.\ for \(c_-<\cexID\).
Furthermore, while the lower ends of the \xiw ranges of both branches lie below \(c_-^2/c_+\) (\(c_-^2\leq\xi_c(c_+)<c_-^2/c_+\)), at the upper end, the \(c_-<\xiw<c_+\) range is only accessible for inverse hybrids, so that the hierarchy at which positive \alphaN first appear differs between them.
The critical value of \(c_-\) at the zero crossing is hence given by the value of \(c_-\) satisfying the condition
\begin{equation}
    \max\limits_{\xiw}
    \left\{\rwinv[c_+]\left(1-\frac{\alphav{c_+}{\xiw}}{\alpha_0}\right)\right\} = 1\,,
    \label{eq:czero}
\end{equation}
where the range of the \xiw over which the maximum is taken differs between inverse detonations and hybrids.
Recall that, although not displayed explicitly, \rwinv also depends on \xiw, cf.~\cref{eq:rw-inverse}, both through the starting point of the rarefaction wave and through \(\vw=\mu(\xiw,c_+)\), but not on \(c_-\).
The second factor in \cref{eq:czero}, on the other hand, decreases with \(c_-\) at fixed \xiw,
\begin{equation}
    \frac{\partial}{\partial(c_-^2)}\left(1-\frac{\alphav{c_+}{\xiw}}{\alpha_0}\right)
    = -\frac{c_+(1+c_+^2)(c_+-\xiw)^2}{\xiw(1-c_+^2)(c_+^2-c_-^2)^2} < 0\,,
    \label{eq:czero-monotonic}
\end{equation}
so that, at fixed \xiw, the sign of \alphaN can change at most once, from negative to positive, as \(c_-\) decreases.
For inverse hybrids, whose \xiw range widens as \(c_-\) decreases, the left-hand side of \cref{eq:czero} therefore increases monotonically with decreasing \(c_-\), and its solution is unique.

At the upper end of the \alphaP range, i.e.\ along the \(\vp=c_+\) line, \(\alphashift[N]=\rwinv[c_+]\,\alphashift[+](c_+,\xiw)\), cf.\ \cref{eq:alpha-shift}.
The factor \alphashift[+] depends on \xiw only through \(v_*(\xiw)\), cf.\ \cref{eq:ID-plus-lower-bound,fn:alpha-bar-plus}.
For \(\xiw<c_-\), \(v_*\) decreases with \xiw, so that \alphashift[+] increases, and it is maximal at \(\xiw=c_-\).
The slope of the projection at \(\xiw=c_-\) is therefore set by the enthalpy ratio alone.
Taking into account that \(\vw=\mu(\xiw,c_+)\) also varies, \cref{eq:dlogrw-cp,eq:lambda-def} gives
\begin{equation}
    \frac{\dd\ln\rwinv}{\dd\xiw}\bigg|_{\vp=c_+}
    = \frac{1+c_+^2}{c_+(1-\xiw^2)}\left[1-\lambda(c_+,\xiw)\,\frac{c_+-\xiw}{2\xiw}\right].
    \label{eq:dlogrw-dxiw-sonic}
\end{equation}
We define \(\xi_r(c_+)\) as the position of the minimum of \(\rwinv[c_+]\), where \cref{eq:dlogrw-dxiw-sonic} vanishes, i.e.
\begin{equation}
    \lambda(c_+,\xi_r(c_+)) = \frac{2 \xi_r(c_+)}{c_+-\xi_r(c_+)}\,,
    \label{eq:xi-r-definition}
\end{equation}
which we find to be unique for all values of \(c_+\) we have tested.
For \(\xi_r<c_-\), i.e.\ for \(\lambda(c_+,c_-) < 2 c_- / (c_+-c_-)\), the projection is still increasing at \(\xiw=c_-\).
Since \(0<\lambda<1\), this holds for all \(c_-\geq c_+/3\).
Close to the zero crossing, however, \(c_-<c_+/3\), and \(\lambda\) has to be evaluated numerically.
Numerically, we find that \(\czeroID(c_+)>\xi_r(c_+)\), that at \(c_-=\czeroID\) the projection increases monotonically on the whole \xiw range of the `(\(+\))' branch, \(\xi_c<\xiw<c_-\), and that its value at \(\xiw=c_-\) is negative for all \(c_->\czeroID\).
Together with \cref{eq:czero-monotonic}, the `(\(+\))' branch of inverse detonations hence first reaches \(\alphaN=0\) at the endpoint \(\xiw\to c_-\), where \(\vpm=c_\pm\), and remains at \(\alphaN<0\) for all \(c_->\czeroID\).
There, \(\alphav{c_+}{c_-}=(c_+-c_-)^2/[3c_-^2(1-c_+^2)]\) and
\(1-\alphav{c_+}{c_-}/\alpha_0=2c_+(1-c_+c_-)/[(1-c_+^2)(c_++c_-)]\),
which yields the threshold condition \cref{eq:czero-ID}.

\begin{figure}
    \centering
    \includegraphics[width=\textwidth]{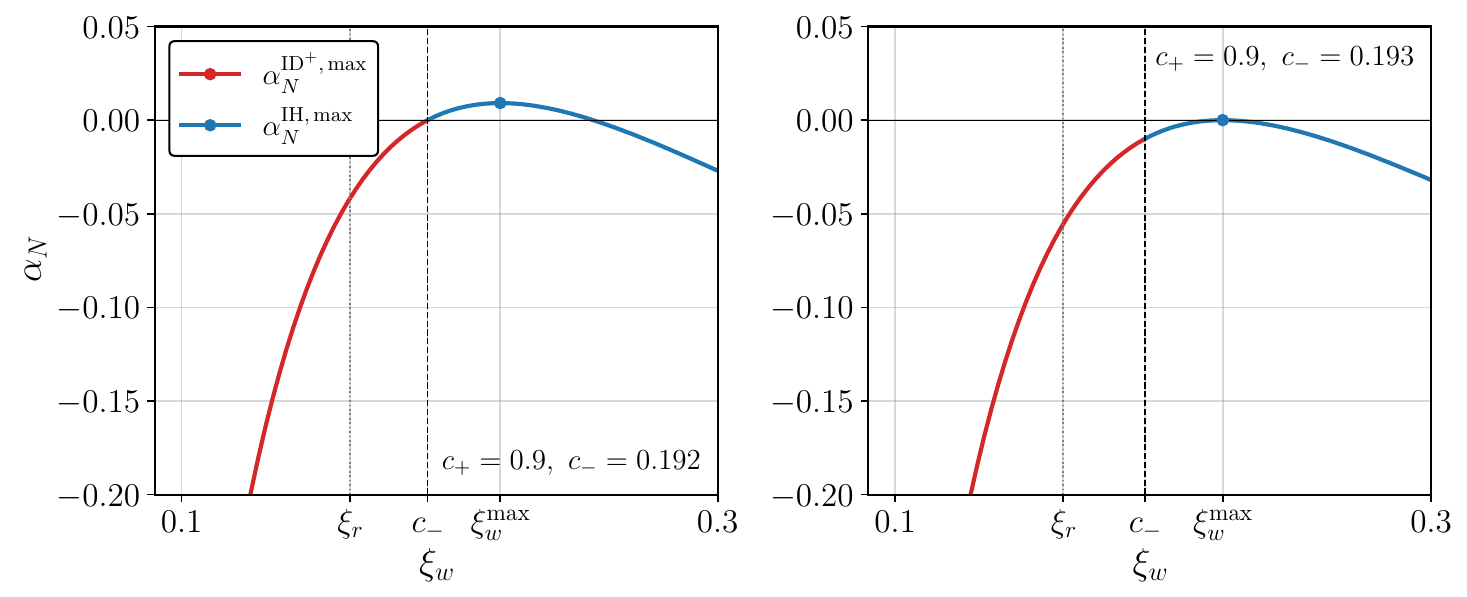}
    \caption{%
        Projection of the \(\vp\to c_+\) sonic endpoint, \(\alpharw{\alphav{c_+}{\xiw}}{\rwinv[c_+]}\), for \(c_+=0.9\) at the two thresholds \(c_-=\czeroID\simeq0.192\) (left) and \(c_-=\czeroIH\simeq0.193\) (right).
        This is the upper \alphaN bound of inverse hybrids for all \xiw, cf.\ \cref{eq:inv-hyb-merged}, and for \(\xiw<c_-\) (red) also the upper bound \alphaNmaxID of the inverse-detonation `\((+)\)' branch wherever it is positive, cf.\ \cref{eq:alphaN-fold-min-IDplus}.
        Since \(c_->\xi_r\), the minimum of \(\rwinv[c_+]\), the projection is still increasing at \(\xiw=c_-\).
        It reaches \(\alphaN=0\) at the upper end of the \xiw range of inverse detonations, \(\xiw\to c_-\) (left), while its maximum, which sets \czeroIH, lies at \(\xiw^{\max}>c_-\), outside the \xiw range of inverse detonations but within the inverse-hybrid range (right).
    }
    \label{fig:zero-crossing-maximum}
\end{figure}

At \(c_-=\czeroID\), the projection is still increasing at \(\xiw=c_-\), so that it is already positive just above \(c_-\), where it corresponds to the upper \alphaN bound of inverse hybrids.
Consequently, \(\czeroIH>\czeroID>\xi_r\), and, as the inverse-detonation part of the projection remains negative for \(c_->\czeroID\), the maximum of the projection at \(c_-=\czeroIH\) lies at \(\xiw>c_-\), see \cref{fig:zero-crossing-maximum}.
This justifies restricting the maximization in \cref{eq:czero-IH} to \(c_-<\xiw<c_+\).
For \(\xiw\to c_+\), the profile ahead of the wall becomes trivial, \(\rwinv\to1\) and \(\alphav{c_+}{\xiw}\to0\), so that the expression in \cref{eq:czero} tends to one for all \(c_-\).
Expanding with \cref{eq:dlogrw-dxiw-sonic} and \(\lambda\to1\), which follows from \(\xiw/c_+<\lambda<1\), the first-order terms cancel and the projection approaches zero from below,
\begin{equation}
    \alphaN = -\frac{(c_+^2+3c_-^2)(c_+-\xiw)^2}{12c_+^2c_-^2(1-c_+^2)} + \mathcal{O}\big((c_+-\xiw)^3\big)\,.
\end{equation}
With a positive slope at \(\xiw=c_-\), the maximum in \cref{eq:czero-IH} hence lies within \(c_- < \xiw < c_+\).

Deeper in the hierarchical regime, \(c_-<\xi_r<\czeroID\), the projection is decreasing at \(\xiw=c_-\), and its maximum can lie below \(c_-\), where it is shared by both branches.
This is the case for the benchmark of \cref{fig:hydro-fold-overlap}, where \(c_-=0.16<\xi_r\simeq0.163\) and the maximum lies at \(\xiw\simeq0.158\).

%
\section{Code Validation and Performance}
\label{app:code-cross-validation}
%

In this appendix we compare \hydrodex with other public codes that provide similar calculations of first-order phase-transition hydrodynamics and of the corresponding sound-wave gravitational-wave signal, in order to validate the physical implementation of \hydrodex and to compare their numerical performance on the same physical problems.
We consider
\texttt{PTtools}~\cite{pttools-git},
\texttt{CosmoGW}~\cite{cosmogw-git},
\texttt{HydroGrav}~\cite{Linton:2026fpj,hydrograv-git},
and \texttt{SNOBEX}~\cite{Mateos:2026moc}.
For the hydrodynamic comparison we also include \texttt{HydrodynamicsFOPT}~\cite{hydrodynamicsfopt-git}.
The independent codes are found to agree across all regimes in which they implement the same physics.
Among the codes considered here, \hydrodex offers the broadest coverage of hydrodynamic solution branches and their gravitational-wave spectra within the constant-speed-of-sound framework.%
\footnote{It does not yet support arbitrary equations of state, though, unlike \texttt{HydroGrav} and \texttt{SNOBEX}.}
In the like-for-like benchmarks presented below, in which all codes are run on the same hardware using all available CPU cores, \hydrodex achieves shorter execution times than \texttt{CosmoGW}, \texttt{HydroGrav}, and the pure-Python version of \texttt{PTtools}; only the Numba-compiled version of \texttt{PTtools} is faster, by a factor of about \(1.5\), in the
infinite-duration comparison.%
\footnote{%
    All benchmarks were performed on a workstation equipped with an Intel Core i5-12500 processor (3.0~GHz base, 4.6~GHz maximum turbo frequency, 6 physical cores, 12 logical threads), 16~GB of RAM, running Ubuntu 24.04, with \python 3.12.3.
    No thread limit was imposed, so the parallelised parts of each code (OpenMP in \hydrodex, Numba in \texttt{PTtools}) used all 12 logical threads.%
}

\subsection{Hydrodynamic comparison}
\label{app:code-cross-validation-hydro}

The hydrodynamic implementations differ significantly in scope.
\hydrodex implements the complete CSS classification discussed in this work, including direct and inverse solutions, independent sound speeds \(c_+\) and \(c_-\), and the mixed-hybrid branches.
Among the other codes considered here, only \texttt{HydrodynamicsFOPT} also contains inverse and
mixed-hybrid solutions.
The remaining implementations are restricted to the standard direct deflagration, hybrid, and detonation branches.

There are further differences in the equations of state that can be used.
In particular, \texttt{CosmoGW} allows a constant sound speed different from \(1/\sqrt{3}\), but its native hydrodynamic solver assumes the same sound speed in the two phases.
It therefore cannot directly reproduce a general CSS configuration with \(c_+\neq c_-\).
All comparisons below are accordingly performed only where the physical assumptions of the codes overlap.

\begin{figure}
    \centering
    \includegraphics[width=\linewidth]
    {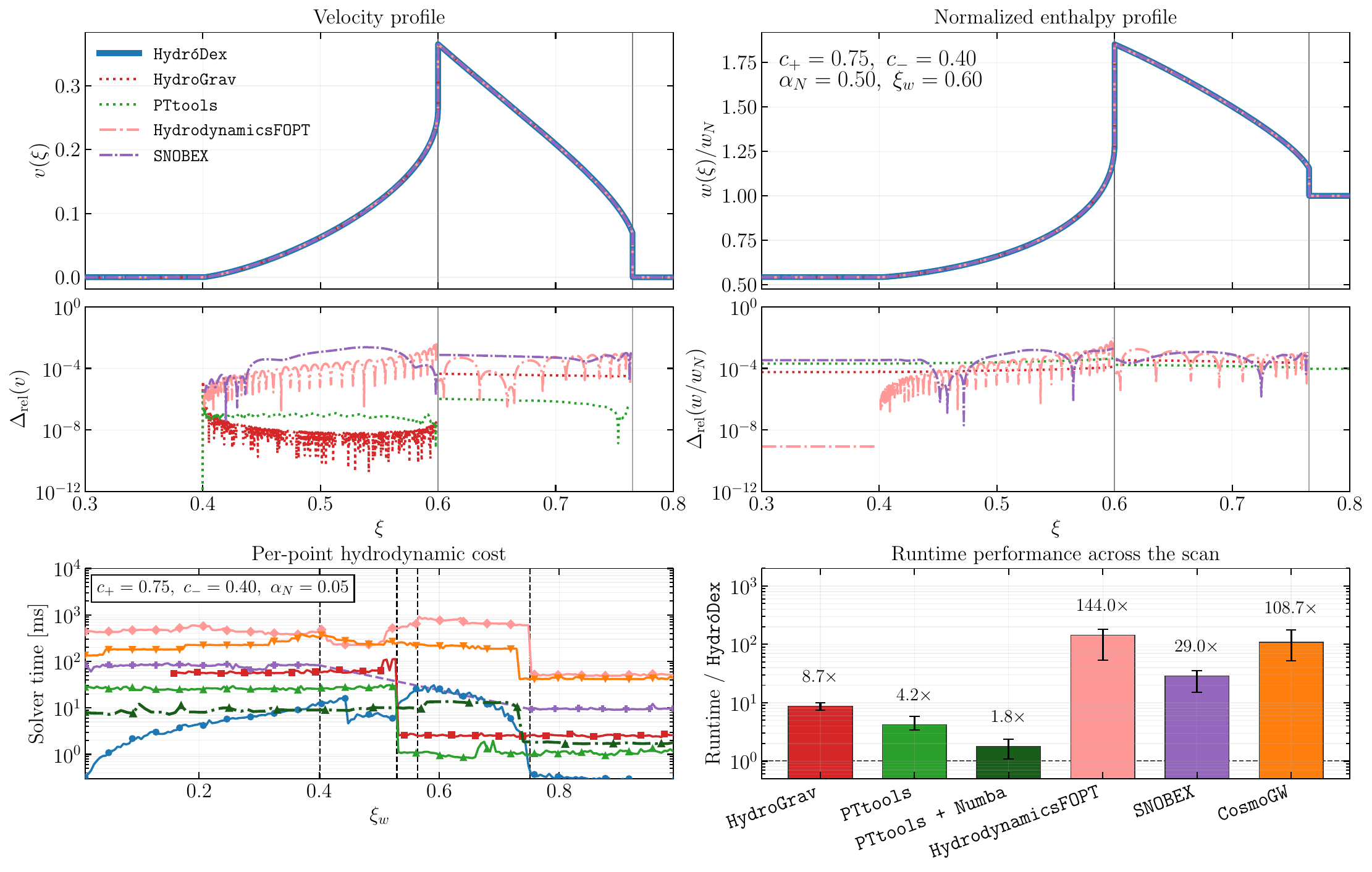}
    \caption{%
        The upper rows compare a representative hybrid profile and its pointwise relative difference with respect to \hydrodex.
        The lower row shows the runtime along a scan in \(\xiw\) at fixed \(c_+=0.75\), \(c_-=0.40\), and \(\alphaN=0.05\).
        Black dashed lines mark changes of the hydrodynamic branch in \hydrodex, while dashed coloured segments indicate failed scan points.
        The \texttt{CosmoGW} timing scan is performed separately at \(c_+=c_-=0.40\).
        The final panel shows the median runtime ratio \(t_{\rm code}/t_{\hydrodex}\), with error bars spanning the \SIrange{25}{75}{\%} interquartile range.
}
    \label{fig:code-comparison-hydro}
\end{figure}

\Cref{fig:code-comparison-hydro} summarizes the hydrodynamic tests.
The upper two rows show a point-by-point comparison for a representative hybrid solution.
For the unequal-sound-speed benchmark displayed in the figure we choose \(c_+=0.75\), \(c_-=0.40\), \(\alphaN=0.50\), and \(\xiw=0.60\).
All codes that support this configuration produce the same velocity and enthalpy profiles within their numerical accuracy.

To quantify the agreement we interpolate the profiles onto a common \(\xi\)-grid and define
\begin{align}
    \Delta_{\rm rel}(v)
    &=
    \frac{
        \left|v_{\rm code}(\xi)-v_{\hydrodex}(\xi)\right|
    }{
        \max_\xi |v_{\hydrodex}(\xi)|
    },
    \qquad&
    \Delta_{\rm rel}(w/w_N)
    &=
    \frac{
        \left|
        (w/w_N)_{\rm code}-(w/w_N)_{\hydrodex}
        \right|
    }{
        \max_\xi
        \left|
        (w/w_N)_{\hydrodex}-1
        \right|
    }.
\end{align}
Small regions around the wall and shock discontinuities are excluded from this pointwise error measure, because the different codes discretize the one-sided states differently and a point-by-point comparison exactly at a discontinuity is not meaningful.
Away from these regions, the agreement is typically many orders of magnitude better than the size of the physical perturbation.

The lower row tests a more demanding parameter slice.
We fix \(c_+=0.75\), \(c_-=0.40\), and \(\alphaN=0.05\), and scan the wall velocity over essentially the full interval \(0<\xiw<1\).
This scan crosses several hydrodynamic branches and, in particular, contains a mixed-hybrid region. 
The vertical black dashed lines in the lower left panel indicate the wall velocities at which the solution switches between physical branches according to \hydrodex.
A dashed coloured segment instead indicates a region where the corresponding code failed to return a solution for one or more scan points.

This distinction is important when interpreting the timing curves.
\texttt{PTtools} and \texttt{HydroGrav}, for example, can appear faster than \hydrodex inside the mixed-hybrid region simply because they do not implement that branch and continue to inconsistently return an ordinary detonation instead.
The two codes are therefore no longer solving the same physical problem in that interval.
Such points are displayed in the per-point timing plot (lower left panel) for completeness, but are not used as like-for-like accuracy or performance comparisons (lower right panel).

Since \texttt{CosmoGW} cannot use \(c_+\neq c_-\), its timing scan is performed separately with \(c_+=c_-=0.40\), at the same fixed \alphaN.
It therefore provides a performance benchmark, but not a point-by-point comparison with the unequal-sound-speed scan.

The performance summary in the lower-right panel shows the ratio of the runtime of the respective reference codes to the runtime of \hydrodex.
For each code we evaluate this ratio on the successful scan points where a like-for-like comparison is available.
The height of each bar is the median value of \(r_t \equiv t_{\text{code}}/t_\hydrodex\), while the error bar shows the interquartile range\footnote{%
    The interquartile range spans the central \SI{50}{\%} of the runtime ratios across the scan, from the \(25^\mathrm{th}\) percentile \(Q_1\) to the \(75^\mathrm{th}\) percentile \(Q_3\).
    It therefore measures how much the relative performance varies with \(\xiw\), rather than an uncertainty on the median itself.%
}, from the \SI{25}{\%} to the \SI{75}{\%} percentile.
Thus a ratio of 1 corresponds to the \hydrodex runtime, while a runtime ratio of 10 means that the other implementation is ten times slower.

The hydrodynamic comparison shows excellent agreement between \hydrodex and the different reference codes, whenever the codes describe the same branch and the same equation of state, in both velocity and enthalpy profiles, with relative differences generally below the per-mille level.
The main differences are instead in scope and performance. 
\hydrodex covers the complete direct and inverse CSS phase space and outperforms other implementations in computational speed in the corresponding like-for-like calculations on the hardware and benchmark points tested.
The only independent code providing a comparison for mixed-hybrid and inverse branches, \texttt{HydrodynamicsFOPT}, reproduces the same solutions but at a considerably larger computational cost.

\subsection{Gravitational-wave comparison}
\label{app:code-cross-validation-gw}

The gravitational-wave comparison involves fewer codes and requires a more careful separation of the physical approximations.
Among the public implementations considered here, \texttt{PTtools}, \texttt{CosmoGW}, and \texttt{HydroGrav} provide a sound-shell model gravitational-wave calculation.
However, they do not implement the same time treatment.

\texttt{PTtools} provides the long-lived, or effectively infinite-duration approximation, whereas \texttt{HydroGrav} implements the full finite-duration momentum convolution.
In principle one could approximate the long-duration limit in \texttt{HydroGrav} by taking a very large source duration, but this would be both numerically inefficient and unnecessary, since the infinite-time limit admits a much simpler semi-analytic treatment. 
\texttt{CosmoGW} and \hydrodex implement both the infinite-duration limit and full finite-duration calculations.
We consequently compare each code only in the regime for which it provides a native implementation.

Another relevant distinction concerns the cosmological background.
The external codes considered here evolve the acoustic perturbations in a radiation-dominated background, \(\omega \equiv {p_{\rm tot}}/{\rho_{\rm tot}} = 1/3\), even when the acoustic sound speed is allowed to differ from \(1/\sqrt{3}\).
The sound speed controlling the acoustic perturbations and the equation-of-state parameter controlling the expansion history are therefore conceptually different quantities.
For the direct comparison we keep \(\omega=1/3\), although \hydrodex in principle allows for a more general treatment  where the acoustic fluid can have \(c_s=c_-\) while the expansion history is specified independently.
One may therefore choose, for example, a background dominated by the same fluid with \(\omega=c_-^2\), or a different dominant component with an equation of state unrelated to that of the plasma undergoing the transition.

\begin{figure}
    \centering
    \includegraphics[width=\linewidth]
    {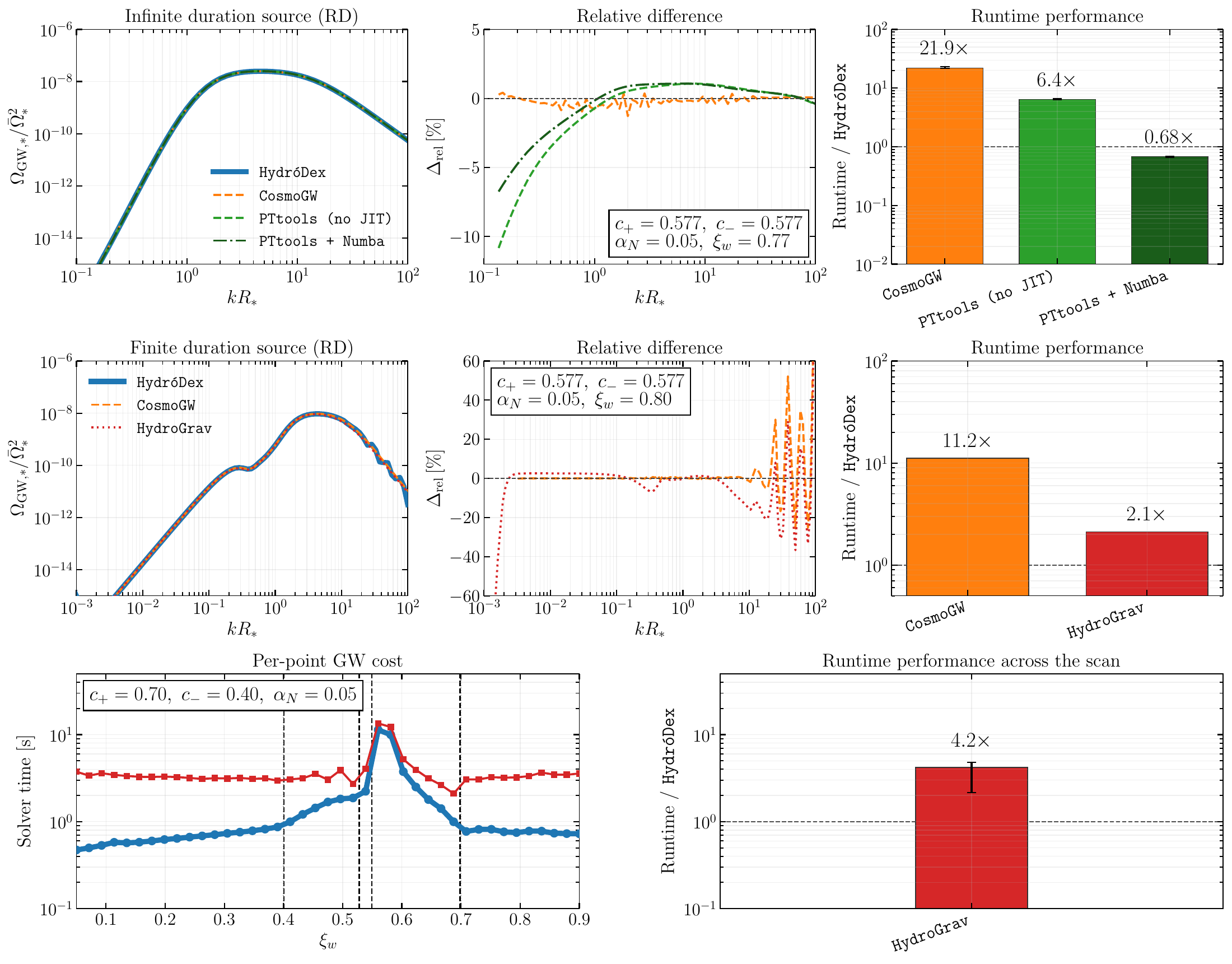}
    \caption{
        The upper row compares the infinite-duration calculation in radiation domination for \(c_s=1/\sqrt{3}\), \(\alphaN=0.05\), and \(\xiw=0.77\) between \hydrodex (solid, blue), \texttt{CosmoGW} (dashed, orange), and \texttt{PTtools} without (dashed, green) and with Numba compilation (dash-dotted, dark green); the runtime ratios are medians over five consecutive runs, with error bars showing the \numrange{25}{75}{\%} interquartile range.
        The middle row compares the full finite-duration calculation with \texttt{CosmoGW} (dashed, orange) and \texttt{HydroGrav} (dotted, red), after the corrections described in the text.
        The lower row shows the runtime for an unequal-sound-speed scan with \(c_+=0.70\), \(c_-=0.40\), and \(\alphaN=0.05\), together with the median runtime ratio \(t_{\rm code}/t_{\hydrodex}\) and its \numrange{25}{75}{\%} interquartile range.
}
    \label{fig:code-comparison-gw}
\end{figure}

\Cref{fig:code-comparison-gw} shows the resulting comparison.
The first row considers the infinite-duration limit in a radiation-dominated background, using \(c_+=c_-=c_s=\frac{1}{\sqrt{3}}\), \(\alphaN=0.05\), and \(\xiw=0.77\).
Here we compare the predictions of \hydrodex (solid, blue), \texttt{CosmoGW} (dashed, orange), and \texttt{PTtools}, the latter both without (dashed, green) and with \texttt{Numba}~\cite{Lam:2015bfh} just-in-time compilation (dash-dotted, dark green).
The three codes agree very well.
\texttt{CosmoGW} reproduces the \hydrodex spectrum to better than \SI{1}{\%} over the whole range shown.
\texttt{PTtools} agrees at the \SI{1}{\%} level around the peak and in the ultraviolet, while towards the infrared the difference grows to \(\simeq \SIrange{7}{10}{\%}\) at \(kR_*\simeq 0.15\).%
\footnote{%
    By default, \texttt{PTtools} adds an additional term \(w v^2/w_N\) to the energy-density perturbation \(\lambda=(e-e_N)/w_N\), which approximates the kinetic contribution \(w\gamma^2 v^2\) to \(T^{00}\) and is of second order in the fluid velocity.
    This term modifies the sine transforms at low \(\chi\), which then propagates to the resulting gravitational-wave spectrum.
    To achieve agreement between the two codes at the percent-level, we modified \texttt{PTtools} to disregard this correction.%
}

The second row compares the full finite-duration calculation in radiation domination.
Here the relevant public implementations are \texttt{CosmoGW} (dashed, orange) and \texttt{HydroGrav} (dotted, red).
Both codes agree with \hydrodex at the percent level over the physically relevant part of the spectrum in their common conformal benchmark, confirming that all three implementations carry out the same finite-duration sound-shell calculation.%
\footnote{%
    To achieve agreement between the results, we needed to account for the different choice of the sound-speed in the gravitational-wave calculation of \texttt{HydroGrav}, which evaluates the acoustic source and the gravitational-wave kernel with \(c_s=c_+\), together with the corresponding prefactor \(3\bar\Gamma_+^2 = 3(w_N/e_N)^2=3(1+c_+^2)^2\), whereas we, since the sound waves propagate in the broken phase after the collisions, instead use \(c_s=c_-\) throughout, i.e.\ \(3\bar\Gamma_-^2=3(1+c_-^2)^2\) in the normalization of \cref{eq:ssm-kinetic-normalization}.%
}

The lower row tests the computational cost over an extended range of wall velocities using an unequal-sound-speed configuration, \(c_+=0.70\), \(c_-=0.40\), \(\alphaN=0.05\), and \(\omega=1/3\), and compares \hydrodex (blue) with \texttt{HydroGrav} (red).
The left panel shows the execution time along the scan, while the right panel reports the median ratio \(t_{\rm code}/t_{\hydrodex}\), with the \SI{25}{75}{\%} interquartile range as an error bar.
The scan confirms that \hydrodex remains faster while simultaneously solving the more general \(c_+\neq c_-\) hydrodynamics.

In summary, \hydrodex provides a fast and accurate implementation of the calculation of hydrodynamic profiles and the corresponding gravitational-wave spectra within the sound-shell model that combines, within a single framework,
\begin{enumerate*}[label=(\roman*)]
    \item the complete direct and inverse CSS hydrodynamic classification, including mixed hybrids;
    \item independent sound speeds \(c_+\) and \(c_-\);
    \item both the infinite-duration and full finite-duration sound-shell model calculations;
    \item an acoustic sound speed that can be specified independently of the background expansion parameter \(\omega\);
    and
    \item an expansion history that need not be radiation dominated.
\end{enumerate*}

\bibliographystyle{JHEP}
{\footnotesize
\bibliography{bibliography}}

\end{document}